\documentclass[12pt,preprint]{aastex}

\newcommand{\as}{^{\prime\prime}}
\newcommand{\am}{^{\prime}}

\shorttitle{VLA-COSMOS SURVEY. II}
\shortauthors{Schinnerer et al.}
 
\begin{document}

\title{The VLA-COSMOS Survey: \\
       II. Source Catalog of the Large Project}
 
\author{E. Schinnerer\altaffilmark{1},
V. Smol{\v c}i{\' c}\altaffilmark{1},
C. L. Carilli\altaffilmark{2},
M. Bondi\altaffilmark{3},
P. Ciliegi\altaffilmark{4},
K. Jahnke\altaffilmark{1},
N.Z. Scoville\altaffilmark{5,6},
H. Aussel\altaffilmark{7,8},
F. Bertoldi\altaffilmark{9},
A.W. Blain\altaffilmark{5},
C.D. Impey\altaffilmark{10},
A.M. Koekemoer\altaffilmark{11},
O. Le Fevre\altaffilmark{12},
C.M. Urry\altaffilmark{13}}

\altaffiltext{1}{Max-Planck-Institut f\"ur Astronomie, K\"onigstuhl 17, 
                 D-69117 Heidelberg, Germany}
\altaffiltext{2}{National Radio Astronomy Observatory, P.O. Box O, 
                 Socorro, NM 87801-0387, U.S.A.}
\altaffiltext{3}{INAF-Istituto di Radioastronomia, Via Gobetti 101,
                 I-40129, Bologna, Italy}
\altaffiltext{4}{INAF-Osservatorio Astronomico di Bologna, Via Ranzani 1, 
                 I-40127 Bologna, Italy}
\altaffiltext{5}{California Institute of Technology, MC 105-24, 1200 East
                 California Boulevard, Pasadena, CA 91125, U.S.A.}
\altaffiltext{6}{Visiting Astronomer, University of Hawaii, 2680 Woodlawn Dr., 
                 Honolulu, HI 96822, U.S.A.}
\altaffiltext{7}{Institute for Astronomy, University of Hawaii, 
                 2680 Woodlawn Dr., Honolulu, HI 96822, U.S.A.}
\altaffiltext{8}{Service d'Astrophysique, CEA/Saclay, 91191 Gif-sur-Yvette, 
                 France}
\altaffiltext{9}{Argelander Institut f\"ur Astronomie, Universit\"at Bonn, 
                 Auf dem H\"ugel 71, D-53121 Bonn, Germany}
\altaffiltext{10}{Steward Observatory, University of Arizona, 
                  933 North Cherry Avenue, Tucson, AZ 85721, U.S.A.}
\altaffiltext{11}{Space Telescope Science Institute, 3700 San Martin Drive, 
                 Baltimore, MD 21218, U.S.A.}
\altaffiltext{12}{Laboratoire d'Astrophysique de Marseille, BP 8, Traverse
                  du Siphon, 13376 Marseille Cedex 12, France}
\altaffiltext{13}{Department of Physics, Yale University, P.O. Box 208101, 
                  New Haven, CT 06520-8121, U.S.A.}

 \begin{abstract}

   The VLA-COSMOS large project is described and its scientific
   objective is discussed. We present a catalog of $\sim 3,600$ radio
   sources found in the $\rm 2\,deg^2$ COSMOS field at 1.4\,GHz. The
   observations in the VLA A and C configuration resulted in a
   resolution of 1.5$^{\prime\prime}$$\times$1.4$^{\prime\prime}$ and
   a mean rms noise of $\rm \sim 10.5(15)\,\mu Jy/beam$ in the central
   $\rm 1(2)\,deg^2$. 80 radio sources are clearly extended consisting
   of multiple components, and most of them appear to be double-lobed
   radio galaxies. The astrometry of the catalog has been thoroughly
   tested and the uncertainty in the relative and absolute astrometry
   are 130\,mas and $<$55\,mas, respectively.

\end{abstract}
  
 
\section{Introduction}\label{intro}
  
The radio source counts above the milli-Jansky level are dominated by
radio galaxies and quasars powered by active galactic nuclei (AGN) in
elliptical host galaxies. However, deep radio surveys at 1.4\,GHz show
an upturn in the integrated source counts at sub-mJy levels revealing
the presence of a population of faint radio sources far in excess of
those expected from the high luminosity radio galaxies and quasars
which dominate at higher fluxes
\citep*{win85,hop98,cil99,ric00,pra01,hop03,huy05}. While radio
sources with relatively bright optical counterparts are starburst
galaxies \citep*[e.g.][]{ben93,afo05}, the ones with fainter optical
counterparts are often redder as expected for early type galaxies
\citep*{gru99}. Recent detailed multi-wavelength follow-up of faint
radio sources showed a mixture of active star forming galaxies and AGN
hosts \citep{roc02,afo06}. The exact mixture of these different
populations (high-z AGN out to the highest redshifts, intermediate-z
post starburst, and lower-z emission line galaxies) as a function of
radio flux level is not very well established, especially in the
$\mu$Jy regime.

In order to fully investigate the nature and evolution of the $\mu$Jy
population it is necessary to couple deep radio observations with high
quality imaging and spectroscopic data from other wavelengths covering
as much of the electromagnetic spectrum as possible. The international
COSMOS (Cosmic Evolution) survey
\citep{sco06}\footnote{http://www.astro.caltech.edu/$\sim$cosmos}
provides such a unique opportunity. COSMOS is a pan-chromatic imaging
and spectroscopic survey of a $1.4^{\circ}\times 1.4^{\circ}$ field
designed to probe galaxy and SMBH (super-massive black hole) evolution
as a function of cosmic environment. One major aspect of the COSMOS
survey is the HST Treasury project \citep{sco06b}, entailing the
largest ever allocation of HST telescope time. The equatorial location
of the COSMOS field offers the critical advantage of allowing major
observatories from both hemispheres to join forces in this endeavor.
State-of-the-art imaging data at all wavelengths
\citep*[X-ray to
centimeter, e.g.][]{has06,sch06,tan06,cap06,ber06,agu06,sch04} plus
large optical spectroscopic campaigns using the VLT/VIMOS and the
Magellan/IMACS instruments \citep{lil06,imp06,tru06} have been or are
currently being obtained for the COSMOS field. These make the COSMOS field
an excellent resource for observational cosmology and galaxy evolution
in the important redshift range $z \sim 0.5 -3$, a time span covering
$\sim$75\% of the lifetime of the universe.

One major scientific rationale of the COSMOS survey is to study the
relation between the large scale structure (LSS) and the evolution of
galaxies and SMBHs. In a $\Lambda$CDM cosmology, galaxies in the early
universe grow through two major processes: dissipational collapse and
merging of lower mass protogalactic and galactic components. Their
intrinsic evolution is then driven by the conversion of primordial and
interstellar gas into stars, with galactic merging and interactions
triggering star formation and starbursts.  Mergers also can perturb
the gravitational potential in the vicinity of the black hole, thus
initiating or enhancing AGN activity. Several lines of evidence
suggest that galaxy evolution and black hole growth are closely
connected; COSMOS offers the chance to observe this connection
directly. While there is general agreement over this qualitative
picture, the timing/occurrence of these events and their dependence on
the local environment remains to be observationally explored
\citep[e.g.][]{fer00}. To study LSS it is essential to obtain high
spatial resolution data over the entire electromagnetic spectrum
covering a significant area on the sky, like 2\,deg$^2$ as in
the case of the COSMOS survey. Also, surveys of active galactic nuclei
benefit from such a combination of areal coverage and depth.

For the radio observations at 1.4\,GHz, it was essential to match the
typical resolution for optical-NIR ground-based data of $\sim$
1$^{\prime\prime}$ to fully exploit the COSMOS database.  Therefore
observations with the NRAO Very Large Array (VLA) had to be conducted
in the A-array that provides a resolution of about 2$^{\prime\prime}$
(FWHM) at 1.4\,GHz. Mosaicking is necessary to cover the large area of
the COSMOS field. The VLA-COSMOS survey consists of the pilot project
\citep{sch04}, the large project (presented here) and the ongoing deep
project (focusing on the central 1\,deg$^2$; Schinnerer et al., in
prep.). The VLA-COSMOS pilot project tested the mosaicking
capabilities in the VLA A-array at 1.4\,GHz in the wide-field imaging
mode and has provided the initial astrometric frame for the COSMOS
field.

Here we present the source catalog derived from the 1.4\,GHz image of the
VLA-COSMOS large project. The paper is organized as follows: after a
brief description of the survey objective (\S \ref{sec:objective}), the
details of the observations and data reduction are presented in \S
\ref{sec:obs} and \S \ref{sec:red}, respectively. In \S \ref{sec:test},
we discuss our tests for flux and astrometric calibration. The
VLA-COSMOS catalog is described in \S \ref{sec:cat}, while the context
of the VLA-COSMOS survey within the COSMOS project is discussed in \S
\ref{sec:cosmos}.

\section{Survey Objective}\label{sec:objective}

Unlike most existing deep survey fields, the COSMOS field is
equatorial and hence has excellent accessibility from all ground-based
facilities (current and future such as [E]VLA and ALMA). In addition,
it has an extensive multi-wavelength coverage \citep{sco06}. This makes
it an ideal field to analyze the (faint) radio source population as a
function of redshift, environment, galaxy morphology and other properties.
The VLA-COSMOS radio observations were matched to study a range of
important issues related to the history of star formation, the growth
of super-massive black holes, and the spatial clustering of galaxies.
The ongoing spectroscopic surveys within the COSMOS project are also
targeting well-defined samples of radio sources as part of the
overall program. In addition, the VLA-COSMOS radio survey is providing
the absolute astrometric frame for the COSMOS field \citep{aus06}, which is
important given the field's large size.

In this paper we describe in detail the observing procedure, and
various tests on data quality and characteristics \citep[astrometry,
fitted source parameters, etc.; see also the pilot project paper
by][]{sch04}. The completeness tests and the number counts of this
survey are under-way (Bondi et al., in prep.) as well as the
identification of optical counterparts using the space- and
ground-based COSMOS imaging data (Ciliegi et al., in prep.).
The full source catalog is available from the COSMOS archive at
IPAC/IRSA\footnote{http://www.irsa.ipac.caltech.edu/data/COSMOS/}.
Subsequent papers will consider important scientific issues such as:
(i) the evolution of radio-loud AGN as a function of environment,
including comparison to X-ray AGN and clusters \citep[see
also][]{smo06}, and a search for type-II radio QSOs, and (ii) a
dust-unbiased survey of star forming galaxies, as revealed in the
sub-mJy radio source population, including consideration of the
evolution of the radio-FIR correlation out to $z \sim 1$ through
comparison with the Spitzer data, and of extreme, high $z$ starbursts
as seen in the MAMBO 250 GHz COSMOS survey \citep{ber06}. In the
following sections we describe the goals of these two key science
programs in more detail.

\subsection{Survey Area}

The sub-mJy radio source counts provide one of the best indicators of
the effect of cosmic variance: number counts of sub-mJy radio sources
in fields of order of $\sim 10\am$ in diameter show a factor three
variation \citep[e.g.][]{hop03}, indicating that such field sizes are
inadequate to map cosmic large scale structure. Thus to properly
sample the faint radio source population and map out its cosmic
structure to the largest relevant scales, it is necessary to survey a
large area at the same resolution and sensitivity. Proper studies of
source clustering require hundreds to thousands of sources. In order
to enable detailed studies of environmental effects on faint, distant
radio source distributions and properties, all as a function of
redshift, several thousand sources are required as well.

Deep radio imaging of the $\rm 2\,deg^2$ COSMOS field with $\sim
3,600$ sources allows one to probe a - unique and - key area of
parameter space. The combination of high sensitivity and high spatial
resolution over a large area (see Tab. \ref{tab:survey}) bridges the
gap between shallow, wider field surveys, such as FIRST \citep*{bec95}
and NVSS \citep*{con98} with about one million source entries, and
ultra-sensitive ($\le 5-7 \mu$Jy), narrow field (single VLA primary
beam $\sim 30\am$ FWHM) studies of a few hundred sources, such as
those by \cite*{fom06,ric00}. Surveys which are comparable in scope to the
VLA-COSMOS large project are the Phoenix deep field survey (PDS),
undertaken with the ATCA \citep*{hop03}, and the VVDS 02hr field done
with the VLA in B-array \citep*{bon03}. These surveys produce a lower
angular resolution and a slightly higher rms (see Tab.
\ref{tab:survey}).

\subsection{Star Forming Galaxies}

\begin{figure}[ht]
\includegraphics[scale=0.7]{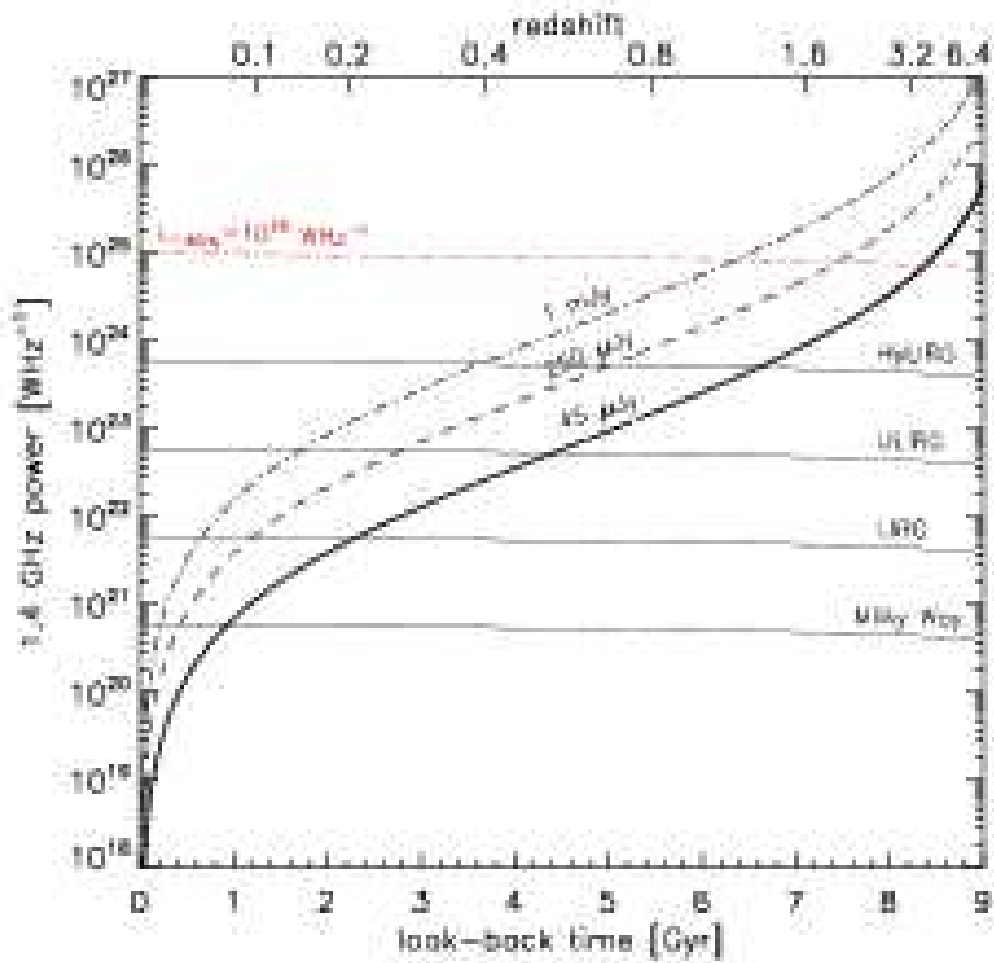}
\caption{The sensitivity limit as a function of (intrinsic) 1.4\,GHz
  luminosity (or power). The limit for the VLA-COSMOS large project
  corresponds to the bold solid line. The expected luminosities for
  various classes of galaxies are indicated by the solid horizontal
  lines. The expected radio power was calculated using the local
  IR-radio relation \citep{con92} and assuming a spectral index of
  $\alpha=0.8$. The horizontal dashed-dotted line corresponds to the
  assumed dividing line between radio-quiet and radio-loud AGN.  (See
  text for details.)  
\label{fig:lum}}
\end{figure}

Tracing the evolution of the cosmic star formation history from
optical surveys bears the large uncertainty of dust corrections
\citep[e.g][]{ste99}. Deep VLA observations of the COSMOS field can
provide a unique, unobscured look at star forming galaxies and highly
extincted galaxies in the full range of environment, especially in
combination with the deep (sub)mm data \citep*{ber06,agu06} and deep
Spitzer infrared imaging \citep{san06} to which the high resolution of
the VLA images provides means to properly identify luminous infrared
galaxies (see Fig. \ref{fig:lum}). The VLA radio data will
particularly be helpful to (a) trace the cosmological star formation
history and (b) test the FIR/radio correlation at high redshifts. The
radio luminosity of local galaxies is well-correlated with their star
formation (SF) rate \citep{con92}, and needs, unlike optical tracers,
no correction for dust obscuration. Thus radio sources with correct
spectral identification (as star forming galaxies) can be
independently used to estimate the SF history (of the luminous
sources).

Recent work by \cite*{haa00} for three deep radio surveys confirms the
trend of rising star formation rate between $z=0$ and $z=1$, however
their calculated star formation rates are significantly larger than
even dust-corrected optically selected star formation rates. A key
uncertainty is the contribution of AGN to the faint ($\rm < 1 mJy$)
radio population, with estimates ranging from 20$\%$ to 80$\%$ for
surveys down to 40\,$\mu$Jy. The (far)IR-radio correlation for star
forming galaxies appears to hold out to high redshift
\citep*{gar02,app04}. However, the number of star forming sources
detected at 1.4\,GHz is small above $z=0.5$. A thorough understanding
of the IR-radio correlation out to higher redshifts is important, as
it has been widely used as a distance measure for sub-mm sources
without any optical counterparts \citep{car00,are05}. Also, an
important question for active star forming galaxies is the role of
mergers, in particular at higher redshift. The FIR imaging alone will
lack sufficient resolution to address this issue, while the optical
imaging will suffer from the standard problem of obscuration in these
very dusty systems. Only arcsecond resolution radio data will allow
the determination of the spatial distribution of star formation in
dusty starbursts on scales relevant for merging galaxies ($\sim 10$
kpc).

\subsection{Active Galactic Nuclei}

Only a large field and deep radio survey can provide information about
the evolution of the currently highly uncertain faint-end of the radio
luminosity function. The fundamental problem in the study of the
evolution of radio-loud AGN has been that samples are drawn from
either very wide field, but very shallow surveys, or very deep, but
very small field surveys. The former are limited at high redshifts to
only extreme luminosity sources, while the latter are plagued by
relatively small number statistics and number variance. The VLA-COSMOS
survey was designed to enable the study of the demographics and
evolution of AGN by encompassing a large cosmological volume and by
providing good statistics on both radio-loud and radio-quiet AGN as a
function of redshift.

Only sub-mJy sensitivities over a wide area are adequate to detect
relatively weak (FRI) radio AGN to very high redshift ($\rm z \sim 6$)
while providing a large number ($\sim 1000$) of AGN sources. At lower
redshift, z $\sim$ 1, a sensitivity of $\rm 1\sigma \approx 10\,\mu
Jy/beam$ is good enough to detect a significant fraction of
radio-quiet, optically-selected QSOs. Moreover, questions regarding
redshift evolution of FRI and FRII sources, their parent galaxy
properties, and environmental dependencies can be addressed
independently for QSOs and radio galaxies. Such observations are
sensitive enough to reach the classic boundary between radio-loud and
radio-quiet AGN (log L$_{1.4\,\rm{GHz}}\,[\mathrm{W\,Hz}^{-1}]$ = 25) at z
$\sim$ 4-5 (depending on the exact spectral index; see Fig.
\ref{fig:lum}). Highly luminous radio-loud objects such as Cygnus A
with log L$_{1.4\,\rm{GHz}}\,[W\,Hz^{-1}] \sim$ 34 \citep*{car96}
should be observable out to their epoch of formation.

\section{Observations}\label{sec:obs}

The goal of the large project of the VLA-COSMOS survey was to image
the entire COSMOS field with an as large as possible uniform rms
coverage while minimizing the observing time required. Since the
observations had to be finished within one configuration cycle,
special requirements arose for the pointing lay-out and the observing
strategy.

\subsection{Lay-out of the Pointing Centers}\label{sec:layout}

\begin{figure}[ht]
\includegraphics[scale=0.7]{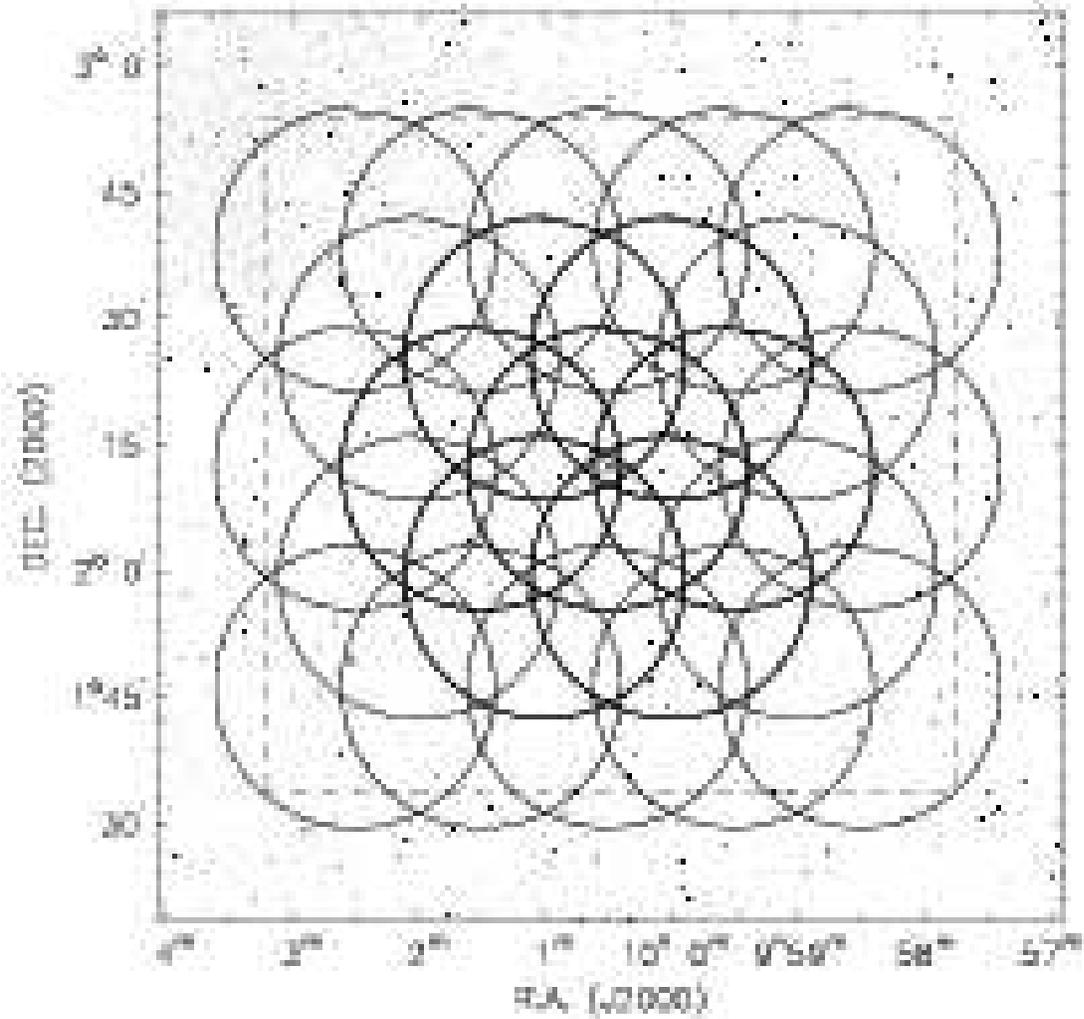}
\caption{The pointing pattern of the VLA-COSMOS Large Project overlaid
  onto a DSS image of the area of the COSMOS field. The heavy-outlined
  circles indicate the pointings observed in the VLA-COSMOS pilot
  project \citep*{sch04}. Each pointing has a radius of 16.8$\am$
  corresponding to the cut-off radius used for making the mosaic. The
  dashed line marks the outline of the COSMOS field covered by ACS
  tiles from the HST-COSMOS survey \citep[see][]{sco06b}.
\label{fig:point}}
\end{figure}

The pointing lay-out was designed to maximize the uniform noise
coverage while minimizing the number of pointings required to limit
overhead due to slewing ($\sim$30\,s slewing time for each change of
pointing). A hexagonal pattern of the pointing centers provides both
a uniform sensitivity distribution and a high mapping efficiency for
large areas \citep[see][]{con98}. To minimize the effect of bandwidth
smearing, we used -- as already tested in the pilot observations
\citep*{sch04} -- a separation of $15\am$ between the individual field
centers. A total of 23 separate pointings was required to fully cover
the $\rm 2\,deg^2$ of the COSMOS field (see Tab. \ref{tab:pos} and
Fig. \ref{fig:point}).

\subsection{Correlator Set-up and Calibrators}\label{sec:set}

We used the standard VLA L-band continuum frequencies of 1.3649 and
1.4351\,GHz and the multi-channel continuum mode to minimize the
effect of bandwidth smearing (in the A configuration). This results in
two intermediate frequencies (IF) with two polarizations, providing 6
useable channels of 3.125\,MHz each, or a total bandwidth of 37.5\,MHz
(observed with both polarizations). (Nominally, 7 channels are
available, however, due to the largely reduced sensitivity in the last
channel, we only used channels 1 to 6.)

The quasar 0521+166 (3C\,138) served as flux and bandpass calibrator
and was observed at the beginning of each observation. To allow for
good correction of atmospheric amplitude and phase variations, we
selected the quasar 1024-008 which was already used in the pilot
observations \citep*{sch04}. 1024-008 is about 6.1$^{\circ}$
away from the COSMOS field center and has a flux of about 1 Jy at
1.4\,GHz. Its positional accuracy is better than $0.01\as$ (VLA
Calibrator Manual 2003); the positional difference is less than
$0.001\as$ between coordinates listed in the VLA Calibrator Manual and
its ICRF \citep[International Celestial Reference Frame;][]{fey04}
position.

The quasar 0925+003 at a distance of about 9$^{\circ}$ from the COSMOS
field center was observed to test the absolute astrometric accuracy of
the observations. Its positional accuracy is known to better than
$0.002\as$, and its 1.4\,GHz flux is similar to the one of 1024-008. It
was also used to test the flux calibration (see Section
\ref{sec:test}).

\subsection{Observing Strategy}

This project holds the status of a VLA Large Project, as it required
240\,hrs of observing time in the A configuration alone. The
observations were scheduled in blocks of 6\,hrs centered at the Local
Siderial Time (LST) of 10:00\,hr. This ensured that the COSMOS field
was always above 40$^{\circ}$ elevation during our observations to
keep the system temperature of the L-band receivers low. These
observing blocks were scheduled over 42 days between September 23th,
2004 and January 9th, 2005 for the A configuration, and between August
26th, 2005 and September, 25, 2005 for the C configuration. The
observing time for the C configuration consisted of 4 observing blocks
each 6\,hrs long, except for the last observation that was 1.5\,hrs
longer.

In order to minimize the impact of varying observing conditions --
especially during the A array observations -- onto the mosaic we
adopted the following scheme: (a) all 23 pointings were observed with
about 6.5 minutes integration time twice each day, (b) the starting
pointing was changed each time, (c) the flux calibrator 0521+166 was
only observed at the beginning (since interpolation between days in
case of a loss was acceptable\footnote{During the observations this
  happened only once, and the flux of the phase calibrator 1024-008
  was fairly stable through the curse of observations (see \S
  \ref{subsec:flux}).} (d) the phase calibrator 1024-008 was observed
every 28 to 35 minutes, and (e) the test calibrator 0925+003 was
observed twice each day after about one-third and two-thirds of the
available observing time. The rotation of the pointings with observing
days also resulted in a more complete $uv$ coverage, and therefore a
rounder synthesized (i.e. DIRTY) beam.

\section{Data Reduction and Imaging}\label{sec:red}

\subsection{Data Reduction}

The data reduction was done using the Astronomical Imaging
Processing System \citep[AIPS;][]{gre03} following the standard routines as
described in the VLA handbook on Data Reduction.  For the flux
calibration and the correction of the atmospheric distortions we used
the pseudo-continuum channel. Before and after this calibration, $uv$
points (of the two calibrators 0521+166 and 1024-008) affected by
radio frequency interference (RFI) were flagged by hand using the AIPS
task 'TVFLAG'. As the data were obtained in the multi-channel
continuum mode, a bandpass calibration was performed on the 'Line'
data after the flux and phase calibration of the pseudo-continuum
channel had been transfered to the 'Line' data. In order to exclude
remaining RFI in the source data (i.e. the individual COSMOS fields),
we checked all channels (per IF and polarization) for RFI using
'TVFLG' and flagged affected points accordingly. During all A-array
observations, significant RFI (affecting $\sim$ 15\% of the data) was
found to be present on IF2 in channel 4 to 6. In addition, all $uv$
data points in the A-array data above an amplitude of 0.4\,Jy were
clipped, since no such strong source is present in any individual
field. The C-array observations were affected by strong RFI and solar
interference, so that only baselines larger than 2.5 k$\lambda$ and 1
k$\lambda$ were included from the data of the first three days and the
last day of observations, respectively. The clipping level was set to
0.45 Jy for the C-array data. 

\subsection{Imaging}\label{subsec:imag}

\begin{figure}[ht]
\includegraphics[angle=-90,scale=0.65]{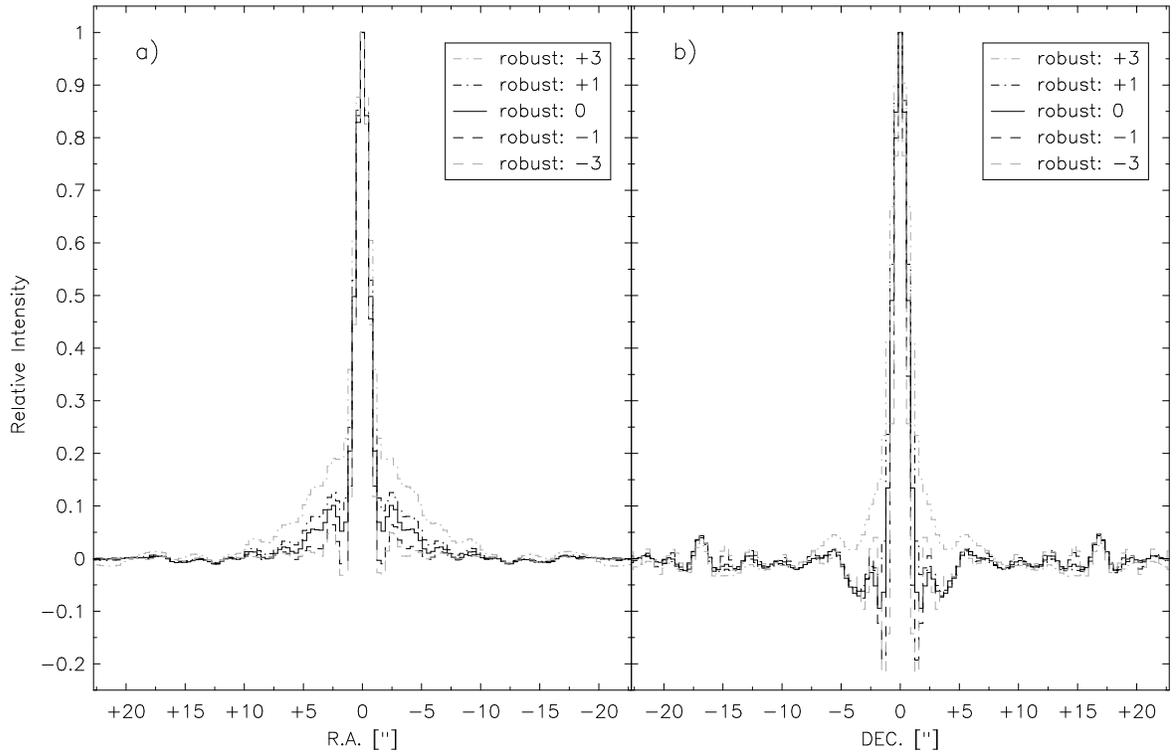}
\caption{Cuts along the x- ({\it a}) and y-axis ({\it b}) of the
  synthesized (i.e. DIRTY) beam for different values of the robust
  weighting: +3 ({\it grey dashed dotted line}), +1 ({\it dashed
    dotted line}), 0 ({\it solid line}), -1 ({\it dashed line}), and
  -3 ({\it grey dashed line}). A value of 0 for the robust parameter
  gave the best compromise between synthesized beam shape and rms
  noise (see text for details).
\label{fig:robust}}
\end{figure}

We performed substantial testing for best imaging quality including
the application of self-calibration on the COSMOS fields themselves.
It was found that no combination of parameters for the
self-calibration in the task 'CALIB' would yield a significant
improvement of the rms (of $>$ 3\%). A robust weighting of 0 provided
the best compromise for the combined A+C array data between a fairly
Gaussian synthesized beam (Fig. \ref{fig:robust}), and still good
sensitivity, i.e. the deviation from Gaussianity only starts below
$\pm$10\% of the peak. This proved to be especially important for
fields which contained bright sources (with peak fluxes up to
10\,mJy/beam) where tests showed that sidelobe artifacts are lowest
when using a robust weighting of 0. The nominal increase in the noise
compared to natural weighting is 1.265. However, the gain in better
cleaning results around bright sources is larger than this nominal
increase. Thus in order to achieve an uniform as possible rms across
the entire COSMOS field, a robust weighting of 0 is used.

\begin{figure}[ht]
\includegraphics[angle=-90,scale=.8]{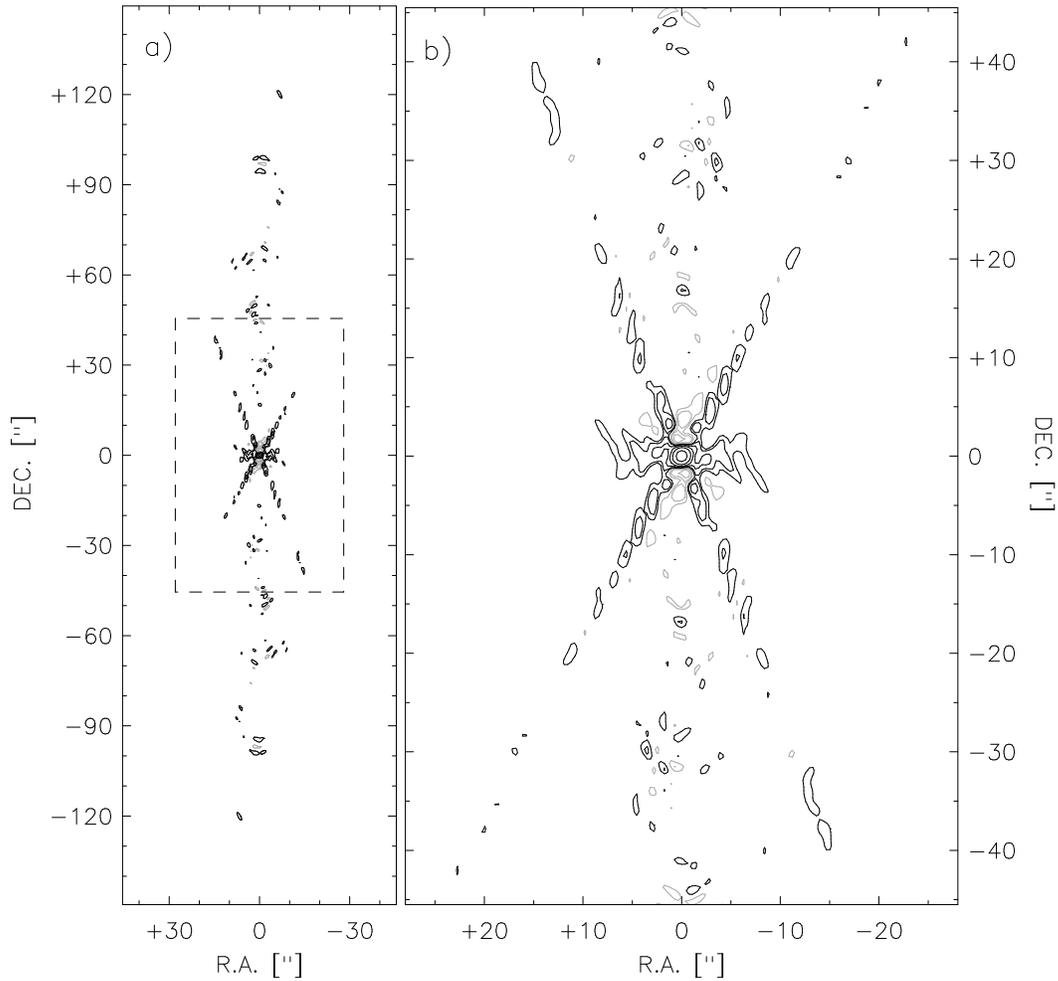}
\caption{Representative synthesized beam belonging to pointing field
  12 for a robust weighting of 0. {\it a)} Large field view with
  contours of 2.5, 5, 10, 20, 40 and 80\% of the maximum. The dashed
  box outlines the area shown in {\it b)}. {\it b)} Zoom into the
  central part of the synthesized beam with contours of 2, 4, 8, 16,
  32, and 64\% of the maximum. (The corresponding negative contours
  are shown in light gray.) The first peaks of the sidelobes are
  below 10\% of the maximum, overall the shape of the synthesized beam
  is fairly well-behaved given the declination of the COSMOS field. 
\label{fig:beam}}
\end{figure}

In order to avoid geometric distortions due to the non-planarity of
the wide-field on the sky, each field was divided into 43 facets of
2048$\times$2048 pixels which were imaged using the option DO3DIMAG in
the AIPS task 'IMAGR'.  The pixel scale of $0.35\as$/pixel has been
well matched to the A+C-array beam size of FHWM $1.5\as \times 1.4\as$
(PA $\sim -50^o$) for a robust weighting of 0 (Fig. \ref{fig:robust}
and \ref{fig:beam}). For each field, a contiguous area of about
1$^{\circ}$ diameter was covered by the facets.  Additional smaller
facets of 128$\times$128 pixels were made using the task 'SETFC' for
positions of NVSS sources with peak fluxes above 0.1 Jy and within a
radial distance of 1.5$^{\circ}$ from the pointing center. This
ensured that sidelobes from strong sources outside the central
1$^{\circ}$ were CLEANed as well.
 
Since most of the COSMOS fields are affected by the sidelobes of
radio galaxies with peak fluxes between 1 to 15 mJy/beam, best
CLEANing results were obtained if CLEAN boxes for individual sources
were provided. This ensured that CLEANing of negative or positive
residuals was minimized. In order to derive the CLEAN boxes for each
field, we used the AIPS task 'IMAGR' to interactively select the CLEAN
boxes in all facets where significant sources were present. This
procedure was performed combining the data of all polarizations and
IFs into one single image to obtain the highest possible S/N
image. The resulting list of CLEAN boxes was saved. In addition, we
required that CLEAN components were subtracted from the $uv$ data
after a facet had been cleaned. This way, CLEAN components in
overlapping facets were not treated separately. In addition, this
requirement also reduced the effect of sidelobe bumps from strong
sources in neighboring facets. 

We would like to note at this point that the reduction process of the
VLA-COSMOS Pilot and Large dataset was not exactly identical.  While
self-calibration was applied to the Pilot data, this step was not done
while reducing the Large survey data: after detailed empirical testing
of the improvements due to self-calibration in the VLA-COSMOS Large
project, we concluded that no significant improvement was achieved,
likely due to the lack of sufficiently bright sources in all parts of
the entire COSMOS field. Since self-calibration adjusts the observed
visibility phases to model phases, it has the potential to alter the
position of a given source. However, it is expected that these
effects cancel out when using several sources within a given pointing.

\begin{figure}[ht]
\includegraphics[scale=0.8]{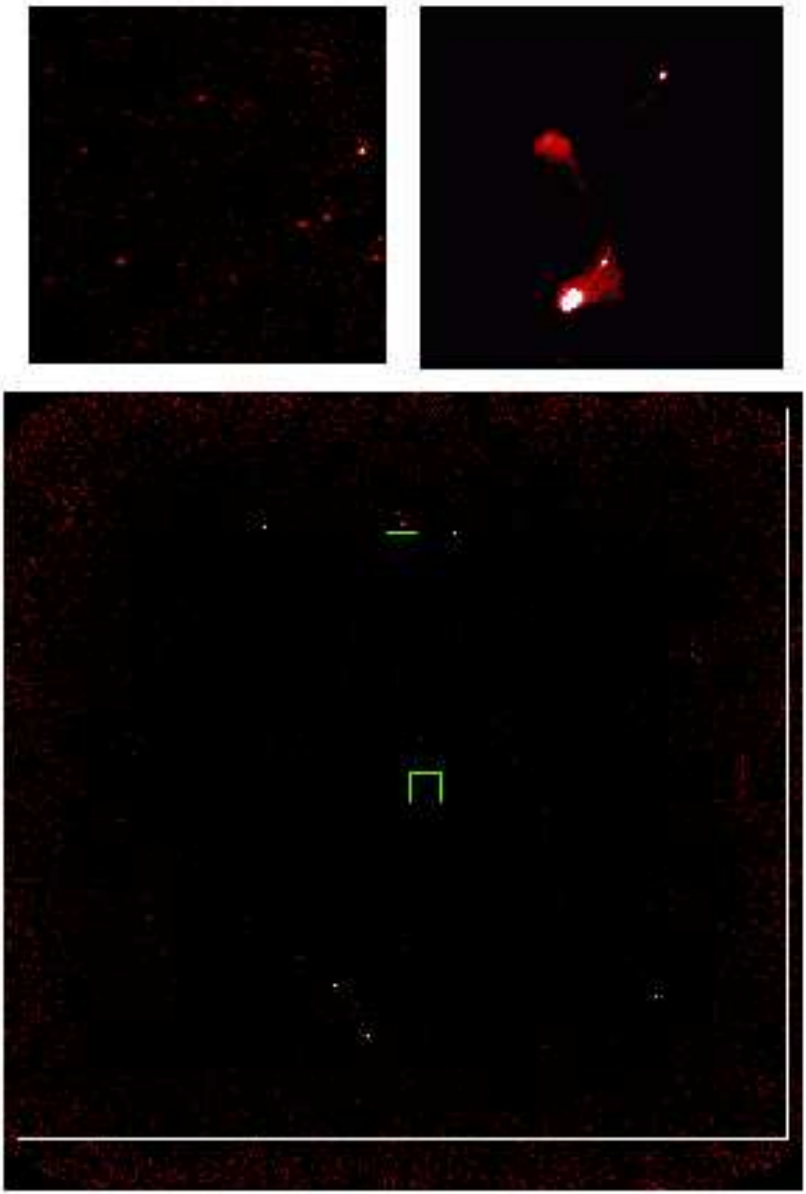}
\caption{The COSMOS field as observed at 1.4\,GHz. {\it Bottom:} The
  2\,deg$^2$ COSMOS field with the ACS coverage
  \citep*[from][]{sco06b} indicated by the gray box. The two green
  boxes outline the regions shown in the top panels. {\it Top:} Two
  regions enlarged demonstrate the quality of the data from the VLA-COSMOS
  large project. The left (right) panel represents the
  lower (upper) green box in the bottom panel. Each panel has a size of
  $2.8\am \times 2.8\am$ corresponding to about 0.1\% of the total area.
\label{fig:map}}
\end{figure}

For the final stage of CLEANing, it turned out that the well known
'beam squint' of the VLA (i.e. slightly different pointing centers for
R and L polarization), and the slightly different frequency coverages
required separate imaging of all polarizations and IF combinations.
The four separate 'IMAGR' runs were performed with the same list of
CLEAN boxes in the automatic mode. The number of iterations was set to
100,000, with a flux limit of 45\,$\mu$Jy/beam ($\sim 1.5\sigma$ in a
single image of a field) and a gain of 0.1 to optimize the CLEANing of
the facets. The 43 facets forming the contiguous area were combined
using the AIPS task 'FLATN'. The four separate images were then
combined using the AIPS task 'COMB' to obtain a single image for each
field. Due to the combination of bandwidth smearing and a significant
drop in sensitivity outside the radius of the Half Power Beam Width,
we decided to use a cut-off radius of 0.4 (corresponding to a radius
of $16.8\am$) when combining the individual fields into the final mosaic
using the task 'FLATN'. The resulting image is shown in Fig. \ref{fig:map}.

\section{Tests}\label{sec:test}

\begin{figure}[ht]
\includegraphics[scale=0.8]{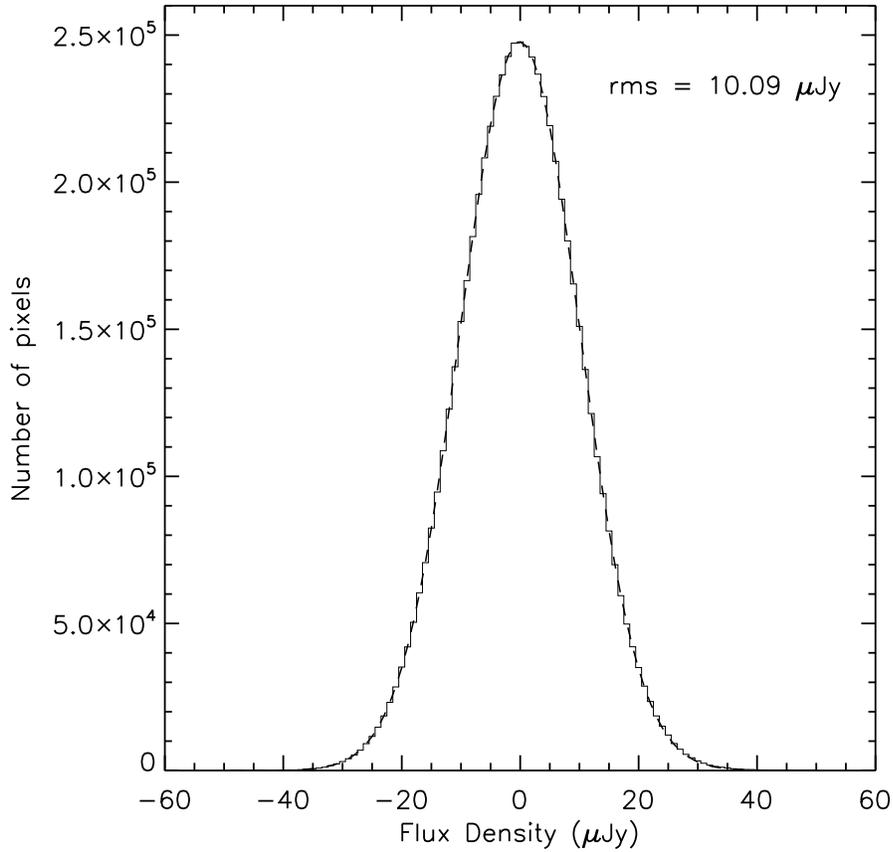}
\caption{Distribution of the noise. Pixel values extracted from a
  $16\am \times 11\am$ box close to the COSMOS field center show a
  Gaussian distribution in agreement with our assumption of Gaussian
  noise. The fitted Gaussian (dashed line) has a rms of $\rm
  10.09\,\mu Jy/beam$ ($\sigma$) (i.e. a FWHM of $\rm 23.76\,\mu
  Jy/beam$). Noise distributions extracted from different boxes located
  through out the uniform part of the field look similar.
\label{fig:noise}}
\end{figure}

We performed a number of tests to evaluate our flux (see \S
\ref{subsec:flux}) and astrometric calibration (see \S
\ref{subsec:astro}) as well as the impact of the CLEAN procedure. For
the last point, we performed a Gaussianity test on the noise. The
noise was extracted from a roughly $16\am \times 11\am$ box close to
the COSMOS field center. The individual noise pixels show a Gaussian
distribution (Fig. \ref{fig:noise}). A Gaussian fit gives an rms of
$\rm 10.09\,\mu Jy/beam$ ($\sigma$) (corresponding to a FWHM of $\rm
23.76\,\mu Jy/beam$). All noise distributions extracted for various
boxes across the part of the field that has an uniform background
showed a Gaussian distribution demonstrating that no artifacts have
been introduced during the CLEAN process.

\subsection{Flux calibration}\label{subsec:flux}

The second phase calibrator 0925+003 was observed twice each day to
allow for assessment of the absolute astrometry and the flux
calibration. Most of the following tests were performed on the A-array
only data, since it covered a wide range in time. We imaged the
calibrator 0925+003 for each day, as well as the two observations per
day separately. All IFs were combined at once, since the source of
interest is at the phase center and any effects due to misalignment
should be negligible. The images were cleaned with $1000$ iterations.
The resulting typical resolution and rms were $1.96\as \times 1.60\as$
(FWHM) and $\sim 870\,\mu$Jy/beam, respectively. The position and flux
of 0925+003 were derived by Gaussian fitting using the AIPS task
'JMFIT' on the individual images.

\begin{figure}[ht]
\epsscale{1.0}
\plotone{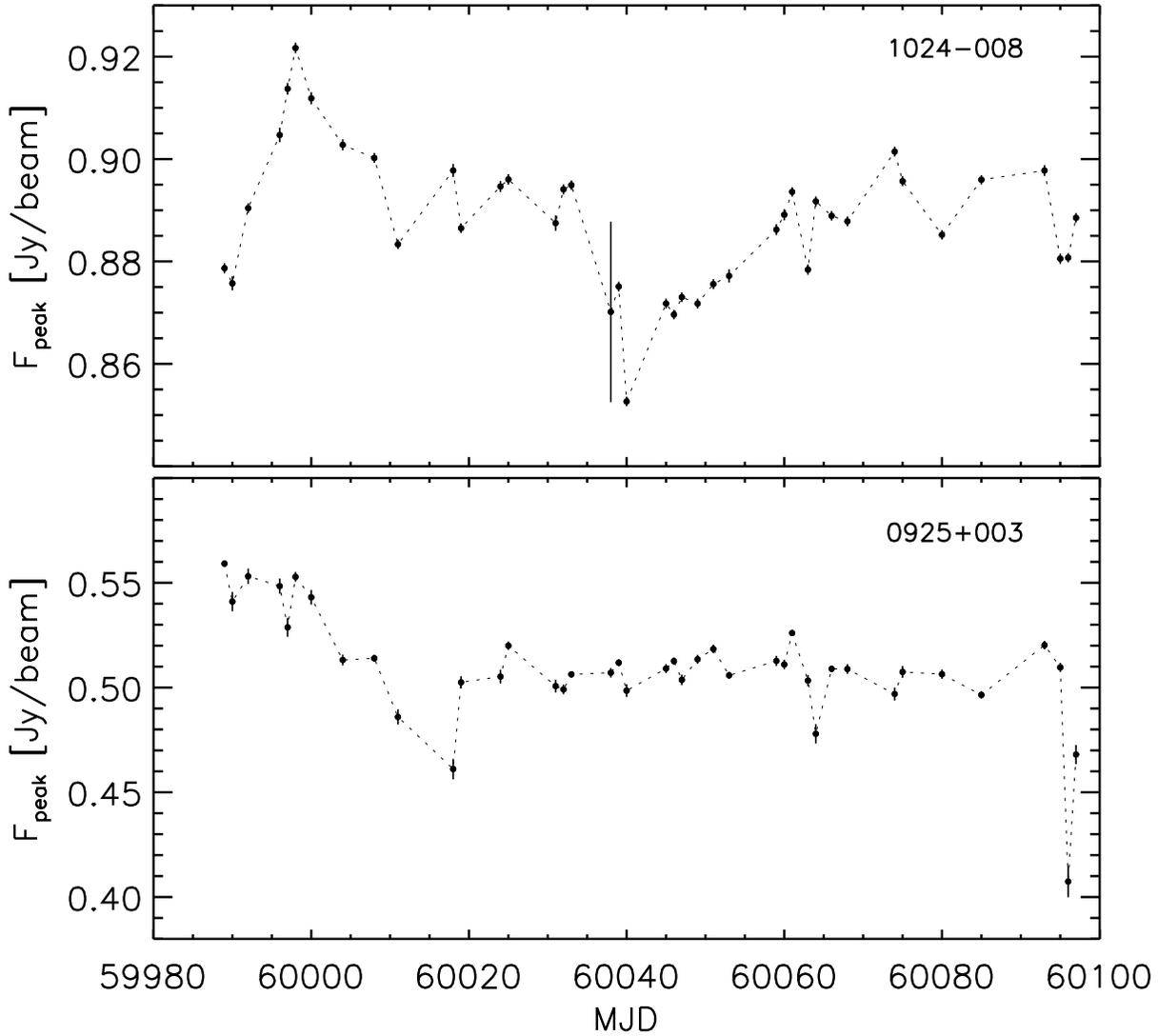}
\caption{Comparison between the flux of the two calibrators 1024-008
  and 0925+003 as a function of observing date. The dots show the peak flux
  density with indicated $3\sigma$ errors.
\label{fig:flux}}
\end{figure}

For most of the days 0521+166 served as the flux calibrator. The
trends of the peak flux of 0925+003 and 1024-008 are not the same over
the course of the observations in the A configuration
(Fig.~\ref{fig:flux}) indicating no systematic effects in the flux
calibration. Note that the error in the flux estimation for calibrator
1024-008 is significantly higher on day MJD 60038 (November 11th, 2004). 
This is due to strong interferences that could not
be entirely removed in the $uv$ data points.

\begin{figure}[ht]
\epsscale{1.0}
\plotone{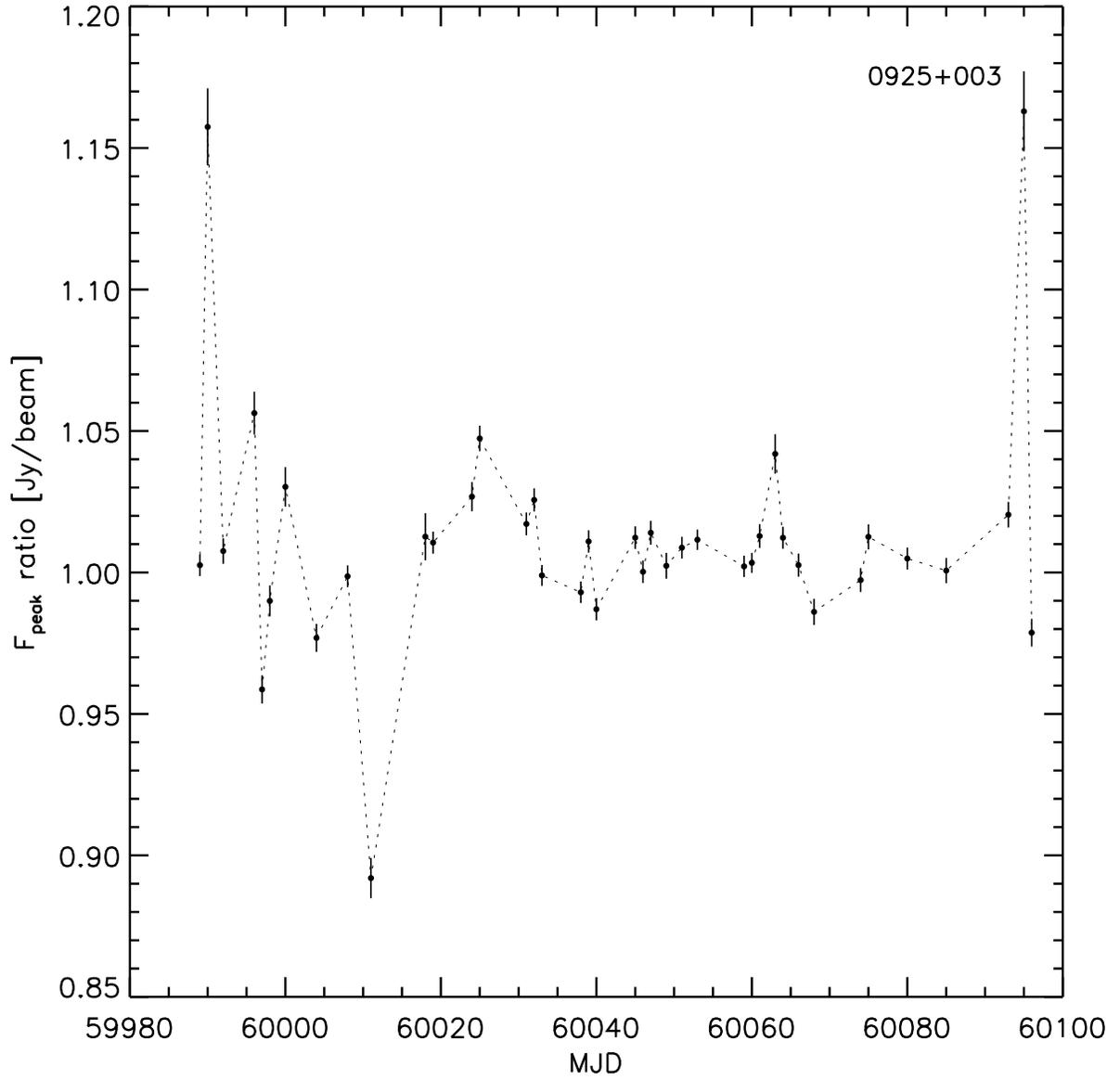}
\caption{The peak flux density variations (dots) of the two
  observations per day for calibrator 0925+003 shown as a ratio of the
  measured peak flux densities. $3\sigma$ errors are indicated.
\label{fig:d_flux}}
\end{figure}

We compared the peak flux density values of 0925+003 of the two
observations per day (Fig.~\ref{fig:d_flux}). The median offset is
4.5\,mJy/beam which corresponds to less than $1\%$ of the total flux
density of 0925+003. The outliers correspond to days MJD 59990
(September 24th, 2004), 60011 (October 15th, 2004) and 60096 (January
8th, 2005). The rms in the maps for those days is about $1.3-2.6$
times the typical rms in the 0925+003 maps. The higher noise is likely
to be caused by worse weather conditions (e.g. it was snowing on
November 13th, 2004) and/or technical problems during observations
(e.g. RFI, intermittent fluctuations of the system temperature
$T_{SYS}$, data corruption on particular antennas). Thus we conclude
that our flux calibration is within the errors expected.

\subsection{Absolute and Relative Astrometry}\label{subsec:astro}

Given the angular resolution of the combined A+C array data of $1.5\as
\times 1.4\as$ (FWHM), we expect to achieve a positional accuracy of
$\sim 0.15\as$ \citep*[corresponding to 1/10th of the beam size;
see][]{fom99} for high S/N sources and $\sim \rm \frac{FHWM}{S/N}$ for
lower S/N cases when extracting the source position within the COSMOS
field.

In order to assess the quality of the absolute astrometric
calibration, all observations of 0925+003 were combined into a single
image. A non-zero offset in RA and DEC of 53~mas and 45~mas,
respectively, has been found relative to the nominal position of
0925+003.  This offset is likely the result of the large angular
separation of 14.5$^{\circ}$ between the two calibrators (i.e.
0925+003 and 1024-008), which could lead to residual phase transfer
errors due to, for example, differential refraction corrections. We
consider this offset as an upper limit to our absolute astrometry
error, since the (center of the) COSMOS field is only 6$^{\circ}$ away
from the phase calibrator 1024-008.

\begin{figure}
\plottwo{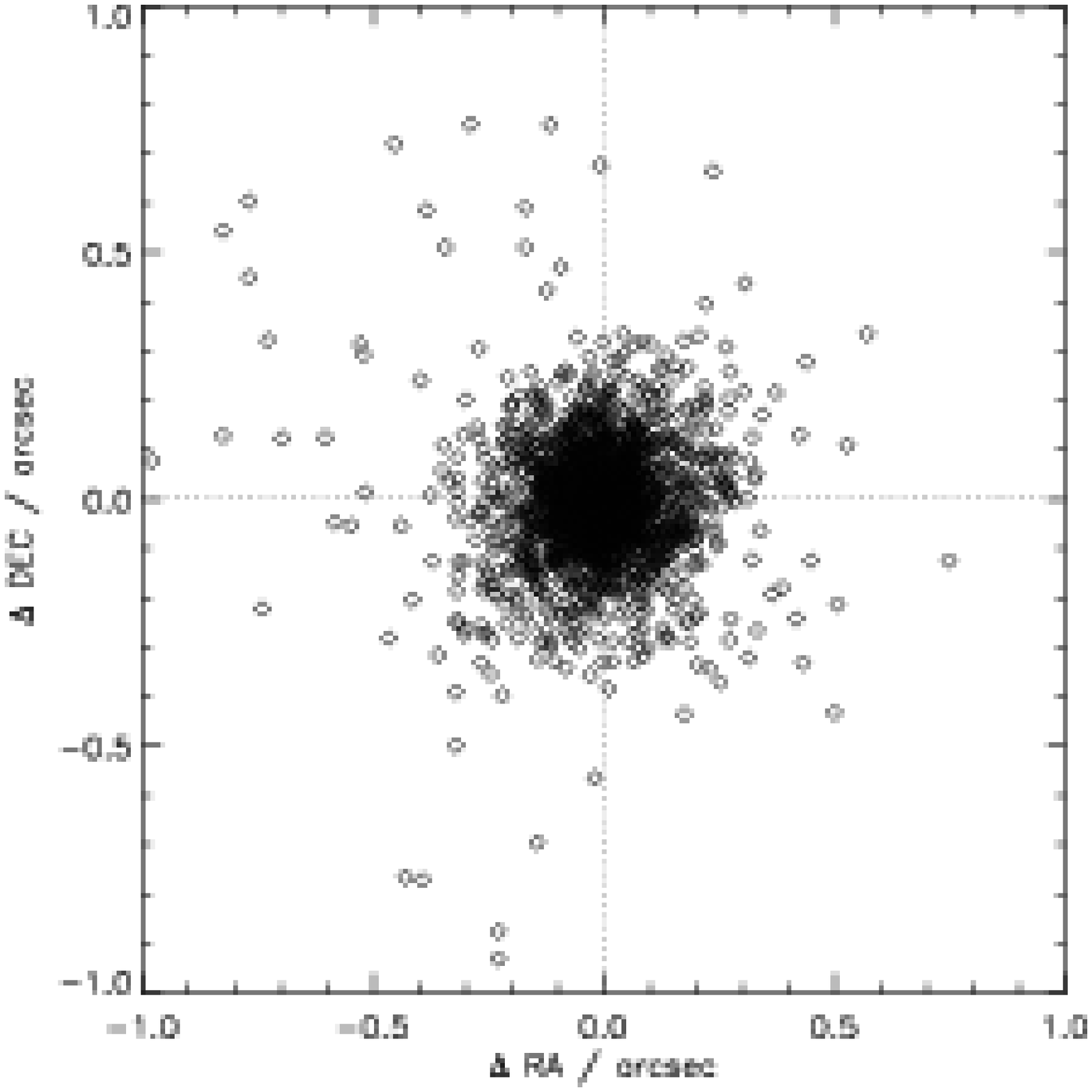}{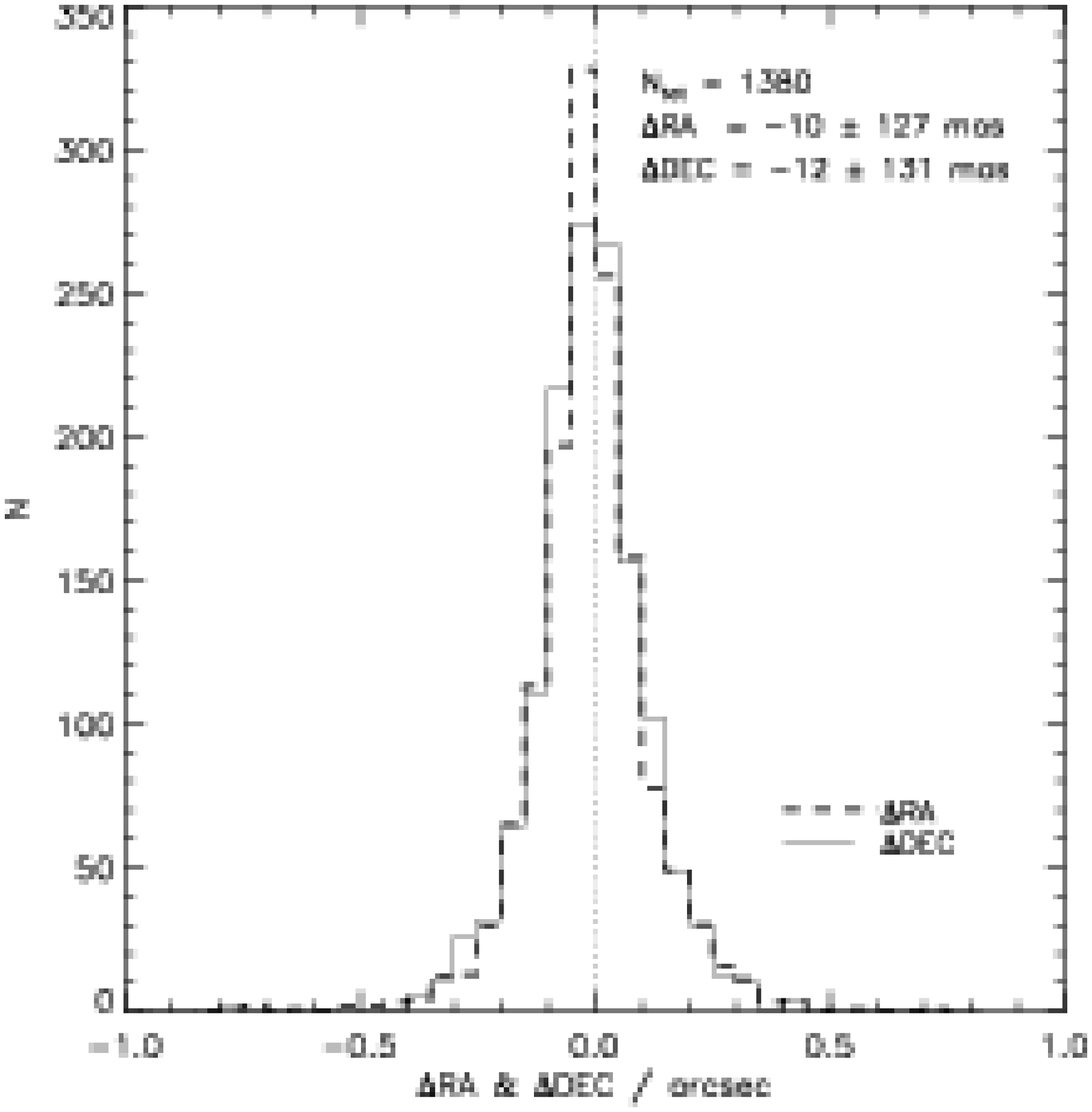}
\caption{The left panel compares the offset in RA ($\Delta RA$) 
with the offset in DEC ($\Delta DEC$)
when positions in single pointings are matched to positions in the
combined mosaic (see text for details). The reference position is the 
one extracted from the mosaic. 
The right panel shows the distributions of $\Delta RA$ (thick dashed line) 
and $\Delta DEC$ (thin solid line). The total number of sources, mean 
and standard deviation of the offsets are indicated.
\label{fig:RDoff0}}
\end{figure}

\begin{figure}
\plotone{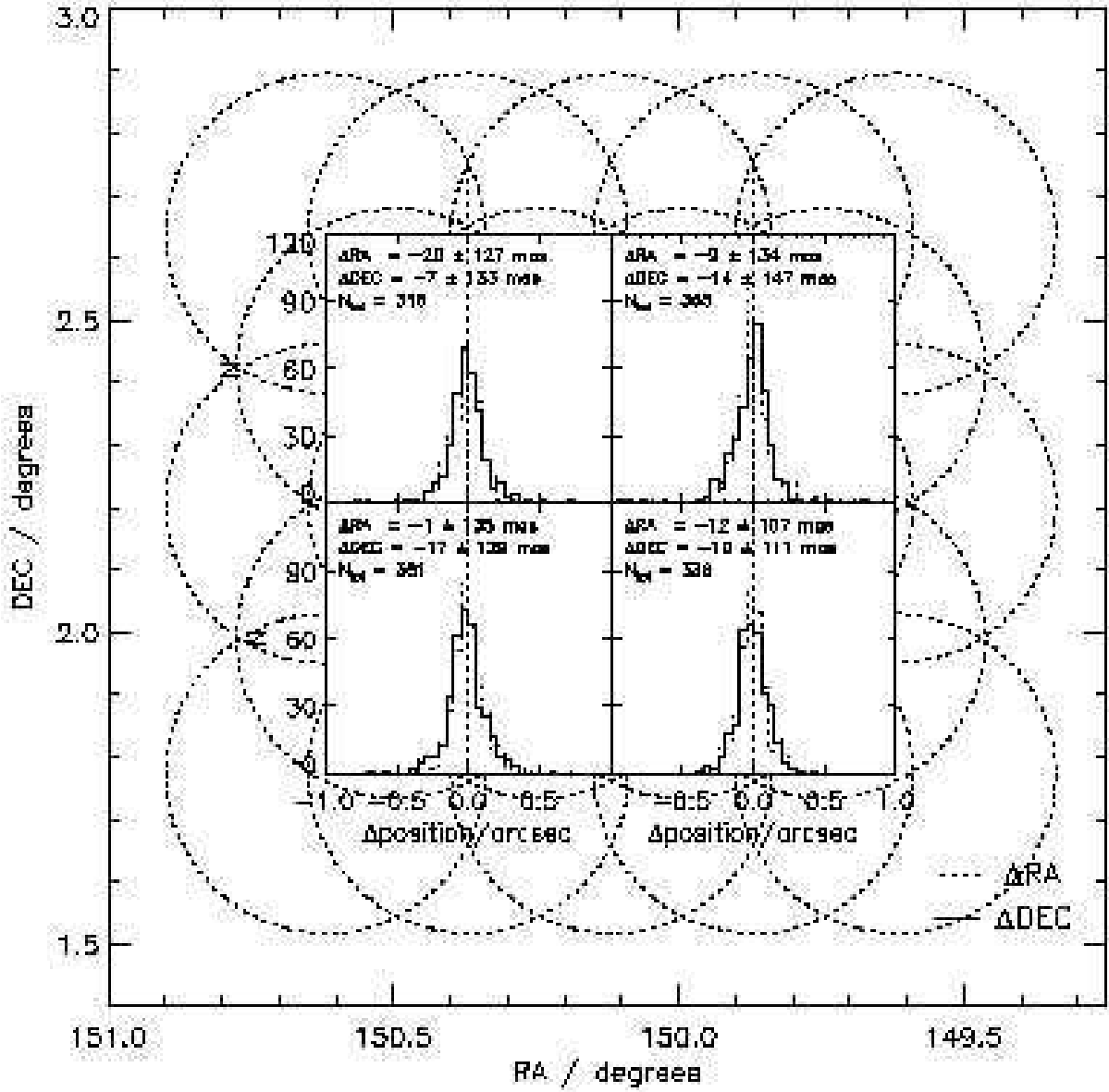}
\caption{Distributions of $\Delta RA$ (thick dashed line) 
and $\Delta DEC$ (thin solid line) for different parts in the inner  
$0.87\Box^\circ$ area. The positions of the four panels in the diagram
correspond exactly to the analyzed area. The mean and
standard deviation of the offsets and the  total number of sources are
indicated in each panel. For clarity the pointing pattern of the VLA-COSMOS
is shown in the background (dotted circles).
\label{fig:RDoff2x2}}
\end{figure}

To test the quality of our relative astrometry, we extracted sources
from each single field and compared their positions to the ones
extracted from the combined mosaic. We searched for sources using the
AIPS task 'SAD' (Search And Destroy). On single fields we ran 'SAD'
searching for sources with fluxes higher than 100\,$\mu$Jy/beam.
'SAD' looks for points above the specified flux limit and merges such
points into contiguous ``islands''. Then it fits components within
these ``islands''. For our astrometric tests, we run 'SAD' rejecting
components within an island with both peak and integrated flux values
lower than 100\,$\mu$Jy/beam which corresponds to $\sim 7\sigma$ in a
single field. On average $\sim 150$ sources were found per pointing.
(In \S\ref{sec:cat} we describe how 'SAD' was run on the mosaic.)
After source extraction we only matched positions of objects which
have a deconvolved major axis of $<3\as$ FWHM and are within a radius
of $\sim17\am$ from the pointing center (which corresponds to our
primary beam cut of 0.4) in the specific field. We analyzed the
offsets in right ascension ($\rm \Delta\,RA$) and declination ($\rm
\Delta\,DEC$) in the central $\rm 0.87\,deg^2$ where the rms noise is
basically uniform. The results are shown in Fig.~\ref{fig:RDoff0}. The
offsets in $\rm \Delta\,RA$ and $\rm \Delta\,DEC$ are $(-10 \pm
127)$\,mas and $(-12 \pm 131)$\,mas, respectively. To search for
possible systematic effects, we analyzed the $\Delta RA$ and $\Delta
DEC$ offsets in different parts of the central $\rm 0.87\,deg^2$ area.
As seen from Fig.~\ref{fig:RDoff2x2}, there are no significant
systematic effects in our relative astrometry as a function of
position within the COSMOS field.

To get a deeper insight into our astrometry we cross-correlated the
COSMOS mosaic source catalog with the VLA FIRST survey catalog
\citep*{bec95}. To minimize the number of spurious matches, we used a
search box size of $2\as$ on a side. Only sources with a major axis
$<3\as$ and COSMOS to FIRST fluxes comparable within 20\%, i.e.  $\rm
0.8<S^{int}_{COSMOS}/S^{int}_{FIRST}<1.2$, were compared. Multiple
component sources and FIRST sources with side lobe flags ($flag=1$)
were excluded. Our final sample of matched sources contains only $28$
objects. The mean offsets and the $1\sigma$ errors for $\rm \Delta RA
= RA_{COSMOS}-RA_{FIRST}$ and $\rm \Delta DEC =
DEC_{COSMOS}-DEC_{FIRST}$ are $(-110\pm273)$\,mas and
$(67\pm232)$\,mas, respectively. Given the low number of matched
sources and the FIRST survey's astrometric accuracy of 500\,mas (or
more) for individual sources \citep*{whi97}, we conclude that the
inferred positional offsets are within the source extraction errors of
both surveys.

In addition, we compared the positions of radio sources extracted from
the VLA-COSMOS Pilot and the Large project. However, we consider this
not a completely independent test, as the same phase calibrator was
used for both projects. We find a median offset of -50\,mas and
90\,mas in $\rm \Delta RA$ and $\rm \Delta DEC$, respectively, while
the rms scatter is 161\,mas and 189\,mas for the first and latter. The
rms scatter is slightly higher than the above derived accuracy of our
relative astrometry ($\sim$130~mas) using only the Large project.
However, this is expected as the rms and the beam size of the Pilot
project is larger: 25\,$\mu$Jy/beam vs. 10\,$\mu$Jy/beam and $1.9\as
\times 1.6\as$ vs. $1.5\as \times 1.4\as$. The derived astrometric
differences between the Pilot and the Large projects are well within
our errors (see \S \ref{subsec:imag} for data reduction difference
between both projects). Hence, we conclude that our relative
astrometric accuracy for the VLA-COSMOS Large project is $\sim$130~mas
and discard this higher rms scatter found from the comparison to the
Pilot data.

Based on arguments presented above, we conclude that the overall
astrometric errors of our derived source positions are dominated by
the uncertainty in the position extraction (due to our beam size) of
$\sim130$\,mas. Our absolute astrometric accuracy is likely to be
better than 55\,mas.

\section{The VLA-COSMOS Catalog}\label{sec:cat}

\subsection{Source Extraction}\label{subsec:find}

\begin{figure}[ht]
\includegraphics[scale=0.75,angle=0]{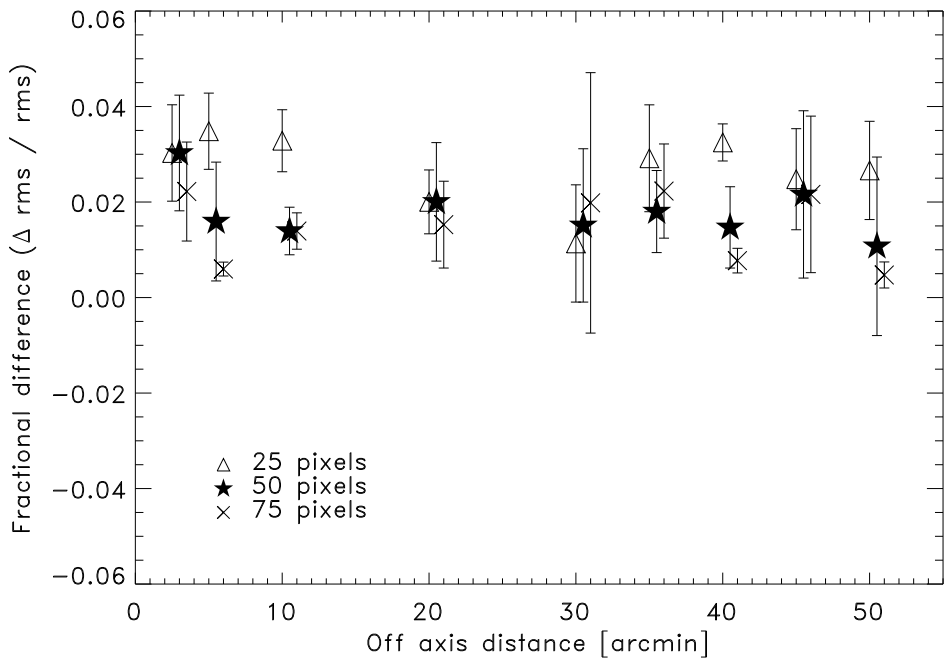}
\caption{Fractional difference between the directly measured rms value in a
  $100\times100$ pixel box and the corresponding value of the
  SExtractor noise map as a function of the radial distance for three
  different noise maps with mesh sizes of 25, 50 and 75 pixels,
  respectively. The x-positions have been shifted by 0.5$\am$
  for clarity. 
\label{fig:mesh}}
\end{figure}

\begin{figure}[ht]
\epsscale{1.0}
\plotone{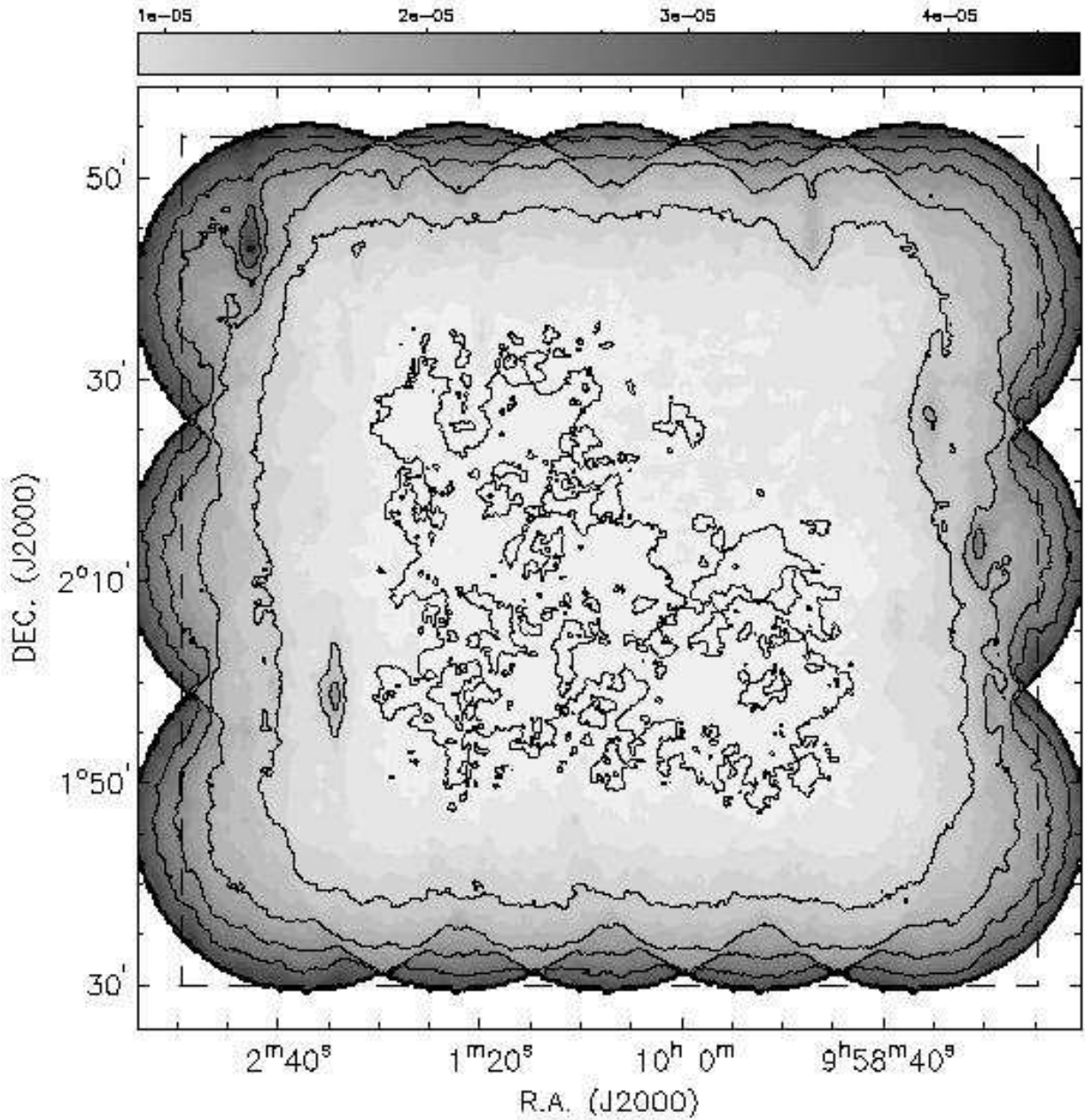}
\caption{Sensitivity map of the area covered by the VLA Large Project
  derived using SExtractor with a mesh size of 50 pixel. The rms is
  fairly uniform except for areas around strong radio sources. Lighter
  shades indicate lower rms noise values. The contours correspond to
  rms levels of 10, 15, 20, 25, 30, and 40 $\mu$Jy/beam. The dashed box
  outlines the area which was searched for radio components.
\label{fig:sens}}
\end{figure}

\begin{figure}[ht]
\epsscale{1.0}
\plotone{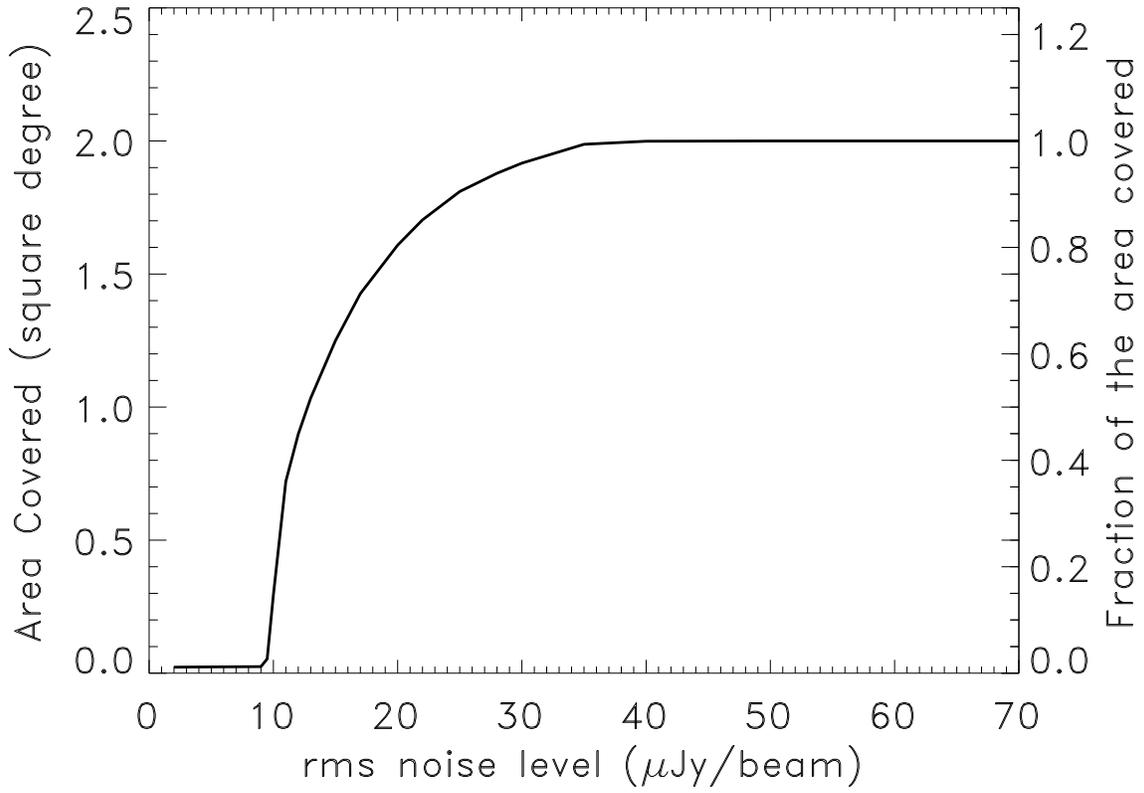}
\caption{Plot of the rms noise level vs. cumulative as well as
  fractional area covered. The full area covered is $\rm 2\,deg^2$ and is
  indicated in Fig. \ref{fig:sens}.
\label{fig:area}}
\end{figure}

In order to select a sample of radio components from the largest
imaged area above a given threshold, defined in terms of the local
signal to noise ratio, we adopted the following approach. First the
software package SExtractor was used to estimate the local background
in each mesh of a grid covering the whole surveyed area
\citep[see][for a general description of SExtractor]{ber96}. Different
noise maps with mesh sizes ranging from 25 to 100 pixels were produced
and examined. The fractional difference between the rms measured in
the SExtractor noise maps and the rms directly measured on the real
map is very small ($\sim$2\%) over the whole map (see Fig.
\ref{fig:mesh}). In the end, we adopted a mesh size of 50 pixels
corresponding to $17.5\as$ which was found to be the best compromise
between closely sampling the variations in rms and avoiding
contamination by larger radio sources. The rms values range from about
9\,$\mu$Jy/beam in the inner regions to about 20\,$\mu$Jy/beam at the
edges of the mosaic with values as high as $30-40\,\mu$Jy/beam around
the few relatively strong sources (see Fig.  \ref{fig:sens}). The mean
rms in the inner $\rm 1\,deg^2$ is 10.5\,$\mu$Jy/beam, the mean rms
over the $\rm 2\,deg^2$ area is 15.0\,$\mu$Jy/beam. The cumulative
area as a function of rms is shown in Fig. \ref{fig:area}.

As a next step, the AIPS task 'SAD' was used to obtain a catalog of
candidate components. 'SAD' attempts to find all the components whose
peaks are brighter than a given flux level. In order to detect radio
components down to the 30\,$\mu$Jy/beam level 'SAD' was run several
times with different search levels (with a decreasing flux limit)
using the resulting residual image each time. We recovered all the radio
components with a peak flux $\rm S_{peak} > 30\,\mu Jy/beam$
(corresponding to roughly 3$\sigma$ in the higher sensitivity
regions). For each component 'SAD' provides peak flux, total flux,
position and size estimated using a Gaussian fit.

However, for faint components the Gaussian fit may be unreliable and a
better estimate of the peak flux (crucial for the selection based on
S/N) can be obtained with a non-parametric second-degree interpolation
using the AIPS task 'MAXFIT'. We ran 'MAXFIT' on all the components
found by 'SAD' and selected only those components for which the peak
flux density found by 'MAXFIT' was greater or equal to 4.5 times the
local rms as derived from the noise map. The (non-parametric) peak
position and flux density as determined by 'MAXFIT' were kept, as the
so derived values should be less affected by assumptions on the real
brightness distribution.

\begin{figure}[ht]
\includegraphics[scale=0.8]{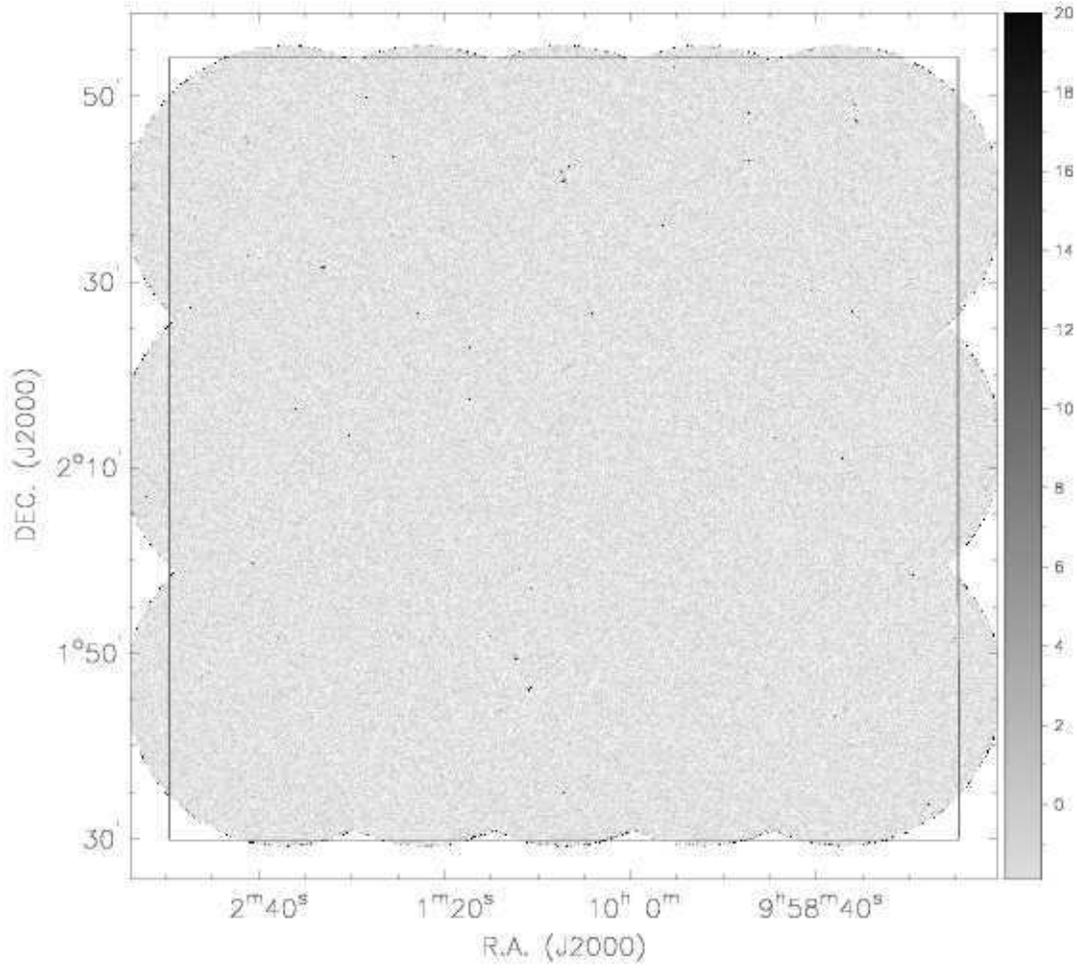}
\caption{Map of the S/N of the VLA-COSMOS Large Project as constructed
  using the SExtractor sensitivity map (Fig. \ref{fig:sens}). Lighter
  shades indicate lower S/N values. The dashed box shows the area
  in which radio sources were identified (see also text).
\label{fig:snr}}
\end{figure}

Finally, we visually inspected the S/N mosaic image (Fig.
\ref{fig:snr}) for components that could have been missed by 'SAD'.
The most likely reason for missing sources is that 'SAD' only recovers
components that can be fitted by a Gaussian fulfilling certain
parameters. Thus, if the fit for a potential component fails, this
component is rejected from the catalogue provided by 'SAD'. Therefore,
the AIPS tasks 'JMFIT' and 'MAXFIT' were run on these potential
components to derive their properties.

In order to exclude 1-pixel wide noise peaks above the detection
threshold (4.5$\sigma$), more scrutiny was used for the 294 components
fitted with both sizes smaller than the CLEAN beam. Only those
components (171) for which JMFIT was able to estimate an upper limit
to the source size greater than the CLEAN beam were kept while the
remaining (123) were identified as noise spikes and excluded from the
catalogue. As a result of the whole procedure a total of 3823
components have been selected (3204 from 'SAD'+'MAXFIT' and 619 from
the S/N image). A more complete analysis on the completeness and
possible biases affecting the catalogue will be described in a future
paper along with the number counts (Bondi et al., in prep).

\subsection{Description of the Catalog}

Some of the components clearly belong to a single radio source (e.g.
jets and lobes of an extended radio galaxy), in other more complex
cases we have also used the optical ground- and space-based images to
discriminate between different components of the same radio source or
separate radio sources. The final catalog (see Tab. \ref{tab:cat}; see
below) lists 3643 radio sources of which 80 are multiple, i.e. better
described by more than a single component. These sources are
identified by the flag 'mult=1' (Tab. \ref{tab:cat}). For these
sources, the listed center is either the one of the radio core or the
optical counterpart when either of these could be reasonably
identified or the luminosity weighted mean position. In addition, we
visually inspected weak ($\le 6\sigma$) sources close to bright
sources with significant sidelobes. A total of 72 sources potentially
lying on sidelobe spikes are flagged with 'slob=1'. 

\begin{figure}[ht]
\epsscale{1.0}
\plotone{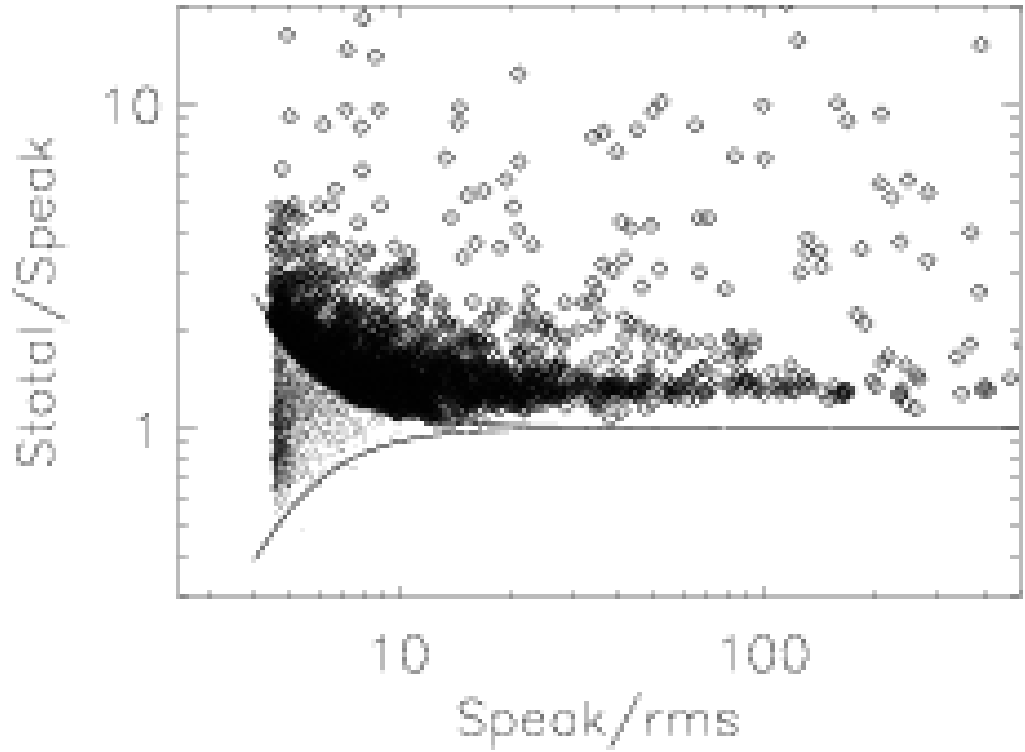}
\caption{Ratio of the total flux $S_T$ to the peak flux $S_P$ as a
  function of the signal-to-noise ratio of the peak flux and the local
  rms. The solid line shows the upper and lower envelopes of the flux
  ratio distribution containing the sources considered unresolved (see
  text). Open symbols show sources considered resolved.  
\label{fig:psf}}
\end{figure}

In Fig. \ref{fig:psf} we plot the ratio of the total integrated flux
density $S_{total}$ and the peak flux density $\rm S_{peak}$ as
functions of the signal to noise ratio S/N ($\rm S_{peak}/rms$) for
all the 3643 sources in the catalog. To select the resolved sources,
we determined the lower envelope of the points in Fig. \ref{fig:psf}
which contains 99\% of the sources with $S_{total}<S_{peak}$, and
mirrored it above the $S_{total}/S_{peak}=1$ line (upper envelope in
Fig.  \ref{fig:psf}). We have considered the 1601 (44\%) sources
laying above the upper envelope resolved. The envelope can be
described by the equation

$$\rm S_{total}/S_{peak} = 1 + [100 / (S_{peak}/rms)^3 ]$$

The resolved sources are flagged in the catalog by 'res=1'. For the
unresolved sources the total flux density is set equal to the peak
brightness and the angular size is undetermined. 

We calculated the uncertainties in the peak flux density $S_{peak}$
and integrated flux $S_{total}$ using the equations given by
\cite*{con97} as outlined in e.g. \cite*{hop03,sch04}. For the
positional uncertainties we used the equations reported in
\cite*[][their equations 4 and 5]{bon03}, using 130 mas as the
calibration error in right ascension and declination \citep*[see
also][their equation 27]{con98}.

\begin{figure}[ht]			 
\resizebox{.9\hsize}{!}{\includegraphics[]{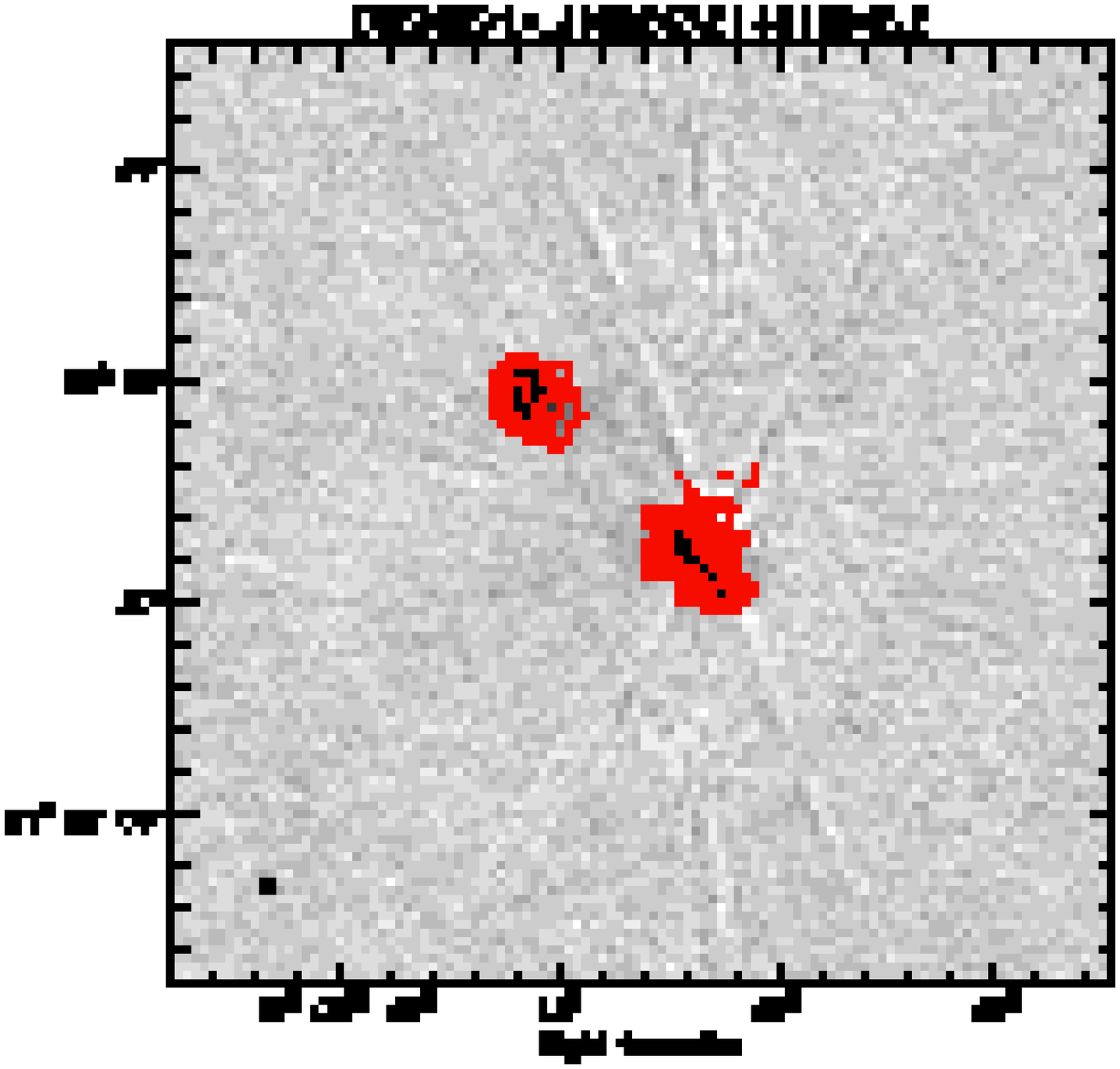}
                      \includegraphics[]{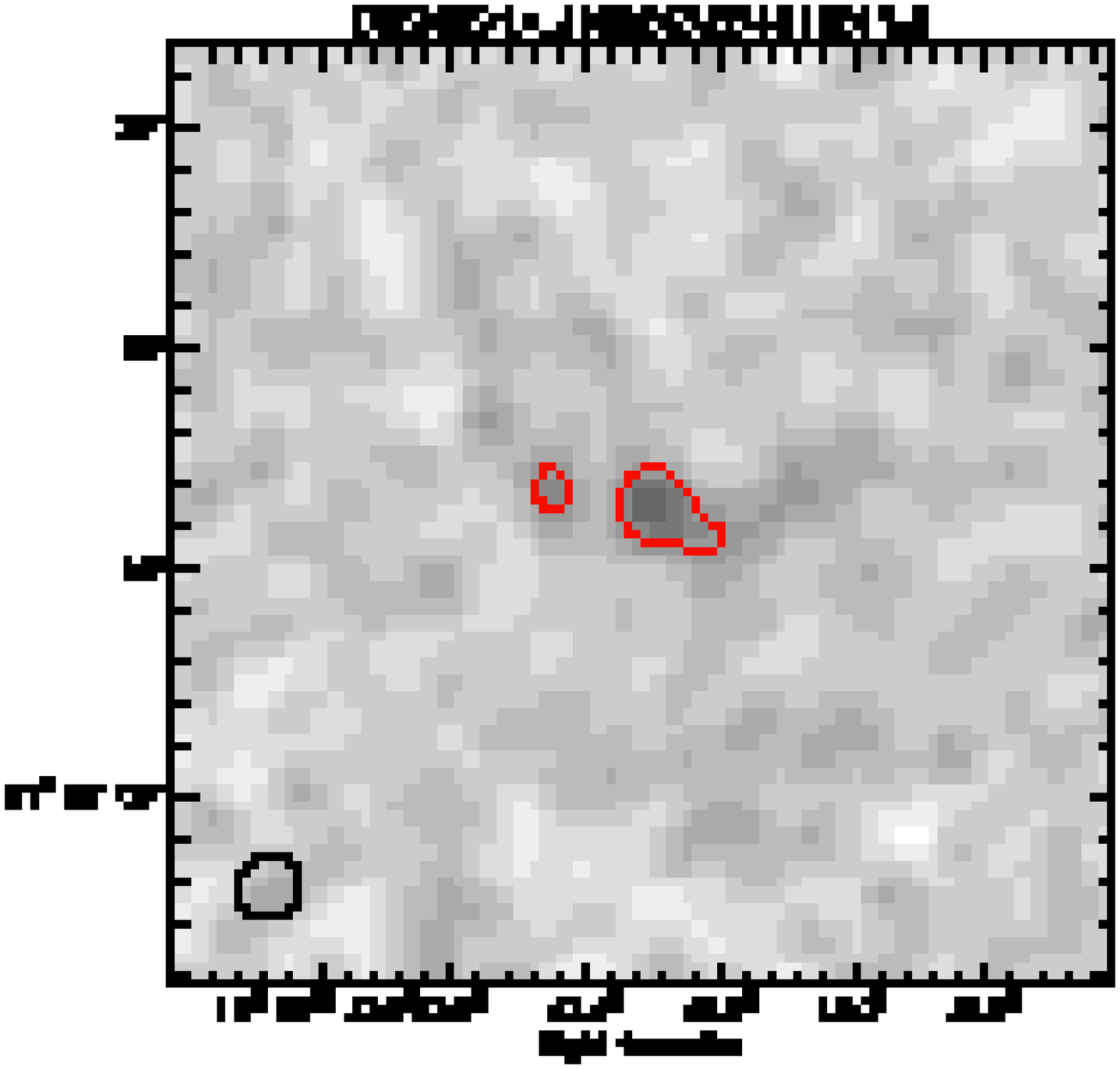}
                      \includegraphics[]{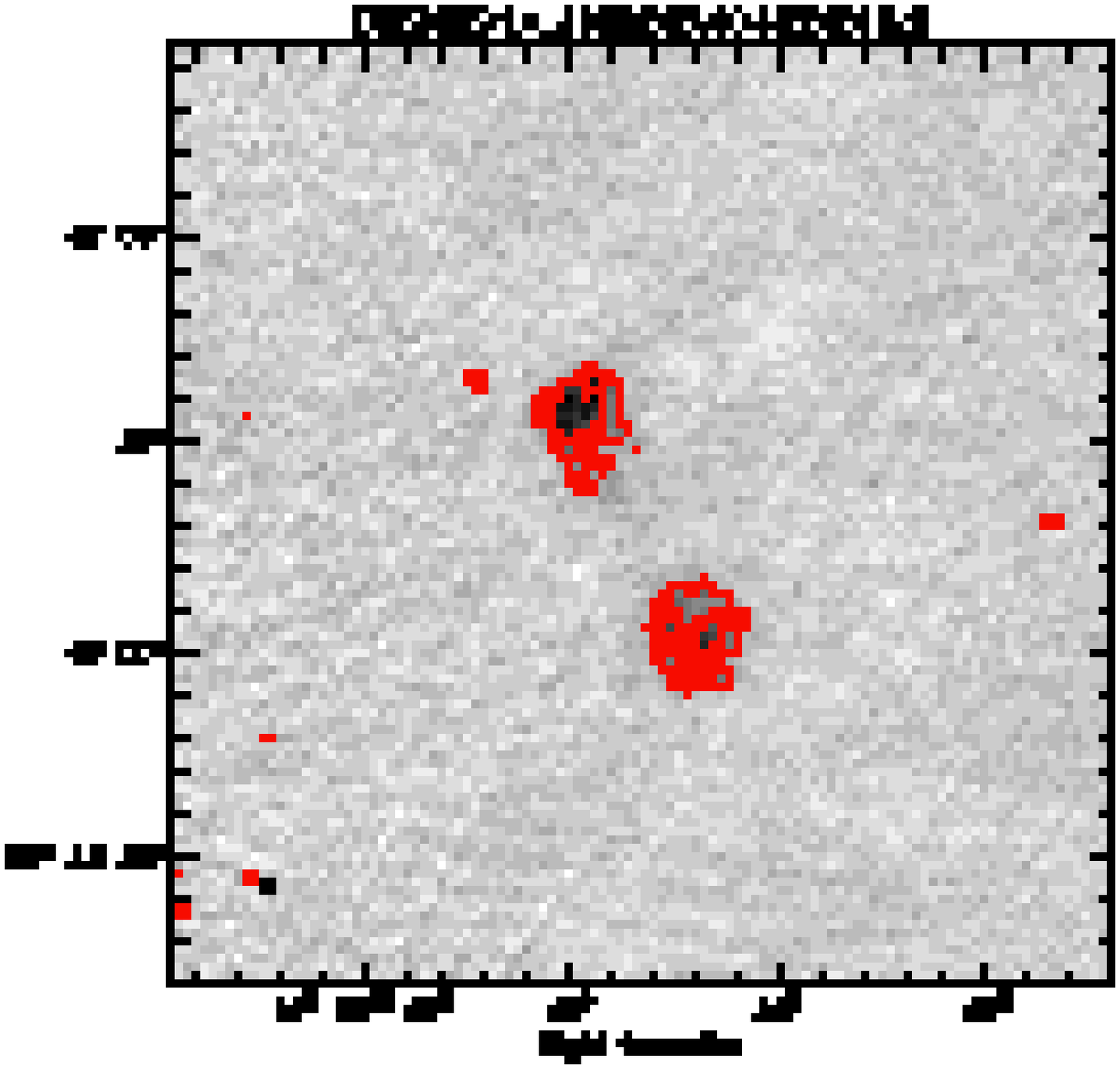}}
\resizebox{.9\hsize}{!}{\includegraphics[]{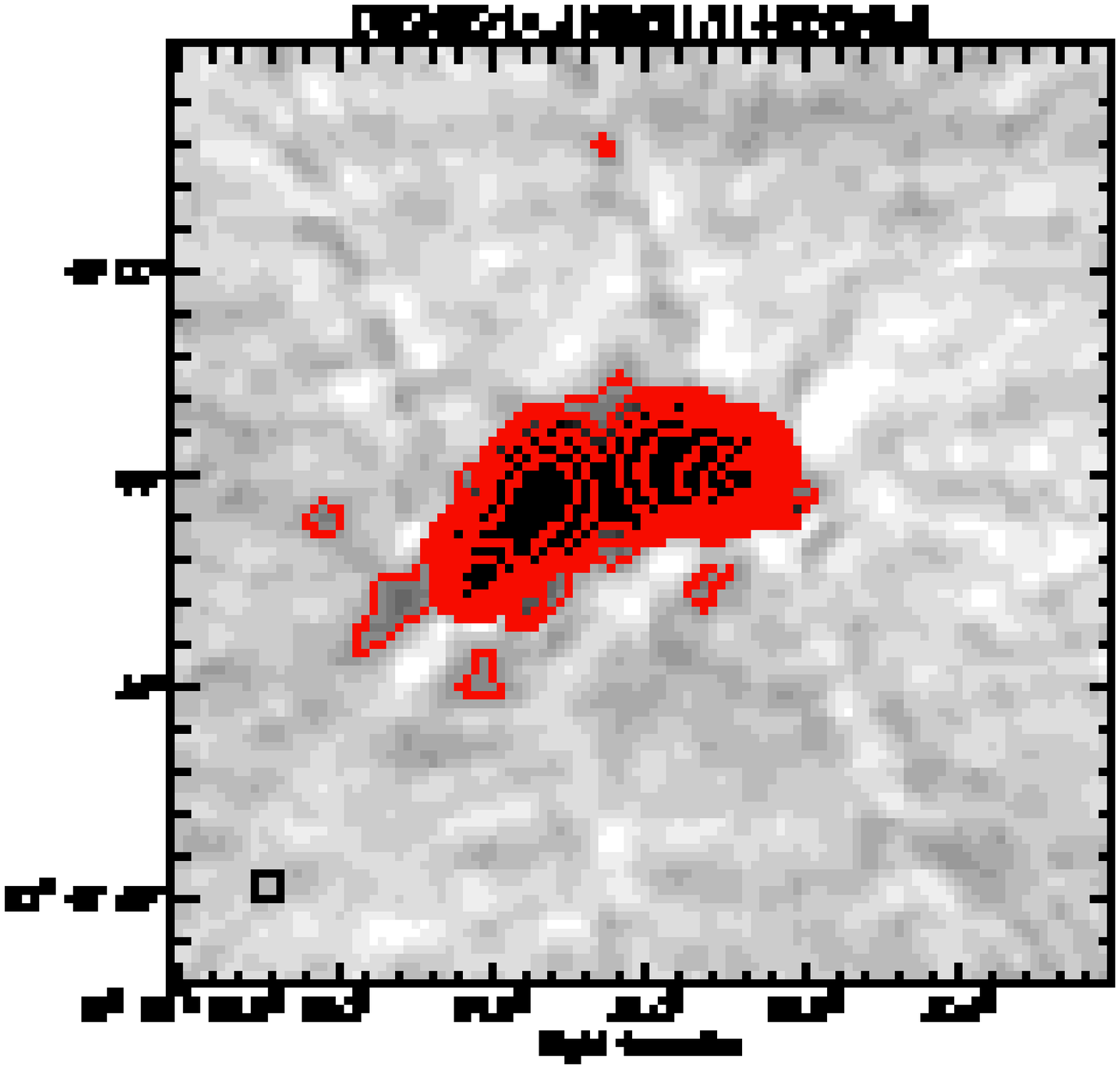}
                      \includegraphics[]{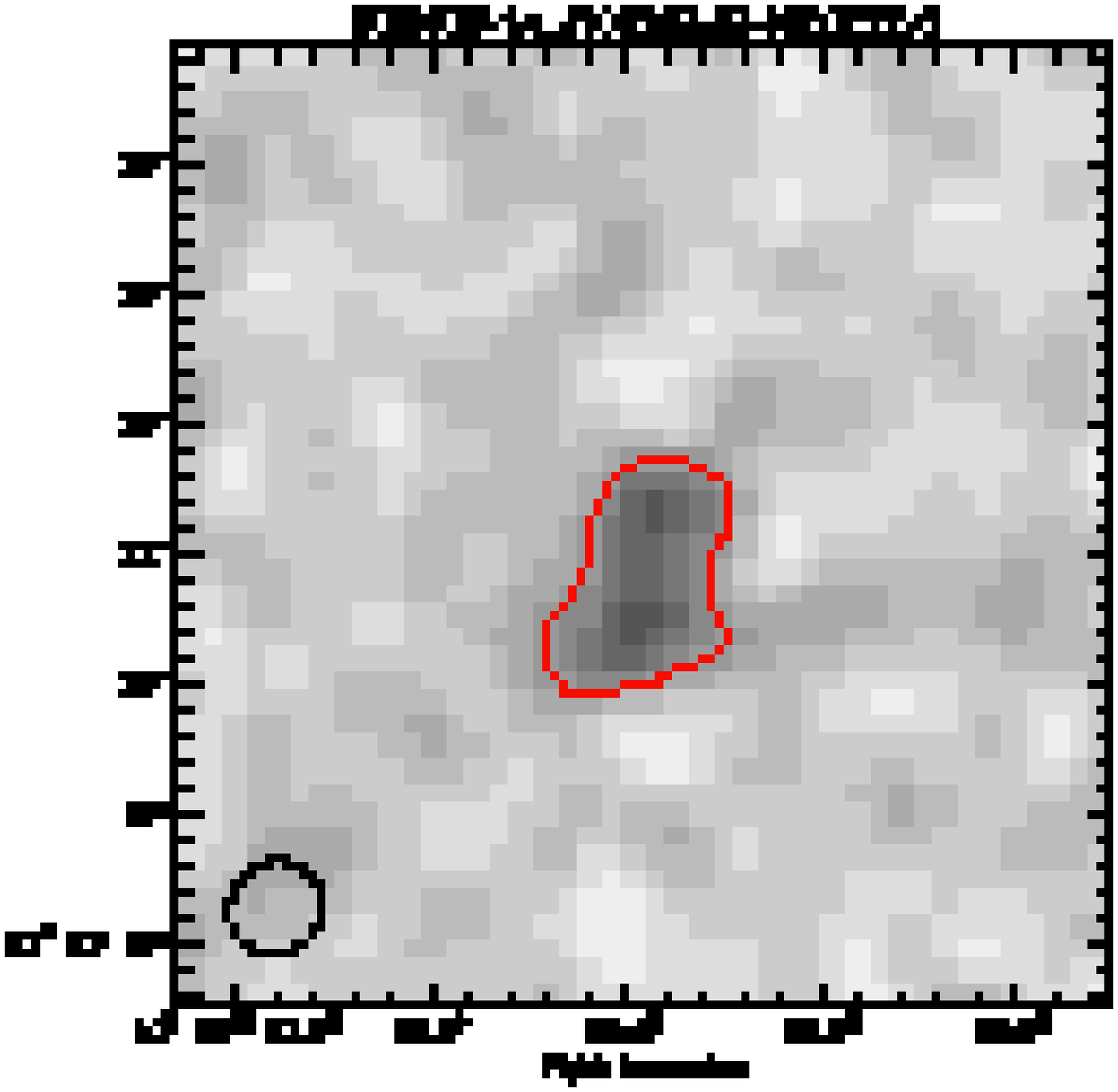}
                      \includegraphics[]{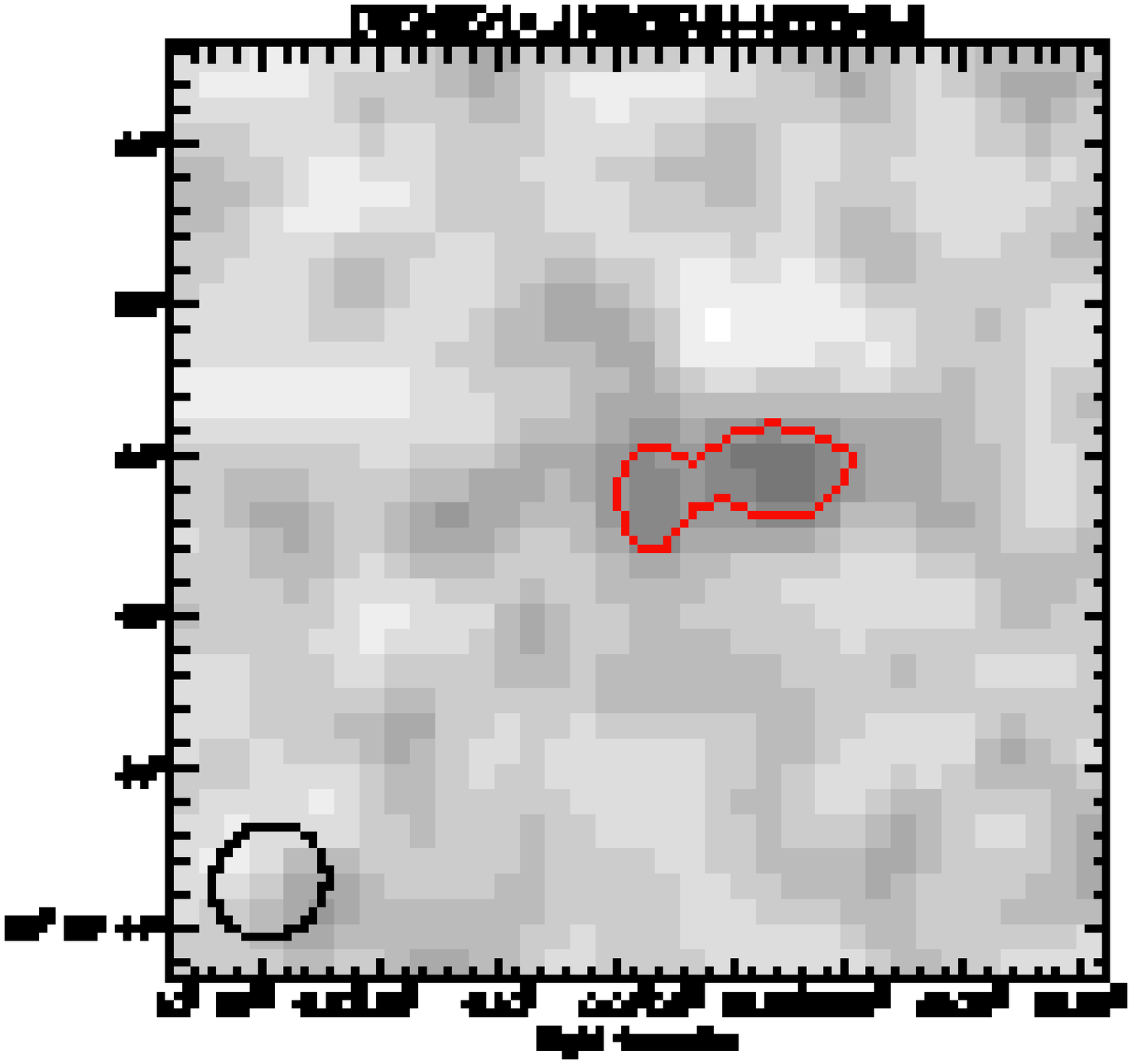}}
\resizebox{.9\hsize}{!}{\includegraphics[]{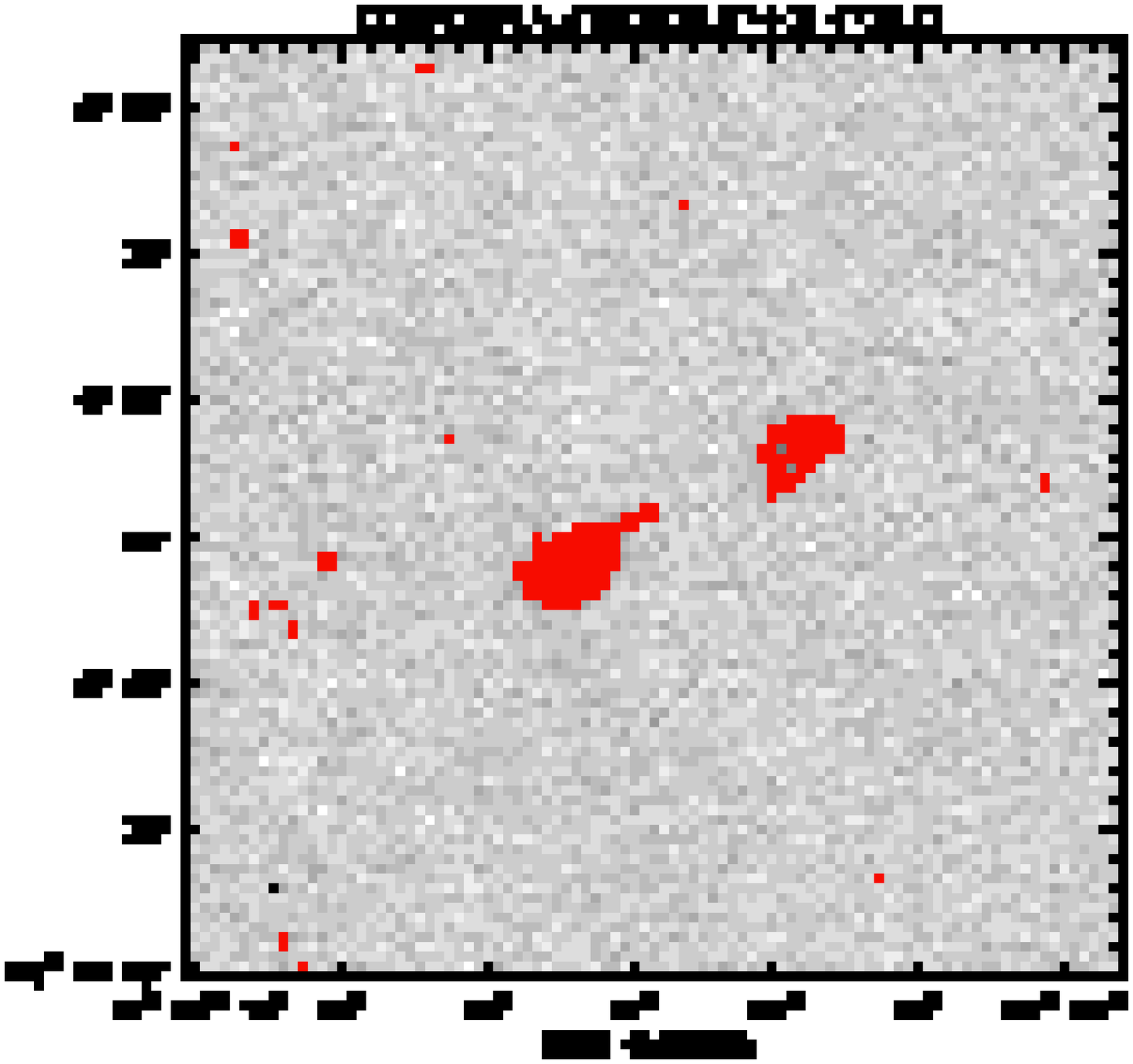}
                      \includegraphics[]{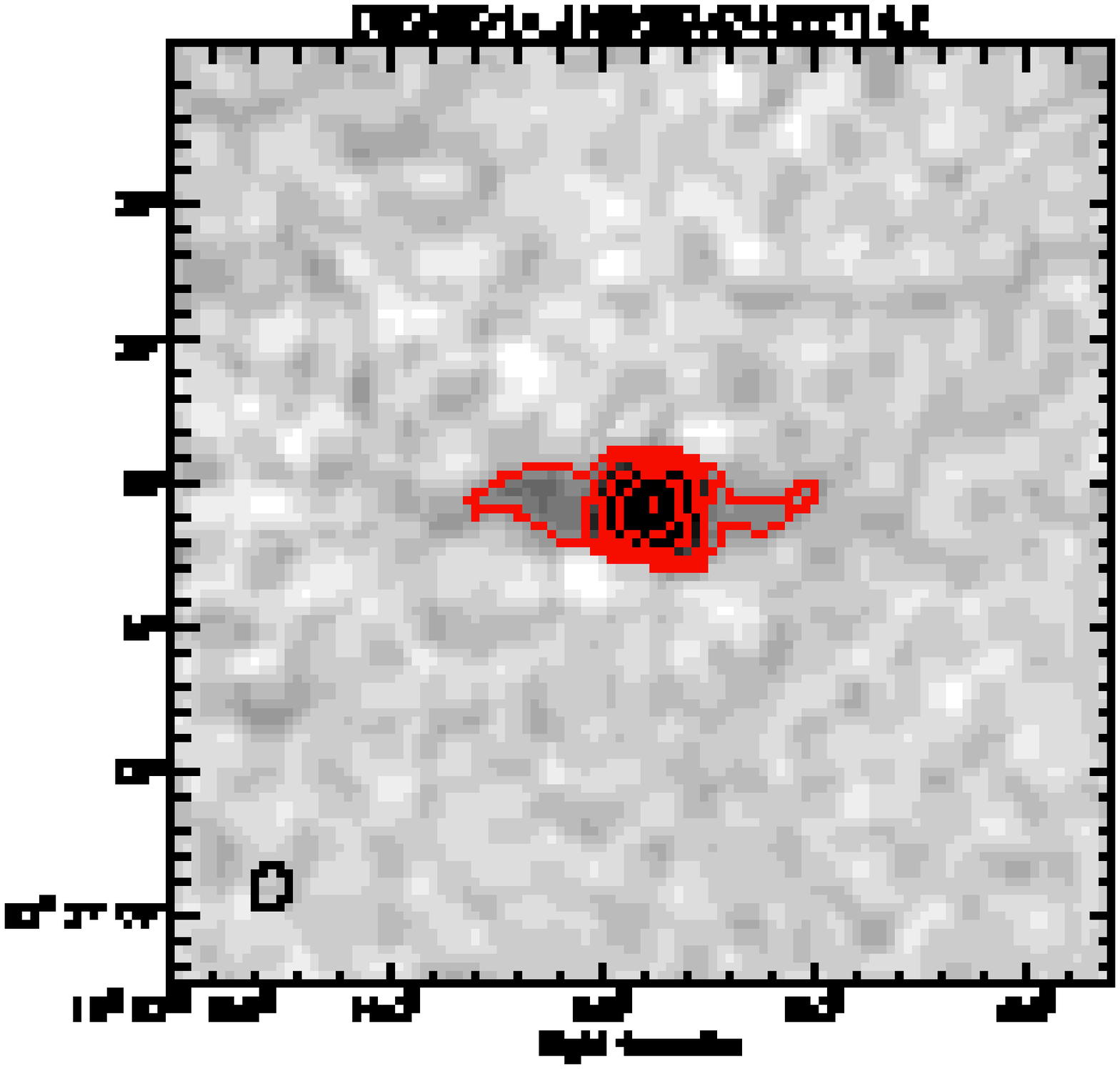}}

                    \caption{Radio sources fitted by multiple Gaussian
                      components and identified as a radio group (see
                      Tab. \ref{tab:multi}). The source name is
                      given at the top of the individual panels.  The
                      grey-scale is from -4$\sigma$ to 10$\sigma$ of
                      the local rms (Tab. \ref{tab:cat}).  The
                      contours start at 4$\sigma$ in steps of $2^n
                      \times \sigma$ with $n=2,3,4,5,\dots$. (The
                      local rms is listed in Tab.  \ref{tab:cat}.) The
                      beam is shown for reference in the bottom left
                      corner. 
\label{fig:multi}}
\end{figure}

For each of the 80 sources fitted with multiple components (see Fig.
\ref{fig:multi}) we list in the multiple source catalog (see Tab.
\ref{tab:multi}) (i) an entry for each of the components identified
with a trailing letter (A, B, C, {\ldots}) in the source name (from
Tab. \ref{tab:cat}), and (ii) an entry for the whole source as it is
listed in the source table (Tab. \ref{tab:cat}). In these cases the
total flux was calculated using the task 'TVSTAT', which allows the
integration of map values over irregular areas, and the sizes are the
largest angular sizes. For these sources the peak flux (at the listed
position) is undetermined and therefore set to a value of '-99.999'.

For each source we list the source name as well as its derived
properties and their uncertainties. All 3643 radio sources are listed
in right ascension order in Tab. \ref{tab:cat} with the following
columns\footnote{Due to bandwidth smearing effects the peak flux and,
  hence, the integrated flux for unresolved sources can be
  underestimated by up to (10-15)\%. An analysis of this will be
  presented in Bondi et al. (in prep.).}:
\\
\\
Column(1): Source name
\\
Column(2): Right ascension (J2000.0)
\\
Column(3): Declination (J2000.0)
\\
Column(4): rms uncertainty in right ascension
\\
Column(5): rms uncertainty in declination
\\
Column(3): Peak flux density and its rms uncertainty
\\
Column(4): Integrated flux density and its rms uncertainty
\\
Column(5): rms measured in the SExtractor noise map
\\
Column(9): Deconvolved source size -- major axis $\theta_{M,dec}$
\\
Column(10): Deconvolved source size -- minor axis $\theta_{m,dec}$
\\
Column(11): Deconvolved source -- position angle $\rm PA_{dec}$
(counterclockwise from North)
\\
Column(12): Flag for resolved (1) and unresolved (0) sources
\\
column(13): Flag for source with multiple (1) or single (0) components
\\
column(14): Flag for potentially spurious source due to sidelobe (1),
otherwise (0)

\begin{figure}[ht]
\plottwo{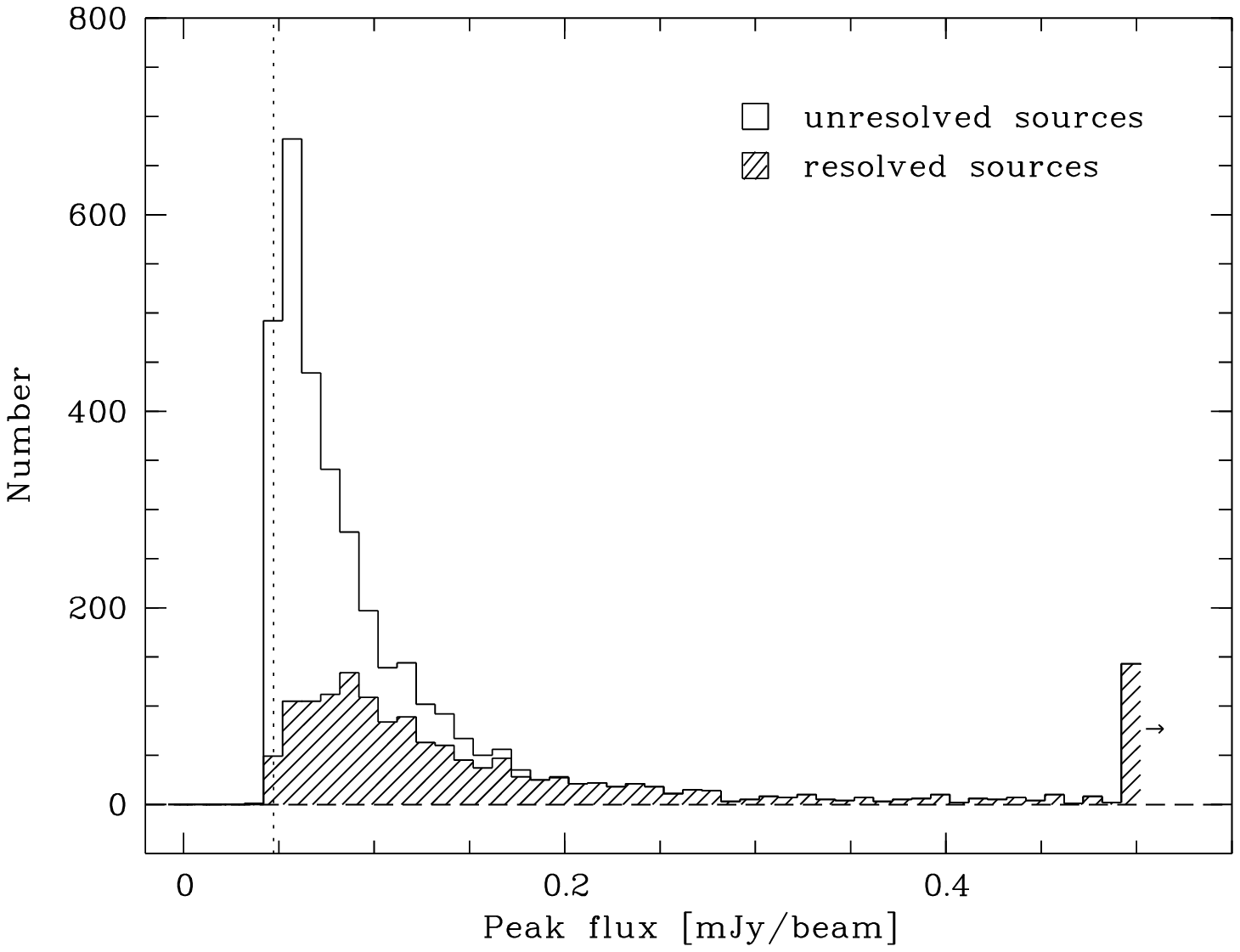}{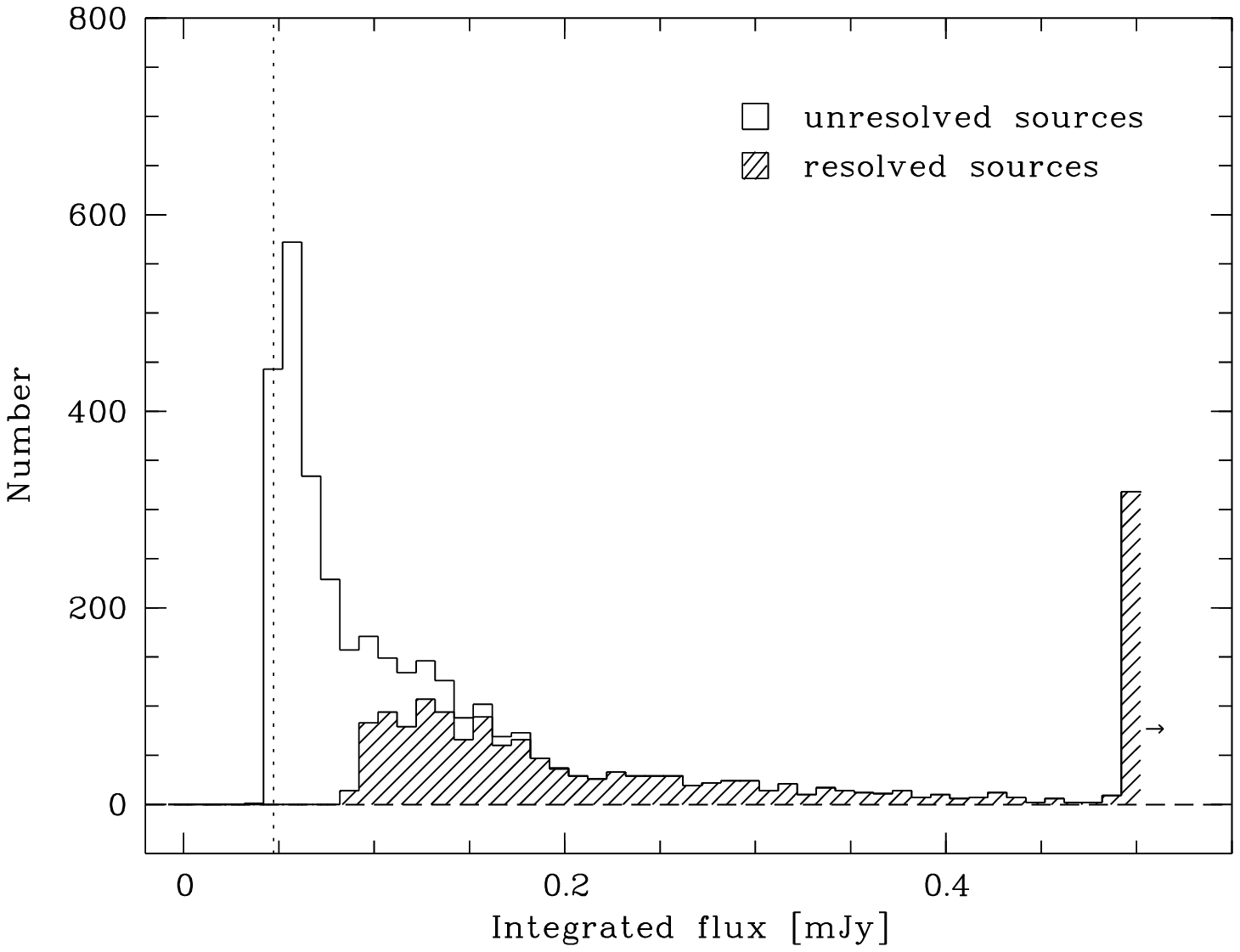}
\caption{Cumulative number distribution of the VLA-COSMOS sources as a
  function of peak ({\it left}) and integrated ({\it right}) flux
  density. The shaded area corresponds to sources that are resolved
  (see text).
\label{fig:cat}}
\end{figure}

The individual components contributing to our multi-component sources
are listed in Tab. \ref{tab:multi}. The columns are the same as for
Tab. \ref{tab:cat}. The (cumulative) peak and integrated flux
distribution of the sources in VLA-COSMOS large project are shown in
Fig. \ref{fig:cat}.

\subsection{Comparison to other Surveys}

We compared the catalog of the VLA-COSMOS large project with the
catalogs of the NVSS, FIRST and VLA-COSMOS pilot project. All three
surveys were also conducted at 1.4\,GHz, however the NVSS and FIRST
surveys used the D- and B-array, respectively \citep{con98,whi97}.

\begin{figure}[ht]
\includegraphics[scale=0.6,angle=-90]{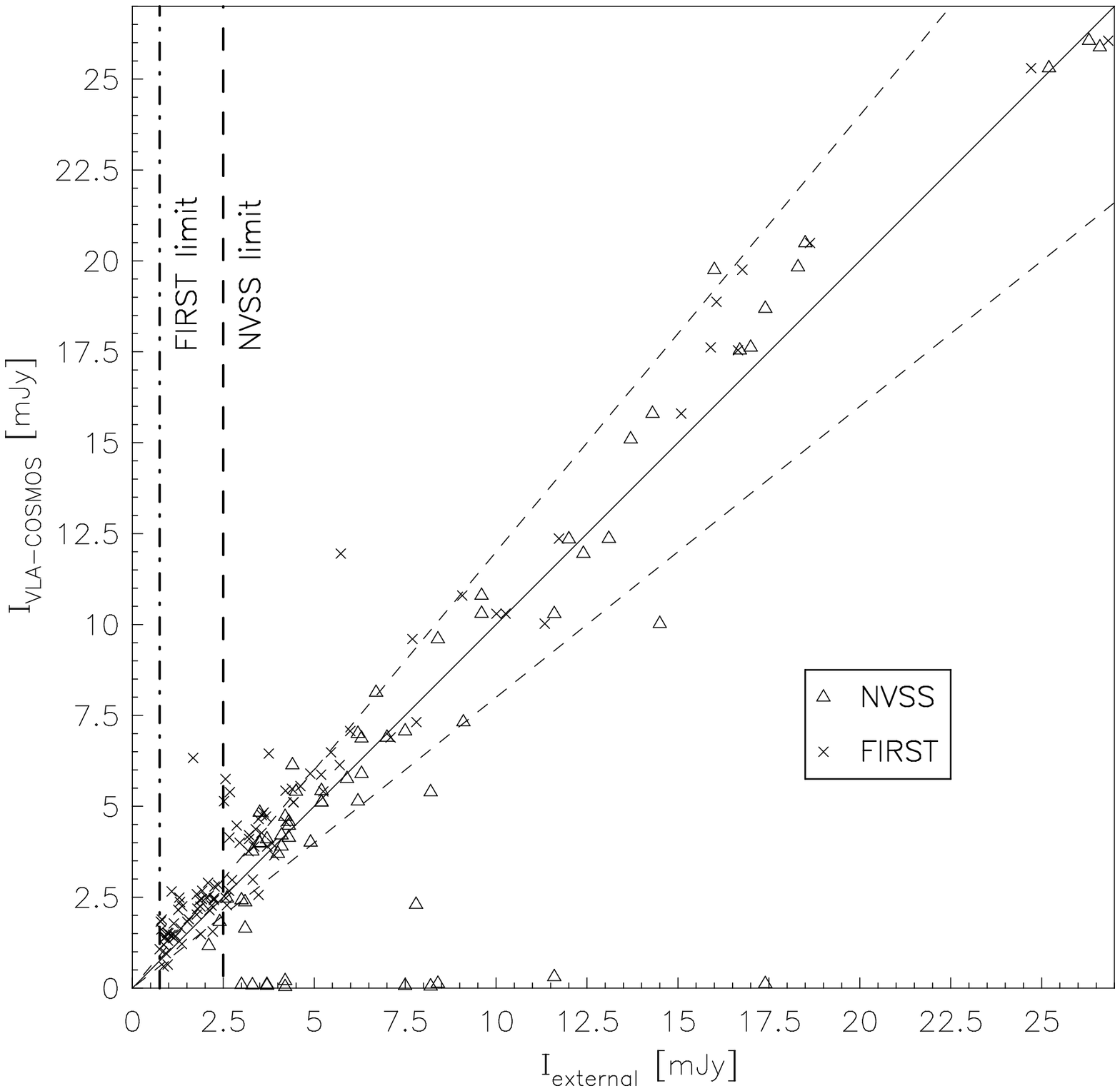}
\caption{Comparison of the derived integrated flux in the VLA-COSMOS
  large project $\rm I_{\rm VLA-COSMOS}$ and the NVSS and FIRST
  surveys $\rm I_{\rm external}$. The solid diagonal line represents a
  flux ratio of unity, while the dashed lines show the $\pm20$\%
  lines. The vertical lines denote the (5$\sigma$) detection limit of
  the NVSS and FIRST surveys \citep{con98,whi97}. The counterparts to
  VLA-COSMOS sources lie within radii of $5\as$ and $1\as$ for the
  NVSS and FIRST survey, respectively. The large discrepancies in the
  derived integrated flux for several NVSS sources is likely due to
  the large difference in resolution (NVSS: $45\as$ FWHM vs.
  VLA-COSMOS:$\sim\,1.5\as$ FWHM), while the discrepancies in the
  integrated flux for the FIRST sources are mainly due to the fact
  that these are part of multi-component VLA-COSMOS sources. 
\label{fig:nvss}}
\end{figure}

Within the area searched for the VLA-COSMOS large project, the NVSS
and FIRST catalogs list 119 and 184 sources, respectively. About 10\%
of the sources in these catalogs have no counterpart in the VLA-COSMOS
survey nor in the other survey, i.e. they are unique to the catalogs
of the NVSS or FIRST survey. Given the sensitivity of the VLA-COSMOS
survey this suggests that these sources are likely false
detections\footnote{The FIRST survey notes on their web-site ({\tt
    http://sundog.stsci.edu/}) that sidelobe flagging near the equator
  is not as reliable as for the northern part of the survey.} as it
seems unlikely that all of them are highly variable sources. We
cross-correlated the NVSS and FIRST catalogs with the catalog of the
VLA-COSMOS large project using a search radius of $5\as$ and $1\as$,
respectively. Figure \ref{fig:nvss} compares the integrated fluxes
derived for the individual sources. The agreement between the values
of the VLA-COSMOS and the NVSS/FIRST survey is fairly good, except for
a number of NVSS sources where our observations have probably resolved
out a large extended flux component. (Note that some of the VLA-COSMOS
multi-component sources consist of more than one FIRST source,
explaining most of the large discrepancies in Fig. \ref{fig:nvss}.)

For 30 sources from the VLA-COSMOS pilot project no counterpart is
present in our catalog of the large project. Given that the
sensitivity of the large project is at least a factor of 2.5 better,
these sources are likely false detections. Thus the fraction of false
detections is about 10\,\% in the pilot catalog. The signal-to-noise
ratio S/N of the sources is below 4.3$\sigma$ of the fitted peak flux
and its calculated error. (This roughly corresponds to a S/N of 5.5
and lower.) This is a factor of 2 more than expected from the
algorithm used which was set to a false detection rate of 5\,\%
\citep{sch04}. As all of the false detection are lying in areas with a
large gradient in the background (i.e.  overlap areas of the
individual pointings at the edge of the field), this strongly suggests
that the local rms was underestimated in these areas and that the used
mesh size of $47\as$ was too large in these areas. (For the large
project a mesh size of $17.5\as$ is used, see \S \ref{subsec:find}.)
We also compared the measured peak and integrated fluxes of both
VLA-COSMOS projects. For sources in the pilot project with significant
detection (S/$\delta$S $>$ 4.5) the measured peak (integrated) flux
agrees within 20\% for about 66\% (50\%) of the sources. However, the
flux measurements agree within the quoted errors for most sources.
The agreement in the integrated flux (also with the error) is lower
for very bright sources ($\rm \ge 1 mJy$). This is very likely due to
the fact that the large project data is more sensitive to low level
extended structure due to its higher sensitivity as well as the
shorter baselines from the C array observations.

\section{The VLA-COSMOS Survey in the COSMOS Context}\label{sec:cosmos}

All data obtained by the COSMOS collaboration will be made available
to the public via the COSMOS archive at IPAC/IRSA. The final reduced
and calibrated data of the VLA-COSMOS pilot project can already be
found there. For the large project of the VLA-COSMOS survey, the final
reduced and calibrated A+C 1.4\,GHz image covering the entire COSMOS
field as well as the source catalogs described here are available as well.

One unique aspect of the overall COSMOS survey is the large ongoing
spectroscopic effort \citep{lil06,imp06}. Given the fortunate timing
of observations, source lists from the VLA-COSMOS survey do provide
target lists for these spectroscopic surveys.  The Magellan-COSMOS
survey \citep{imp06} is targeting potential AGN candidates (from the
X-ray and radio surveys) down to an $\rm i_{AB} = 23.0\,mag$. Most
VLA-COSMOS sources with optical counterparts fulfilling this criteria
are being observed by this survey. At the time of writing, for
over 200 radio sources a spectral classification has already been
obtained, with an expected total of 500 sources \citep*{tru06}. In
addition, the zCOSMOS survey \citep{lil06} is including VLA-COSMOS
sources with optical counterparts down to $\rm B_{AB} = 25.0\,mag$ in
their target lists as compulsory targets.

Therefore, we expect that over 1,500 VLA-COSMOS sources will have
optical spectra, once the spectroscopic surveys are completed. These
spectra do not only provide very accurate redshifts, but also allow a
better classification of the nature of the host galaxy (AGN vs. star
formation). Thus the VLA-COSMOS survey will provide the largest sample
of radio sources with spectral information in the redshift range
$z>0.3$. For comparison, in the local universe, the largest samples of
radio sources with optical spectra are the combined 2dFGRS+NVSS with
757 sources \citep*{sad02} and the combined SDSS+FIRST with 5454
entries \citep*{ive02}. Together with the information available from
the other wavelengths covering the X-ray to mm regime, COSMOS will
provide a unique dataset for the study of the faint radio source
population.

 
\acknowledgments

The National Radio Astronomy Observatory (NRAO) is operated by
Associated Universities, Inc., under cooperative agreement with the
National Science Foundation. We would like to thank the NRAO for their
support during this project with special thanks to Barry Clark and
Joan Wrobel. For fruitful discussions we thank Frazer Owen, Jim
Condon, Bill Cotton, Andrew Hopkins and Jos\'{e} Afonso. We thank the
anonymous referee for constructive comments which helped improving the
paper. VS's visit to NRAO was supported by HST-GO-09822.31-A. CC
thanks the Max-Planck-Gesellschaft and the Humboldt-Stiftung for
support through the Max-Planck-Forschungspreis. CC acknowledges
support through NASA grant HST-GO-09822.33-A. KJ acknowledges support
by the German DFG under grant SCHI 536/3-1. The COSMOS Science meeting
in May 2005 was supported in part by the NSF through grant
OISE-0456439. The Digitized Sky Survey was produced at the Space
Telescope Science Institute under U.S. Government grant NAG W-2166.

{\it Facilities:} \facility{NRAO}

\clearpage 

\begin{deluxetable}{lccccl}
\tabletypesize{\scriptsize}
\tablecaption{Radio Surveys at 1.4\,GHz\label{tab:survey}}
\tablewidth{0pt}
\tablehead{
\colhead{Field} & 
\colhead{Area} &  
\colhead{rms} & 
\colhead{resolution} & 
\colhead{\# of objects} & 
\colhead{Reference} \\
 &
\colhead{[deg$^2$]} &  
\colhead{[$\mu$Jy/beam]} & 
\colhead{[$\as \times \as$]} &  & 
}
\startdata
COSMOS (large)  &     2 &   10.5 & 1.5$\times$1.4 &   3643 & this paper \\
COSMOS (pilot)  & 0.837 &   25 & 1.9$\times$1.6 &    246 & Schinnerer et al. 2004\\
HDFN   &  0.35 &  7.5 &    2.0$\times$1.8 &    314 & Richards 2000\\ 
SSA 13 &  0.32 &  4.8 &     1.8 &  810 & Fomalont et al. 2006\\ \hline
FIRST  &10,000 &  150 &       5 &    1,000,000 & Becker et al. 1995\\
FLS    &     5 &   23 &       5 &   3565 & Condon et al. 2003\\
VVDS   &     1 &   17 &       6 &   1054 & Bondi et al. 2003\\
ATHDFS &  0.35 &   11 & 7.1$\times$6.2 &    466 & Norris et al. 2005, Huynh et al. 2005\\
ATESP  &    26 &   79 &    14$\times$8 &   2960 & Prandoni et al. 2001\\ \hline
PDS    &  4.56 &   12 &    12$\times$6 &   2090 & Hopkins et al. 2003\\
ELAIS\tablenotemark{a} &  4.22 &   27 &      15 &    867 & Ciliegi et al. 1999 \\
Lockman&  0.35 &  120 &      15 &    149 & de Ruiter et al. 1997\\ \hline
NVSS   &34,000 &  350 &      45 &  1,700,000 & Condon et al. 1998\\
\enddata
\tablenotetext{a}{consists of 3 fields of the ELAIS survey: N1, N2, and N3}
\end{deluxetable}

\begin{deluxetable}{lrrl}
\tablecaption{VLA Pointing Centers\label{tab:pos}}
\tablewidth{0pt}
\tablehead{
\colhead{Pointing \#} & 
\colhead{R.A. (J2000)} & 
\colhead{DEC (J2000)} & 
\colhead{Remark}}
\startdata
F01 & 10:02:28.67 & +02:38:19.84 & \\
F02 & 10:01:28.64 & +02:38:19.84 & \\
F03 & 10:00:28.60 & +02:38:19.84 & \\
F04 & 09:59:28.56 & +02:38:19.84 & \\
F05 & 09:58:28.52 & +02:38:19.84 & \\
F06 & 10:01:58.66 & +02:25:20.42 & \\
F07 & 10:00:58.62 & +02:25:20.42 & P1 in pilot project\\
F08 & 09:59:58.58 & +02:25:20.42 & P2 in pilot project\\
F09 & 09:58:58.54 & +02:25:20.42 & \\
F10 & 10:02:28.67 & +02:12:21.00 & \\
F11 & 10:01:28.64 & +02:12:21.00 & P3 in pilot project\\
F12\tablenotemark{a} & 10:00:28.60 & +02:12:21.00 & P4 in pilot project\\
F13 & 09:59:28.56 & +02:12:21.00 & P5 in pilot project\\
F14 & 09:58:28.62 & +02:12:21.00 & \\
F15 & 10:01:58.66 & +01:59:21.58 & \\
F16 & 10:00:58.62 & +01:59:21.58 & P6 in pilot project\\
F17 & 09:59:58.58 & +01:59:21.58 & P7 in pilot project\\
F18 & 09:58:58.54 & +01:59:21.58 & \\
F19 & 10:02:28.67 & +01:46:22.24 & \\
F20 & 10:01:28.64 & +01:46:22.24 & \\
F21 & 10:00:28.60 & +01:46:22.24 & \\
F22 & 09:59:28.56 & +01:46:22.24 & \\
F23 & 09:58:28.52 & +01:46:22.24 & \\
\enddata
\tablenotetext{a}{COSMOS field center}
\tablecomments{Pointing centers for the VLA-COSMOS large project at 1.4\,GHz.}
\end{deluxetable}

\begin{deluxetable}{lrrccrrcrrrccc}
 \tabletypesize{\scriptsize}
\rotate
\tablecaption{1.4\,GHz Source Catalog of the VLA-COSMOS Large Project (abridged)\label{tab:cat}}
\tablewidth{0pt}
\setlength{\tabcolsep}{0.02in}
\tablehead{
\colhead{Name} &
\colhead{R.A.} & 
\colhead{Dec.} & 
\colhead{$\rm \sigma_{R.A.}$} &
\colhead{$\rm \sigma_{Dec.}$} &
\colhead{$\rm S_{peak}$} & 
\colhead{$\rm S_{total}$} &
\colhead{rms} & 
\colhead{$\rm \theta_{M,dec}$} &
\colhead{$\rm \theta_{m,dec}$} &
\colhead{$\rm PA_{dec}$} & 
&\colhead{Flags}&
\\
& \colhead{(J2000.0)} & \colhead{(J2000.0)} & 
\colhead{[$\as$]} & \colhead{[$\as$]} &
\colhead{[mJy/beam]} & 
\colhead{[mJy]} &  
\colhead{[mJy/beam]} &
\colhead{[$^{\prime\prime}$]} & \colhead{[$^{\prime\prime}$]} &
\colhead{[$^o$]} & 
\colhead{res\tablenotemark{a}}&
\colhead{slob\tablenotemark{b}}&
\colhead{mult\tablenotemark{c}}
}
\startdata
COSMOSVLA\_J095738.80+024203.2 & 09 57 38.800 & +02 42 03.19 & 0.19 & 0.19 &  0.112 $\pm$ 0.024 &  0.112 $\pm$ 0.024 & 0.024 & 0.00 & 0.00 &  0.0 & 0 & 0 & 0\\
COSMOSVLA\_J095738.97+021630.3 & 09 57 38.972 & +02 16 30.32 & 0.19 & 0.19 &  0.112 $\pm$ 0.025 &  0.112 $\pm$ 0.025 & 0.025 & 0.00 & 0.00 &  0.0 & 0 & 0 & 0\\
COSMOSVLA\_J095739.10+021503.1 & 09 57 39.097 & +02 15 03.05 & 0.19 & 0.19 &  0.119 $\pm$ 0.024 &  0.129 $\pm$ 0.024 & 0.024 & 0.00 & 0.00 &  0.0 & 0 & 0 & 0\\
COSMOSVLA\_J095739.23+024539.0 & 09 57 39.229 & +02 45 39.02 & 0.19 & 0.19 &  0.126 $\pm$ 0.028 &  0.126 $\pm$ 0.028 & 0.028 & 0.00 & 0.00 &  0.0 & 0 & 0 & 0\\
COSMOSVLA\_J095739.39+023655.5 & 09 57 39.390 & +02 36 55.47 & 0.19 & 0.19 &  0.111 $\pm$ 0.024 &  0.111 $\pm$ 0.024 & 0.024 & 0.00 & 0.00 &  0.0 & 0 & 0 & 0\\
COSMOSVLA\_J095739.44+021850.9 & 09 57 39.441 & +02 18 50.87 & 0.18 & 0.18 &  0.133 $\pm$ 0.027 &  0.133 $\pm$ 0.027 & 0.027 & 0.00 & 0.00 &  0.0 & 0 & 0 & 0\\
COSMOSVLA\_J095739.71+023103.5 & 09 57 39.712 & +02 31 03.53 & 0.13 & 0.13 &  0.124 $\pm$ 0.027 &  0.124 $\pm$ 0.027 & 0.027 & 0.00 & 0.00 &  0.0 & 0 & 0 & 0\\
COSMOSVLA\_J095739.81+013653.4 & 09 57 39.814 & +01 36 53.40 & 0.17 & 0.17 &  0.156 $\pm$ 0.030 &  0.156 $\pm$ 0.030 & 0.030 & 0.00 & 0.00 &  0.0 & 0 & 0 & 0\\
COSMOSVLA\_J095740.60+020145.1 & 09 57 40.602 & +02 01 45.13 & 0.20 & 0.19 &  0.225 $\pm$ 0.035 &  0.377 $\pm$ 0.105 & 0.035 & 2.52 & 0.00 & 55.8 & 1 & 0 & 0\\
COSMOSVLA\_J095740.99+024921.1 & 09 57 40.986 & +02 49 21.13 & 0.18 & 0.18 &  0.154 $\pm$ 0.034 &  0.154 $\pm$ 0.034 & 0.034 & 0.00 & 0.00 &  0.0 & 0 & 0 & 0\\
COSMOSVLA\_J095741.11+015122.6 & 09 57 41.107 & +01 51 22.58 & 0.13 & 0.14 &-99.990 $\pm$-99.990 & 45.620 $\pm$ -99.990 & 0.024 &53.00 & 9.00 &  0.0 & 1 & 0 & 1\\
COSMOSVLA\_J095741.25+024346.2 & 09 57 41.250 & +02 43 46.20 & 0.19 & 0.19 &  0.123 $\pm$ 0.025 &  0.123 $\pm$ 0.025 & 0.025 & 0.00 & 0.00 &  0.0 & 0 & 0 & 0\\
COSMOSVLA\_J095741.34+020346.1 & 09 57 41.338 & +02 03 46.13 & 0.22 & 0.22 &  0.152 $\pm$ 0.031 &  0.152 $\pm$ 0.031 & 0.031 & 0.00 & 0.00 &  0.0 & 0 & 0 & 0\\
COSMOSVLA\_J095741.52+023841.2 & 09 57 41.525 & +02 38 41.21 & 0.18 & 0.17 &  0.116 $\pm$ 0.023 &  0.116 $\pm$ 0.023 & 0.023 & 0.00 & 0.00 &  0.0 & 0 & 0 & 0\\
COSMOSVLA\_J095741.74+025004.0 & 09 57 41.737 & +02 50 03.96 & 0.19 & 0.19 &  0.160 $\pm$ 0.034 &  0.160 $\pm$ 0.034 & 0.034 & 0.00 & 0.00 &  0.0 & 0 & 0 & 0\\
COSMOSVLA\_J095741.89+020426.4 & 09 57 41.895 & +02 04 26.42 & 0.17 & 0.17 &  0.181 $\pm$ 0.031 &  0.181 $\pm$ 0.031 & 0.031 & 0.00 & 0.00 &  0.0 & 0 & 0 & 0\\
COSMOSVLA\_J095742.30+020426.1 & 09 57 42.305 & +02 04 26.07 & 0.13 & 0.13 & 11.371 $\pm$ 0.031 & 20.492 $\pm$ 0.228 & 0.031 & 1.88 & 0.35 & 57.1 & 1 & 0 & 0\\
COSMOSVLA\_J095742.61+022827.8 & 09 57 42.612 & +02 28 27.81 & 0.20 & 0.19 &  0.133 $\pm$ 0.029 &  0.133 $\pm$ 0.029 & 0.029 & 0.00 & 0.00 &  0.0 & 0 & 0 & 0\\
COSMOSVLA\_J095742.71+024540.4 & 09 57 42.711 & +02 45 40.41 & 0.17 & 0.17 &  0.134 $\pm$ 0.026 &  0.134 $\pm$ 0.026 & 0.026 & 0.00 & 0.00 &  0.0 & 0 & 0 & 0\\
COSMOSVLA\_J095743.04+015650.8 & 09 57 43.044 & +01 56 50.82 & 0.15 & 0.15 &  0.425 $\pm$ 0.030 &  0.747 $\pm$ 0.098 & 0.030 & 2.11 & 0.20 &129.1 & 1 & 0 & 0\\
COSMOSVLA\_J095743.23+013851.0 & 09 57 43.228 & +01 38 51.05 & 0.17 & 0.17 &  0.139 $\pm$ 0.025 &  0.139 $\pm$ 0.025 & 0.025 & 0.00 & 0.00 &  0.0 & 0 & 0 & 0\\
COSMOSVLA\_J095743.40+015620.7 & 09 57 43.400 & +01 56 20.72 & 0.34 & 0.19 &  0.183 $\pm$ 0.030 &  0.289 $\pm$ 0.102 & 0.030 & 2.64 & 0.30 & 73.2 & 1 & 0 & 0\\
COSMOSVLA\_J095743.73+014132.5 & 09 57 43.729 & +01 41 32.47 & 0.18 & 0.17 &  0.121 $\pm$ 0.022 &  0.121 $\pm$ 0.022 & 0.022 & 0.00 & 0.00 &  0.0 & 0 & 0 & 0\\
COSMOSVLA\_J095743.87+023038.5 & 09 57 43.872 & +02 30 38.52 & 0.15 & 0.14 &  0.412 $\pm$ 0.026 &  0.727 $\pm$ 0.084 & 0.026 & 1.98 & 0.33 & 57.7 & 1 & 0 & 0\\
\enddata

\tablenotetext{a}{Flag if source is -- according to Fig. \ref{fig:psf} -- resolved (1) or unresolved (0)}
\tablenotetext{b}{Flag if source is potentially spurious due to sidelobe bump (1) or not (0)}
\tablenotetext{c}{Flag if source consists of multiple components (1) or a single component (0)}

\tablecomments{Catalog of radio sources at 1.4\,GHz detected in the
  COSMOS field with a S/N$\ge$4.5 in the VLA-COSMOS large project data
  (see \S \ref{sec:cat}). Radio sources with multiple Gaussian fits
  are flagged ('mult=1'), their multiple components are listed
  separately in Tab. \ref{tab:multi}. The table is available in its
  entirety via the link to a machine-readable version above and/or via
  the COSMOS archive at
  IPAC/IRSA\footnote{http://www.irsa.ipac.edu/data/COSMOS/tables/}. A
  portion is shown here for guidance regarding its form and content. }

\end{deluxetable}

\begin{deluxetable}{lrrccrrcrrrccc}
 \tabletypesize{\scriptsize}
\rotate
\tablecaption{Multi-components of sources in the VLA-COSMOS catalog (abridged)\label{tab:multi}}
\tablewidth{0pt}
\setlength{\tabcolsep}{0.03in}
\tablehead{
\colhead{Name} &
\colhead{R.A.} & 
\colhead{Dec.} & 
\colhead{$\rm \sigma_{R.A.}$} &
\colhead{$\rm \sigma_{Dec.}$} &
\colhead{$\rm S_{peak}$} & 
\colhead{$\rm S_{total}$} &
\colhead{rms} & 
\colhead{$\rm \theta_{M}$} &
\colhead{$\rm \theta_{m}$} &
\colhead{PA} & 
&\colhead{Flags}&
\\
& \colhead{(J2000.0)} & \colhead{(J2000.0)} & 
\colhead{[$\as$]} & \colhead{[$\as$]} &
\colhead{[mJy/beam]} & 
\colhead{[mJy]} &  
\colhead{[mJy/beam]} &
\colhead{[$^{\prime\prime}$]} & \colhead{[$^{\prime\prime}$]} &
\colhead{[$^o$]} & 
\colhead{res\tablenotemark{a}}&
\colhead{slob\tablenotemark{b}}&
\colhead{mult\tablenotemark{c}}
}
\startdata
COSMOSVLA\_J095741.11+015122.6A & 09 57 39.708 & +01 51 41.59 & 0.13 & 0.13 &  1.971 $\pm$ 0.026 &  8.612 $\pm$ 0.202 & 0.026 &  3.36 & 2.56 & 124.8 & 1 & 0 & 0 \\ 
COSMOSVLA\_J095741.11+015122.6B & 09 57 39.858 & +01 51 43.67 & 0.13 & 0.13 &  1.463 $\pm$ 0.026 &  4.289 $\pm$ 0.151 & 0.026 &  2.64 & 1.85 &  84.7 & 1 & 0 & 0 \\
COSMOSVLA\_J095741.11+015122.6C & 09 57 40.100 & +01 51 38.36 & 0.24 & 0.24 &  0.227 $\pm$ 0.026 & 10.694 $\pm$ 1.260 & 0.026 & 13.17 & 7.03 & 133.9 & 1 & 0 & 0 \\
COSMOSVLA\_J095741.11+015122.6D & 09 57 41.107 & +01 51 22.58 & 0.14 & 0.13 &  0.497 $\pm$ 0.025 &  0.754 $\pm$ 0.069 & 0.025 &  1.62 & 0.50 & 114.2 & 1 & 0 & 0 \\
COSMOSVLA\_J095741.11+015122.6E & 09 57 41.686 & +01 51 11.30 & 0.22 & 0.21 &  0.314 $\pm$ 0.024 &  8.037 $\pm$ 0.820 & 0.024 & 11.14 & 5.54 & 130.0 & 1 & 0 & 0 \\
COSMOSVLA\_J095741.11+015122.6F & 09 57 42.166 & +01 51 03.17 & 0.13 & 0.13 &  2.227 $\pm$ 0.024 & 12.488 $\pm$ 0.229 & 0.024 &  3.73 & 2.49 & 134.7 & 1 & 0 & 0 \\
COSMOSVLA\_J095741.11+015122.6  & 09 57 41.107 & +01 51 22.58 & 0.13 & 0.14 &-99.990 $\pm$-99.990 & 45.620 $\pm$-99.990 & 0.024 & 53.00 & 9.00 &   0.0 & 1 & 0 & 1 \\
COSMOSVLA\_J095755.84+015804.2A & 09 57 55.792 & +01 58 05.76 & 0.13 & 0.14 &  0.791 $\pm$ 0.022 &  3.370 $\pm$ 0.155 & 0.022 &  3.51 & 2.07 & 150.5 & 1 & 0 & 0 \\
COSMOSVLA\_J095755.84+015804.2B & 09 57 55.847 & +01 58 01.95 & 0.17 & 0.14 &  0.501 $\pm$ 0.022 &  1.657 $\pm$ 0.151 & 0.022 &  3.85 & 1.86 & 108.6 & 1 & 0 & 0 \\
COSMOSVLA\_J095755.84+015804.2C & 09 57 55.898 & +01 58 04.18 & 0.18 & 0.23 &  0.531 $\pm$ 0.022 &  1.714 $\pm$ 0.214 & 0.022 &  5.89 & 2.06 &  31.4 & 1 & 0 & 0 \\
COSMOSVLA\_J095755.84+015804.2  & 09 57 55.840 & +01 58 04.24 & 0.13 & 0.16 &-99.990 $\pm$-99.990 &  6.450 $\pm$-99.990 & 0.022 & 21.96 & 6.86 &   0.0 & 1 & 0 & 1 \\
COSMOSVLA\_J095756.45+025155.6A & 09 57 56.418 & +02 51 56.26 & 0.34 & 0.25 &  0.170 $\pm$ 0.031 &  0.302 $\pm$ 0.111 & 0.031 &  2.84 & 0.54 & 122.4 & 1 & 0 & 0 \\
COSMOSVLA\_J095756.45+025155.6B & 09 57 56.484 & +02 51 54.91 & 0.19 & 0.18 &  0.167 $\pm$ 0.031 &  0.167 $\pm$ 0.031 & 0.031 &  0.00 & 0.00 &   0.0 & 0 & 0 & 0 \\
COSMOSVLA\_J095756.45+025155.6 & 09 57 56.451 & +02 51 55.59 & 0.20 & 0.43 &-99.990 $\pm$-99.990 &  0.300 $\pm$-99.990 & 0.031 &  3.75 & 1.43 &   0.0 & 1 & 0 & 1 \\
COSMOSVLA\_J095800.80+015857.2A & 09 58 00.619 & +01 58 53.03 & 0.18 & 0.17 &  0.348 $\pm$ 0.019 &  3.684 $\pm$ 0.303 & 0.019 &  7.16 & 2.59 &  51.5 & 1 & 0 & 0 \\
COSMOSVLA\_J095800.80+015857.2B & 09 58 00.798 & +01 58 57.15 & 0.13 & 0.13 &  7.204 $\pm$ 0.019 & 16.624 $\pm$ 0.183 & 0.019 &  1.89 & 1.58 & 156.5 & 1 & 0 & 0 \\
COSMOSVLA\_J095800.80+015857.2  & 09 58 00.798 & +01 58 57.15 & 0.13 & 0.13 &-99.990 $\pm$-99.990 & 18.875 $\pm$-99.990 & 0.019 & 10.00 & 3.00 &   0.0 & 1 & 0 & 1 \\
COSMOSVLA\_J095815.51+014923.7A & 09 58 15.502 & +01 49 24.61 & 0.16 & 0.23 &  0.145 $\pm$ 0.014 &  0.496 $\pm$ 0.083 & 0.014 &  3.49 & 1.62 &   5.7 & 1 & 0 & 0 \\
COSMOSVLA\_J095815.51+014923.7B & 09 58 15.520 & +01 49 22.18 & 0.20 & 0.34 &  0.080 $\pm$ 0.014 &  0.080 $\pm$ 0.014 & 0.014 &  0.00 & 0.00 &   0.0 & 0 & 0 & 0 \\
COSMOSVLA\_J095815.51+014923.7  & 09 58 15.509 & +01 49 23.75 & 0.15 & 0.24 &-99.990 $\pm$-99.990 &  0.500 $\pm$-99.990 & 0.014 &  3.75 & 1.43 &   0.0 & 1 & 0 & 1 \\
\enddata

\tablenotetext{a}{Flag if component is -- according to Fig. \ref{fig:psf} -- resolved (1) or unresolved (0)}
\tablenotetext{b}{Flag if component is potentially spurious due to sidelobe bump (1) or not (0)}
\tablenotetext{c}{Flag if source consists of multiple components (1) or one of its single components (0)}

\tablecomments{ List of individual components that made up the 80
  radio sources that were fitted by multiple Gaussian. These
  multi-component sources are flagged in Tab. \ref{tab:cat} by a
  'mult=1'.  The table is available in its entirety via the link to a
  machine-readable version above and/or via the COSMOS archive at
  IPAC/IRSA. A portion is shown here for guidance regarding its form
  and content. }

\end{deluxetable}

\clearpage
 
\clearpage
\begin{center}
\resizebox{.9\hsize}{!}{\includegraphics[]{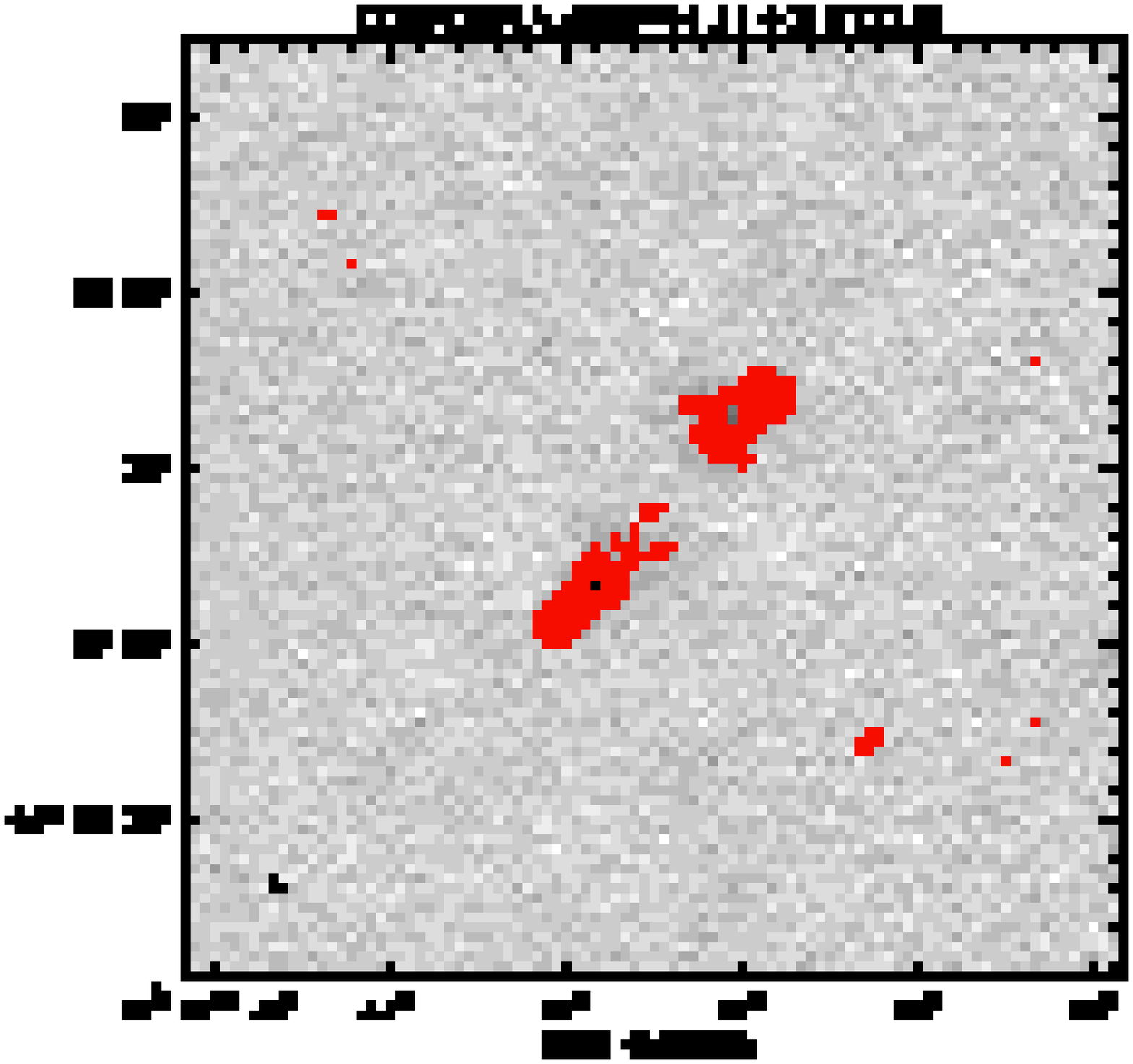}
                      \includegraphics[]{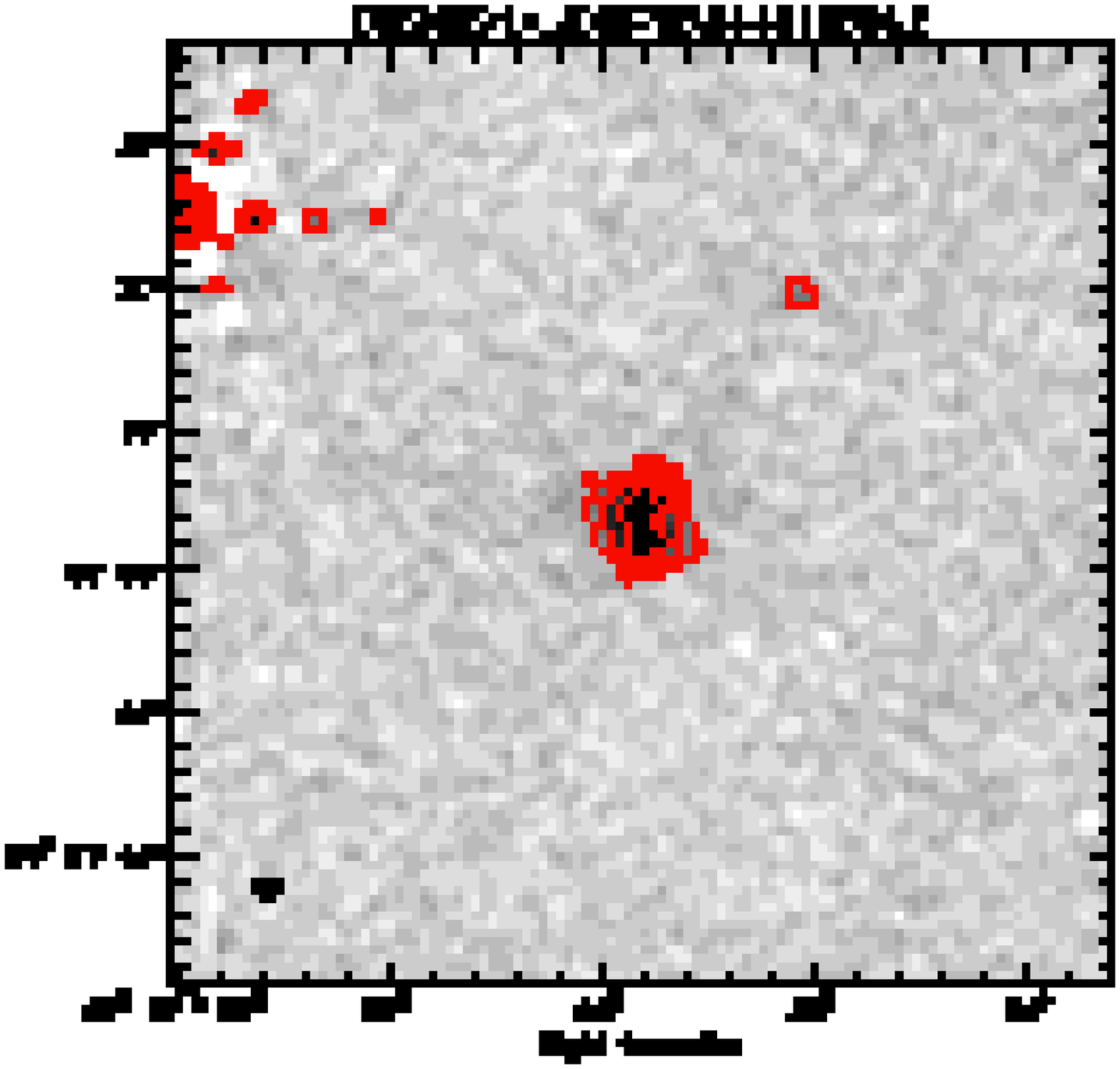}
                      \includegraphics[]{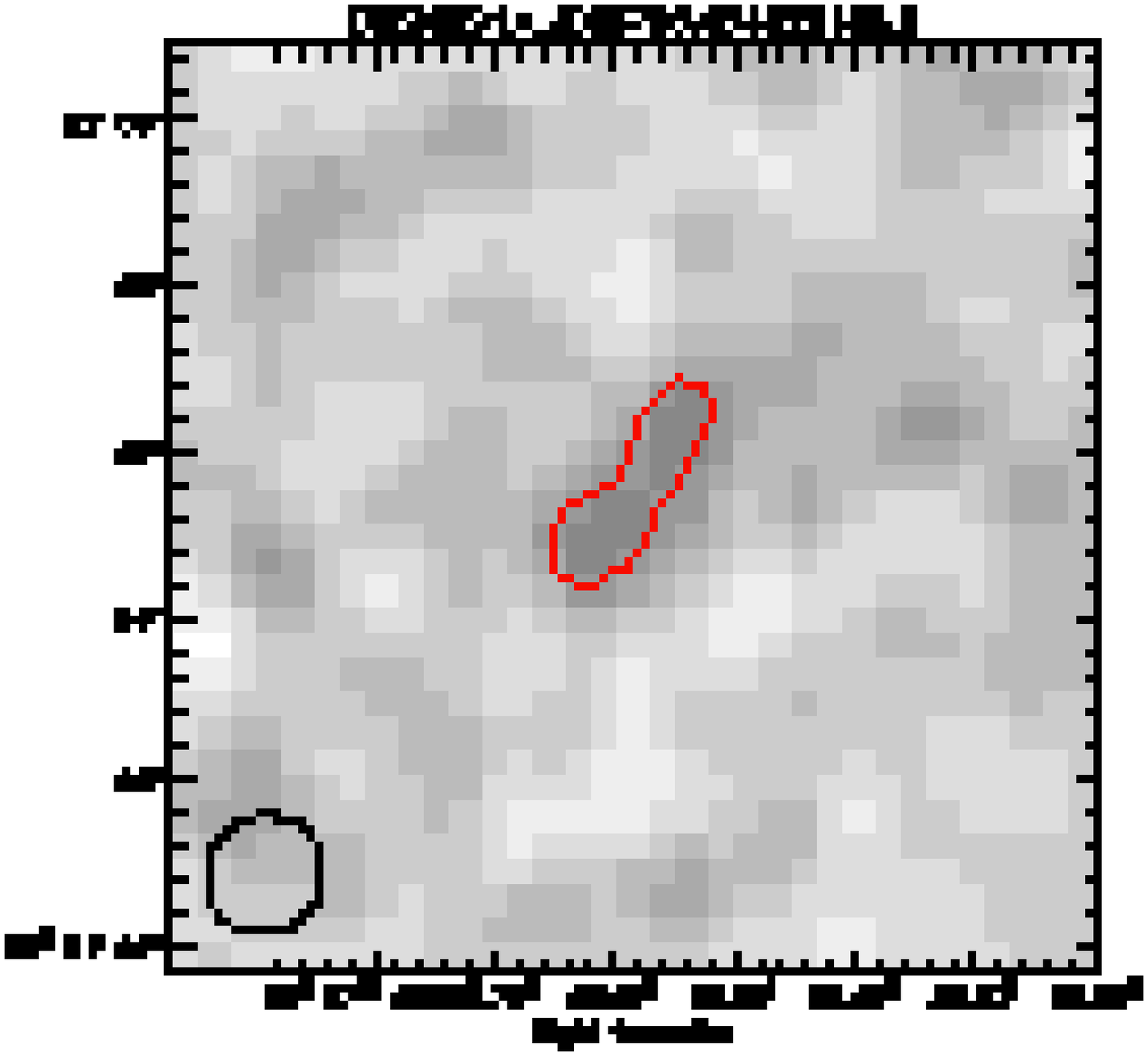}}\\
\resizebox{.9\hsize}{!}{\includegraphics[]{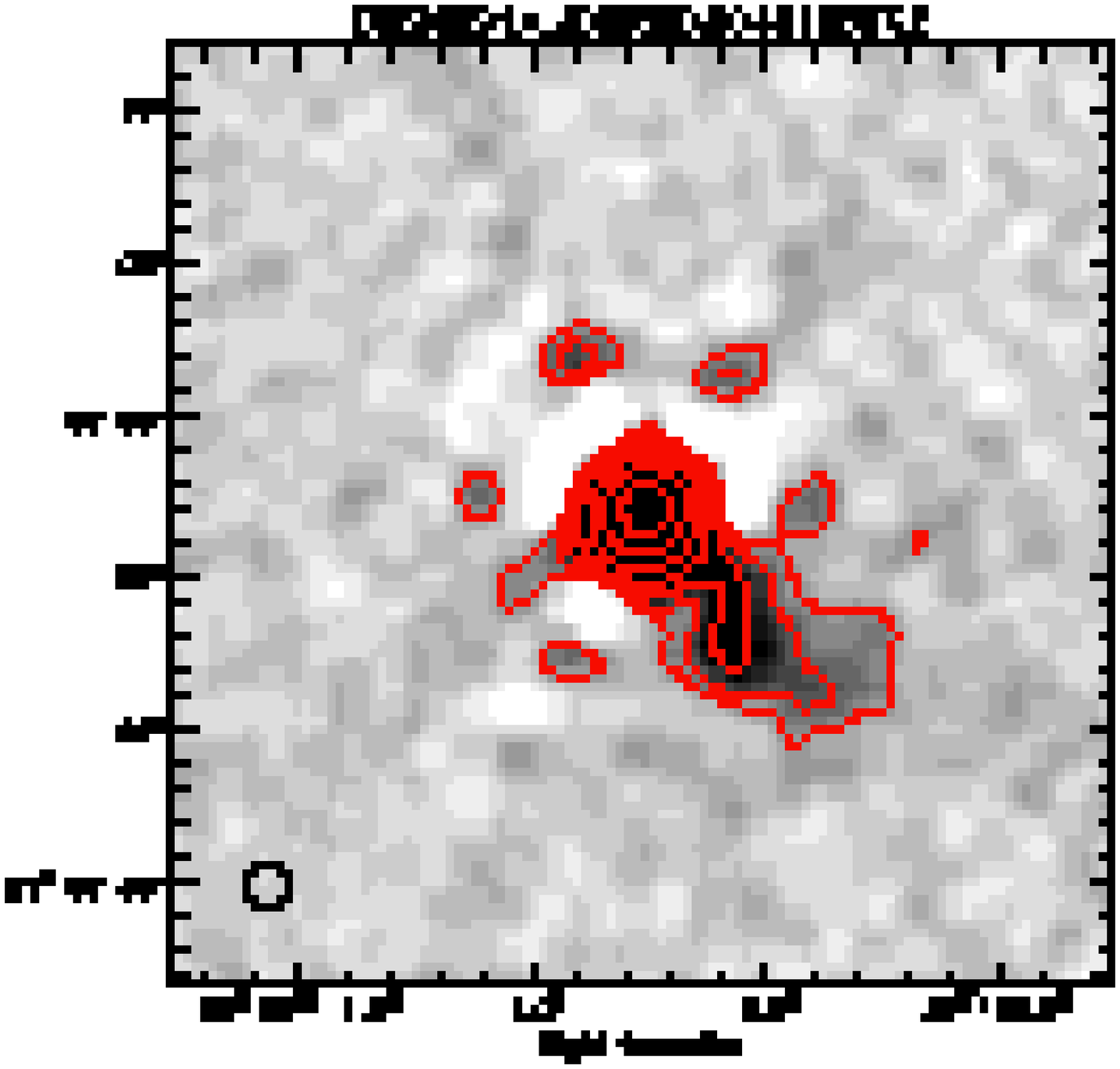}
                      \includegraphics[]{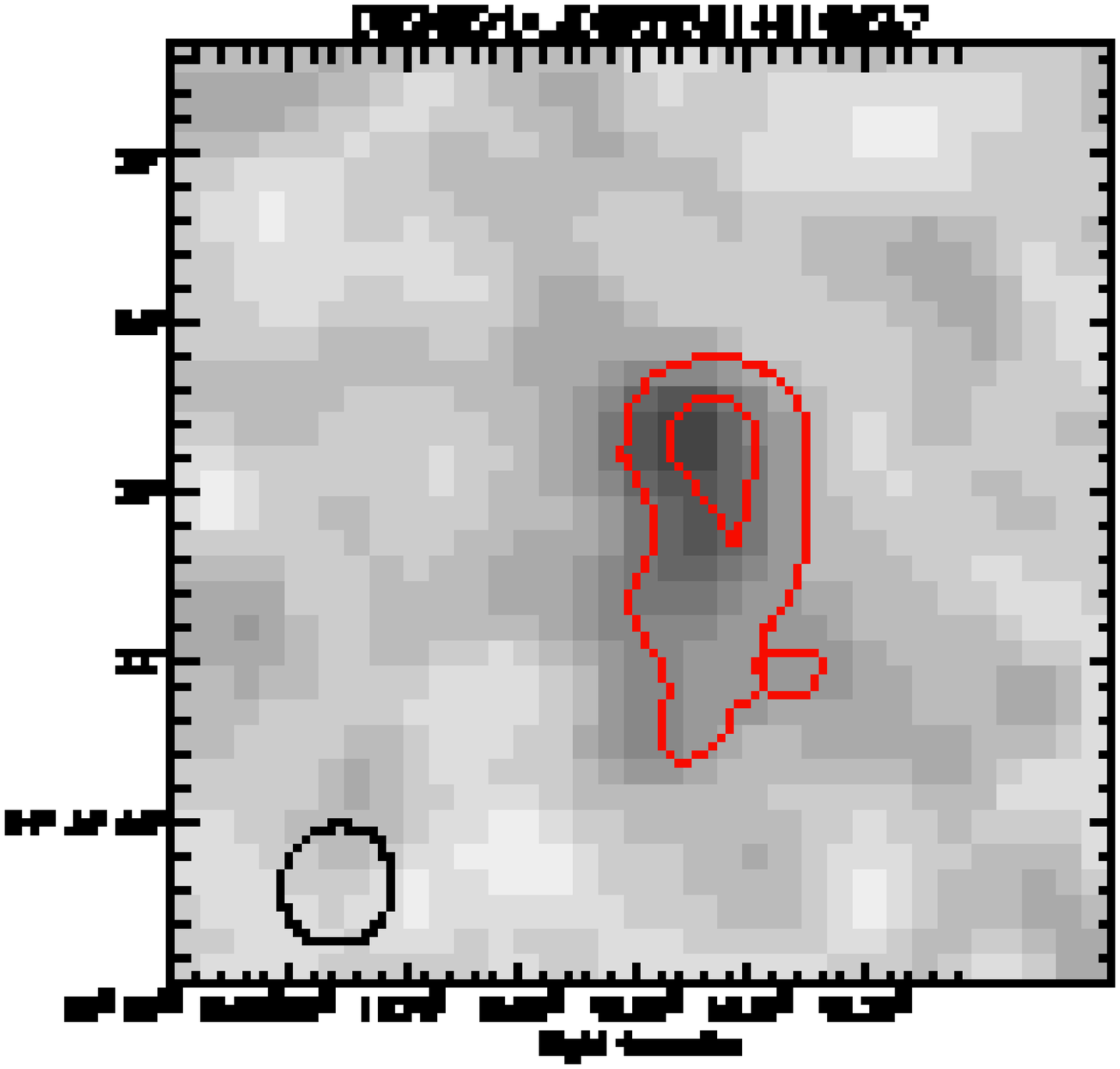}
                      \includegraphics[]{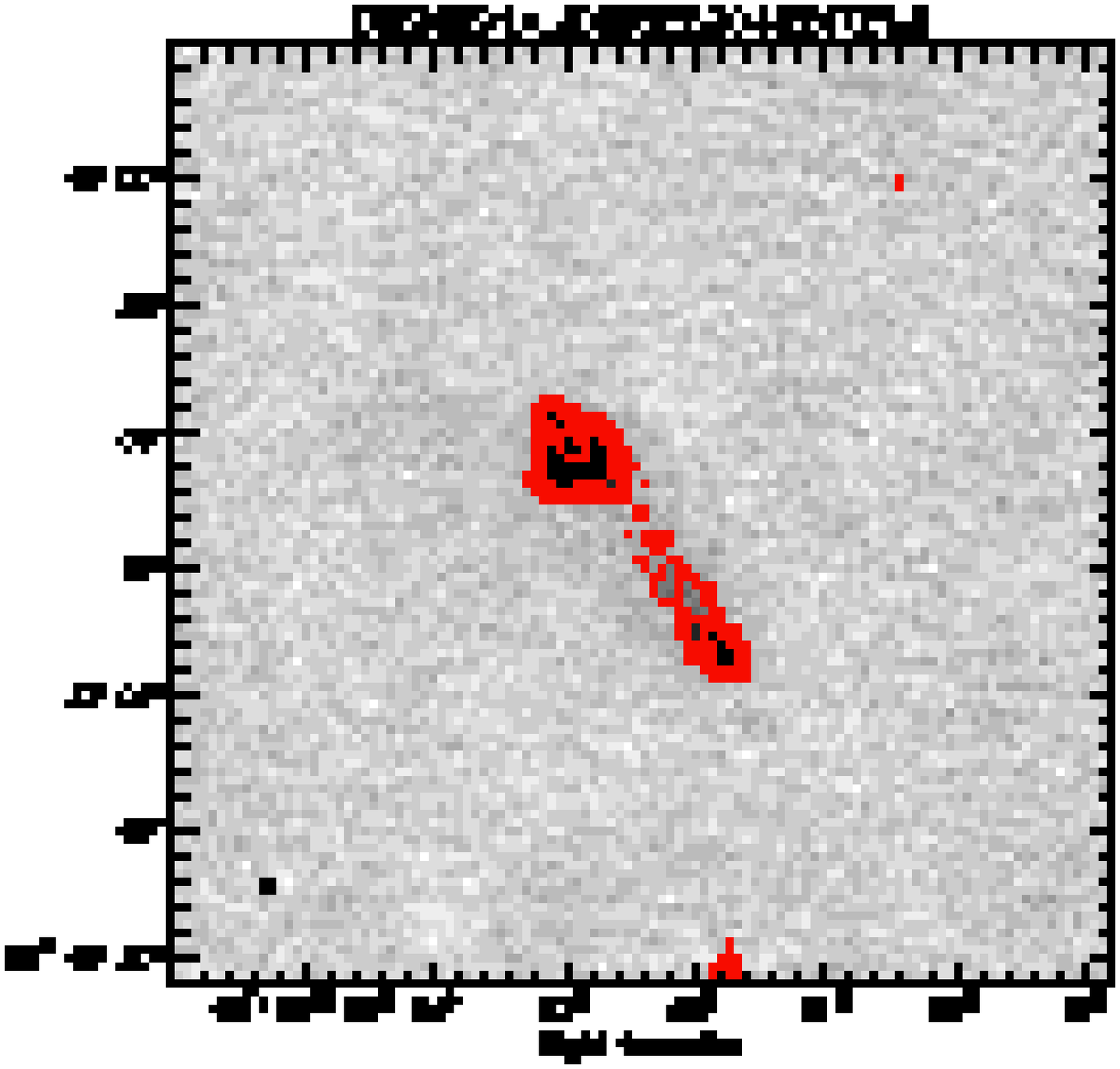}}\\
\resizebox{.9\hsize}{!}{\includegraphics[]{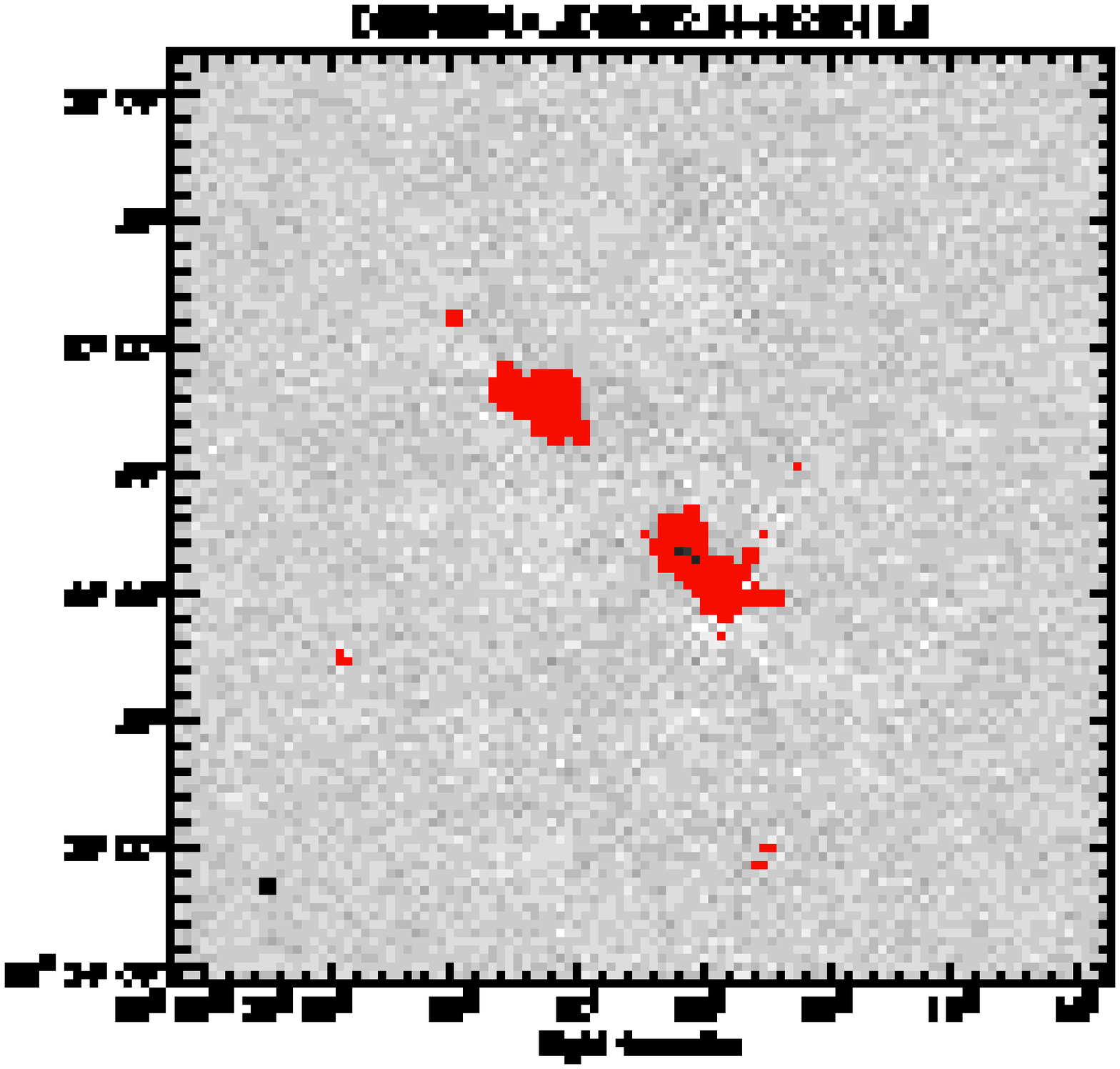}
                      \includegraphics[]{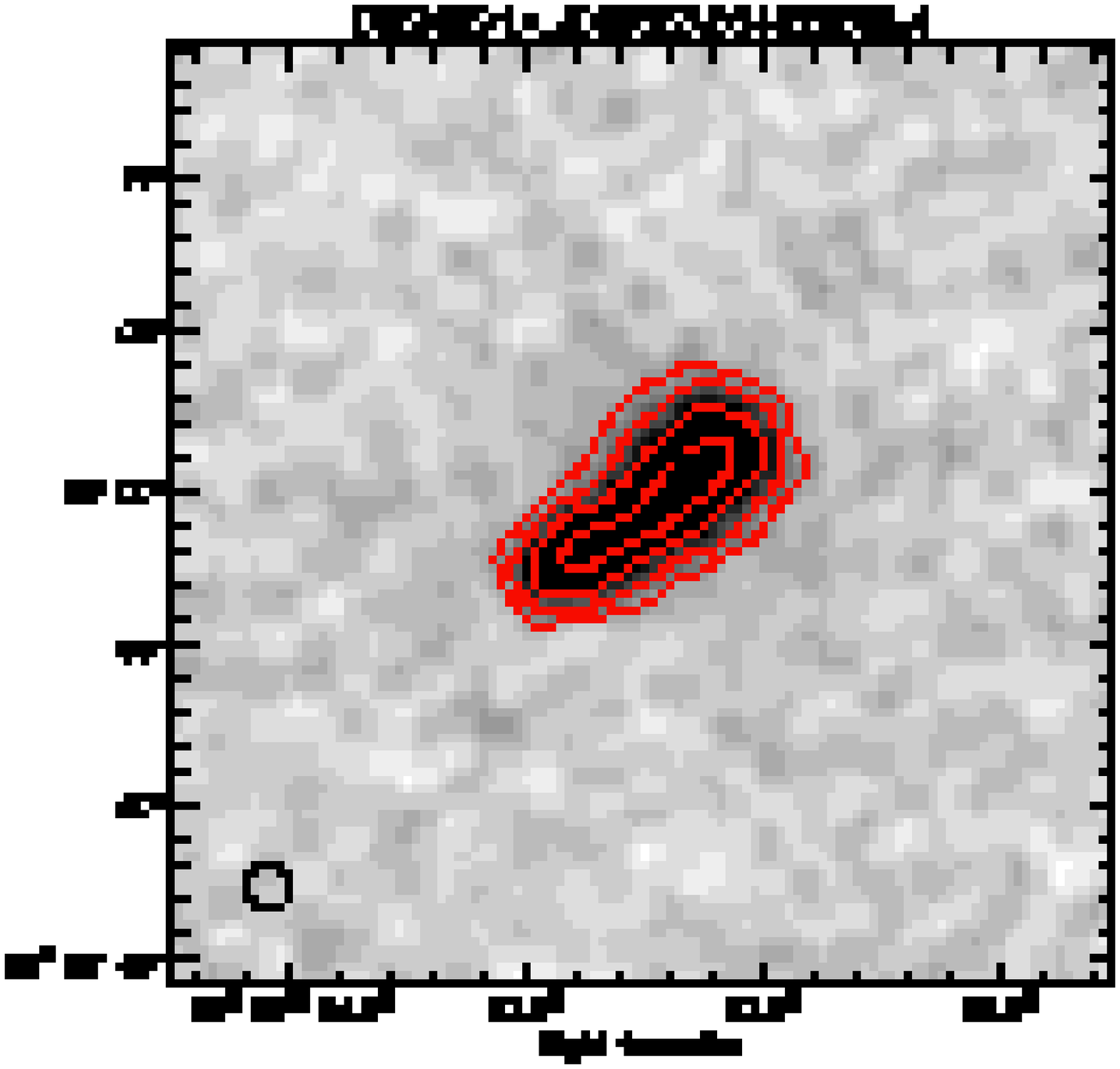}
                      \includegraphics[]{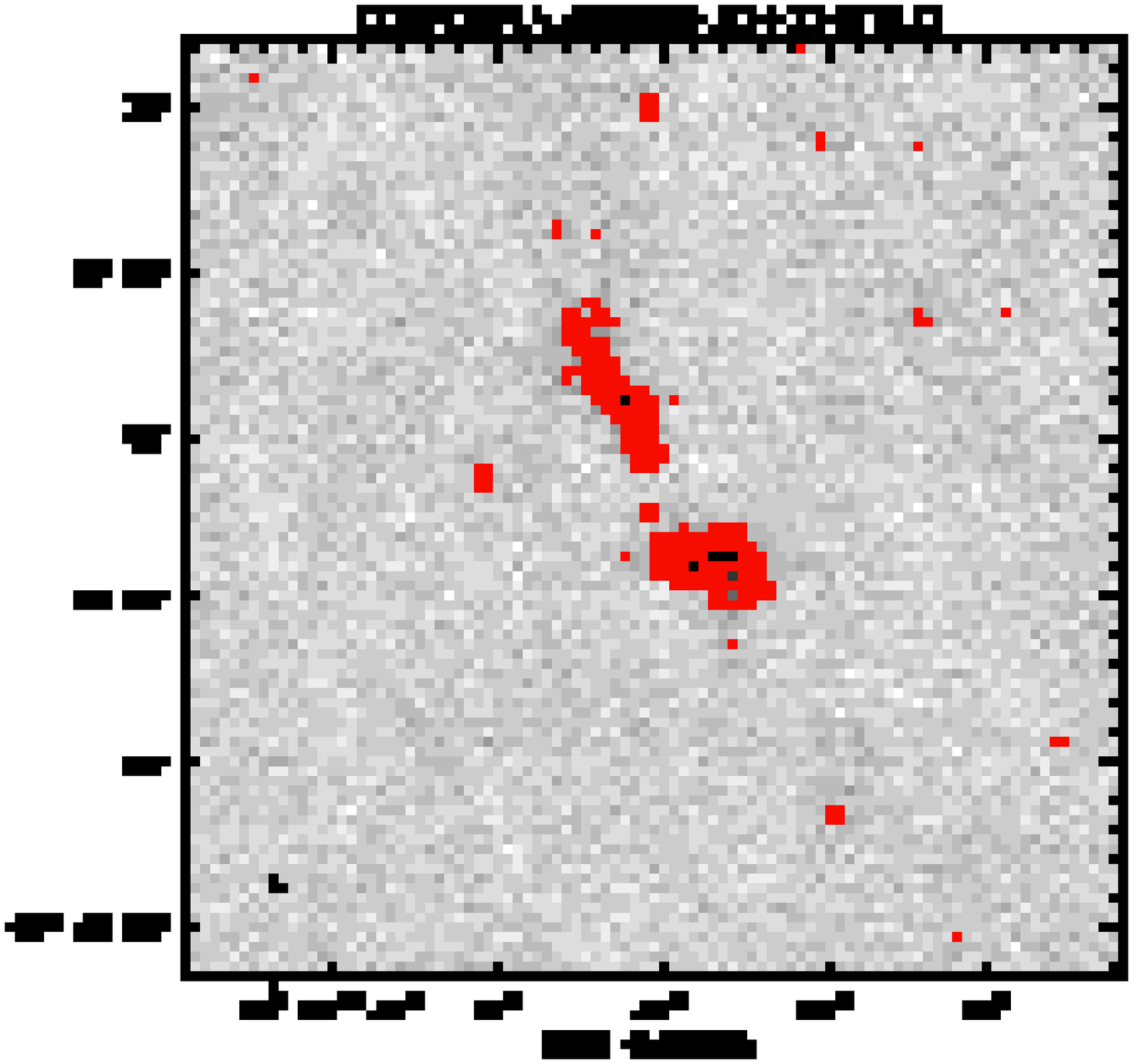}}\\
\resizebox{.9\hsize}{!}{\includegraphics[]{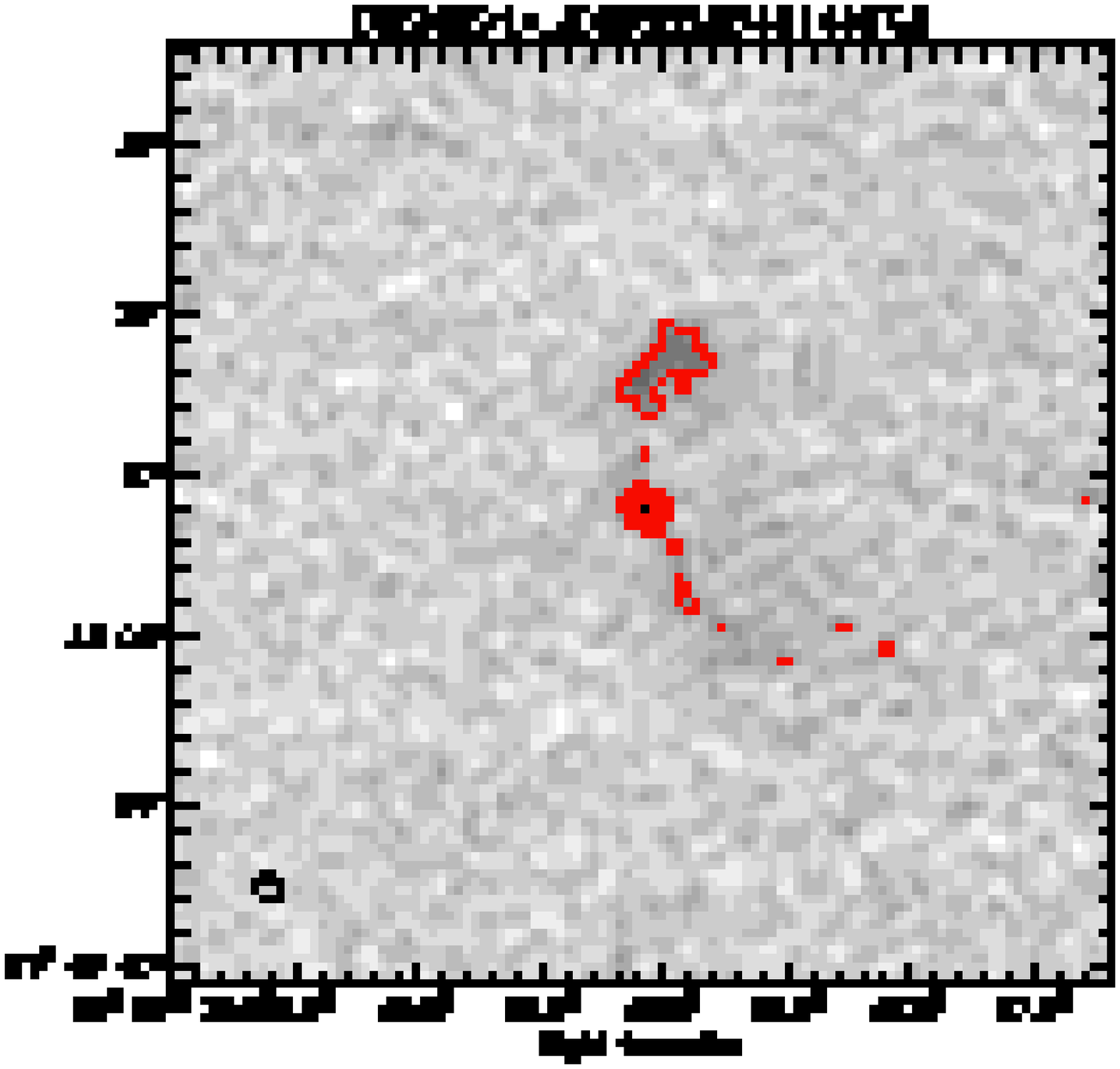}
                      \includegraphics[]{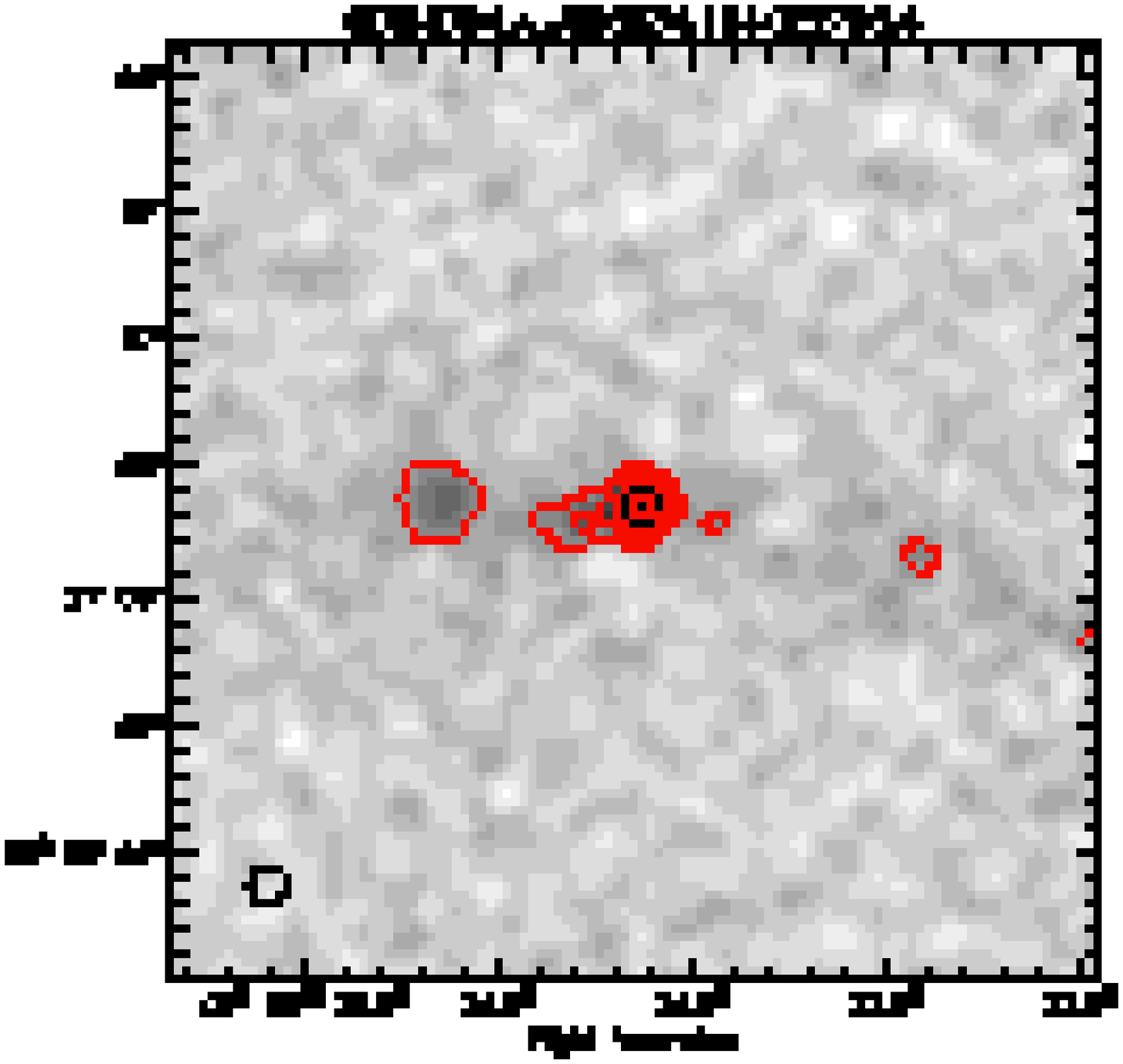}
                      \includegraphics[]{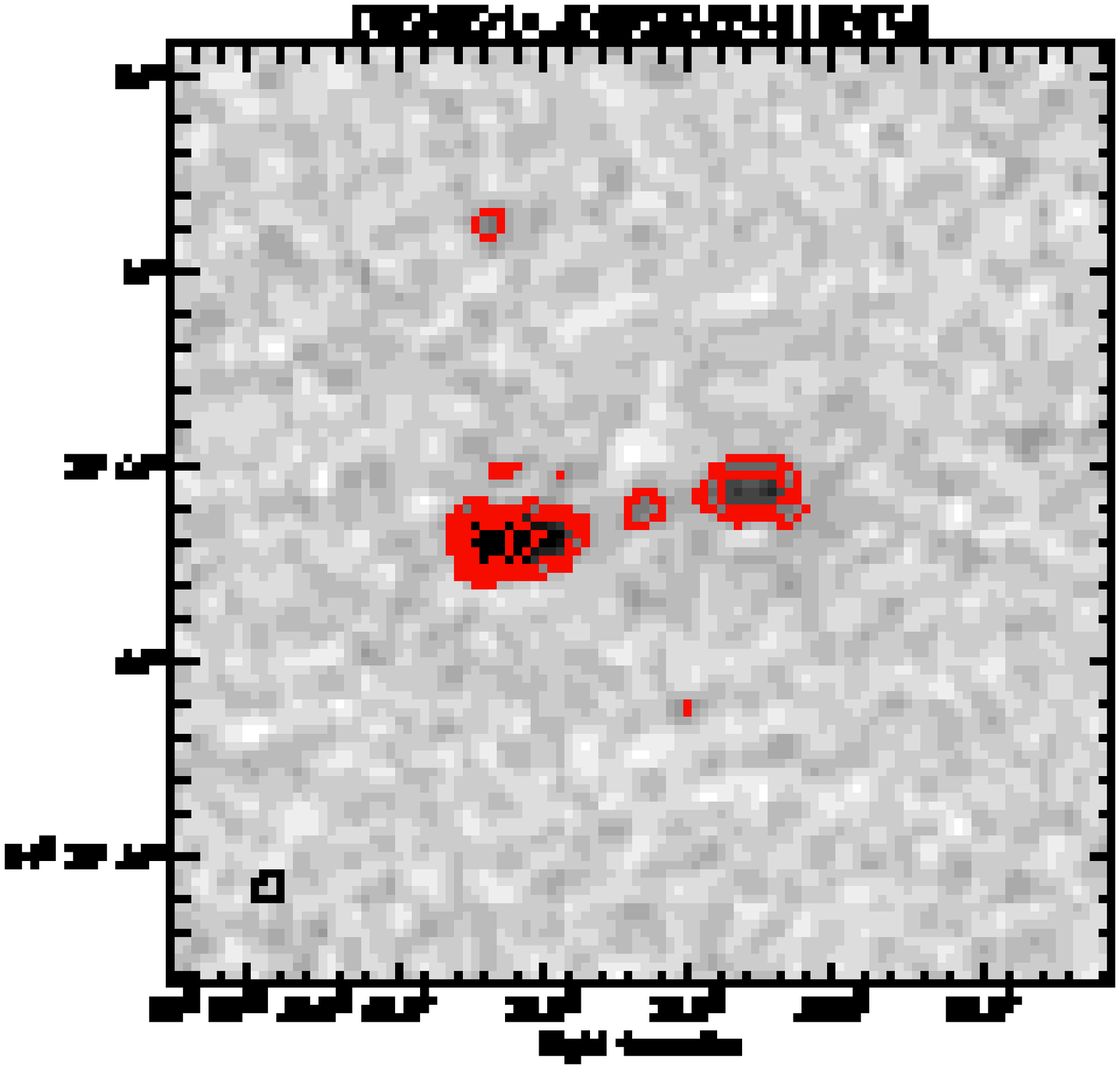}}
\clearpage            
\resizebox{.9\hsize}{!}{\includegraphics[]{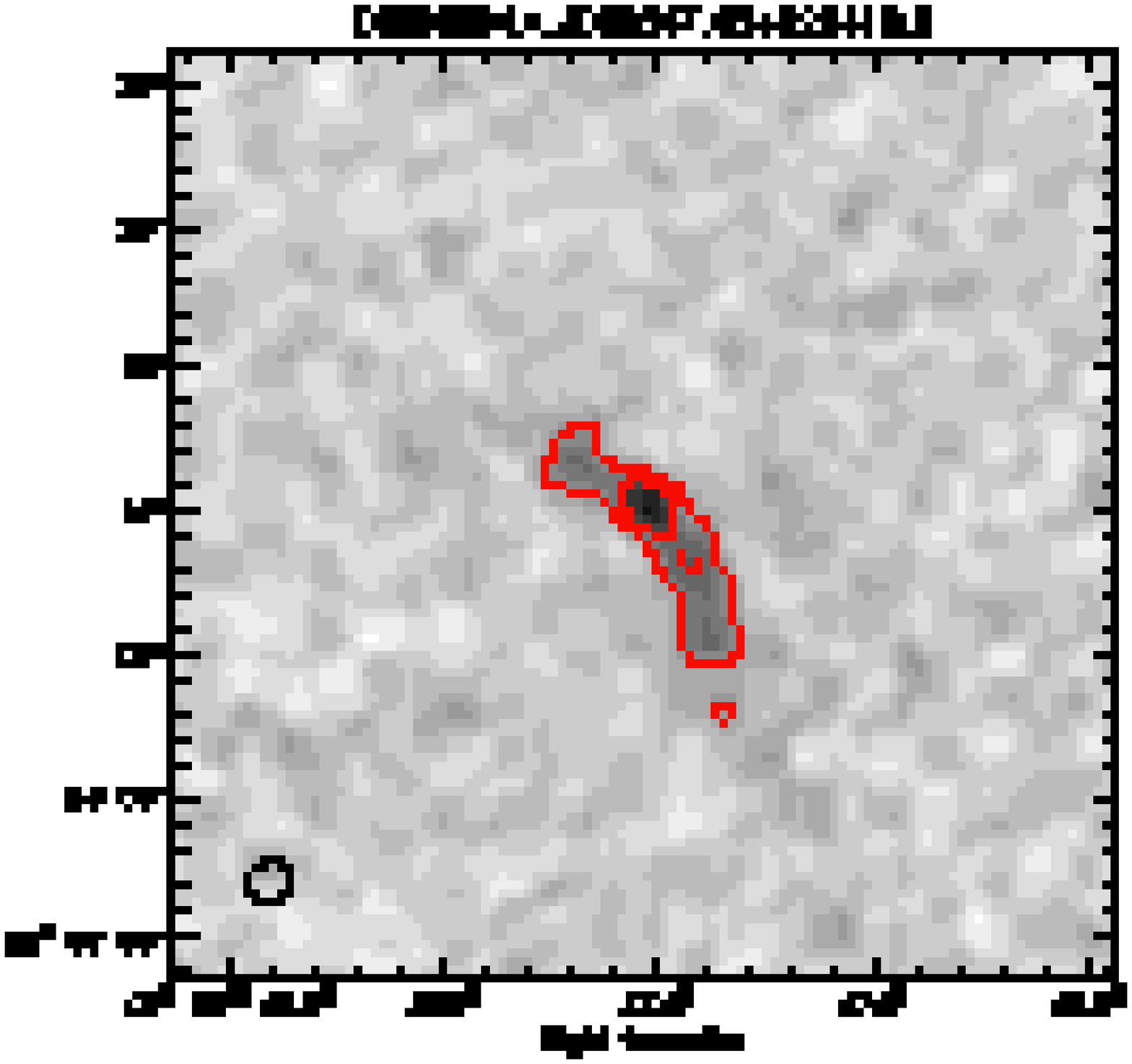}
                      \includegraphics[]{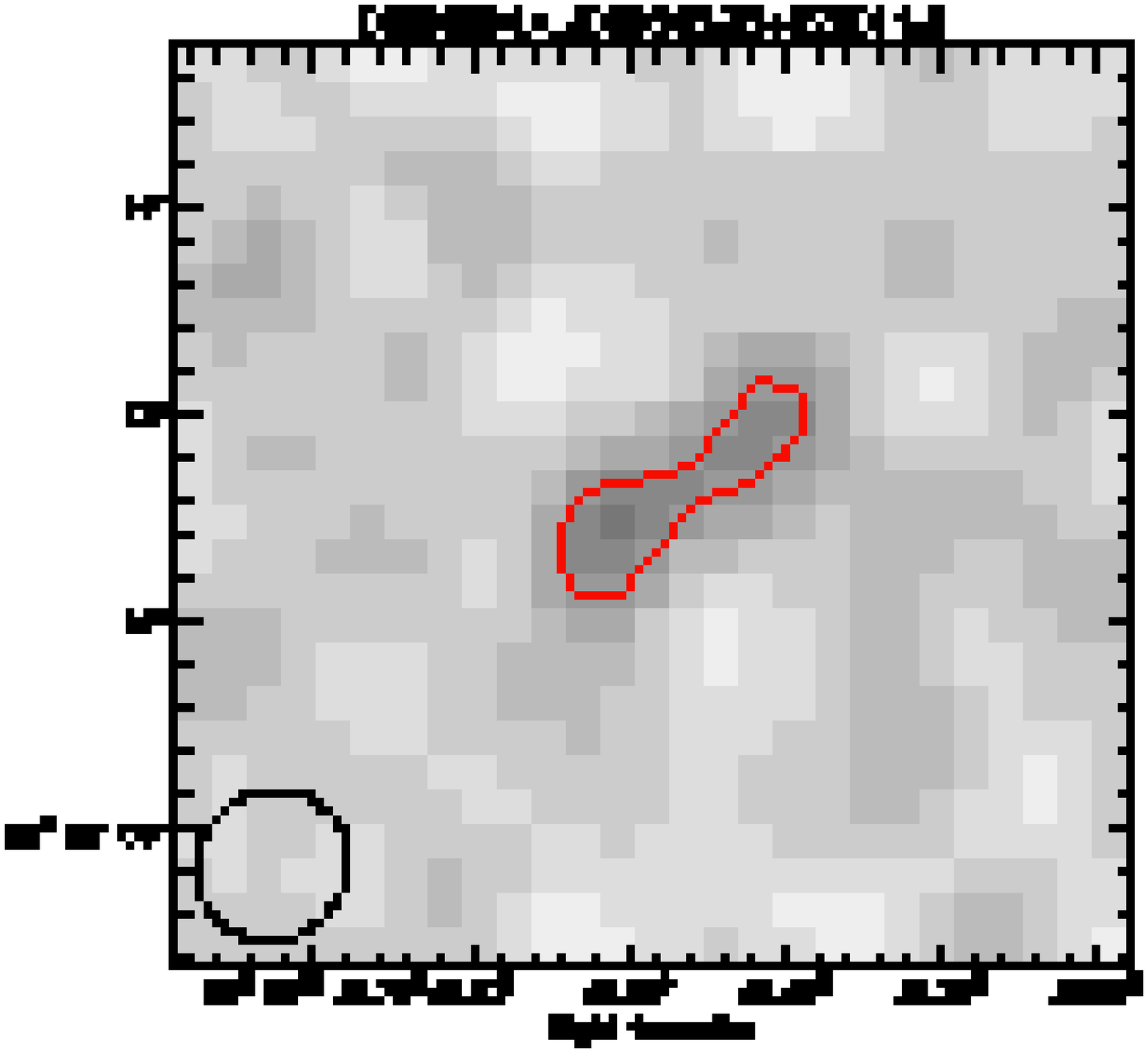}
                      \includegraphics[]{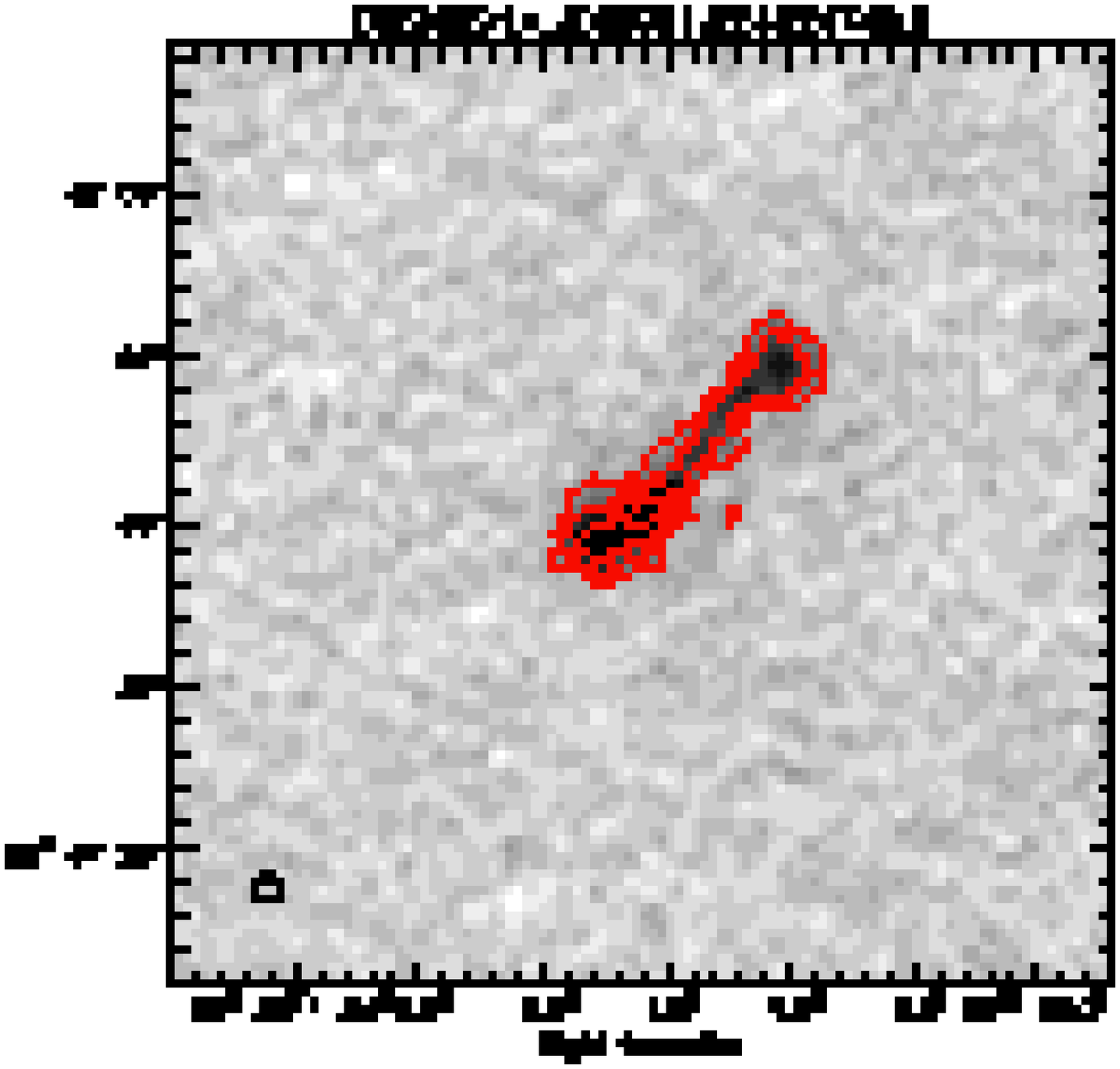}}\\
\resizebox{.9\hsize}{!}{\includegraphics[]{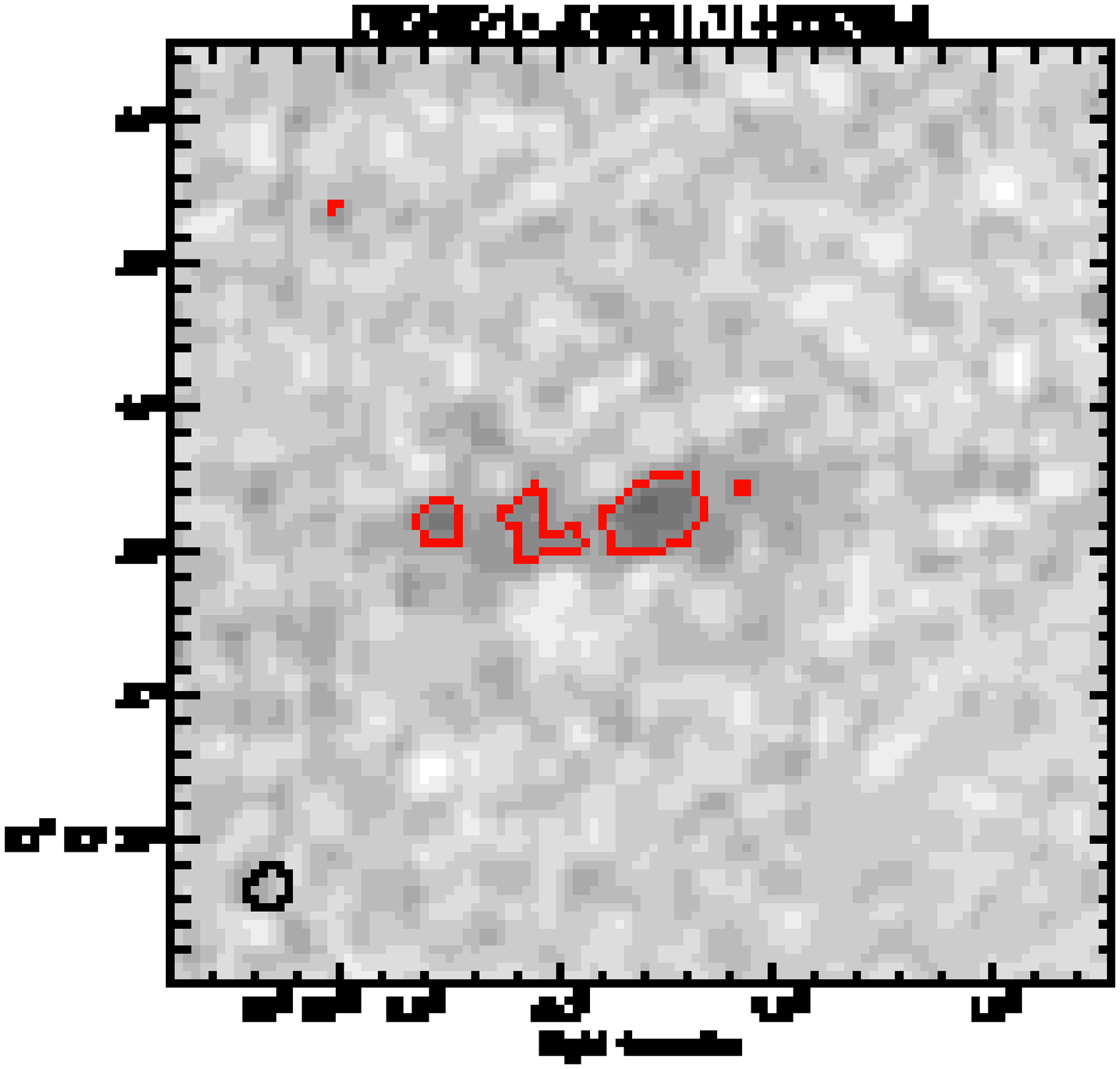}
                      \includegraphics[]{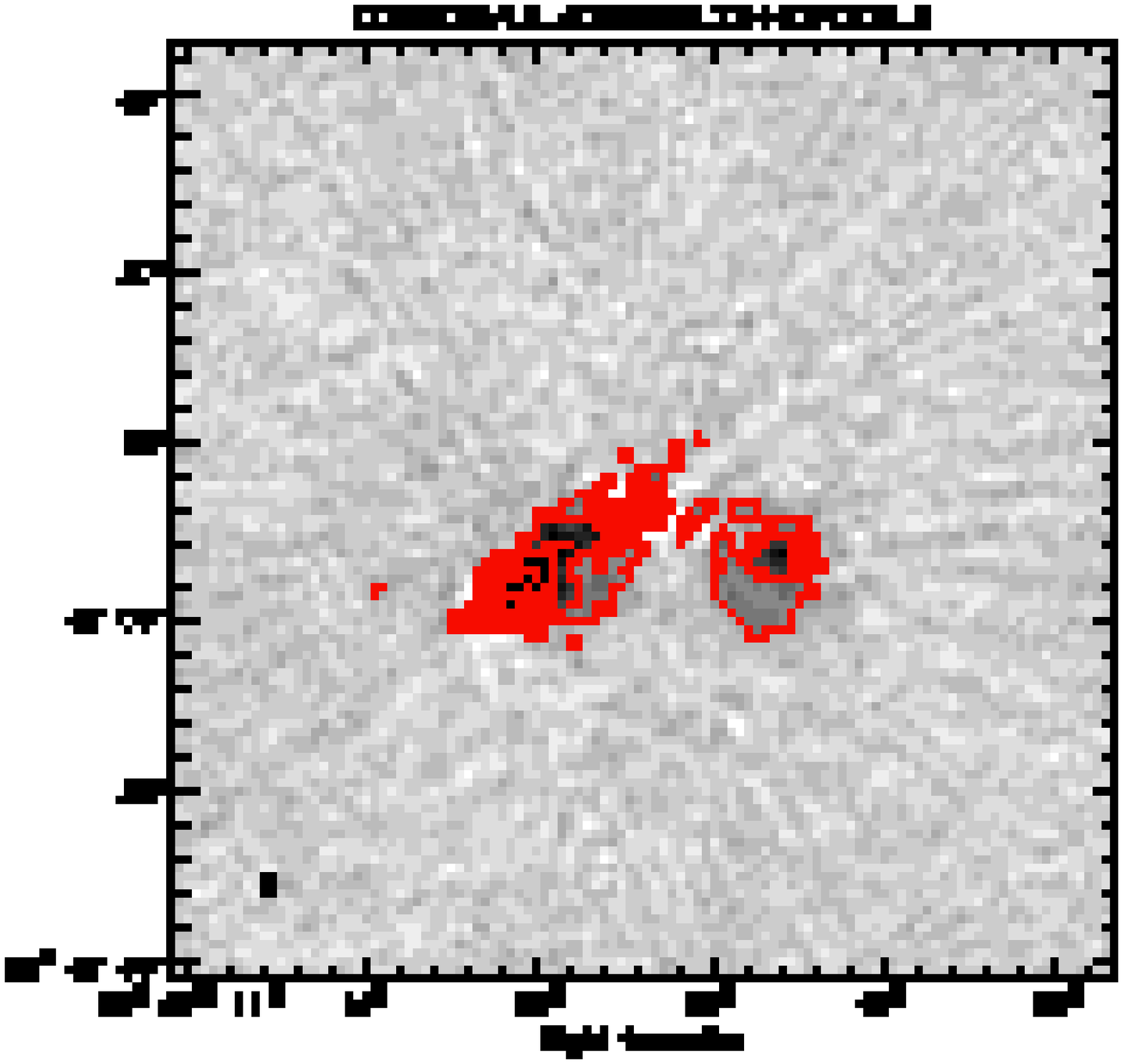}
                      \includegraphics[]{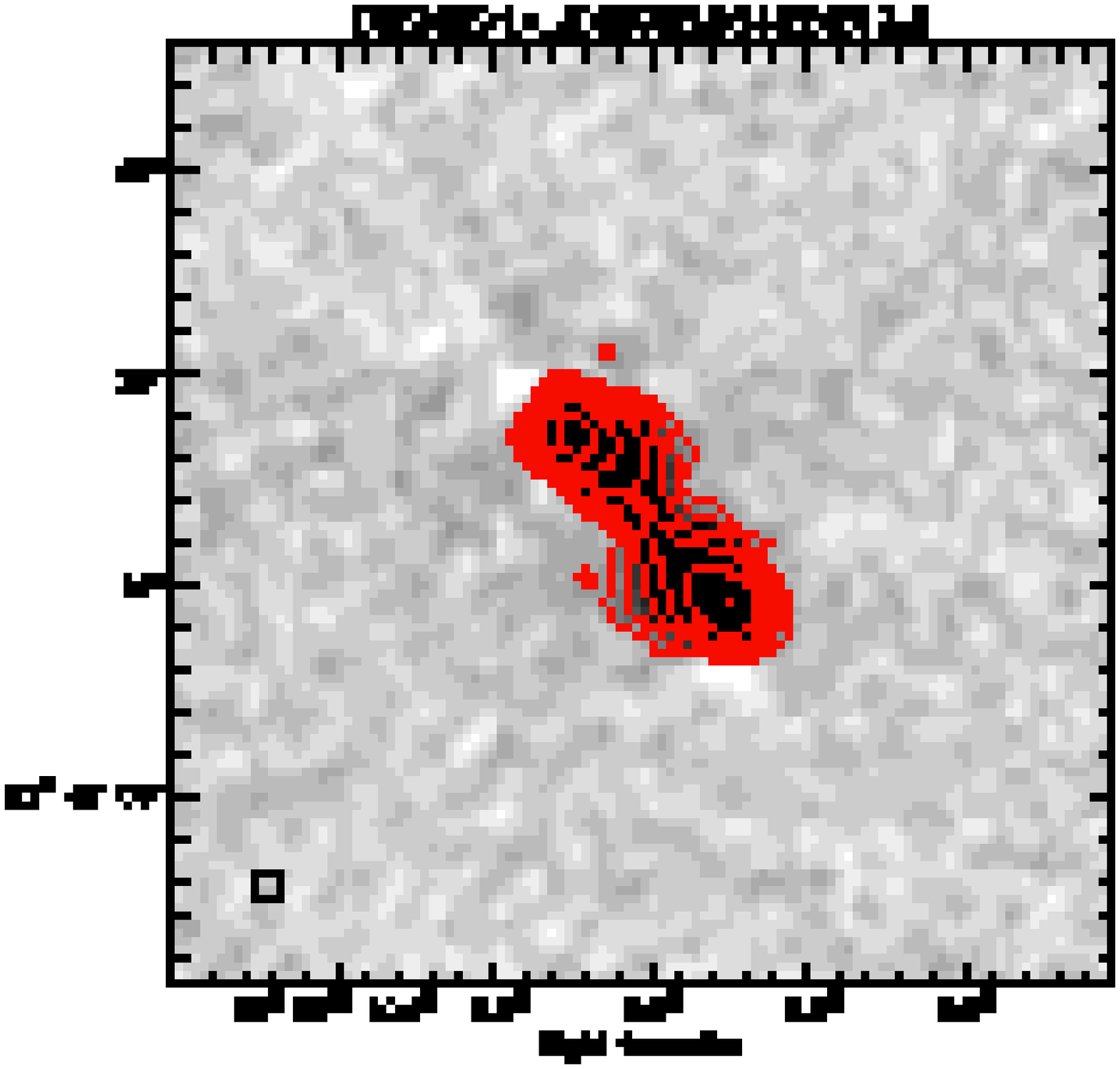}}\\
\resizebox{.9\hsize}{!}{\includegraphics[]{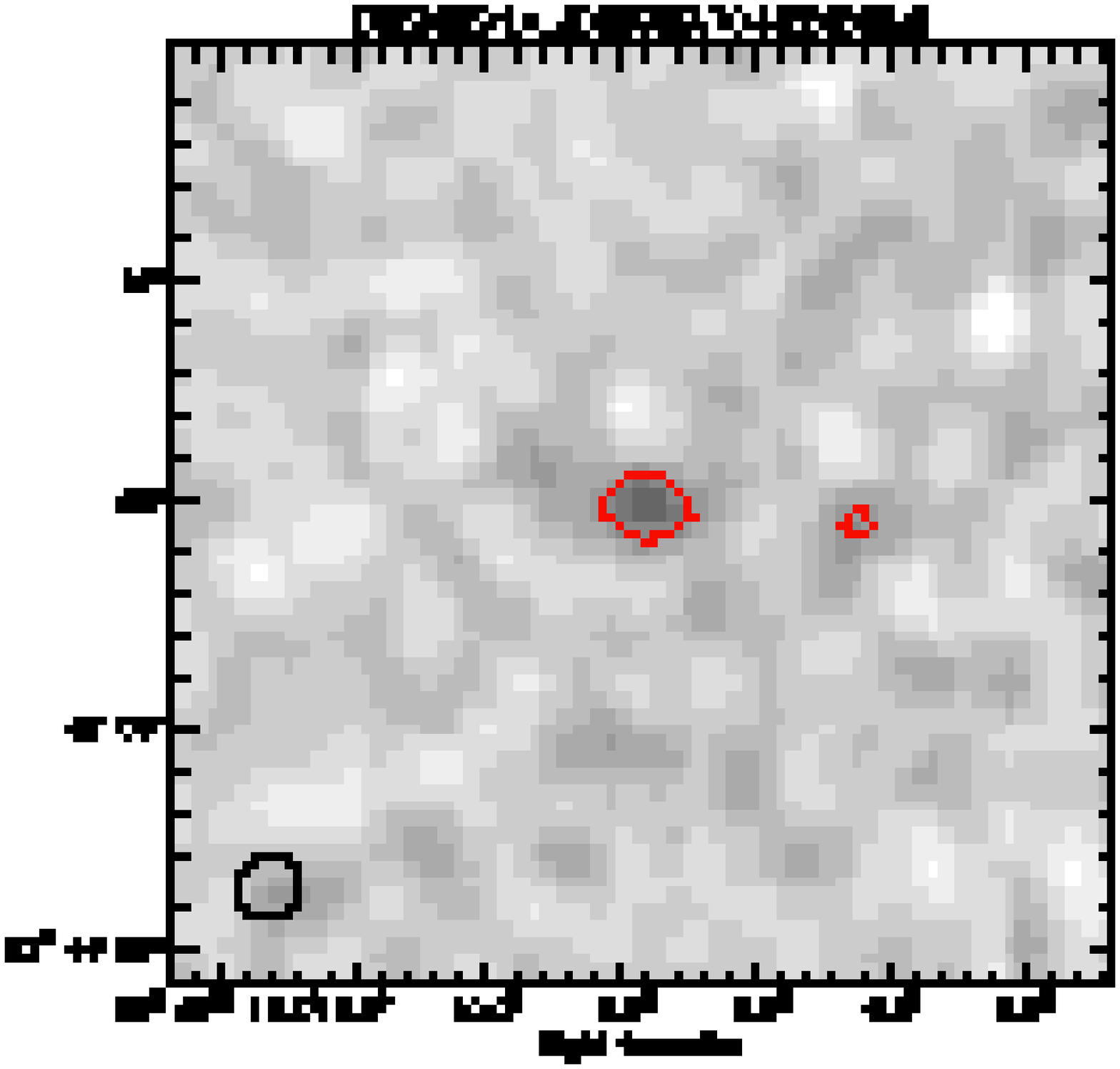}
                      \includegraphics[]{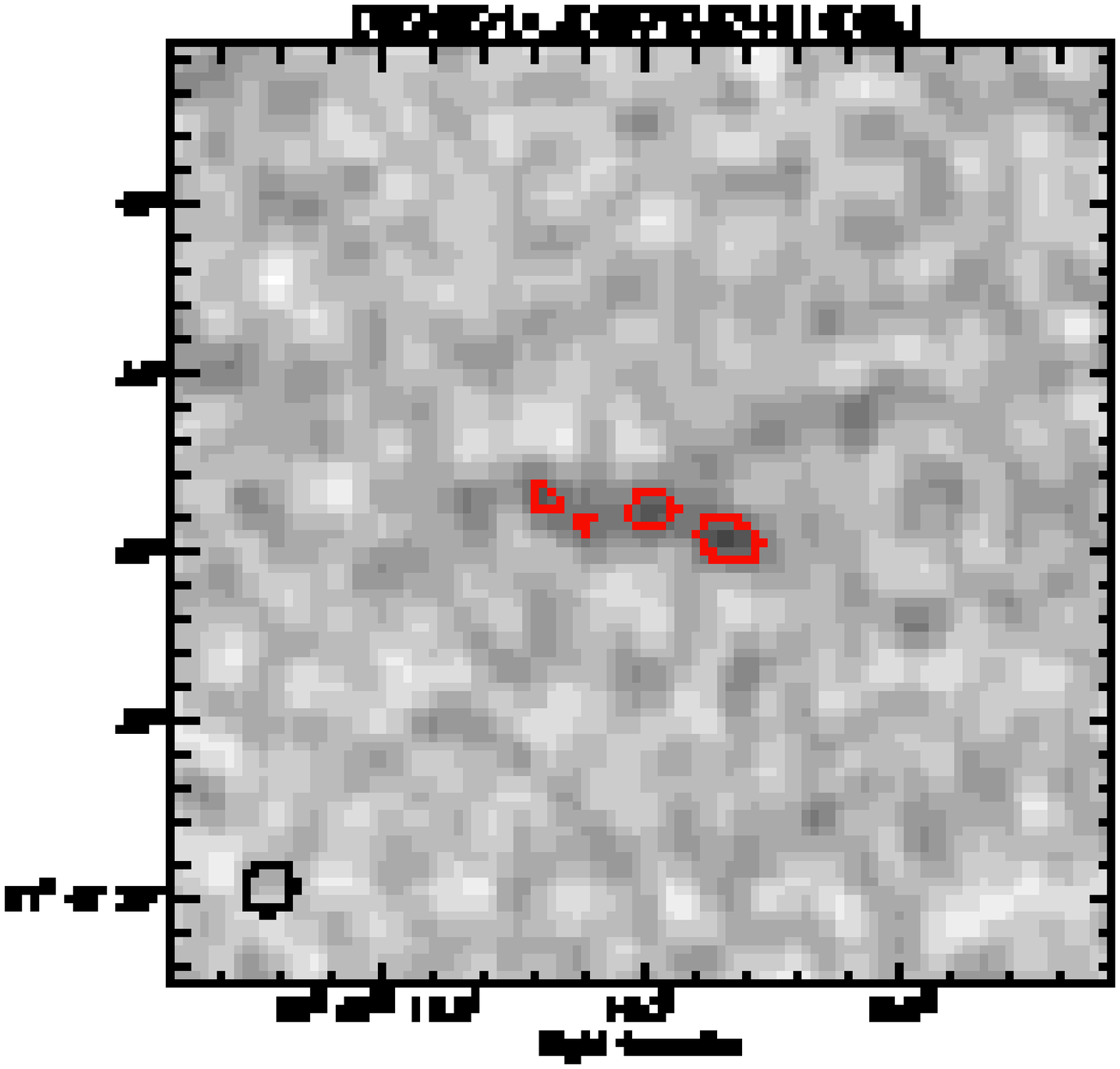}
                      \includegraphics[]{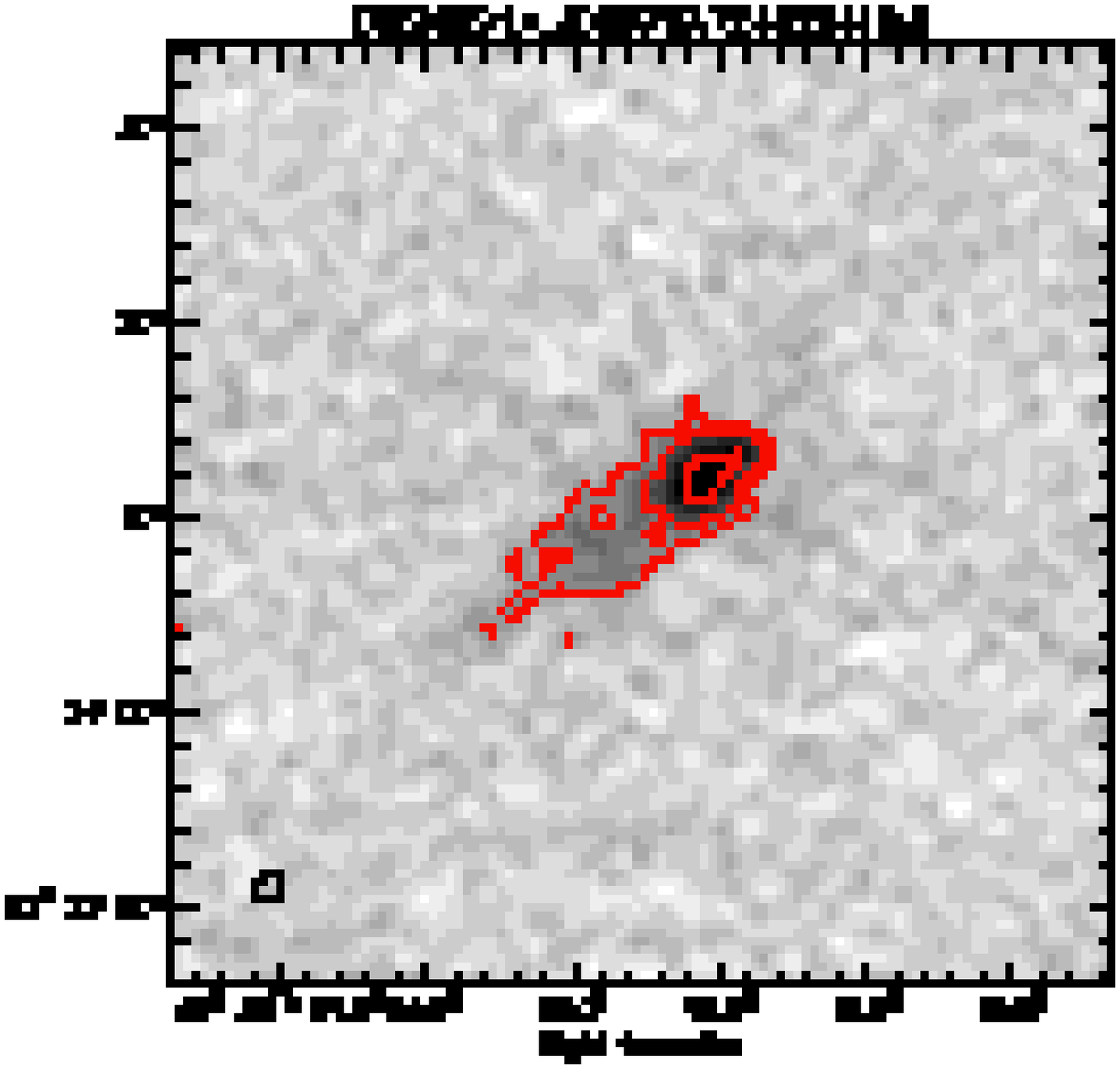}}\\
\resizebox{.9\hsize}{!}{\includegraphics[]{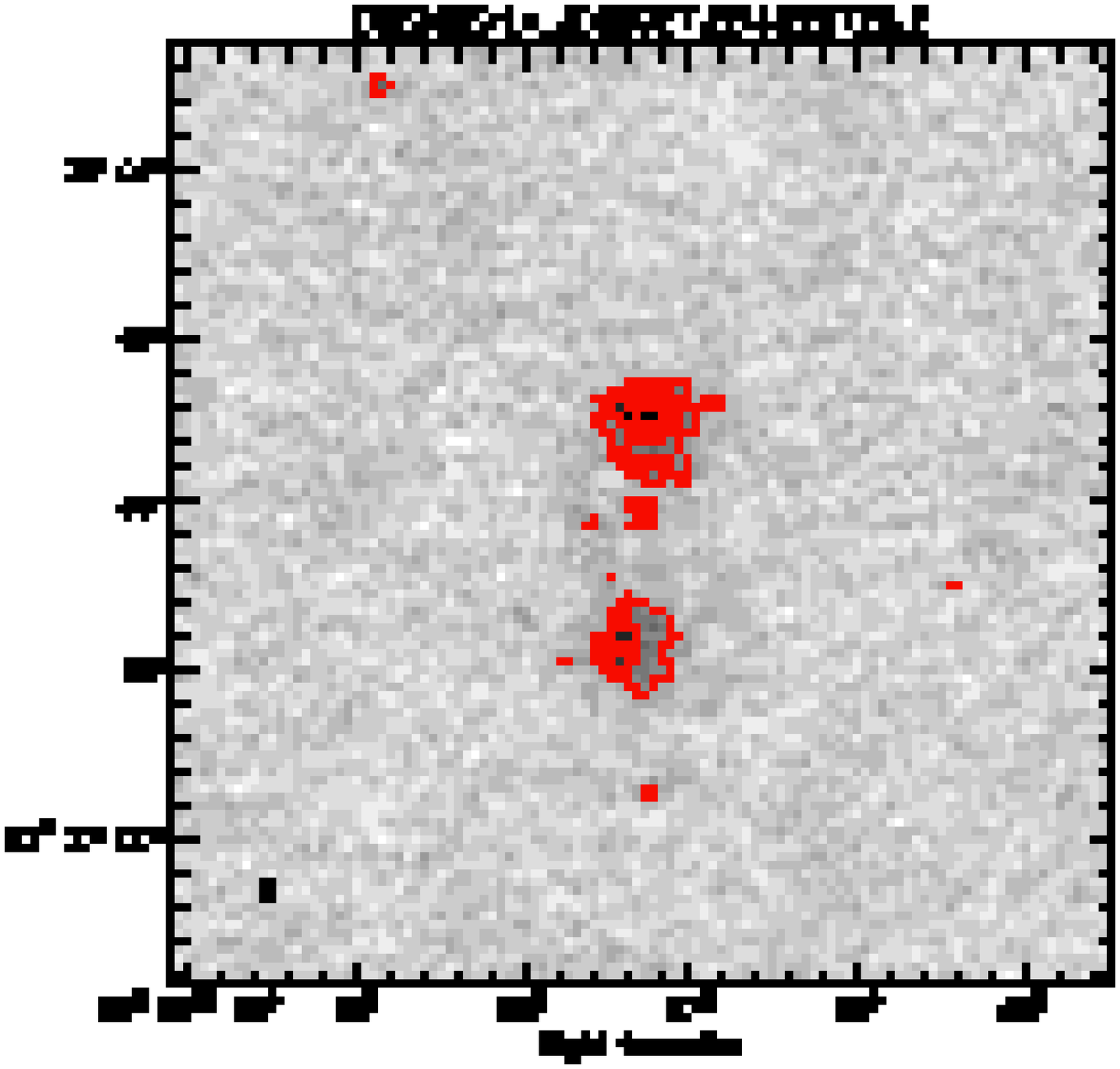}
                      \includegraphics[]{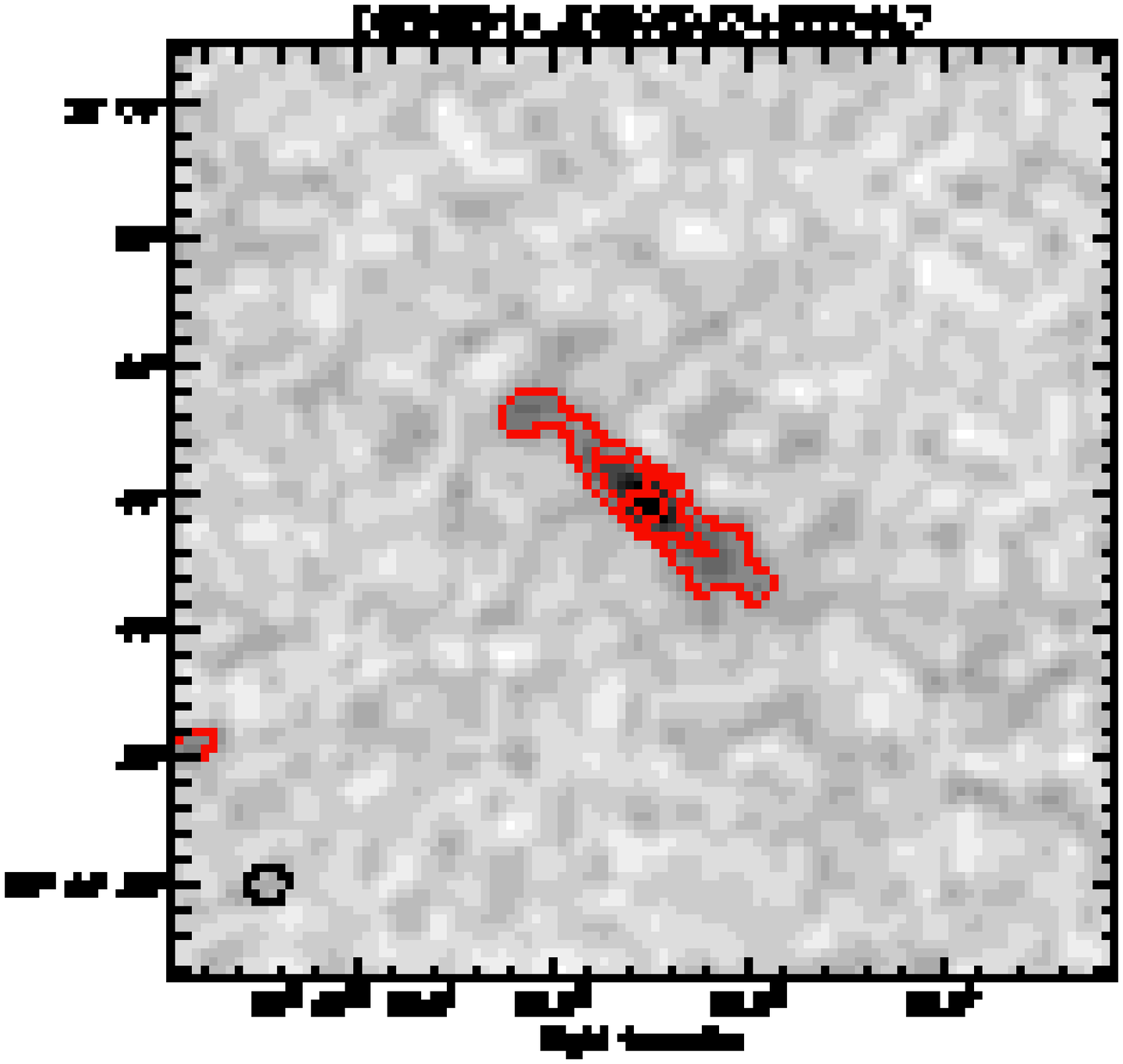}
                      \includegraphics[]{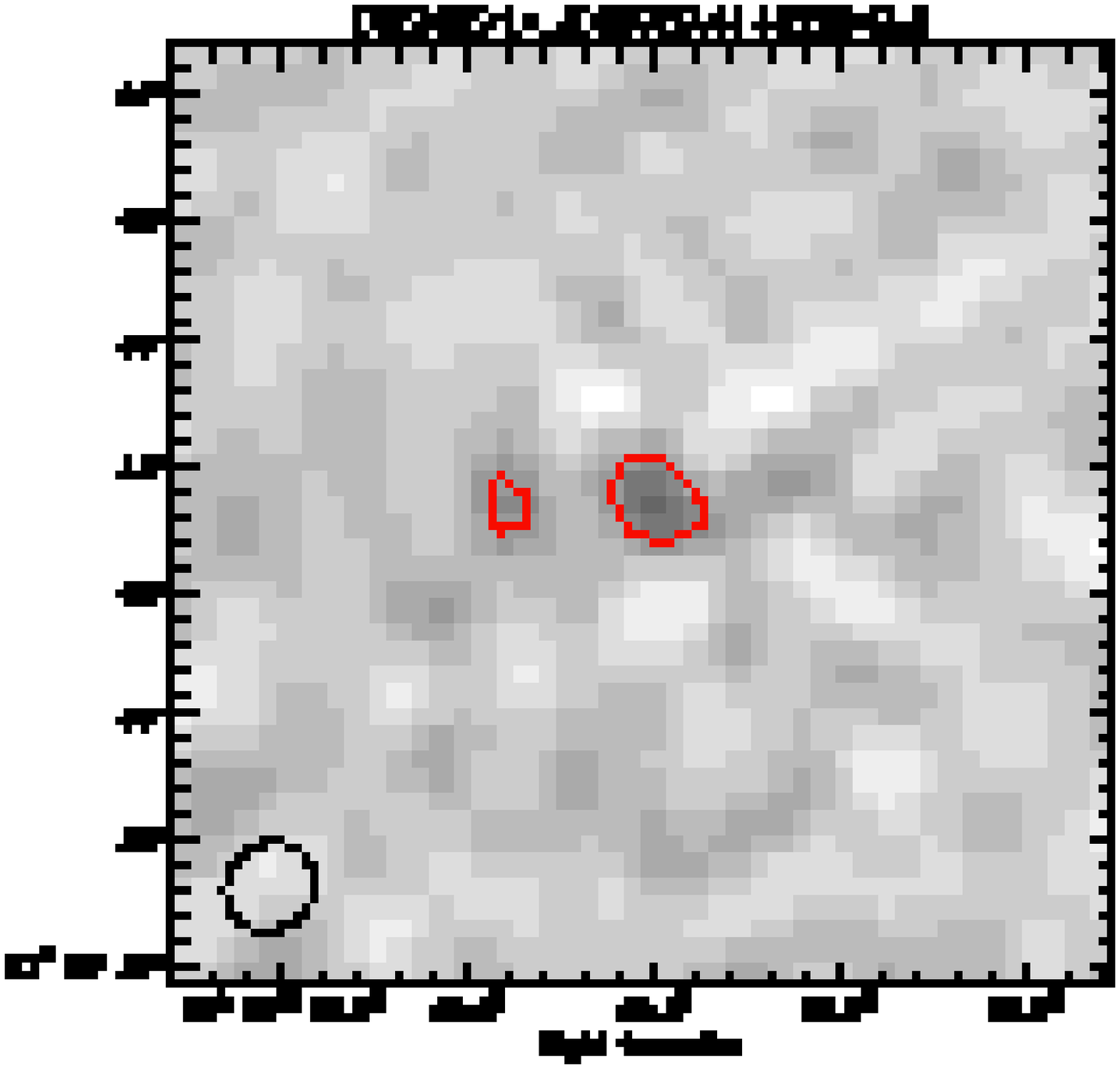}}\\
\clearpage
\resizebox{.9\hsize}{!}{\includegraphics[]{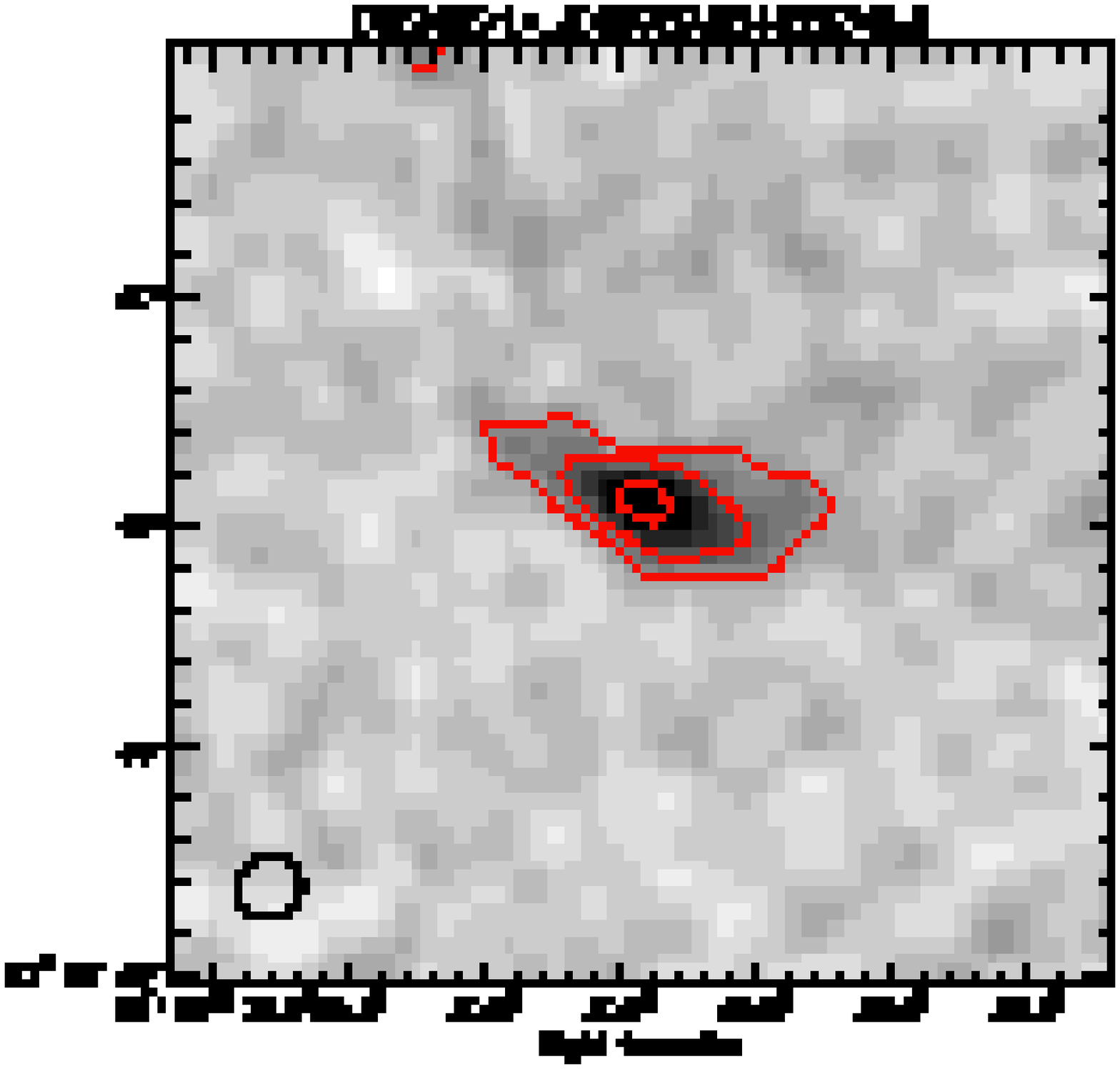}
                      \includegraphics[]{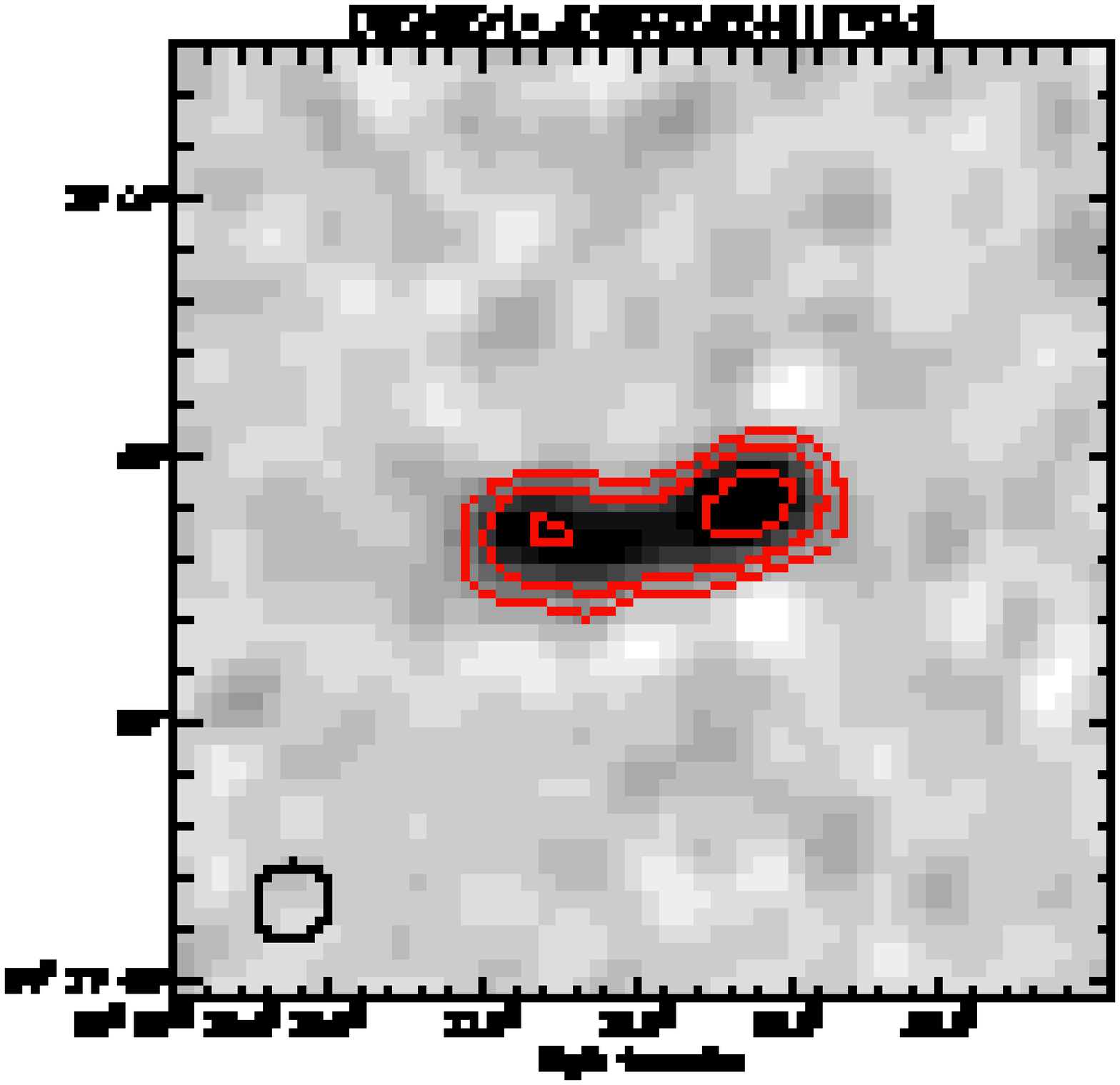}
                      \includegraphics[]{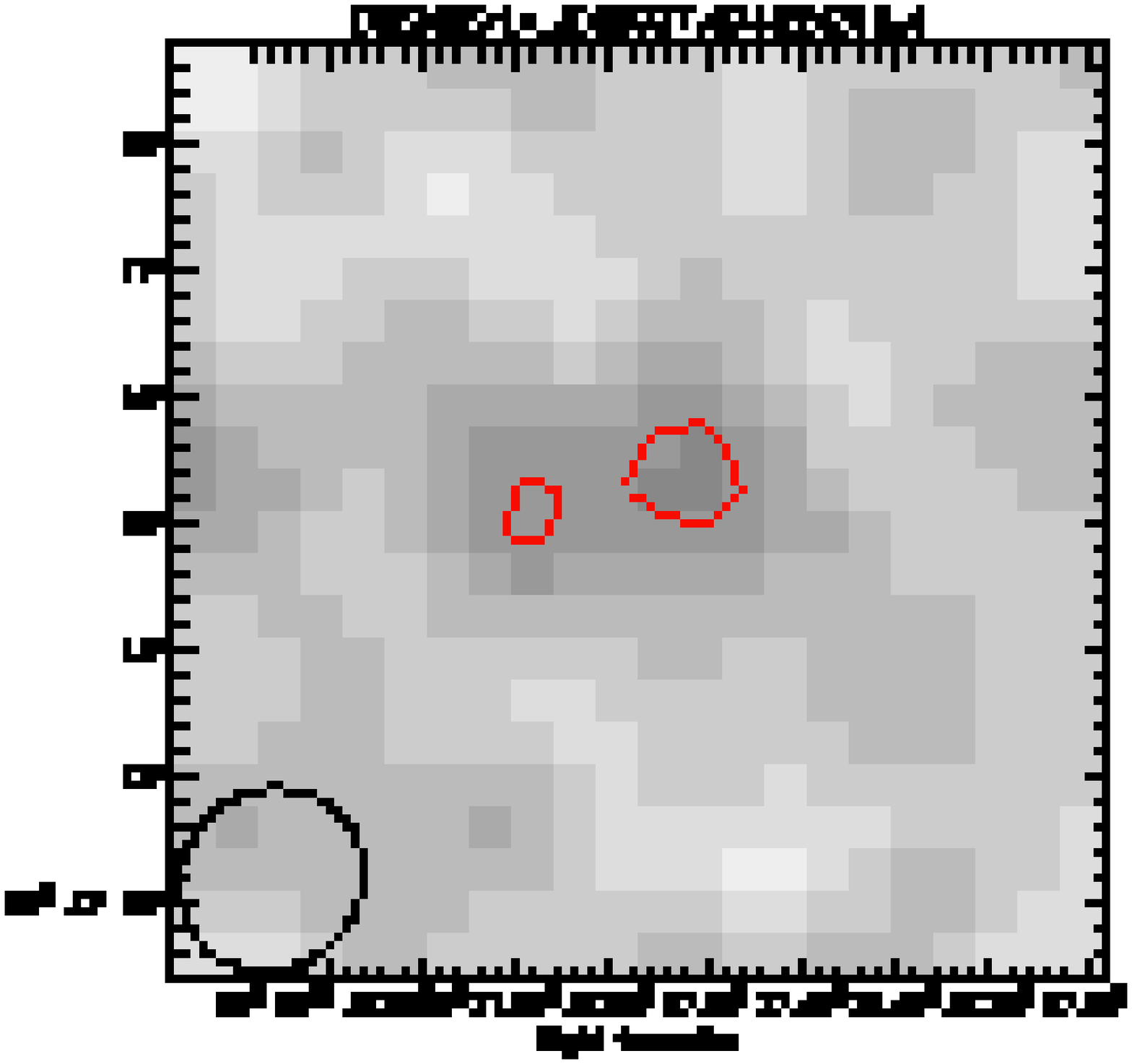}}\\
\resizebox{.9\hsize}{!}{\includegraphics[]{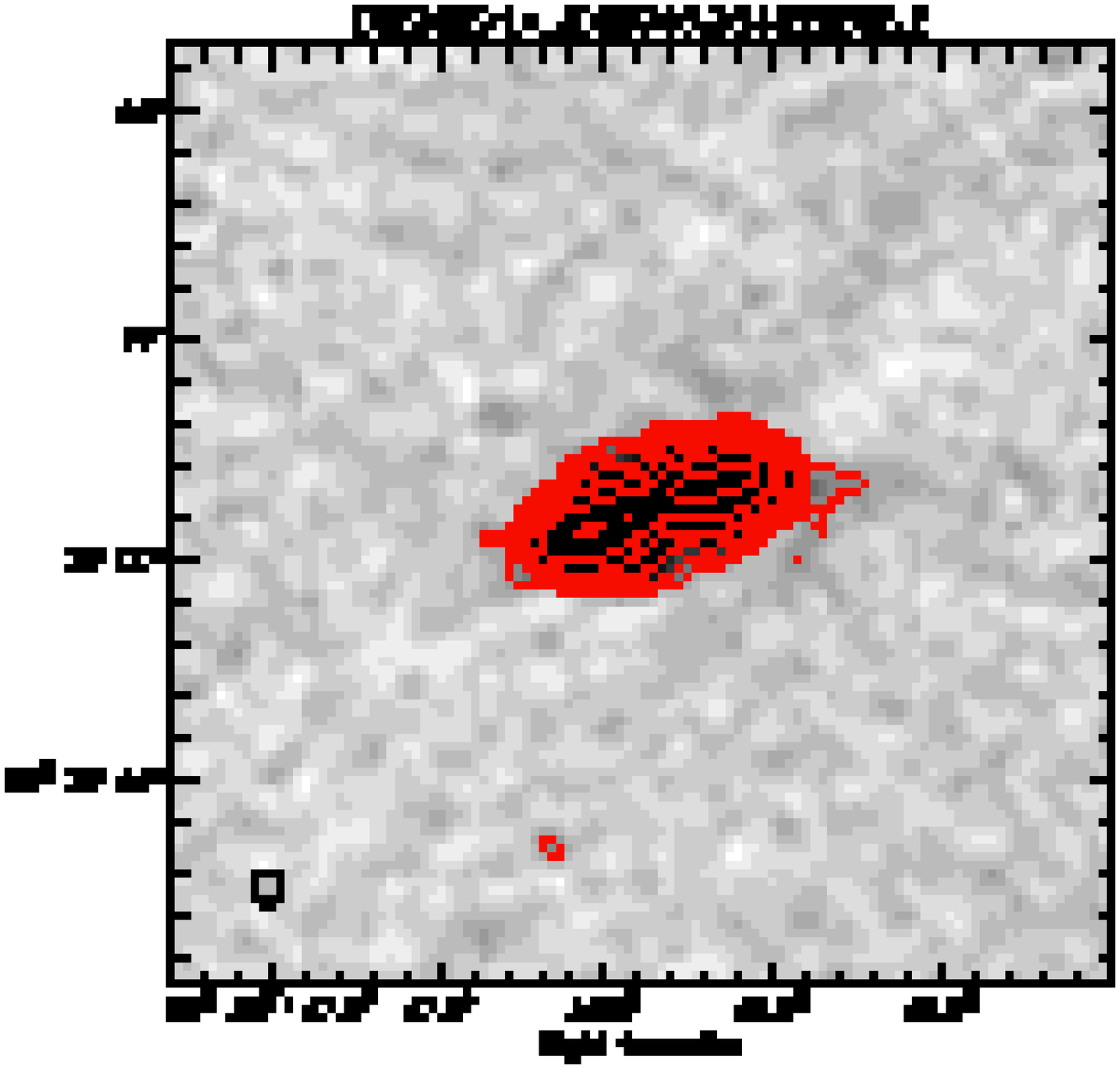}
                      \includegraphics[]{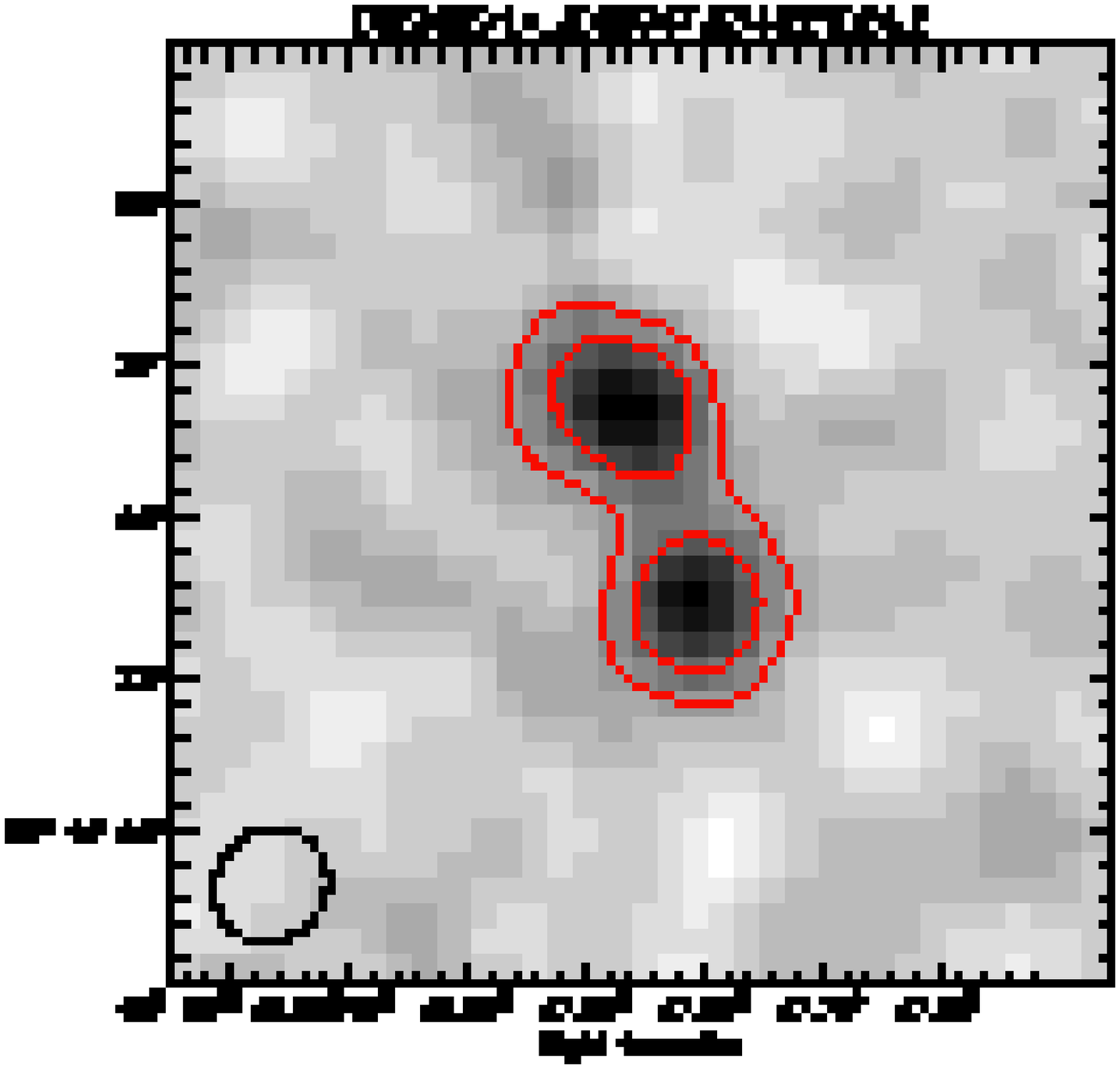}
                      \includegraphics[]{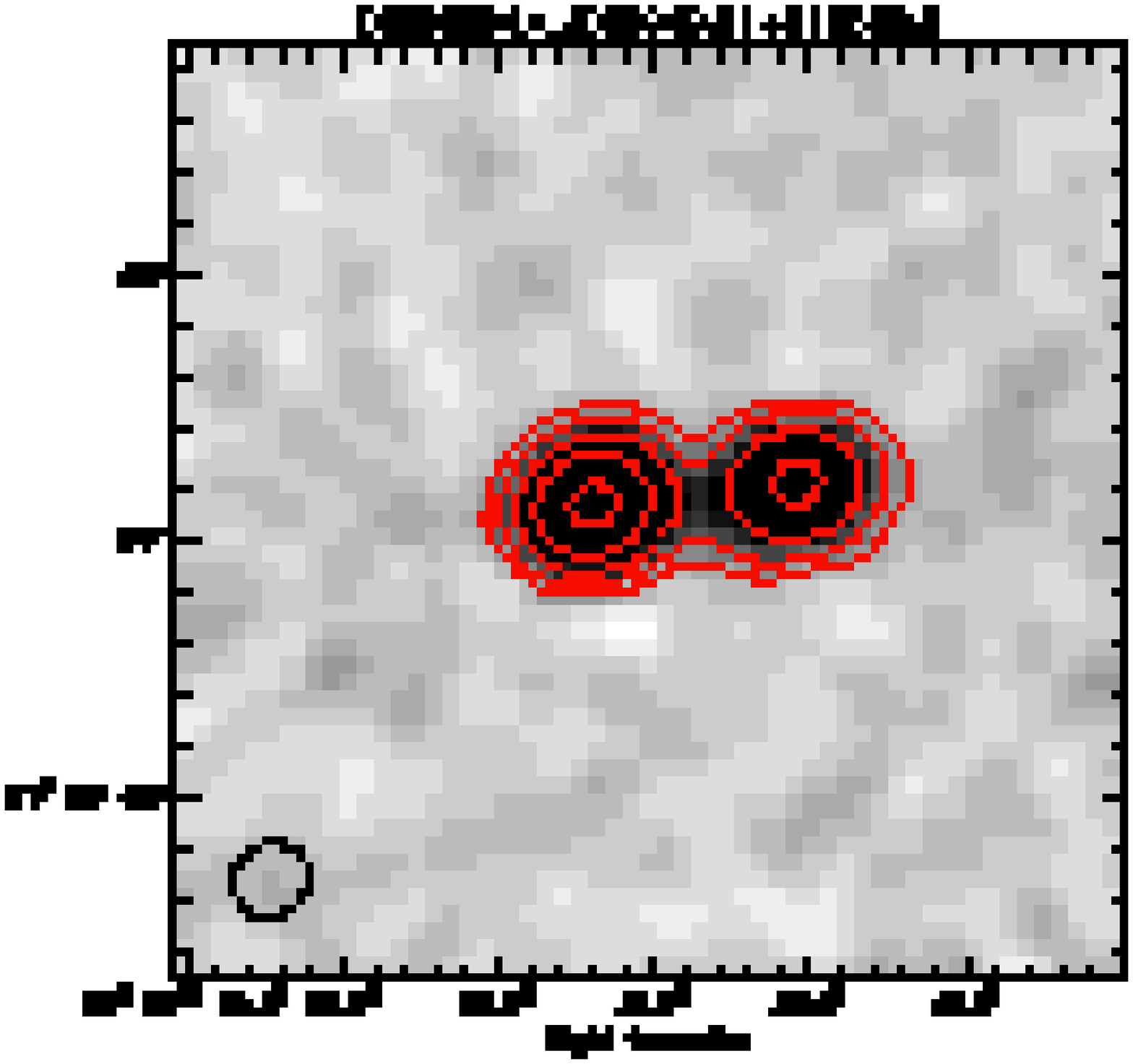}}\\
\resizebox{.9\hsize}{!}{\includegraphics[]{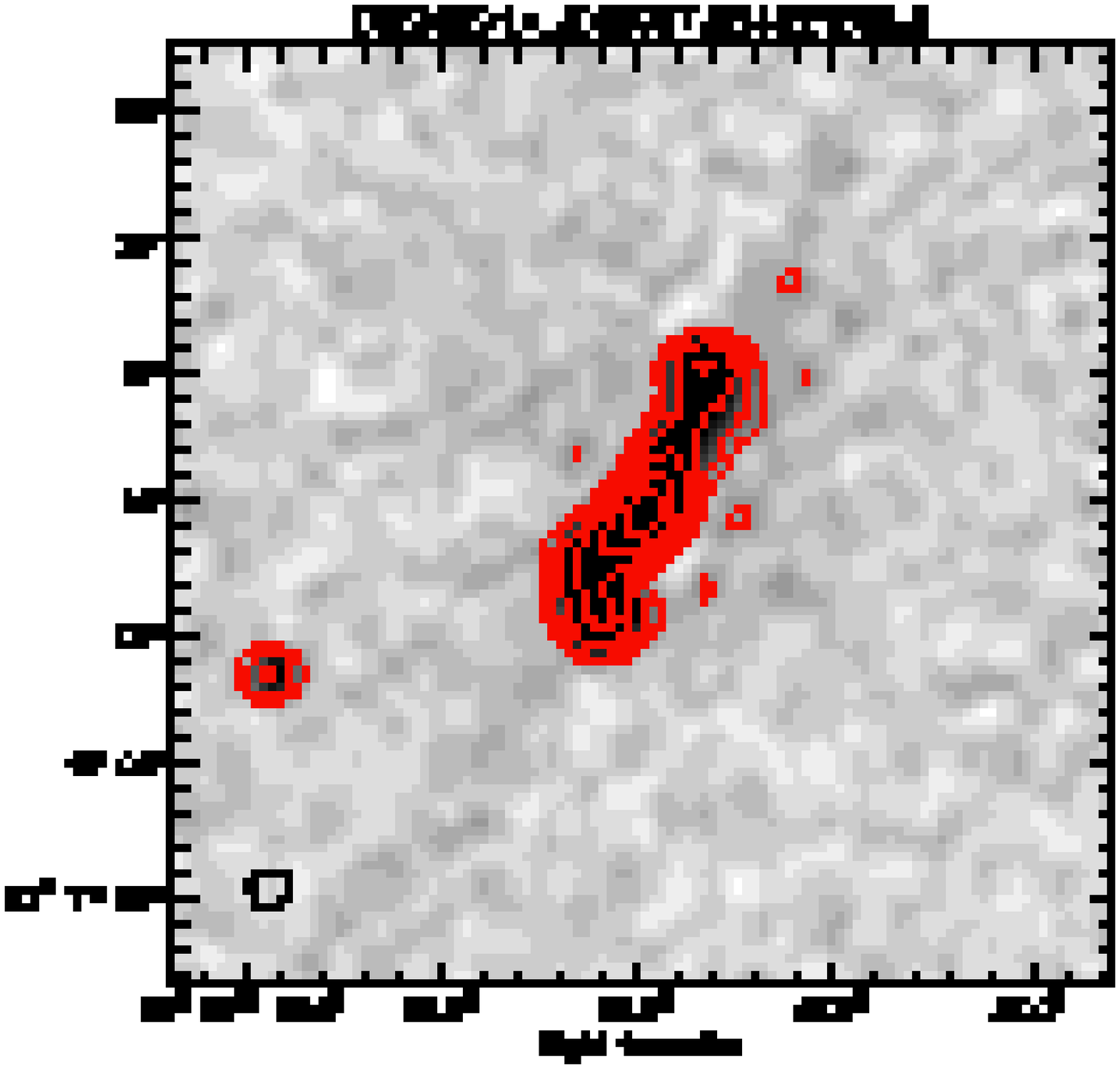}
                      \includegraphics[]{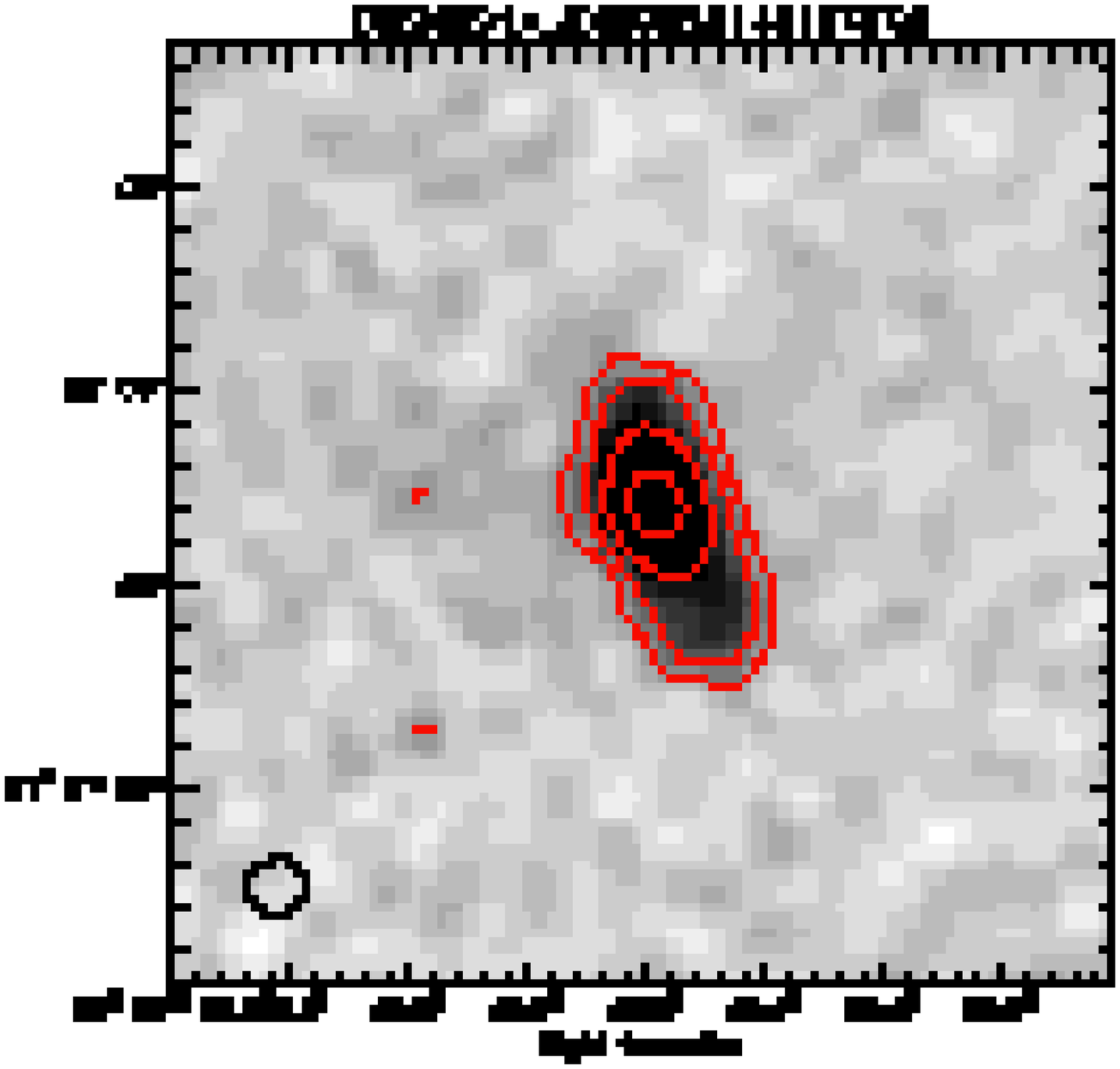}
                      \includegraphics[]{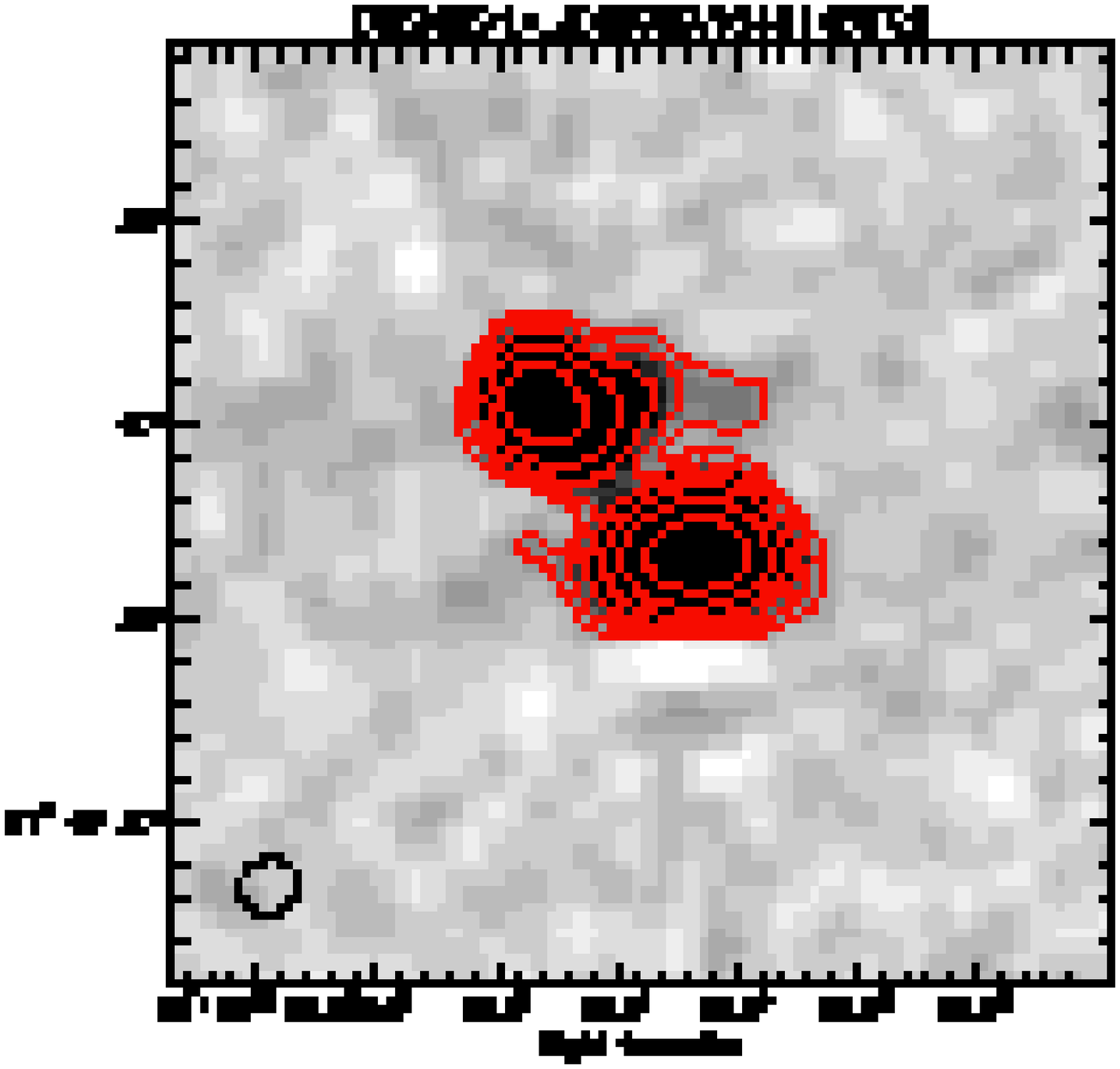}}\\
\resizebox{.9\hsize}{!}{\includegraphics[]{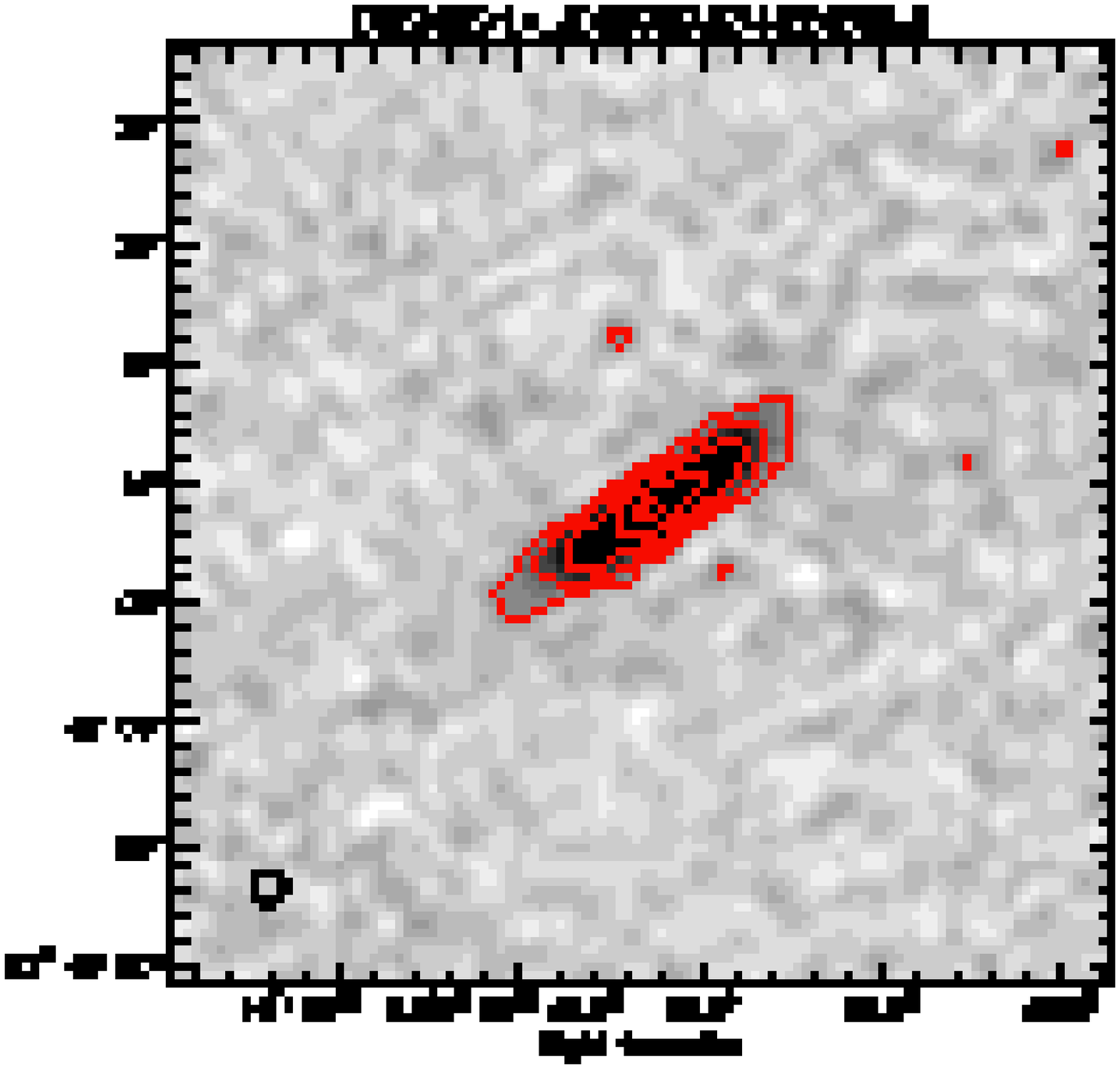}
                      \includegraphics[]{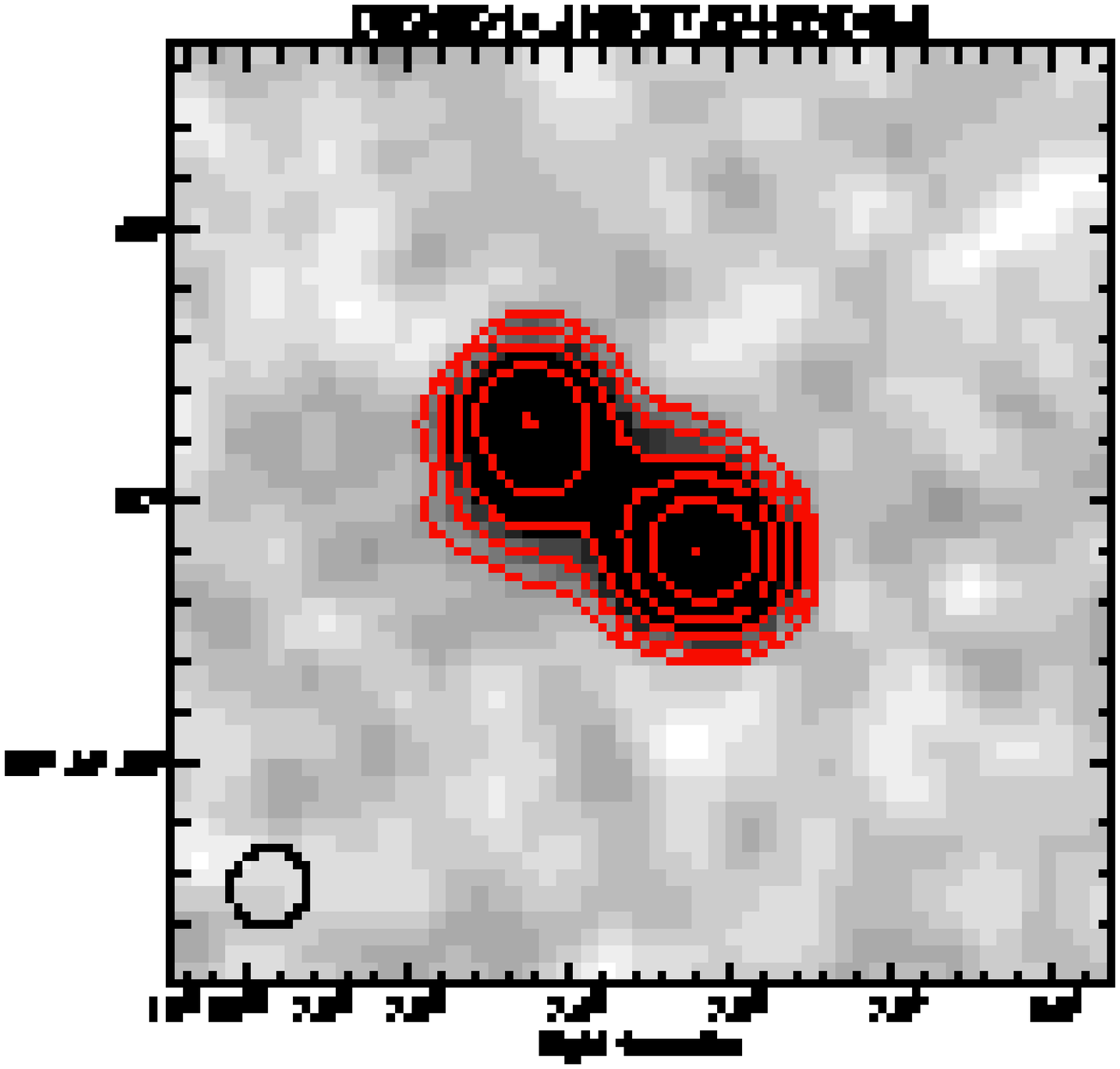}
                      \includegraphics[]{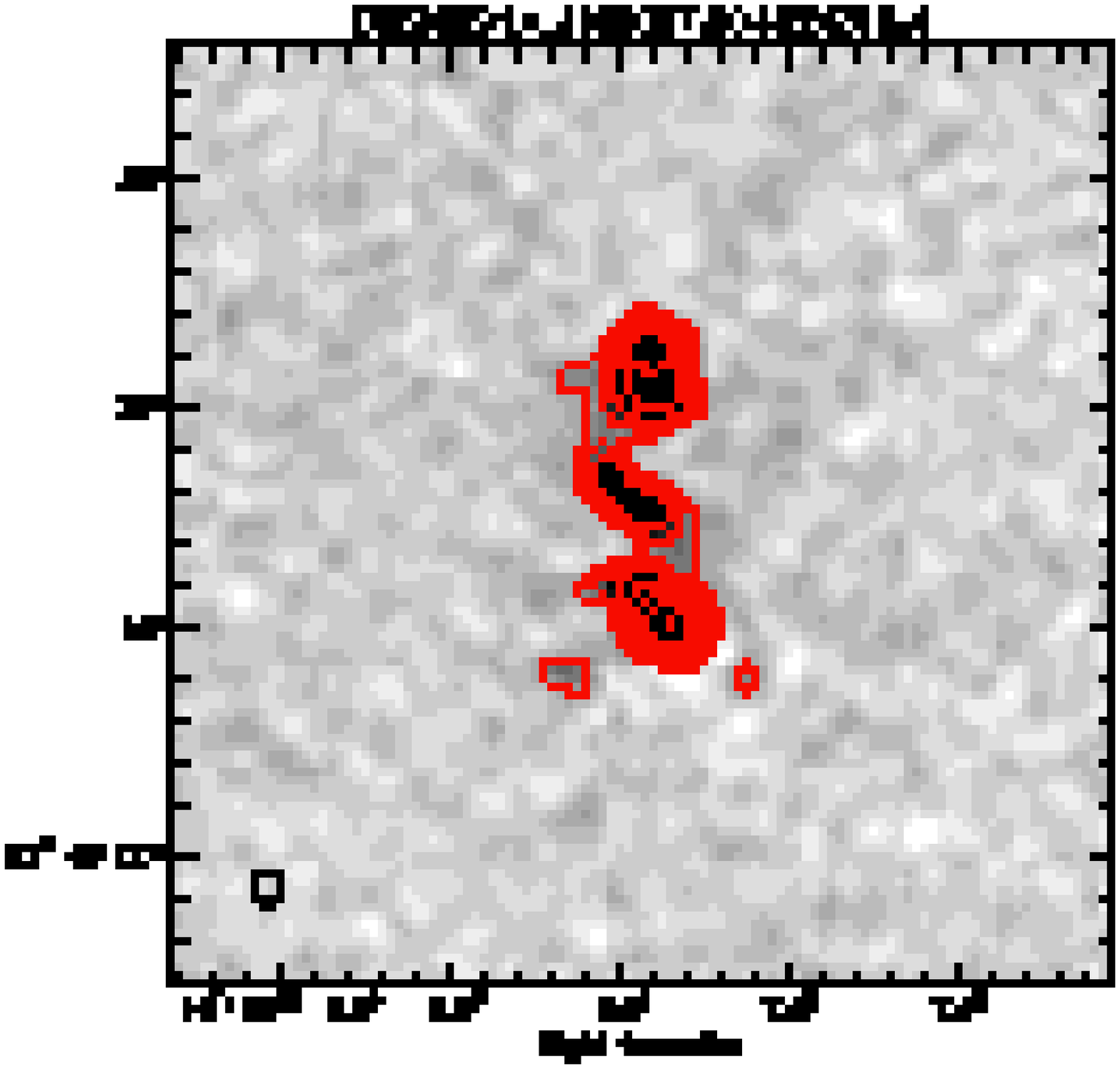}}\\
\clearpage
\resizebox{.9\hsize}{!}{\includegraphics[]{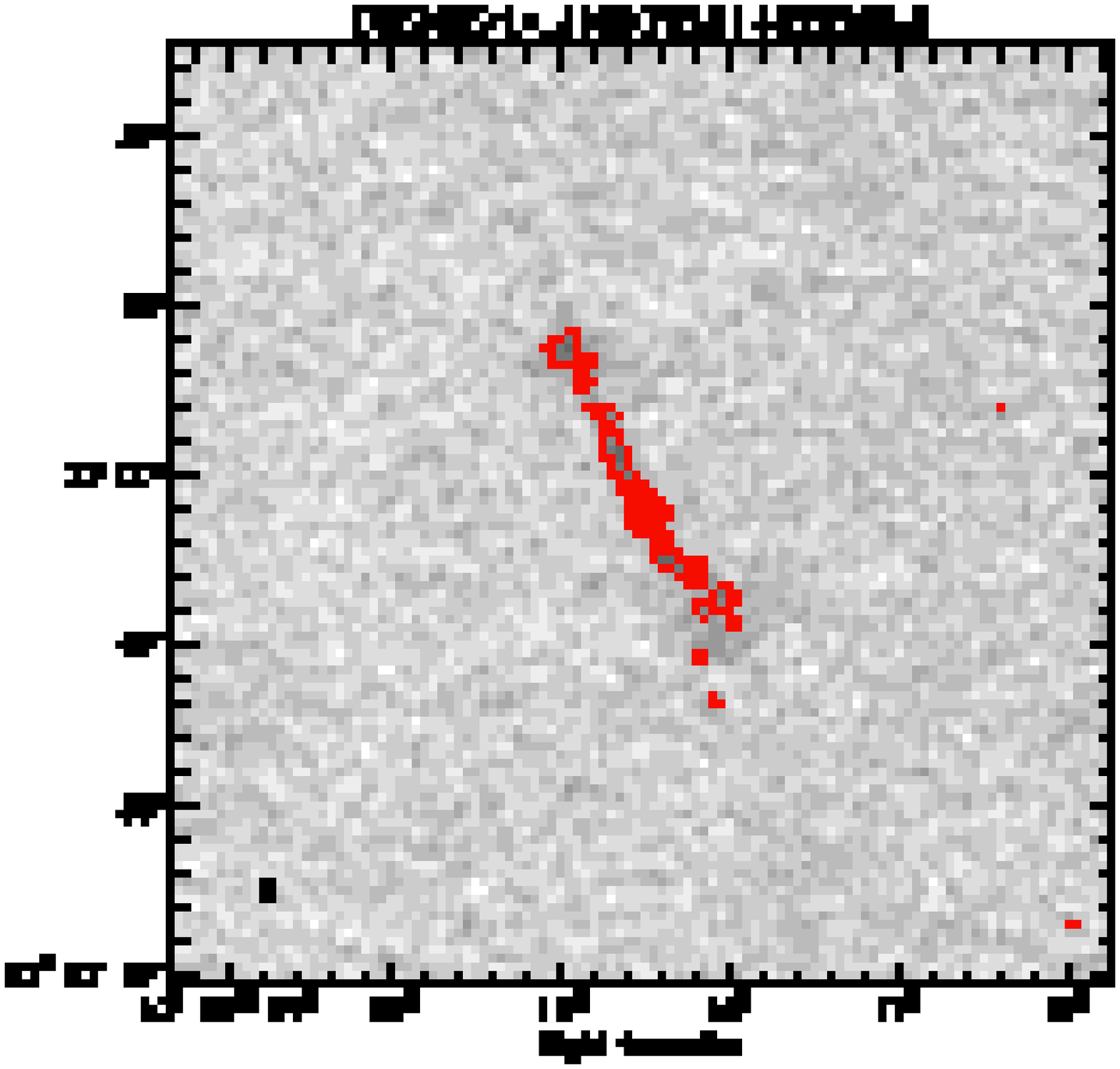}
                      \includegraphics[]{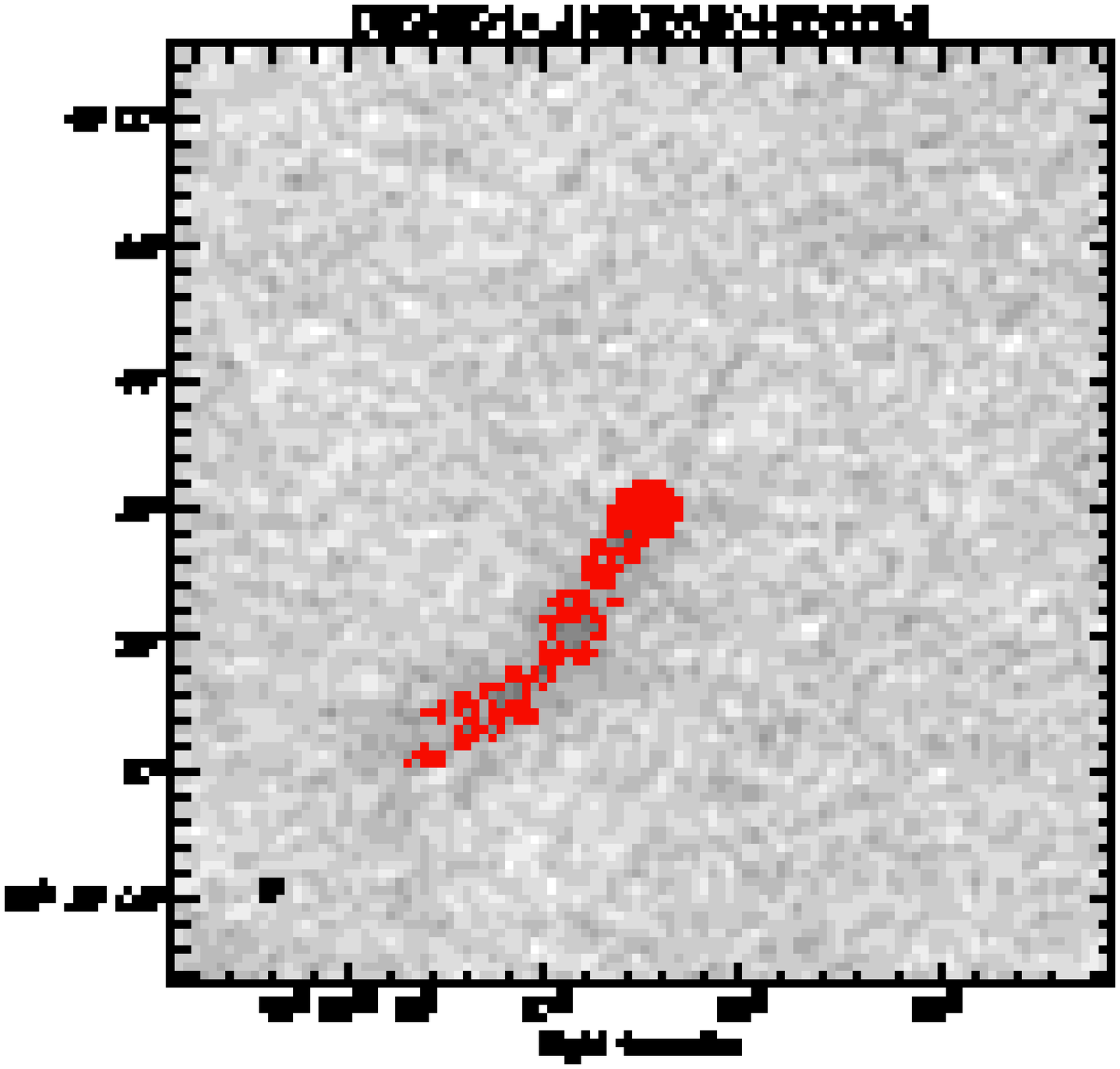}
                      \includegraphics[]{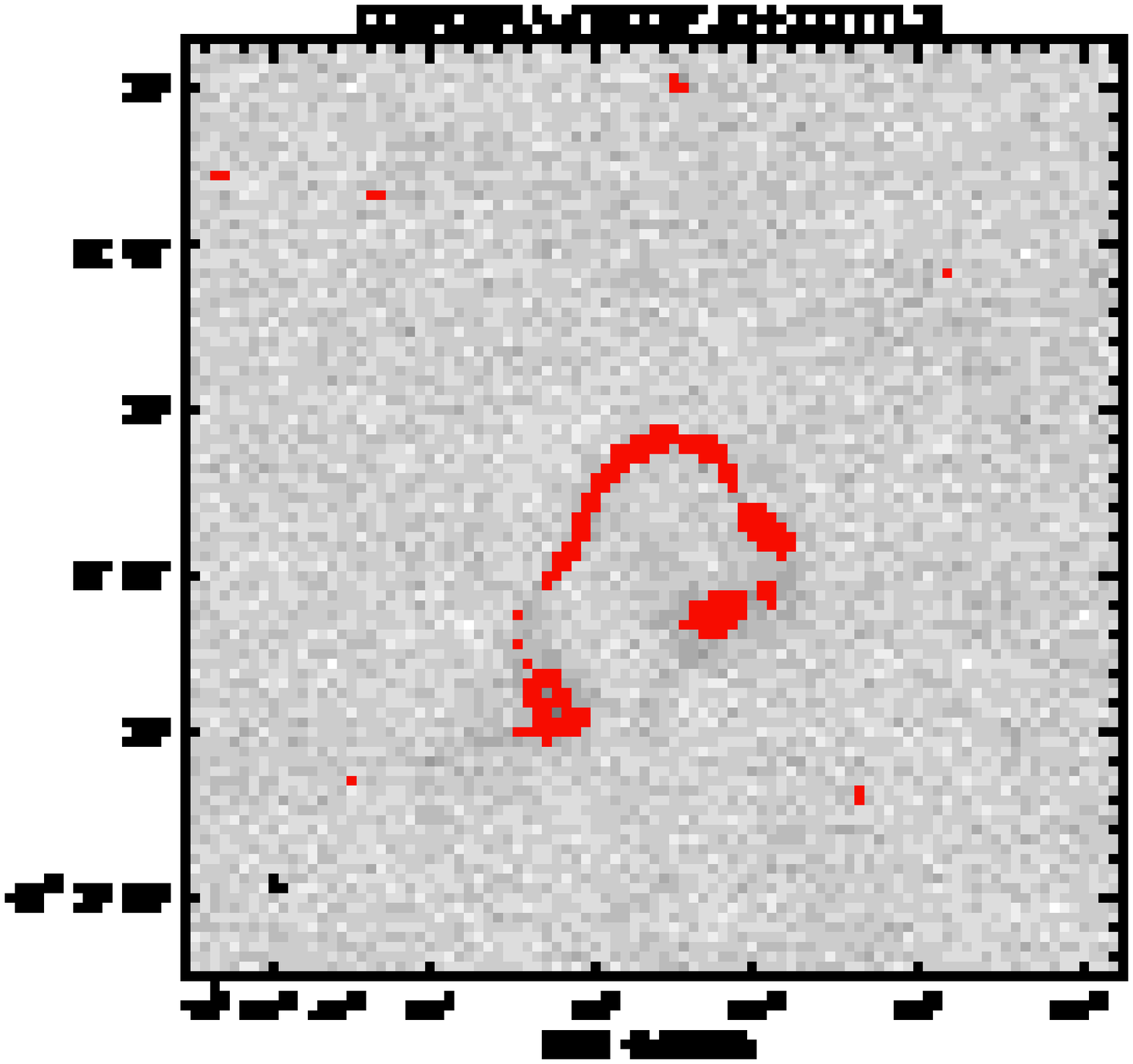}}\\
\resizebox{.9\hsize}{!}{\includegraphics[]{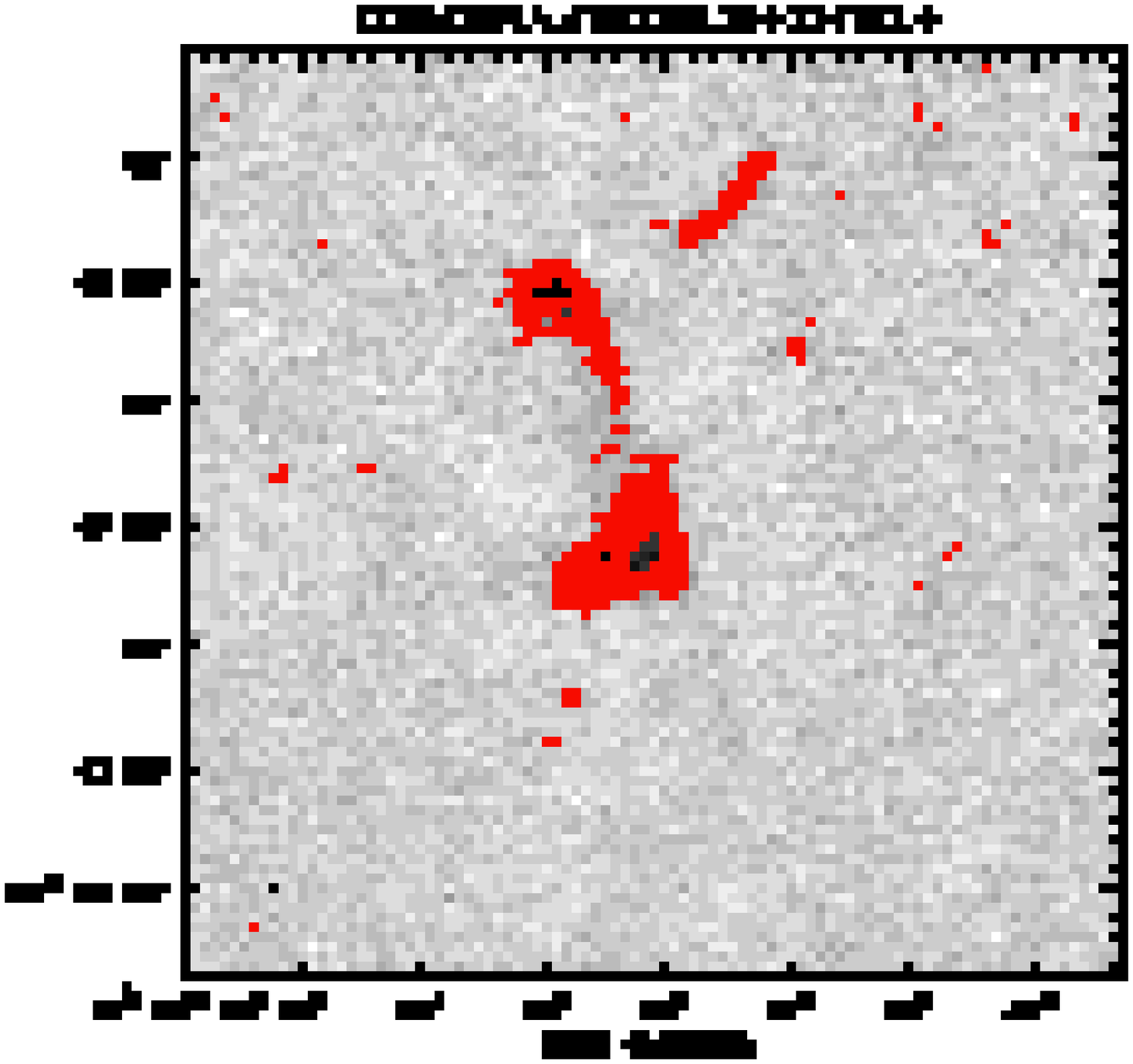}
                      \includegraphics[]{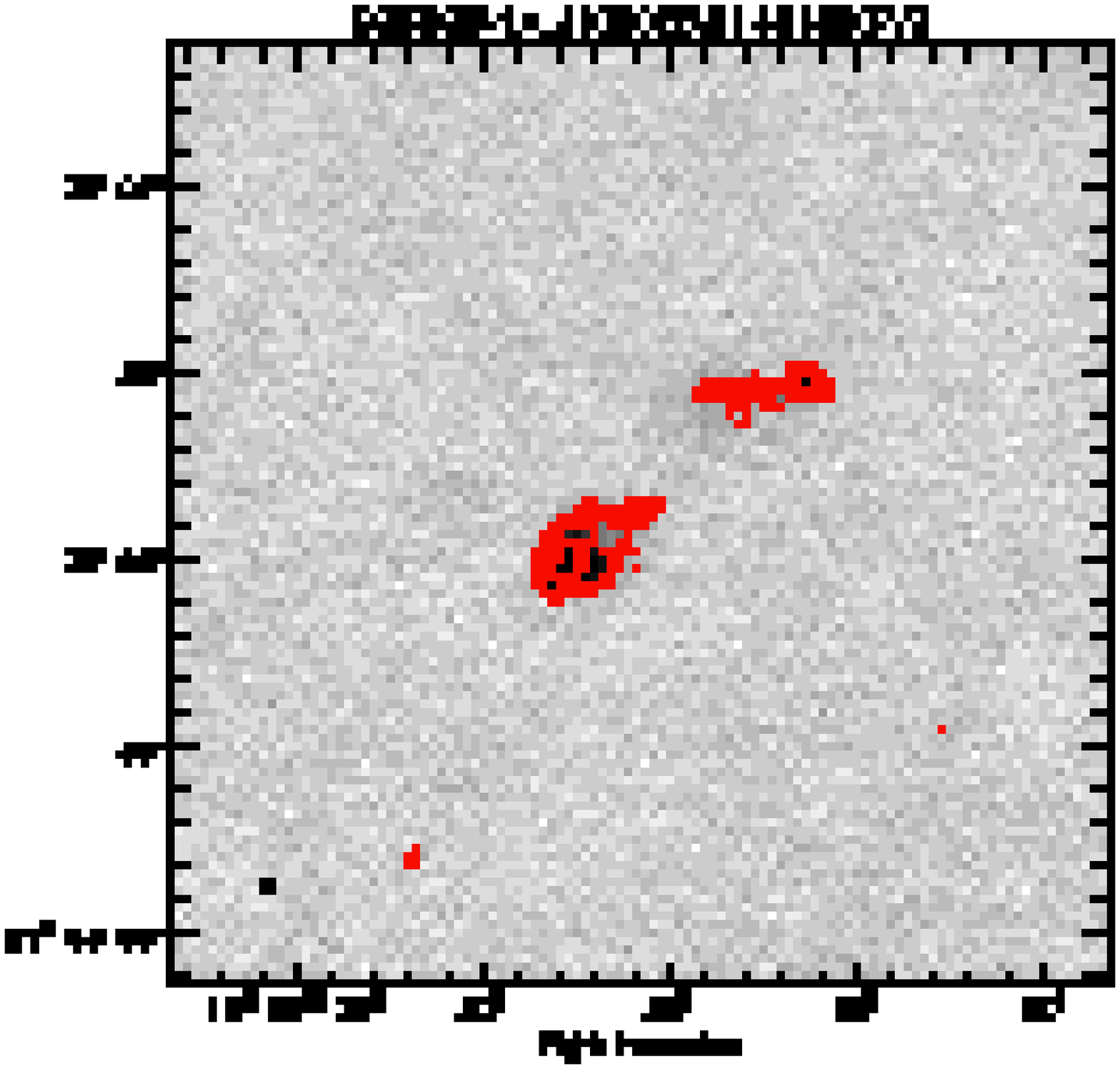}
                      \includegraphics[]{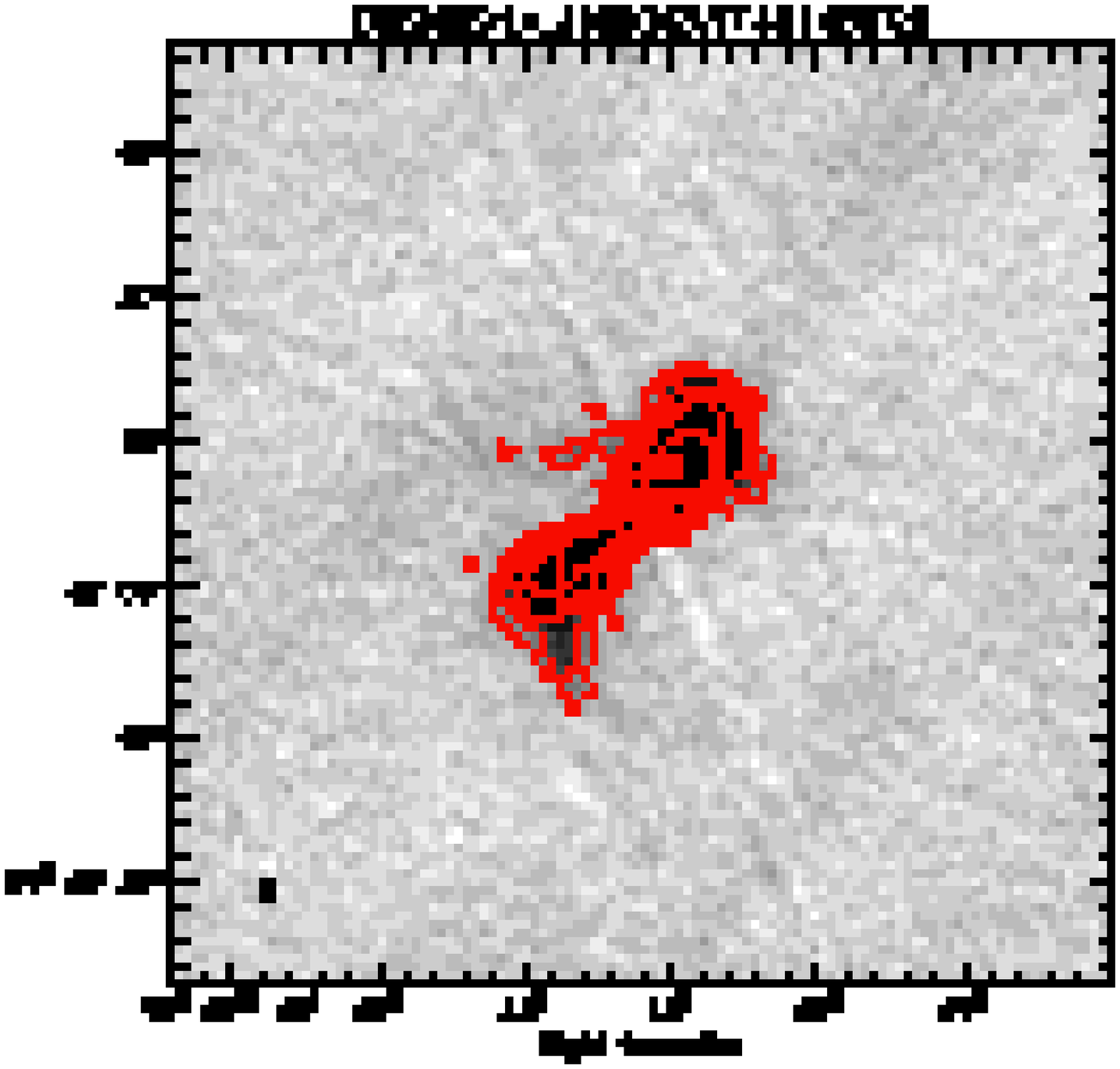}}\\
\resizebox{.9\hsize}{!}{\includegraphics[]{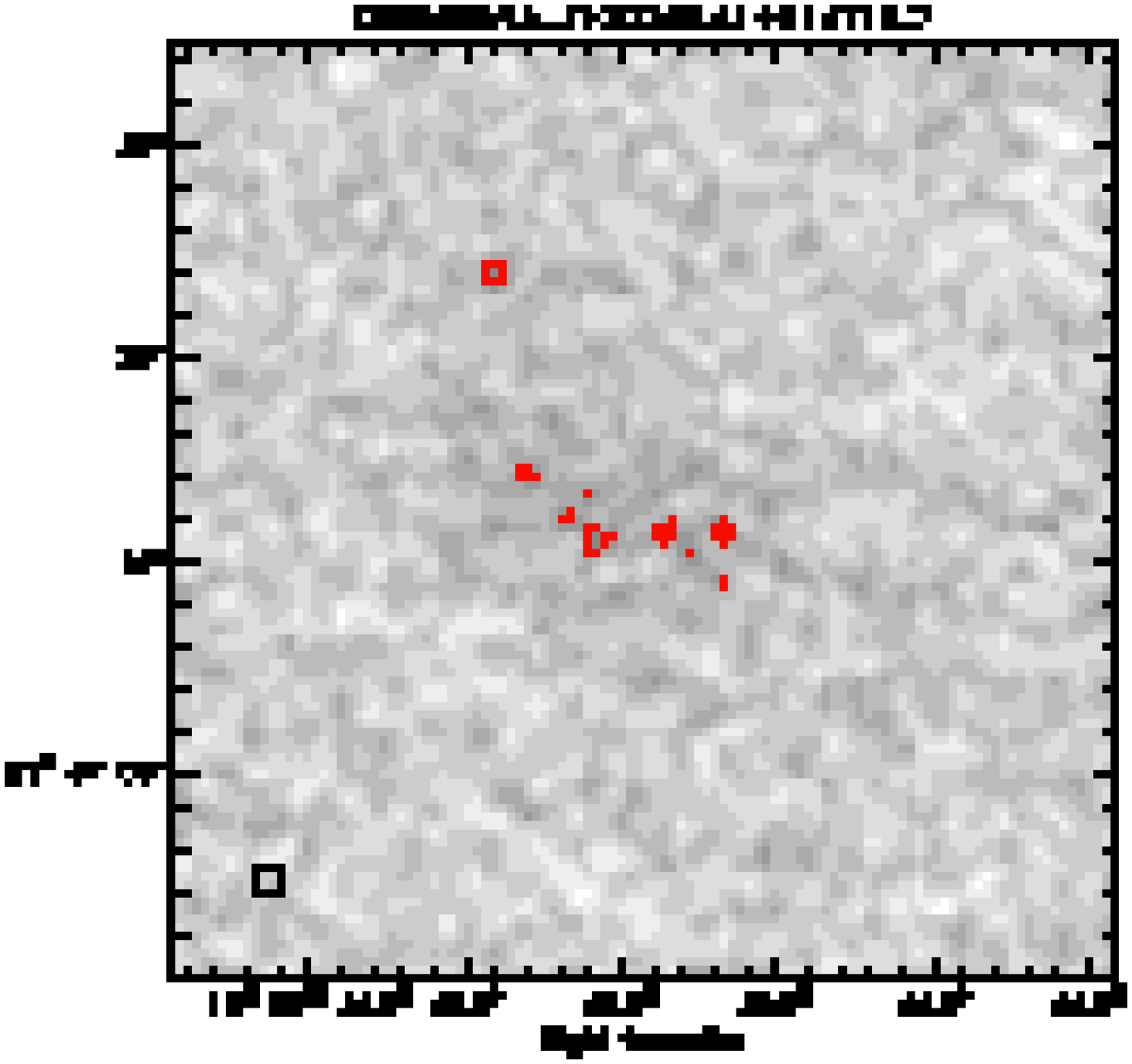}
                      \includegraphics[]{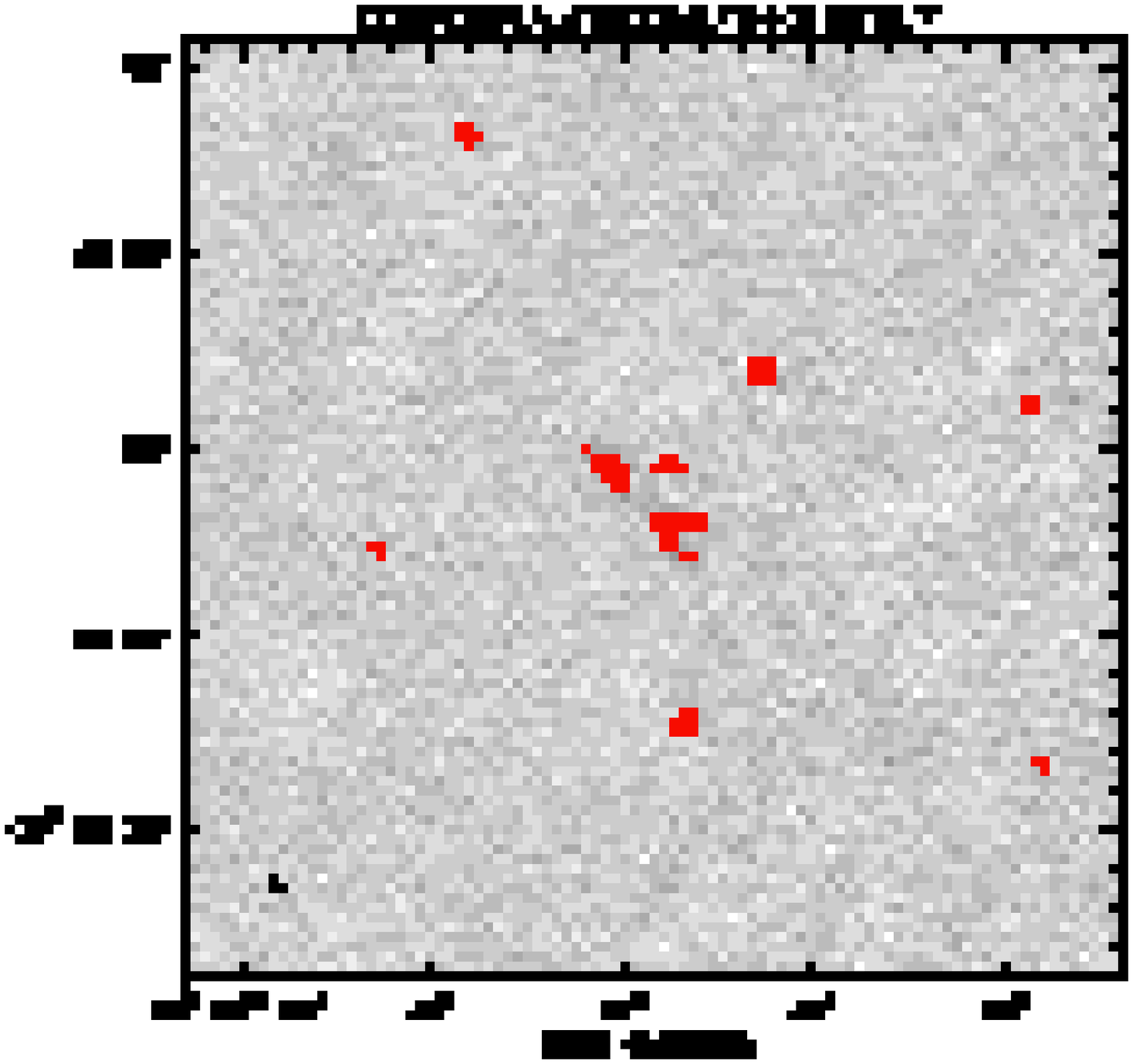}
                      \includegraphics[]{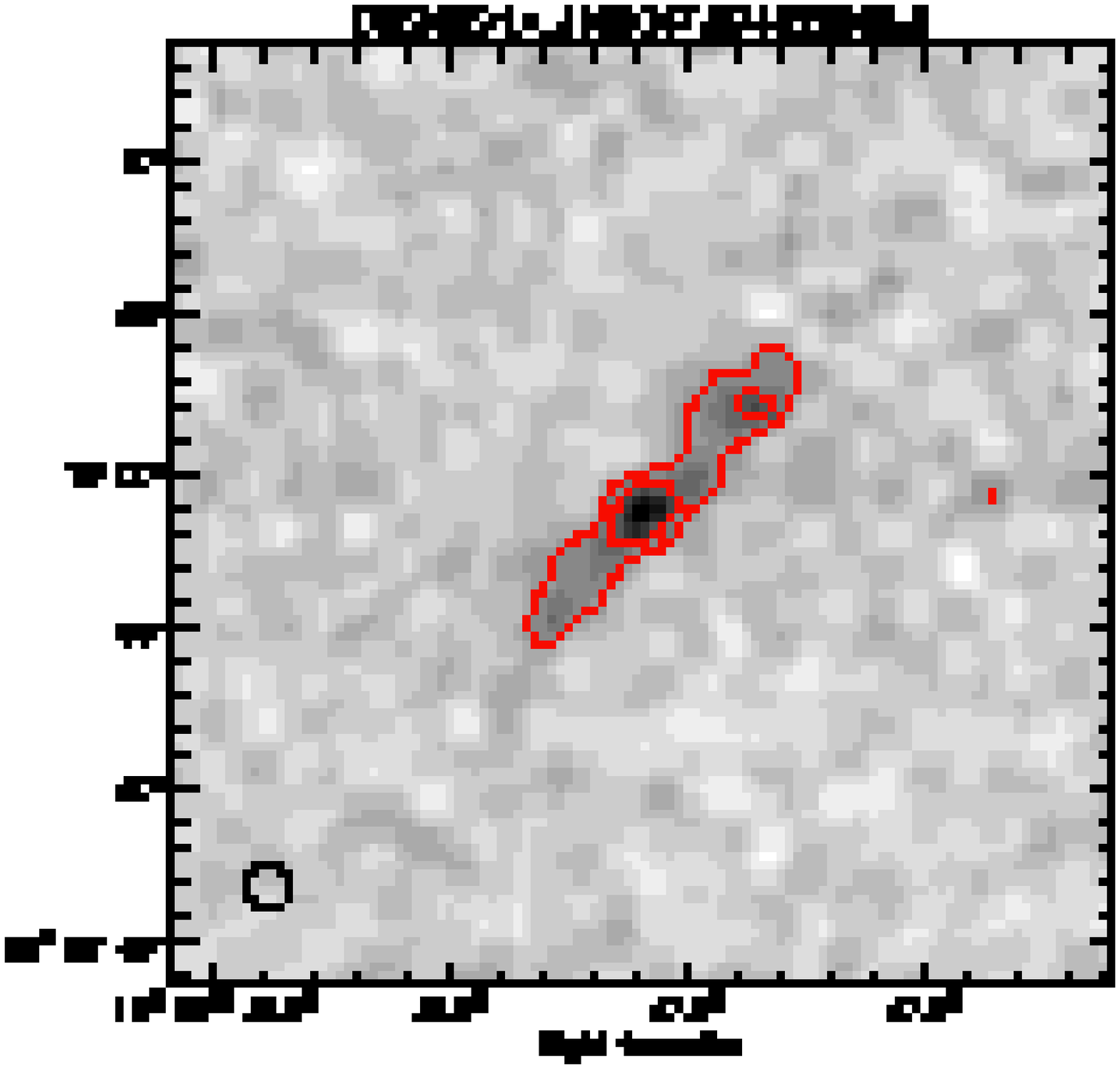}}\\
\resizebox{.9\hsize}{!}{\includegraphics[]{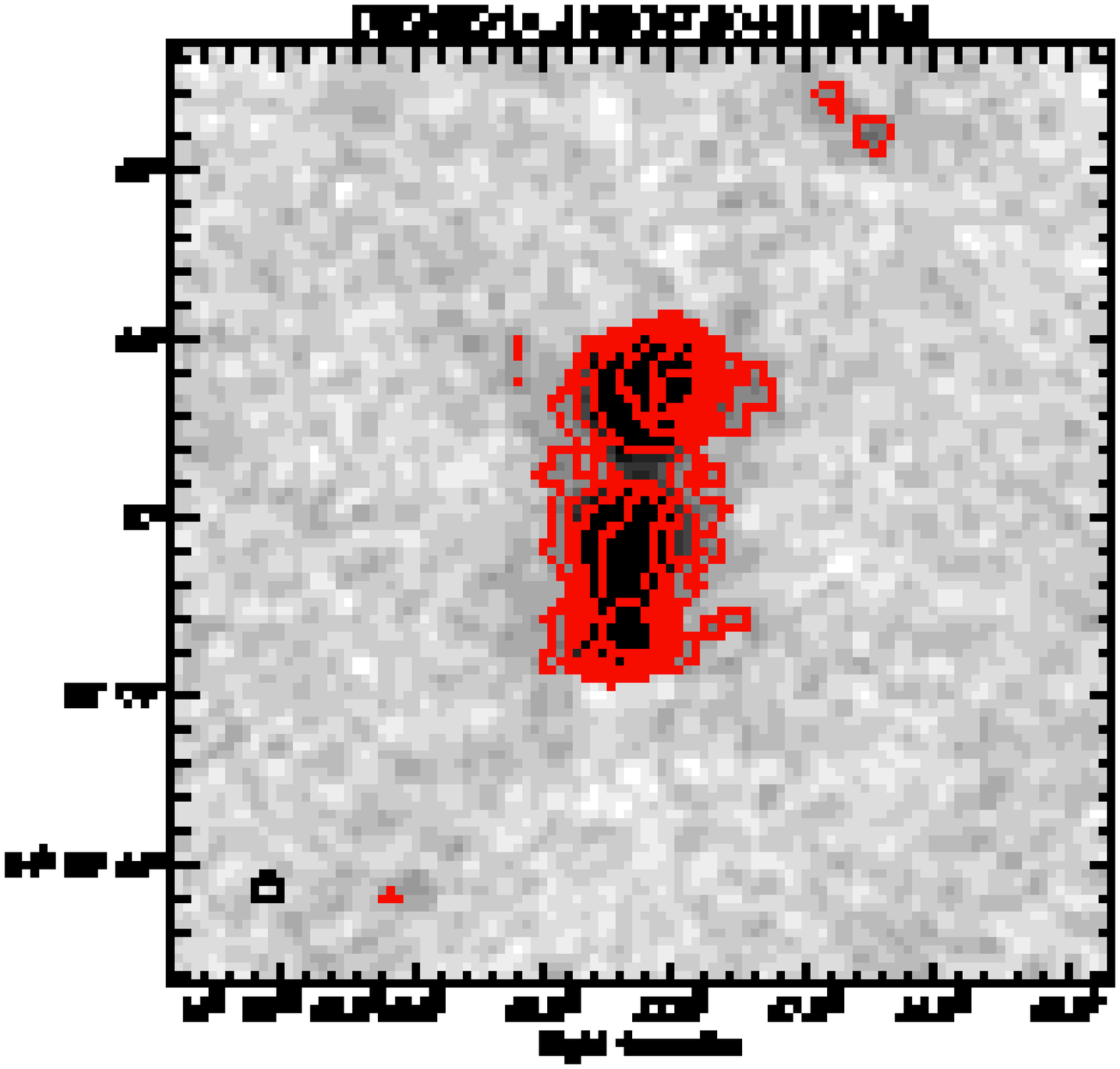}
                      \includegraphics[]{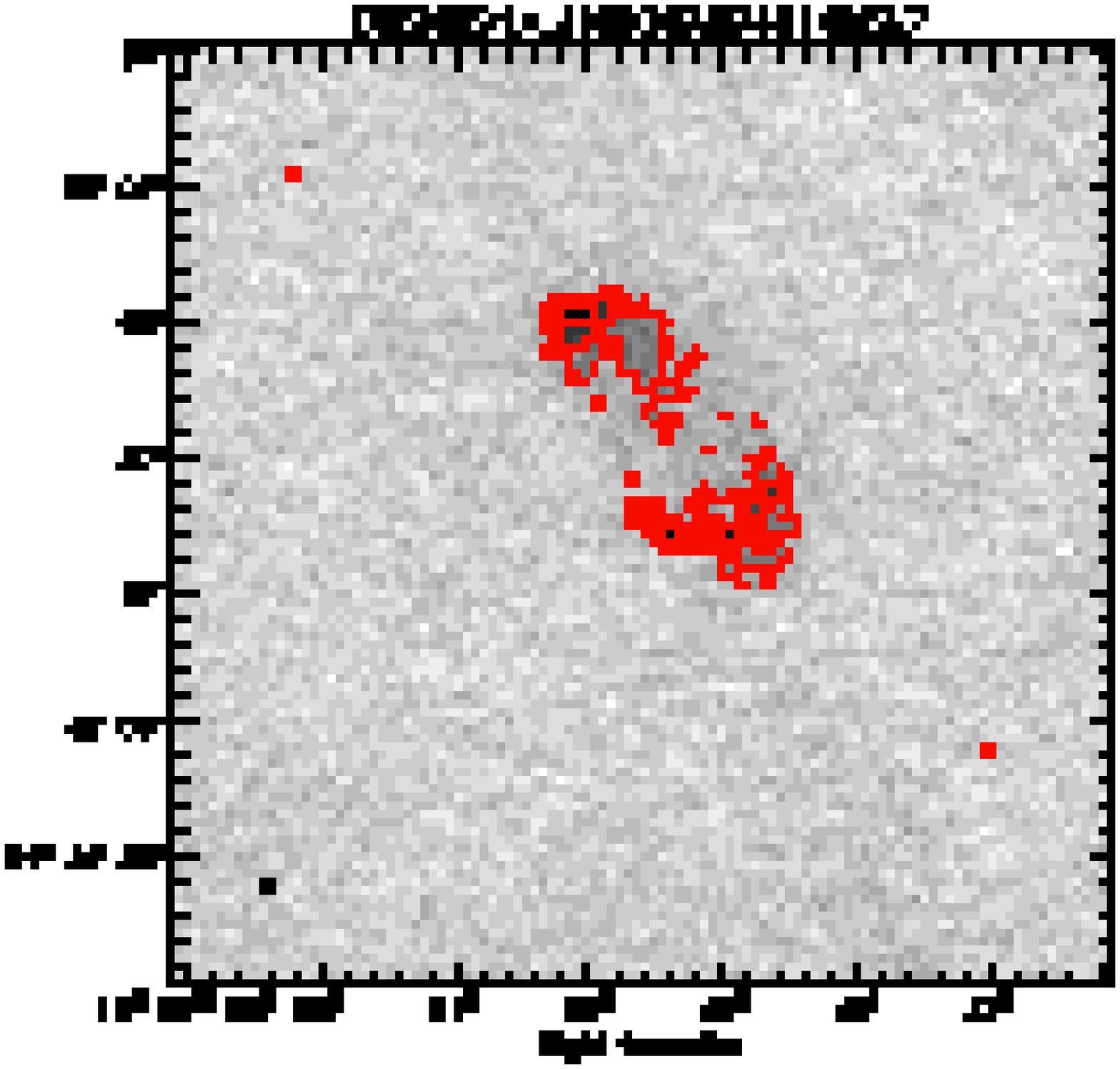}
                      \includegraphics[]{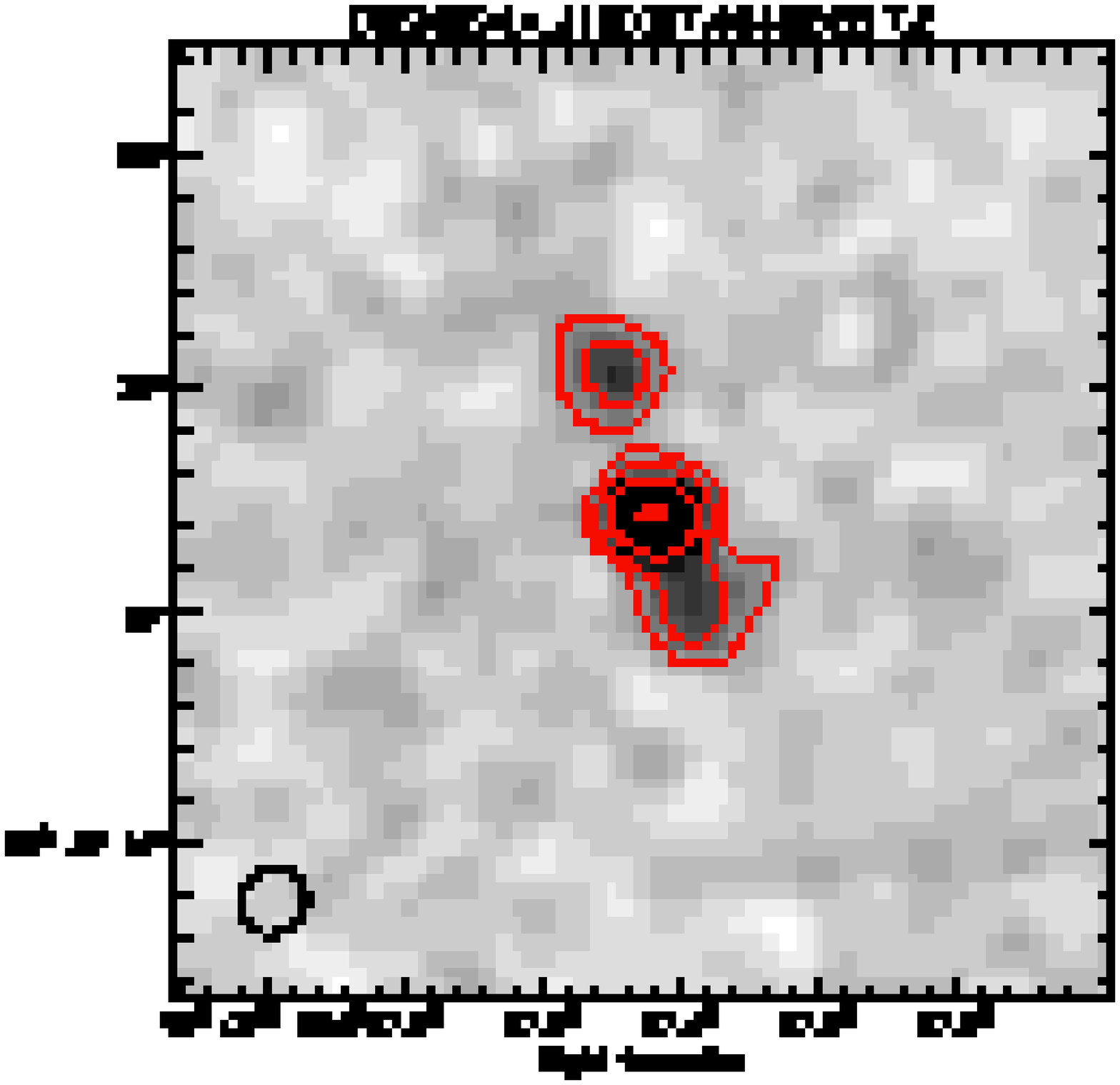}}
\clearpage
\resizebox{.9\hsize}{!}{\includegraphics[]{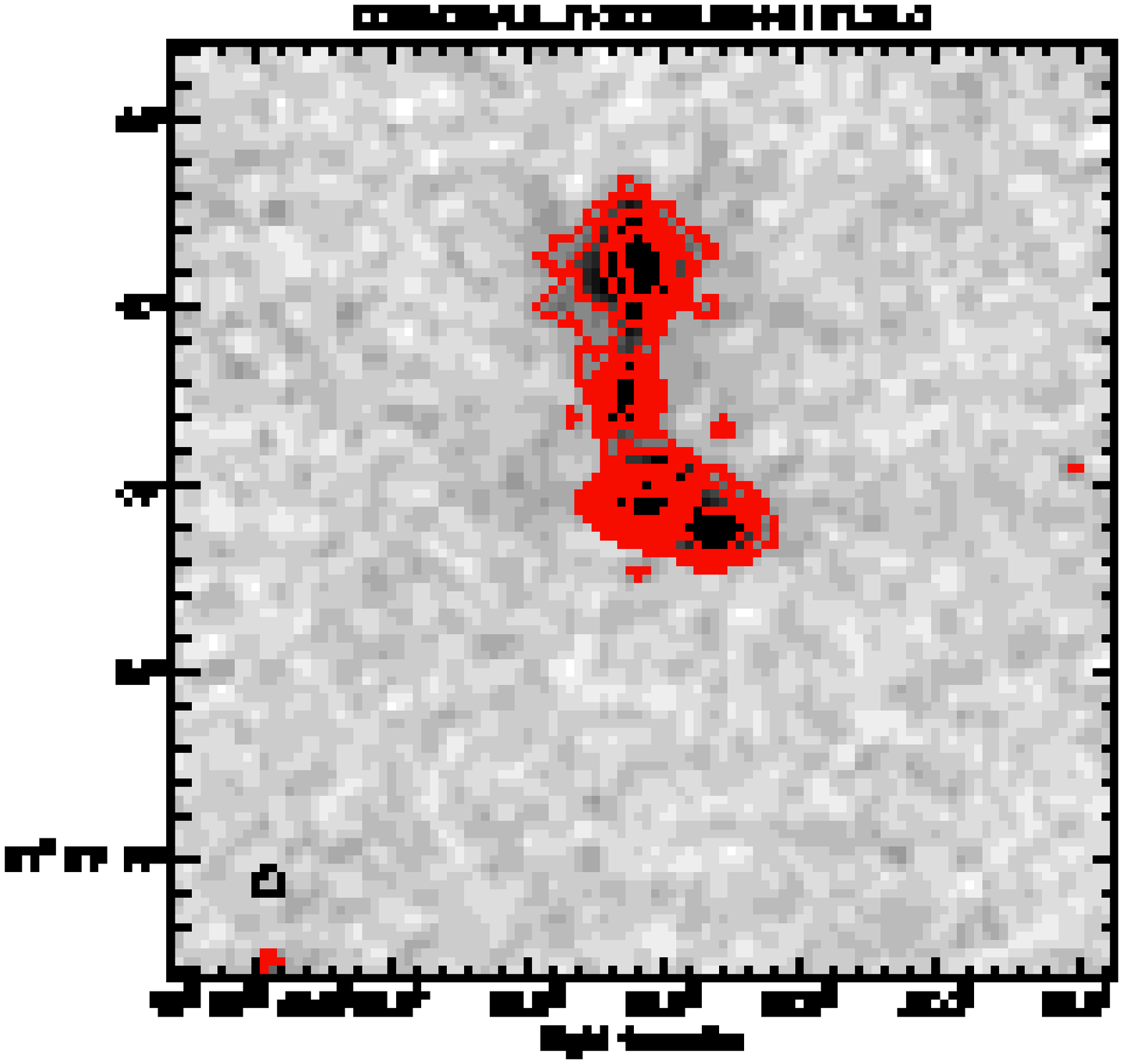}
                      \includegraphics[]{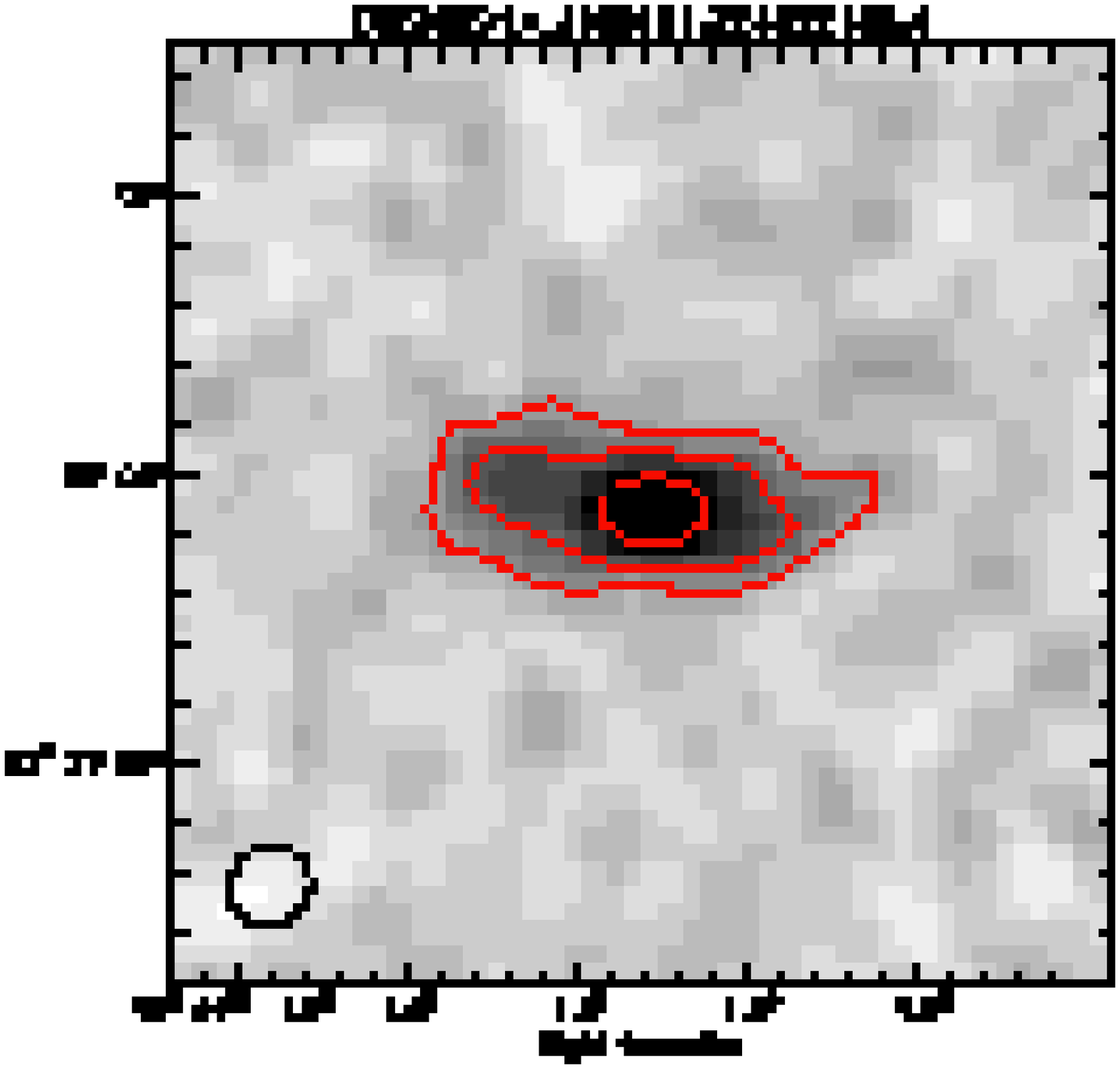}
                      \includegraphics[]{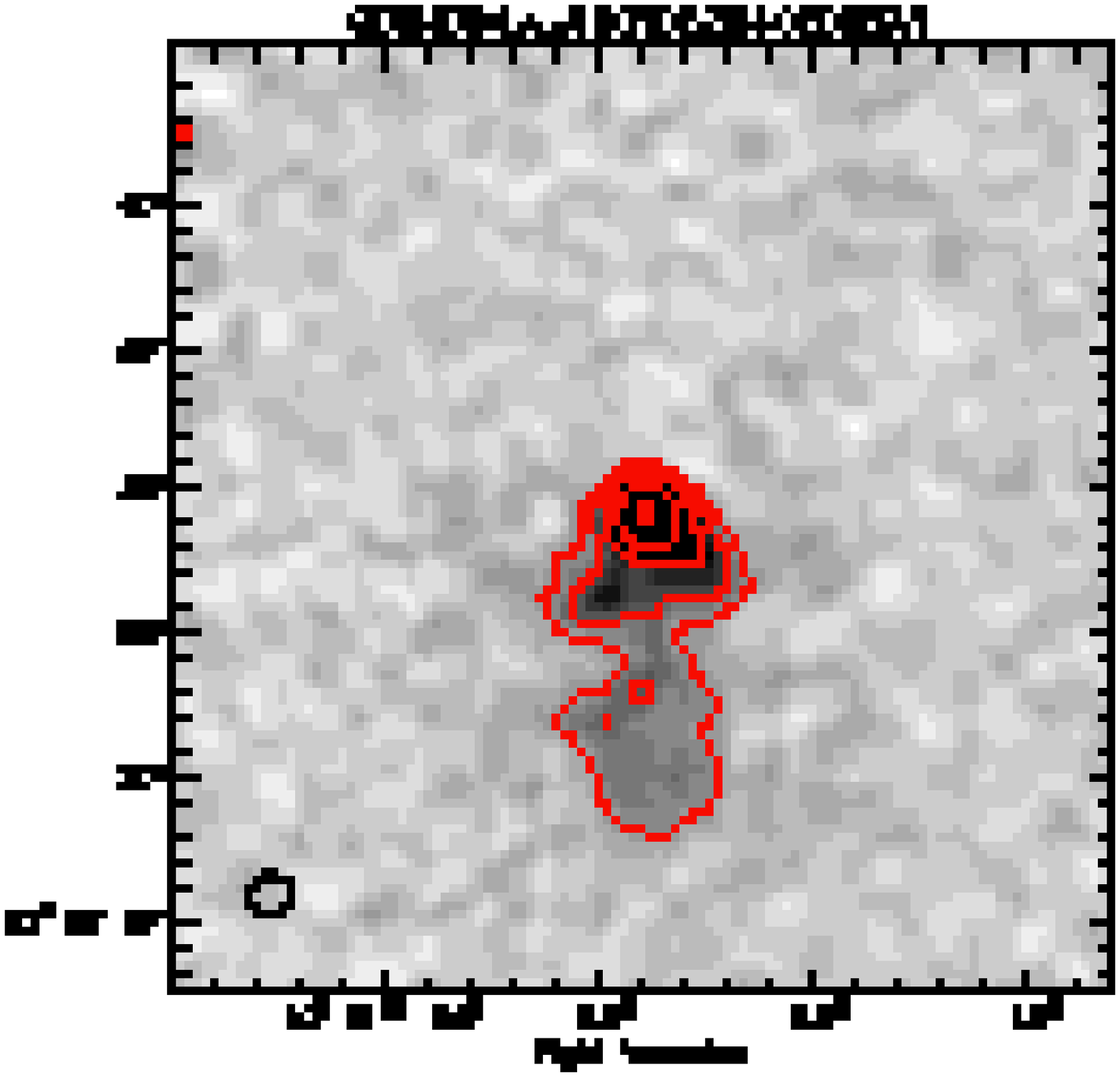}}\\
\resizebox{.9\hsize}{!}{\includegraphics[]{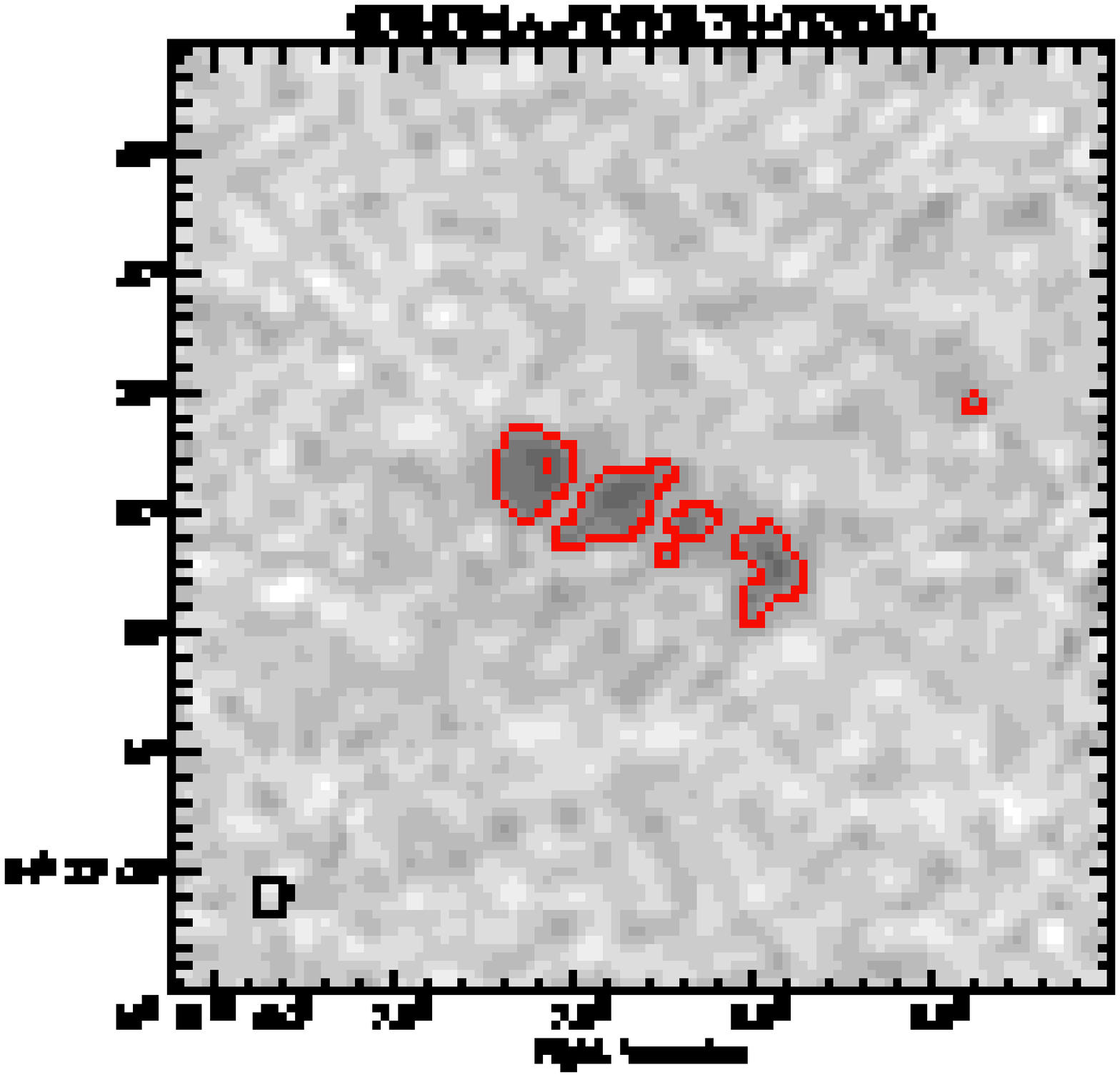}
                      \includegraphics[]{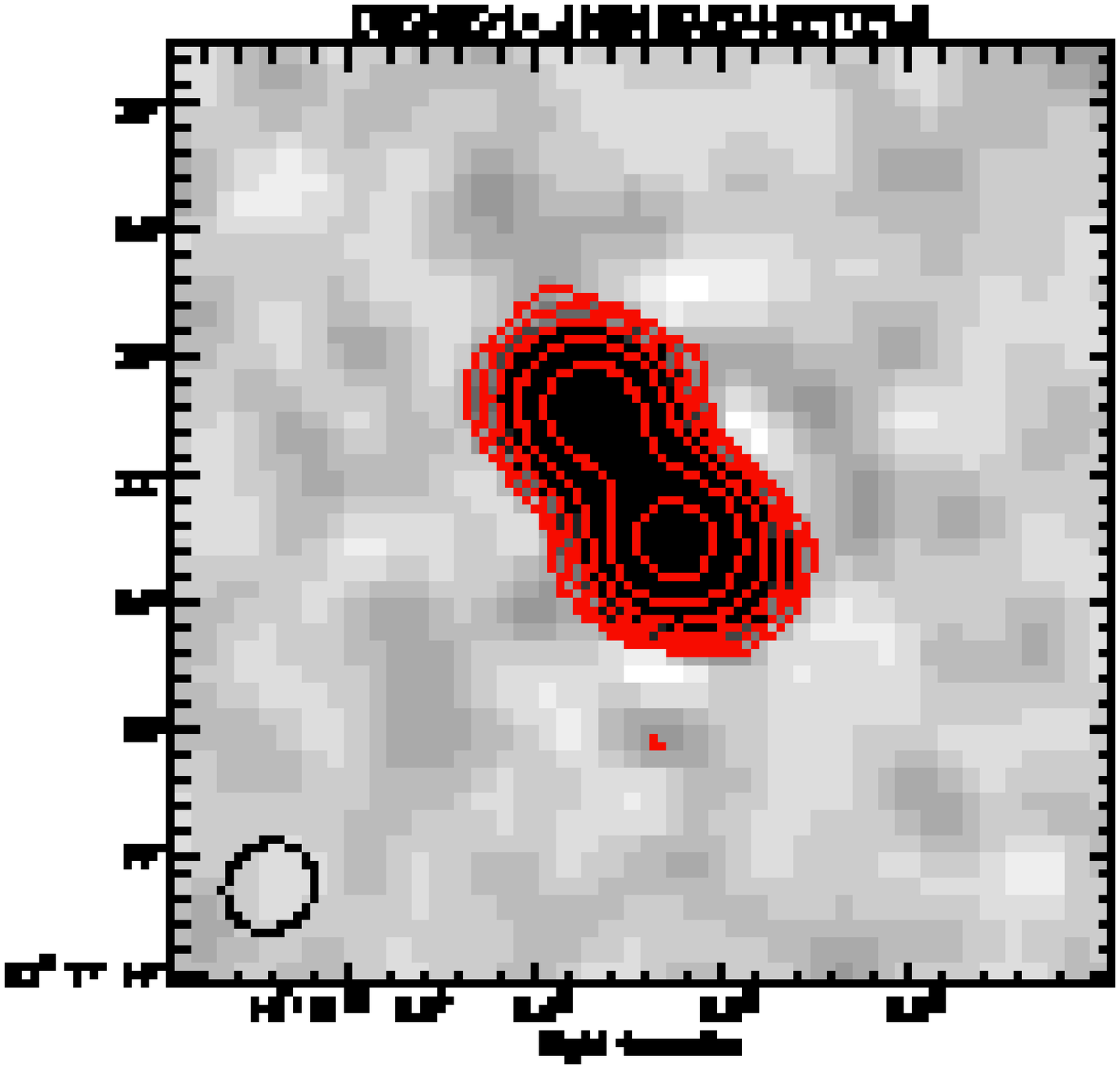}
                      \includegraphics[]{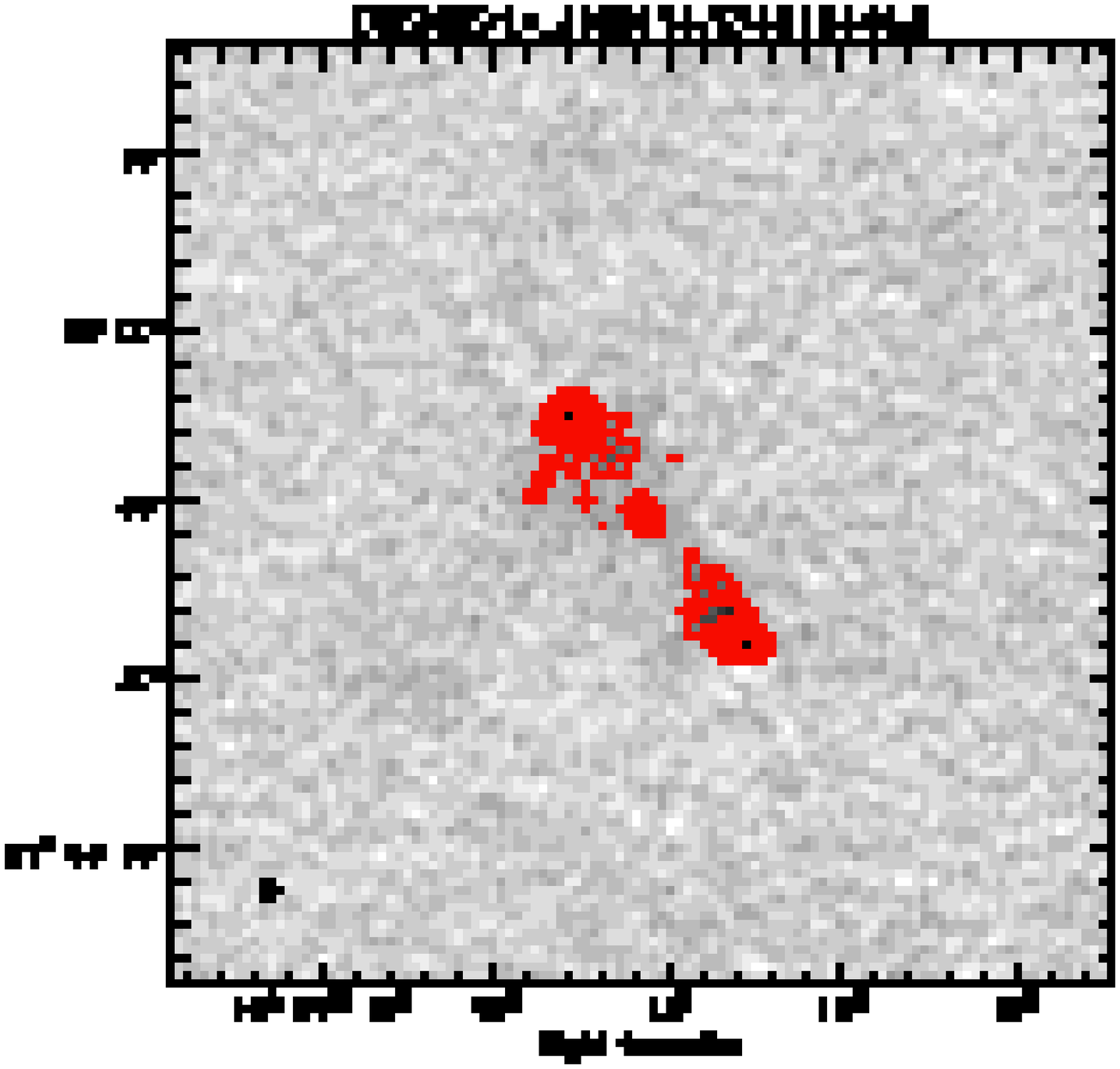}}\\
\resizebox{.9\hsize}{!}{\includegraphics[]{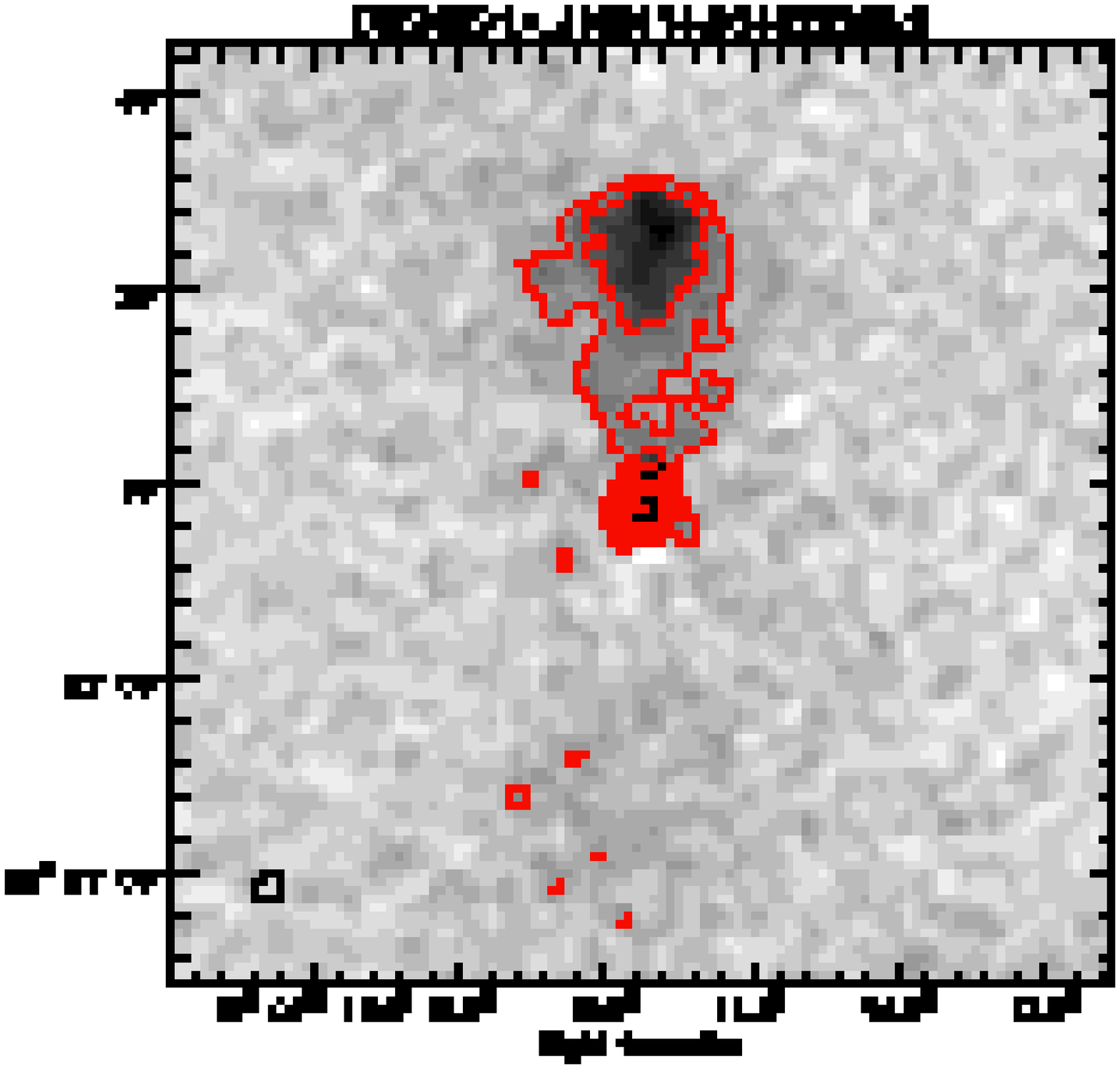}
                      \includegraphics[]{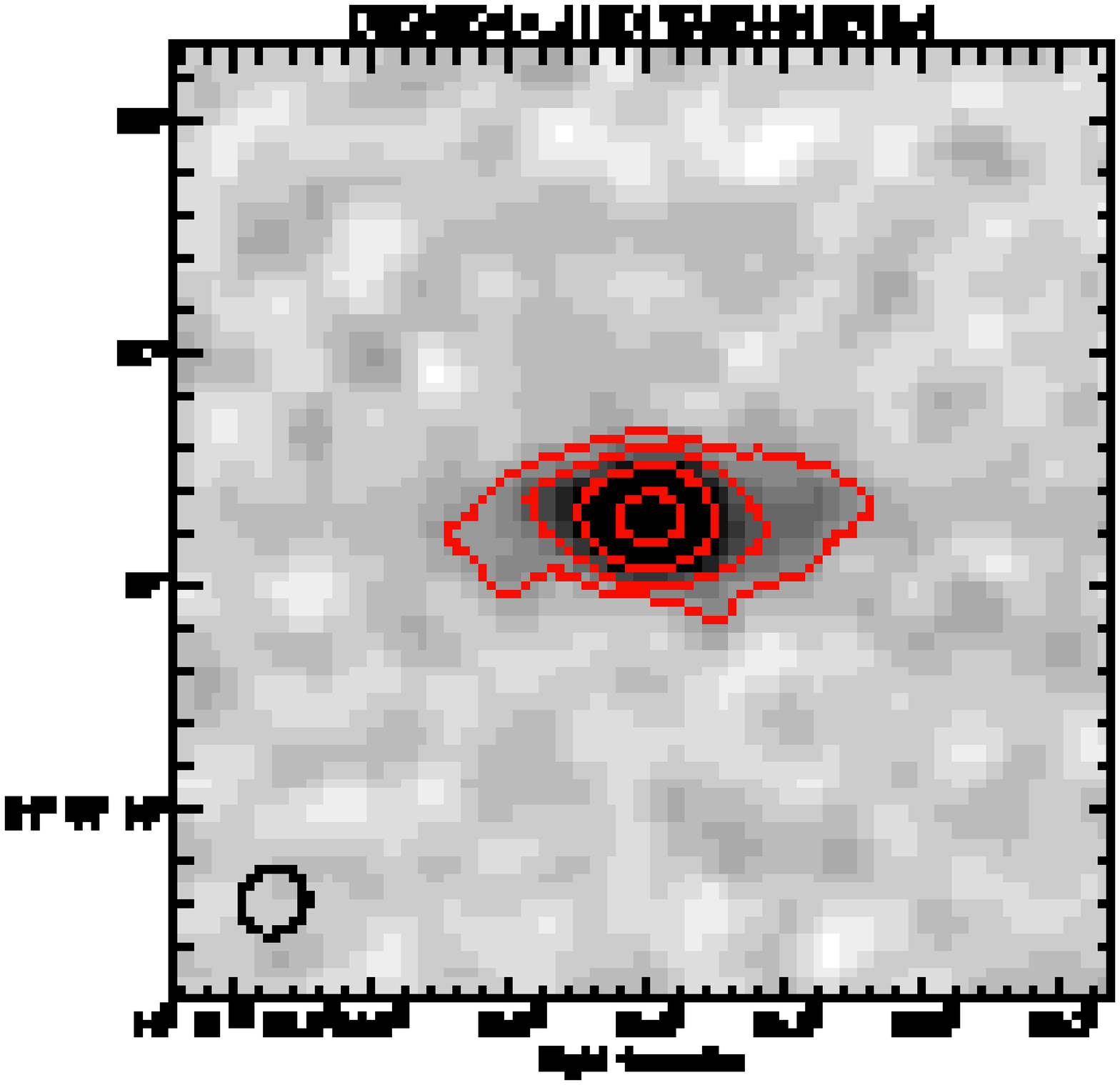}
                      \includegraphics[]{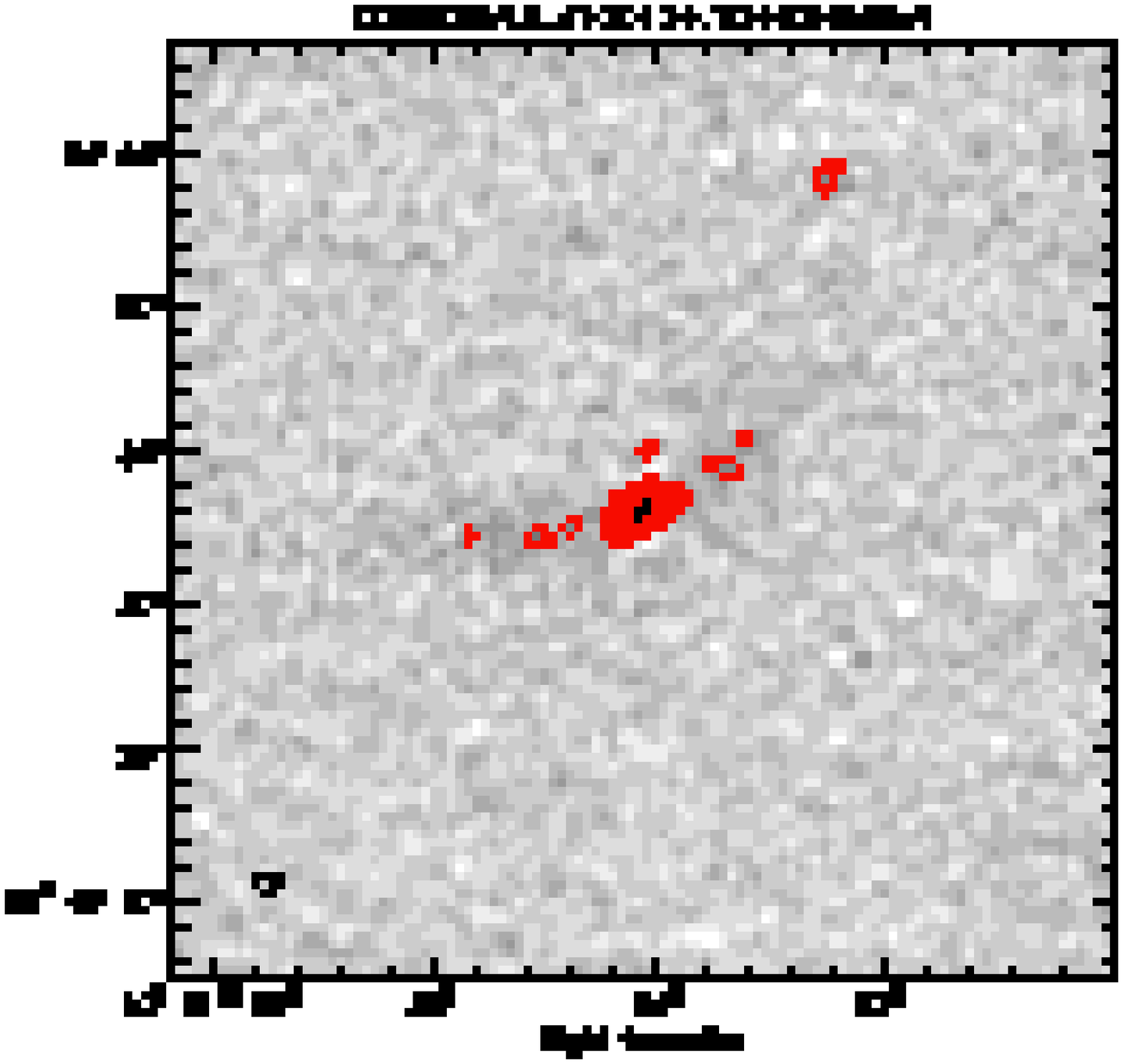}}\\
\resizebox{.9\hsize}{!}{\includegraphics[]{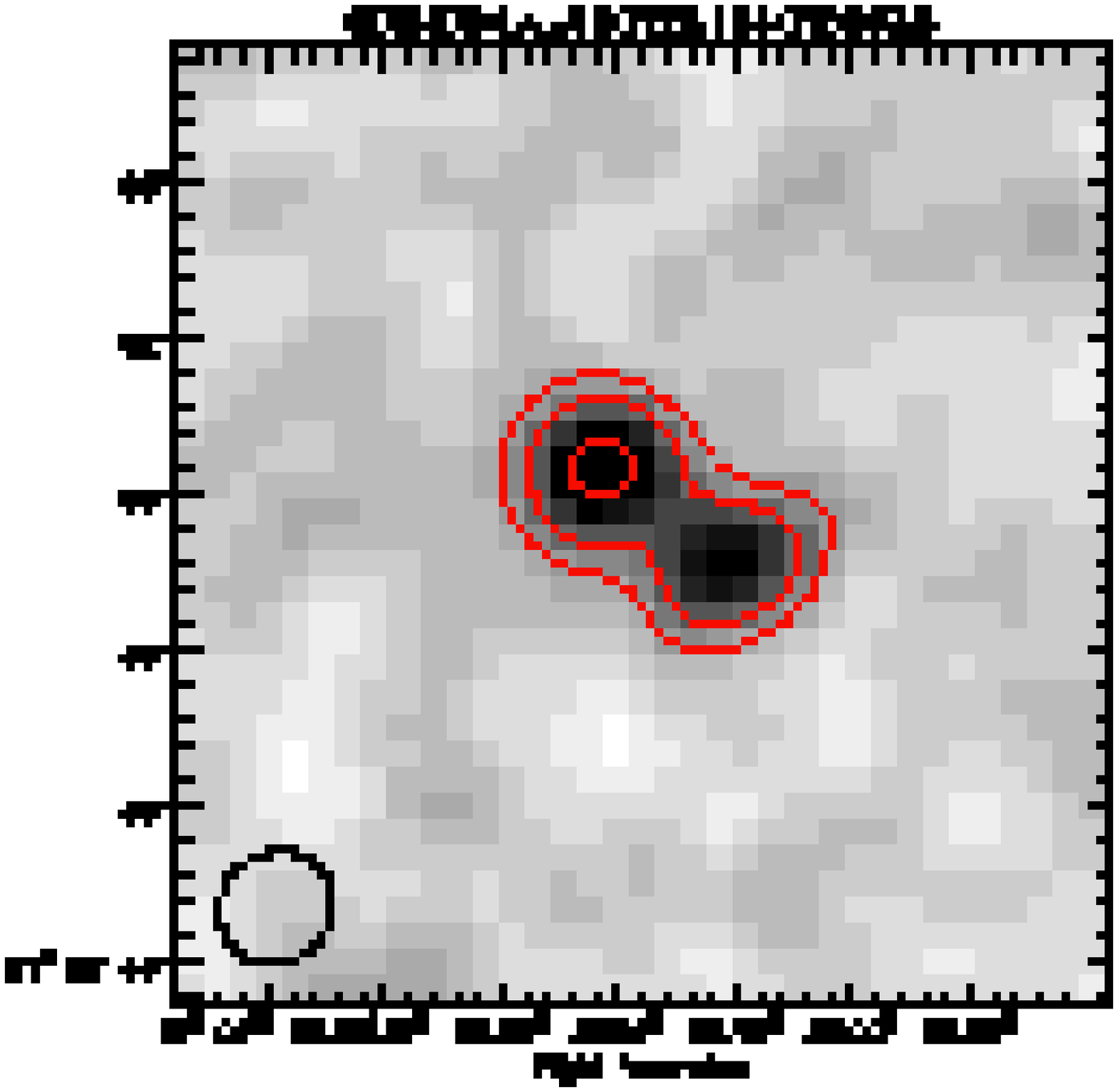}
                      \includegraphics[]{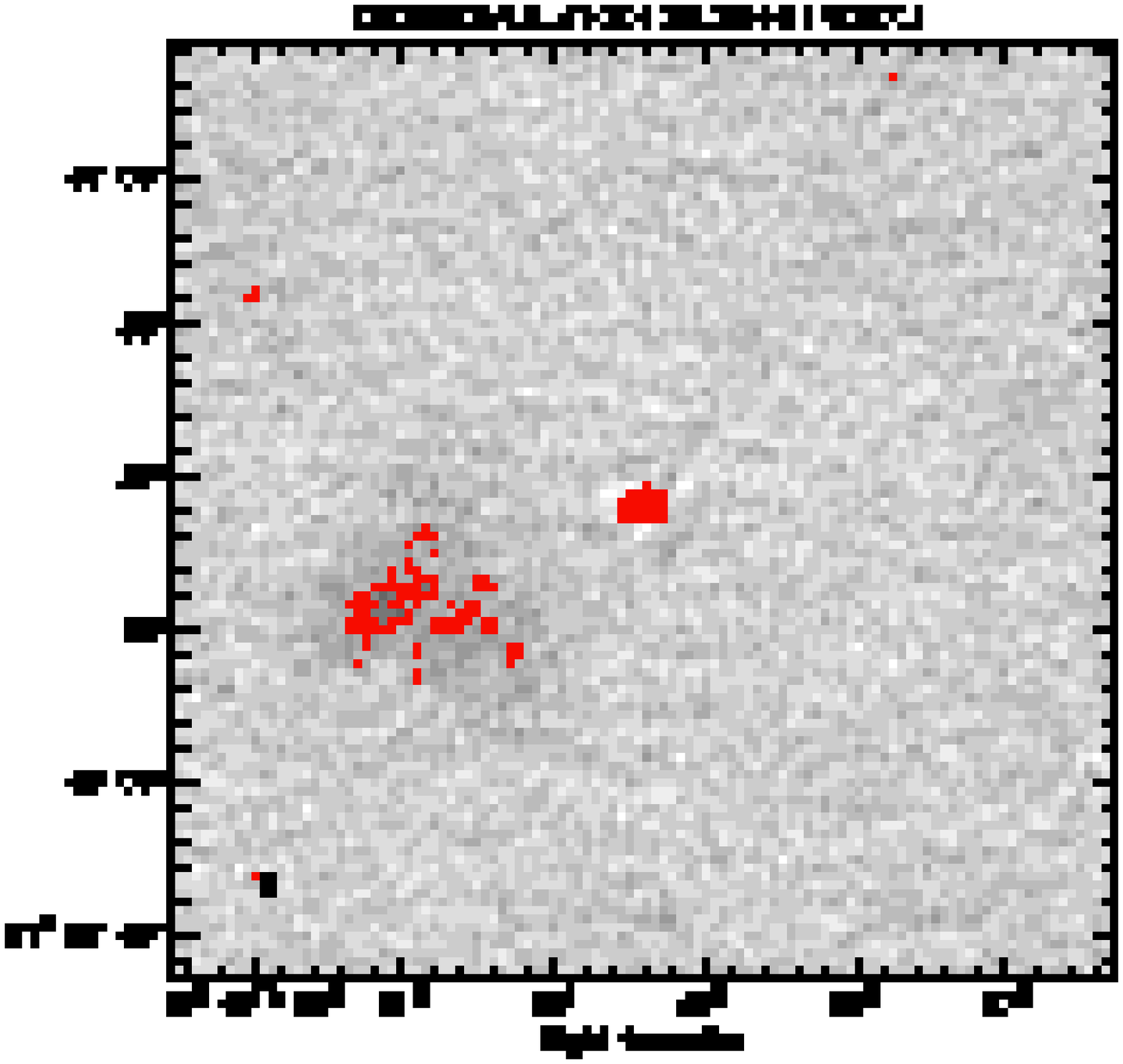}
                      \includegraphics[]{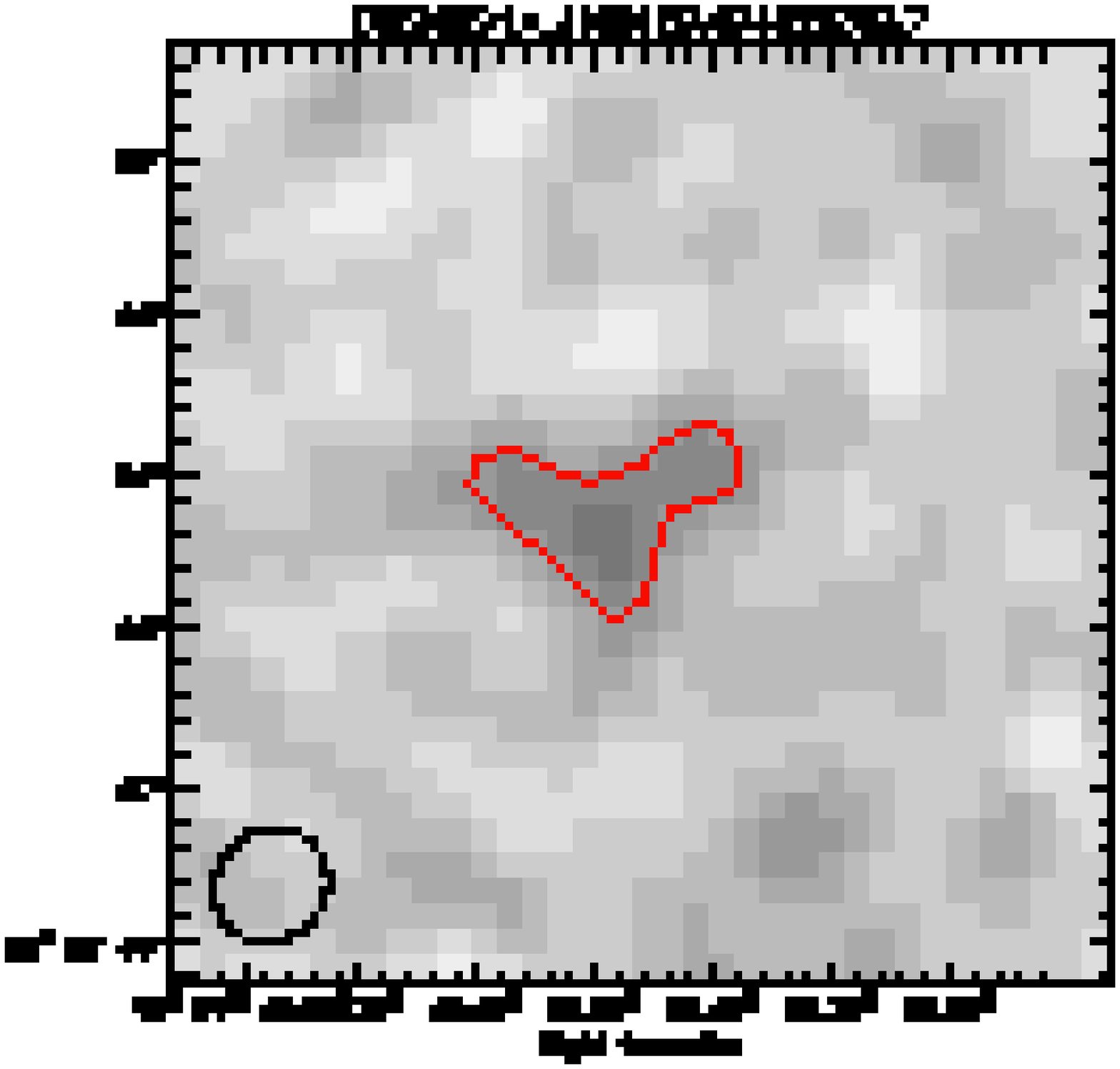}}
\clearpage
\resizebox{.9\hsize}{!}{\includegraphics[]{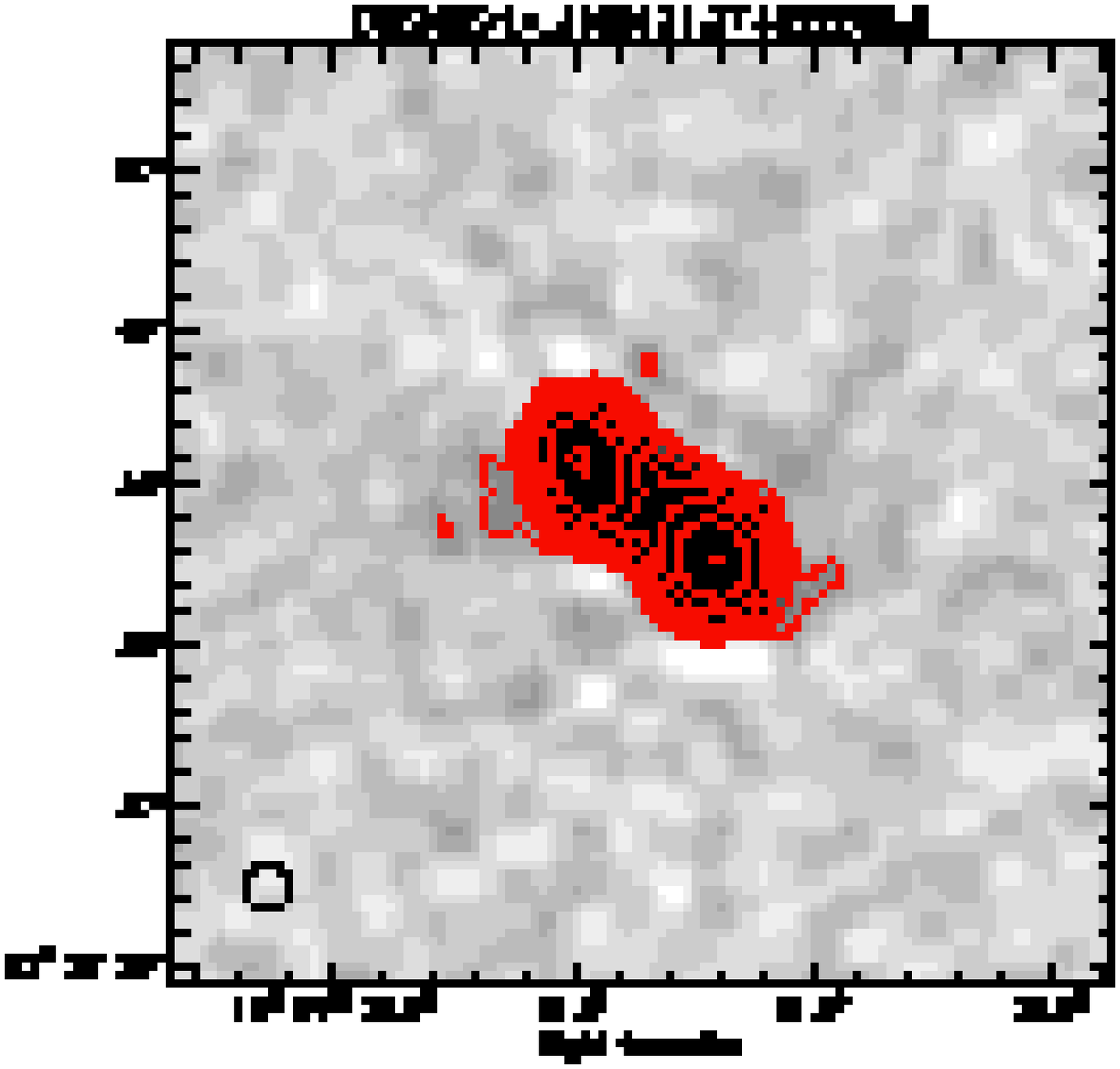}
                      \includegraphics[]{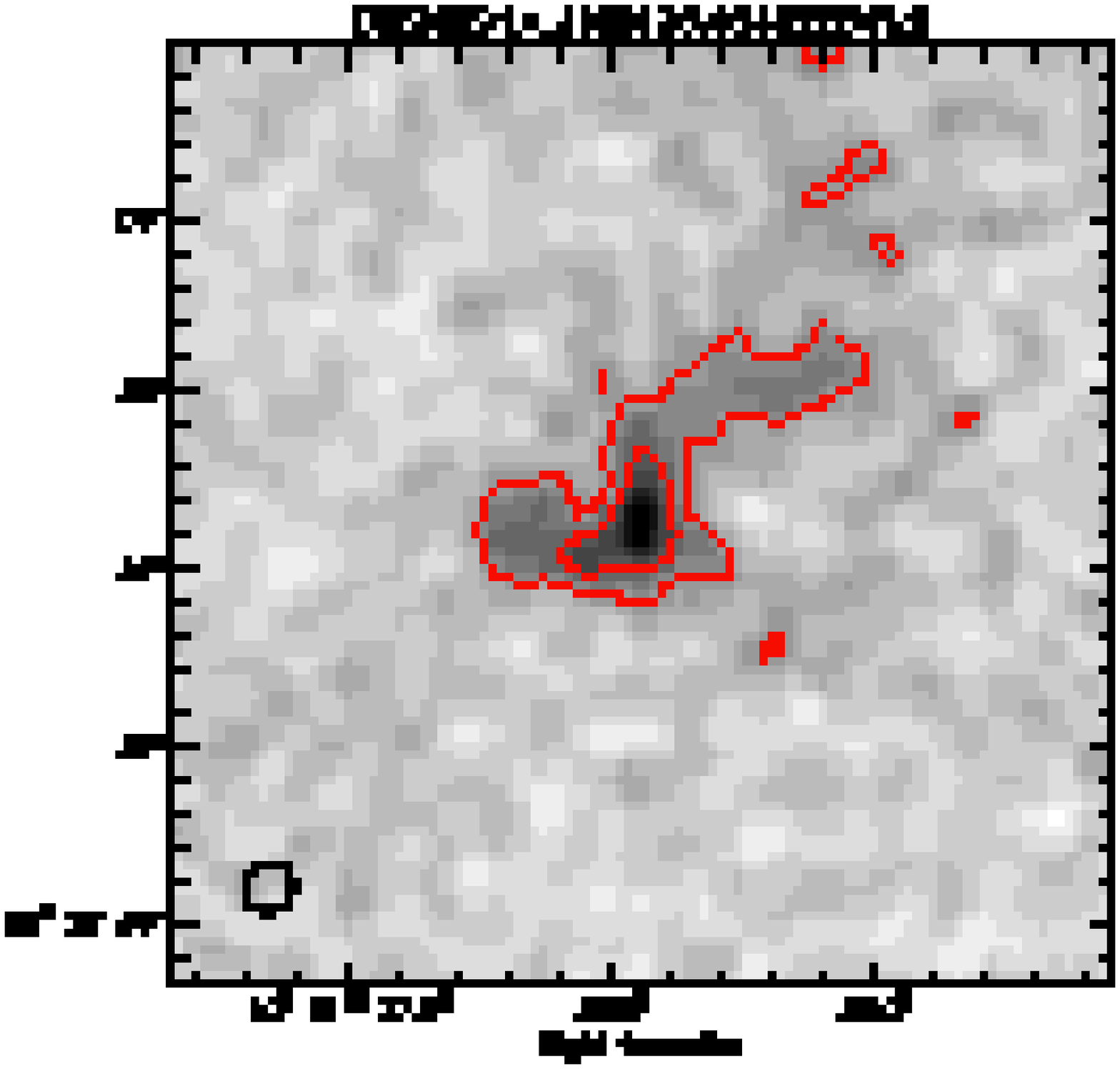}
                      \includegraphics[]{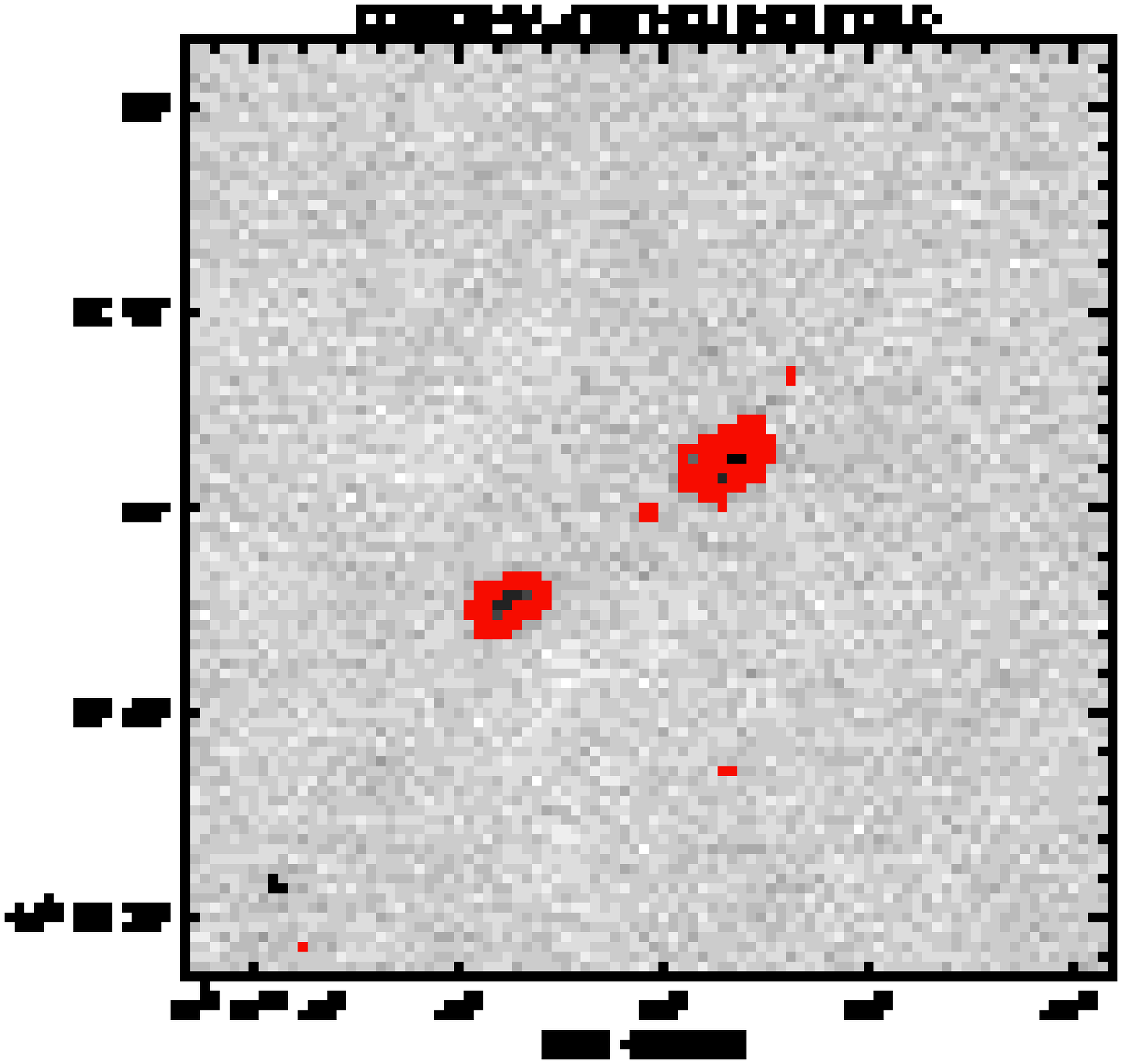}}\\
\resizebox{.9\hsize}{!}{\includegraphics[]{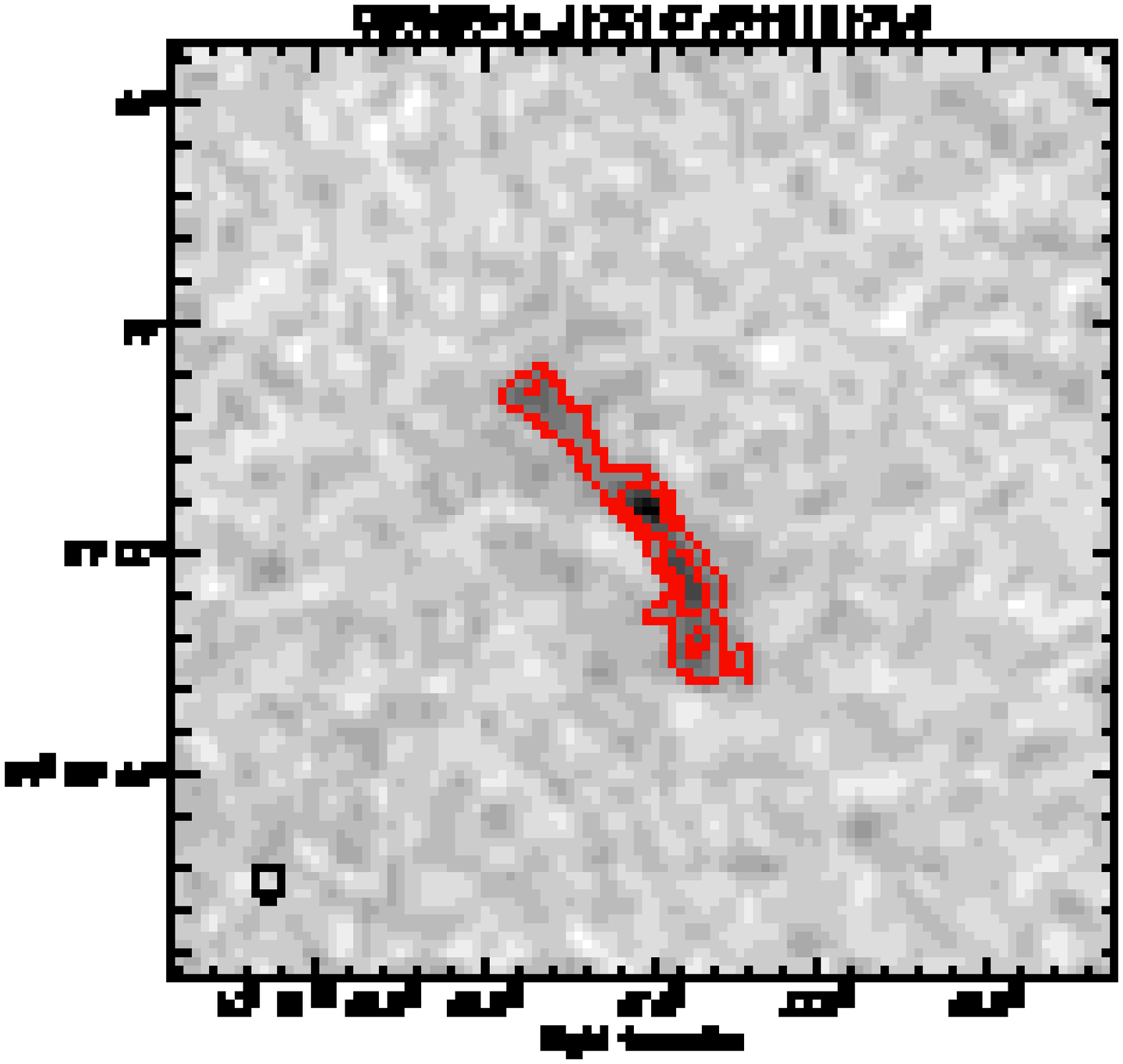}
                      \includegraphics[]{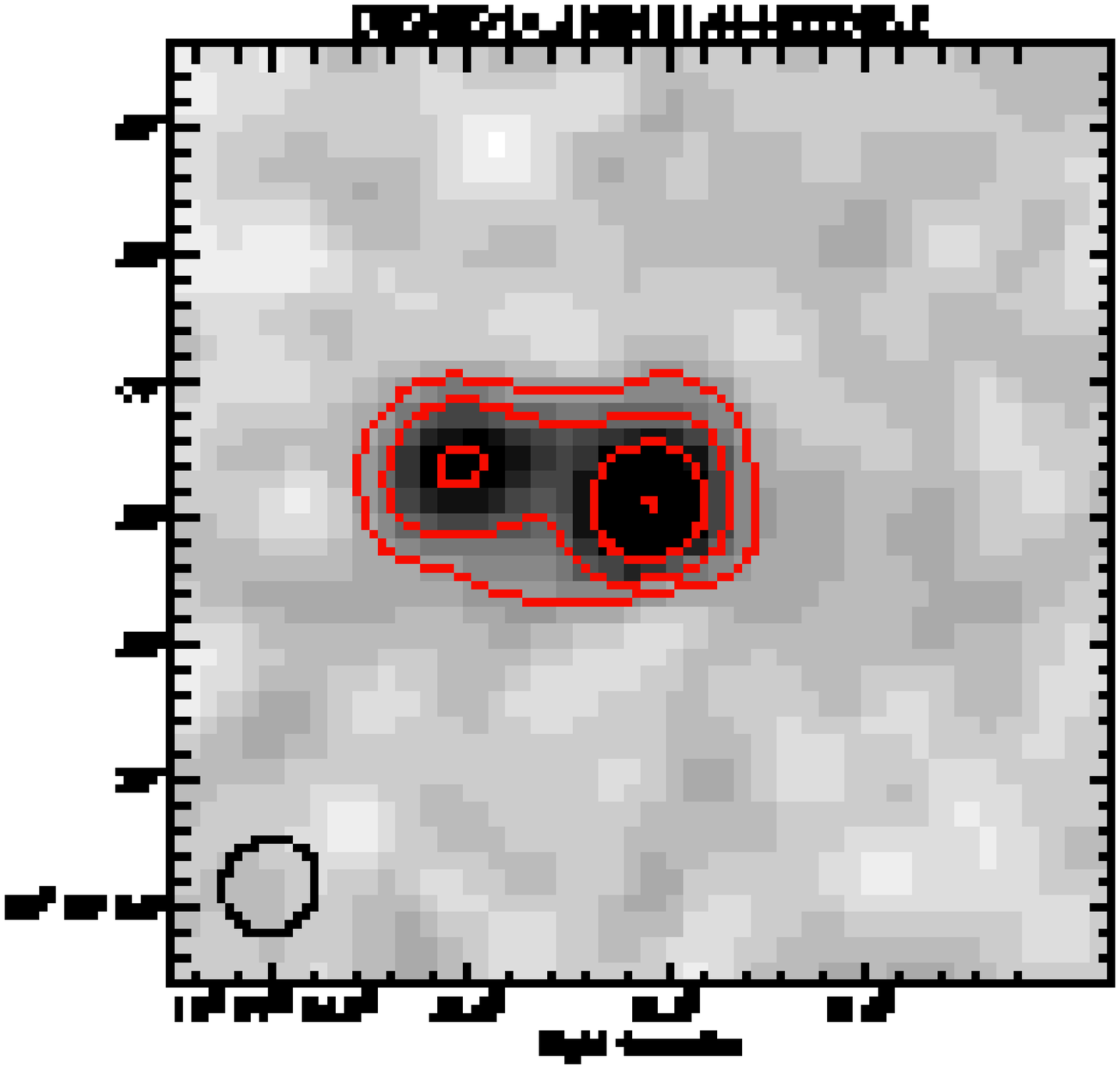}
                      \includegraphics[]{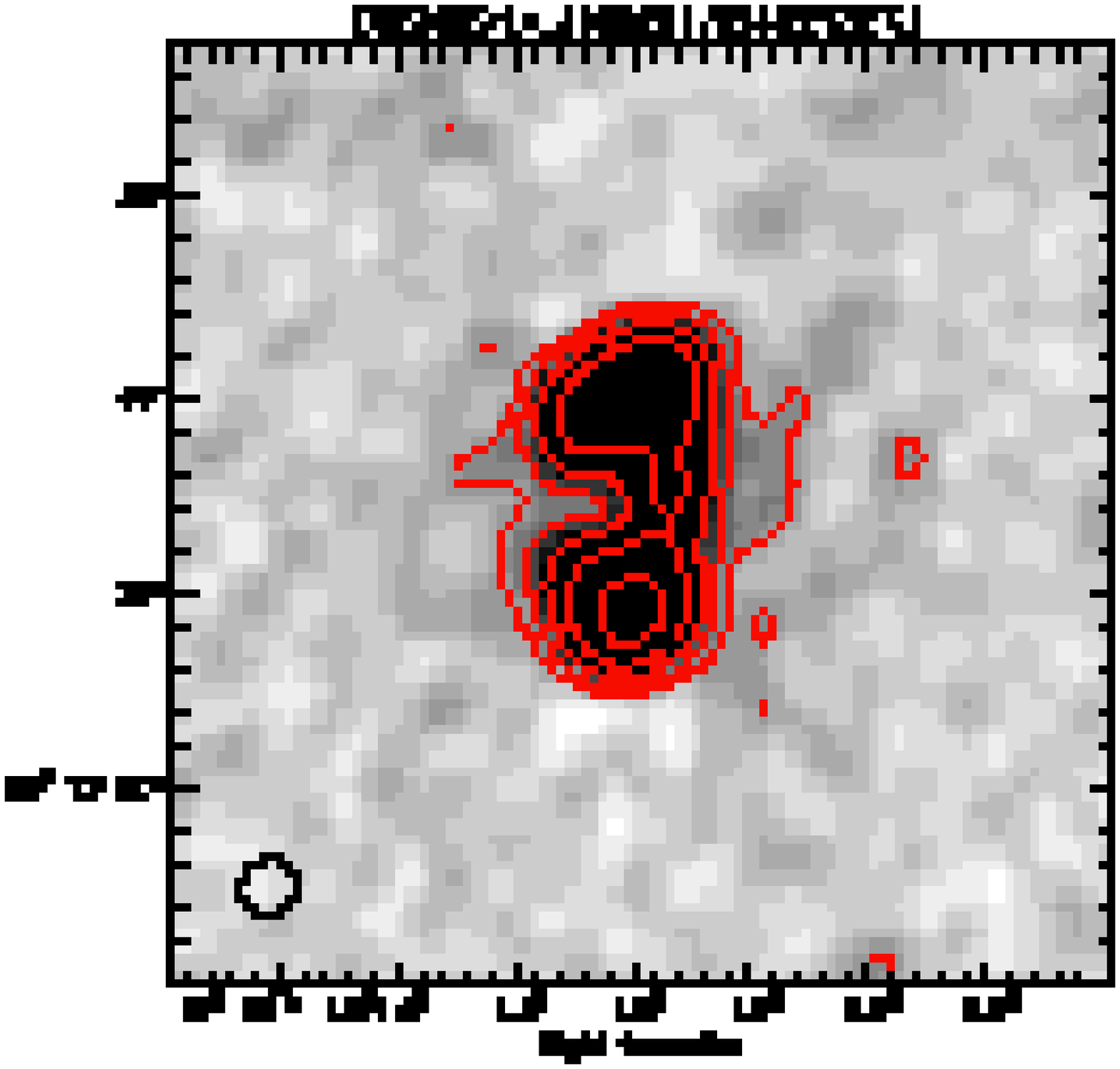}}\\
\resizebox{.9\hsize}{!}{\includegraphics[]{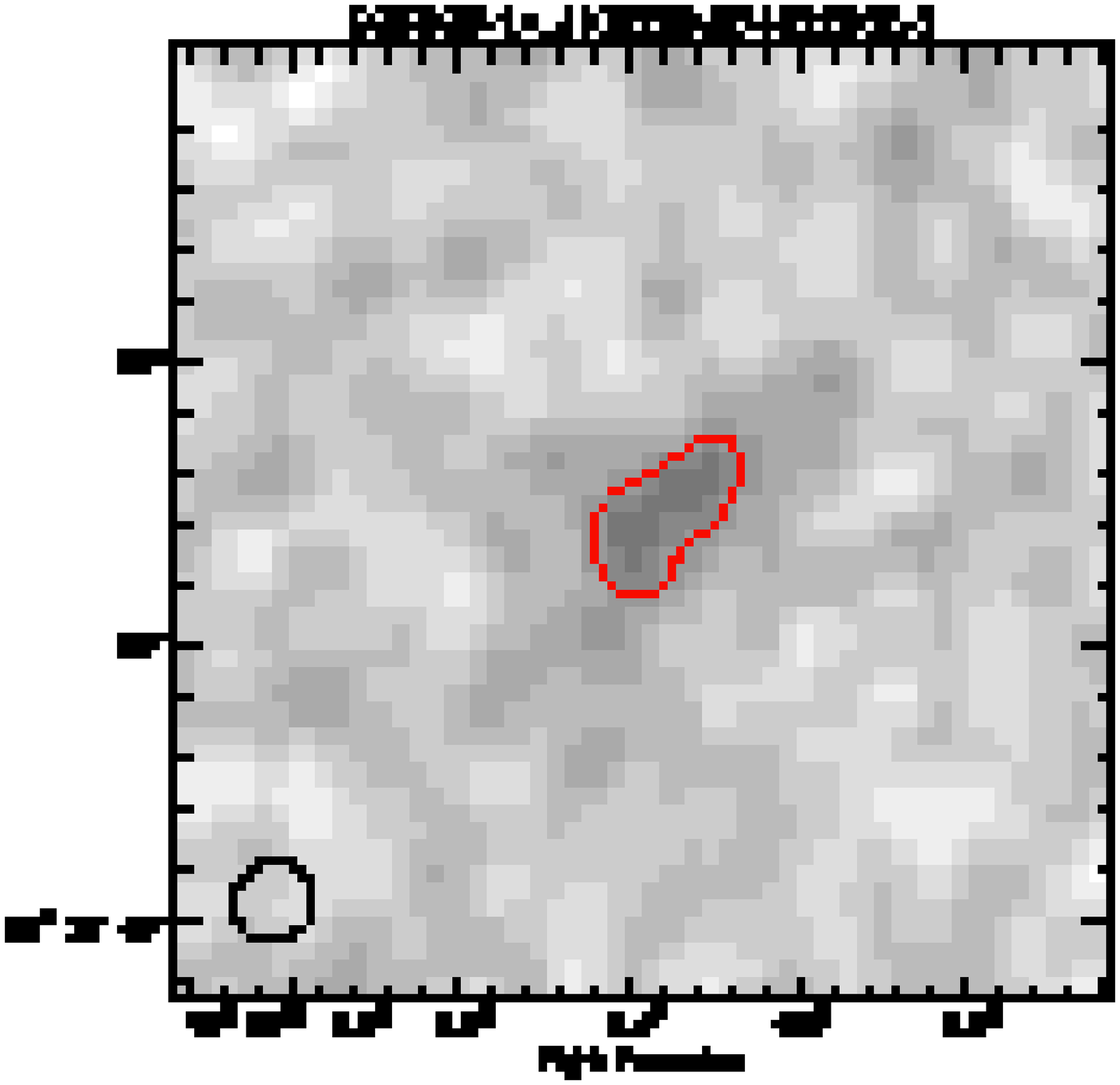}
                      \includegraphics[]{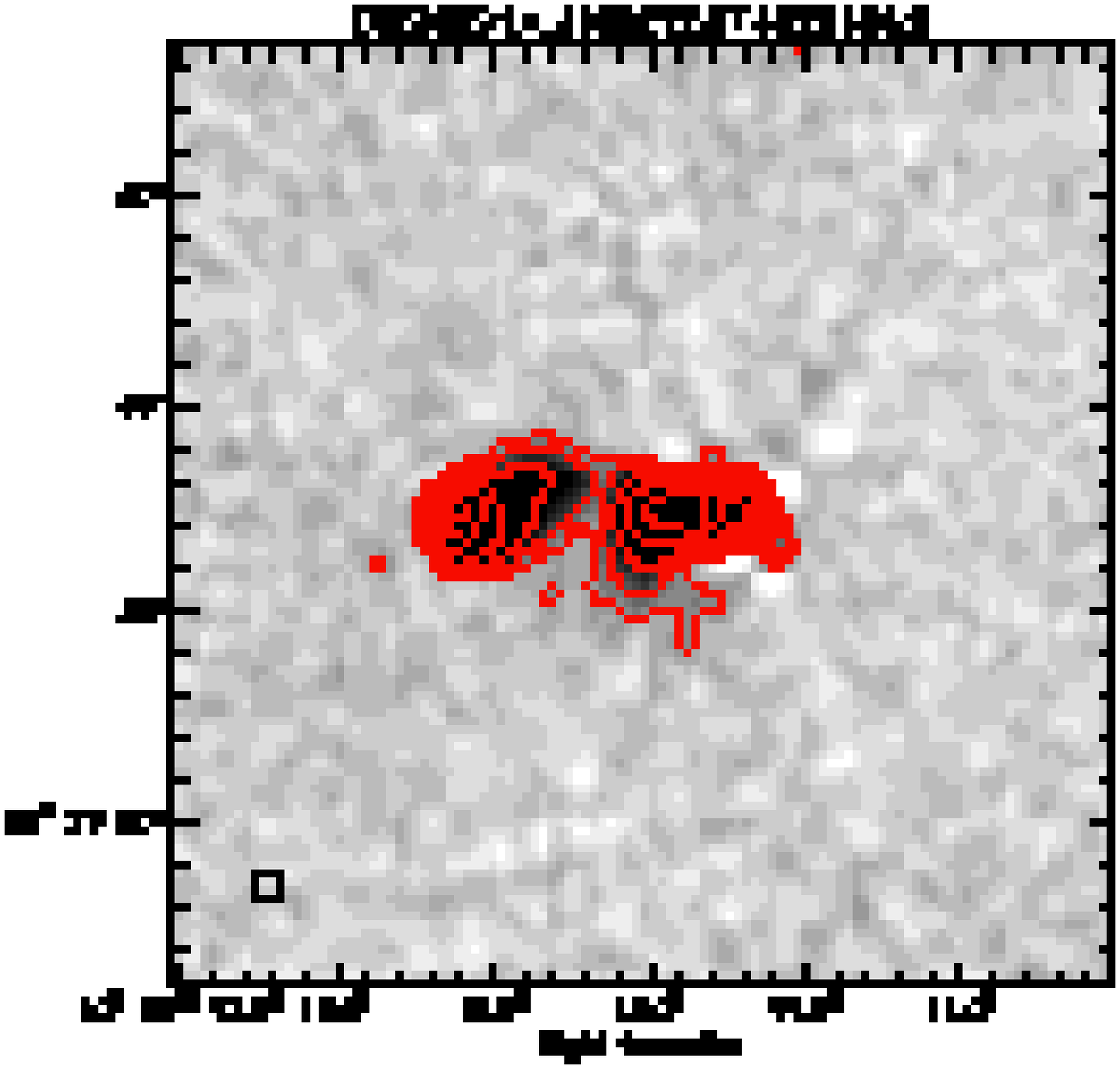}
                      \includegraphics[]{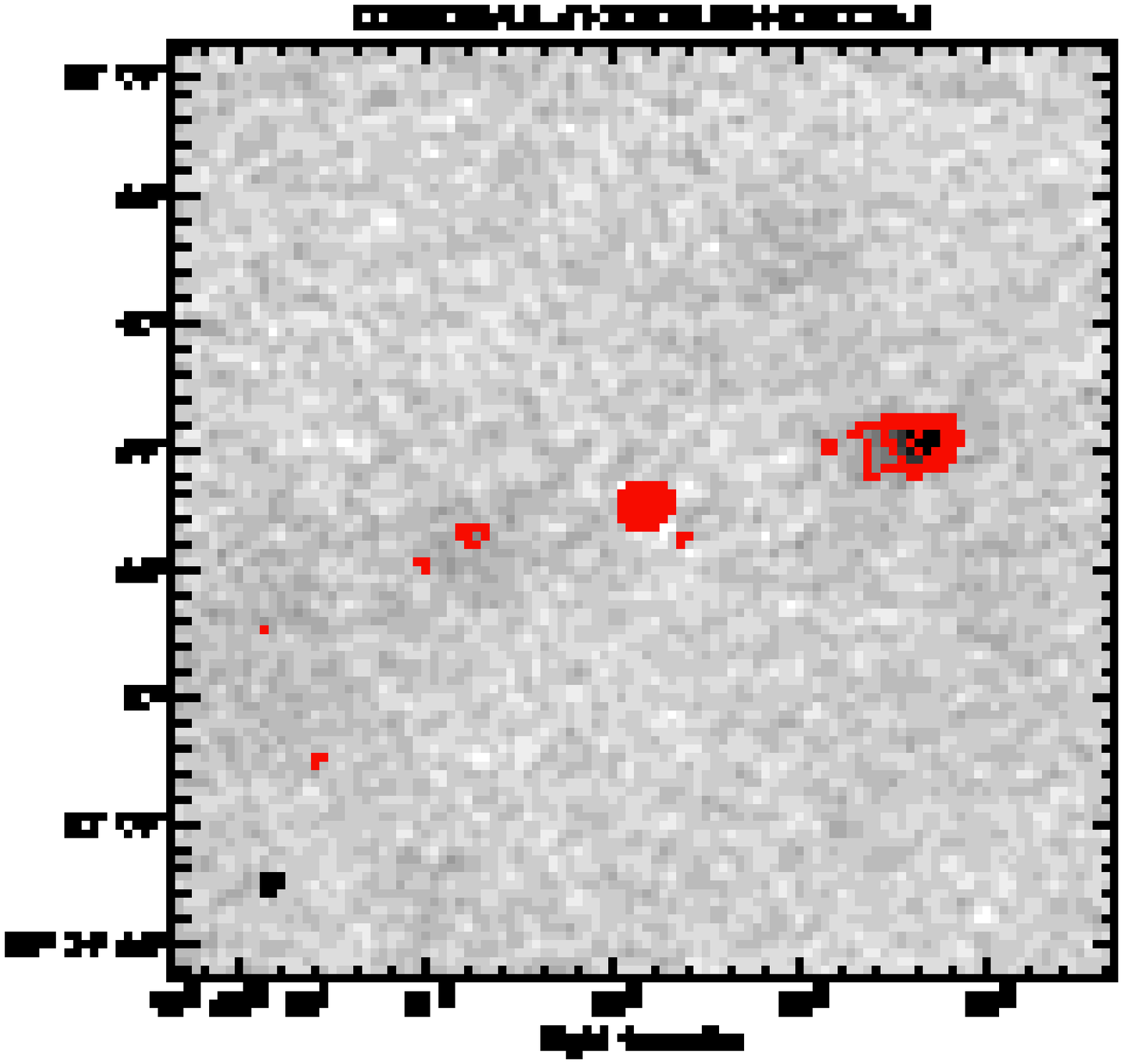}}\\
\resizebox{.9\hsize}{!}{\includegraphics[]{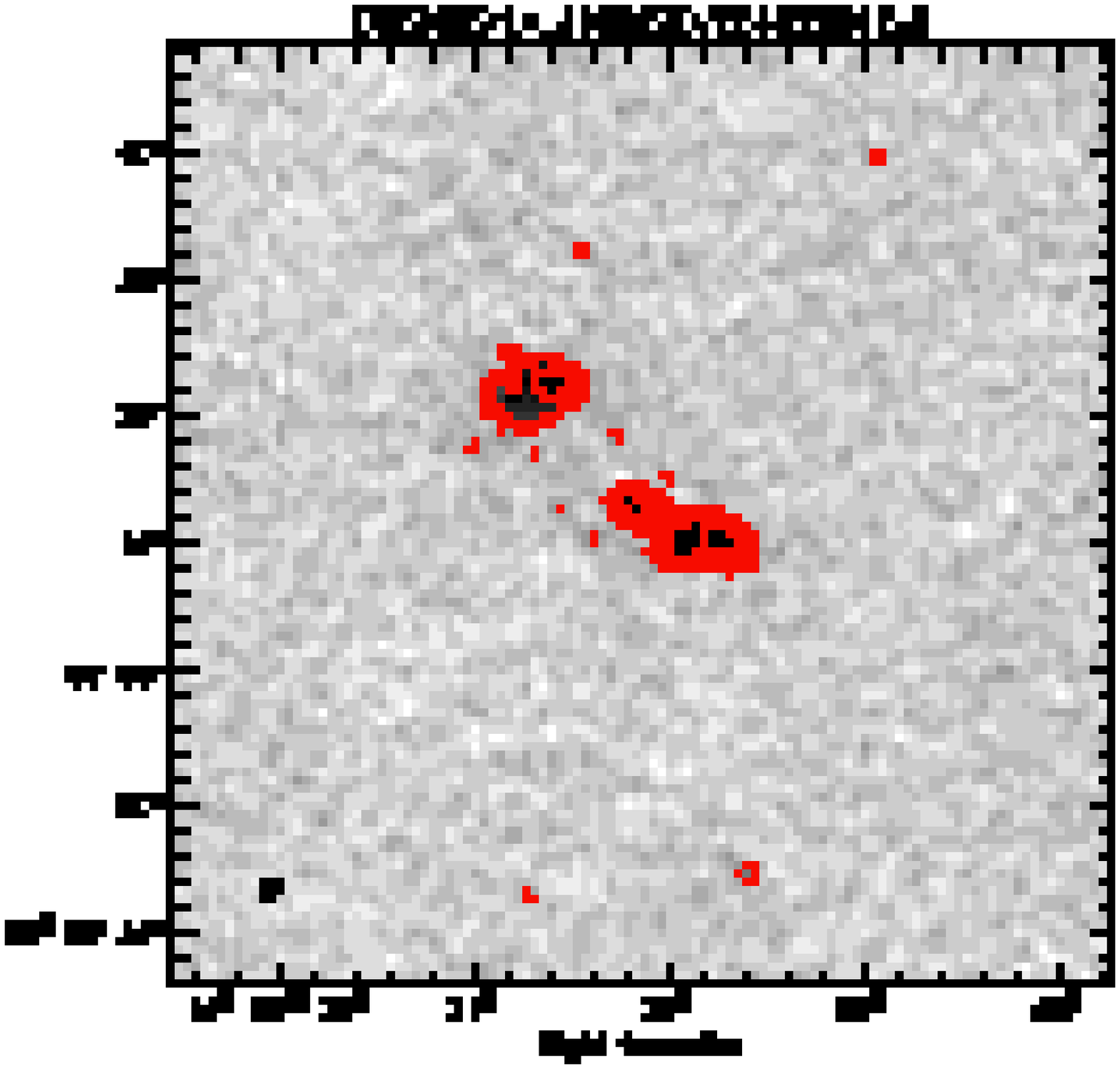}
                      \includegraphics[]{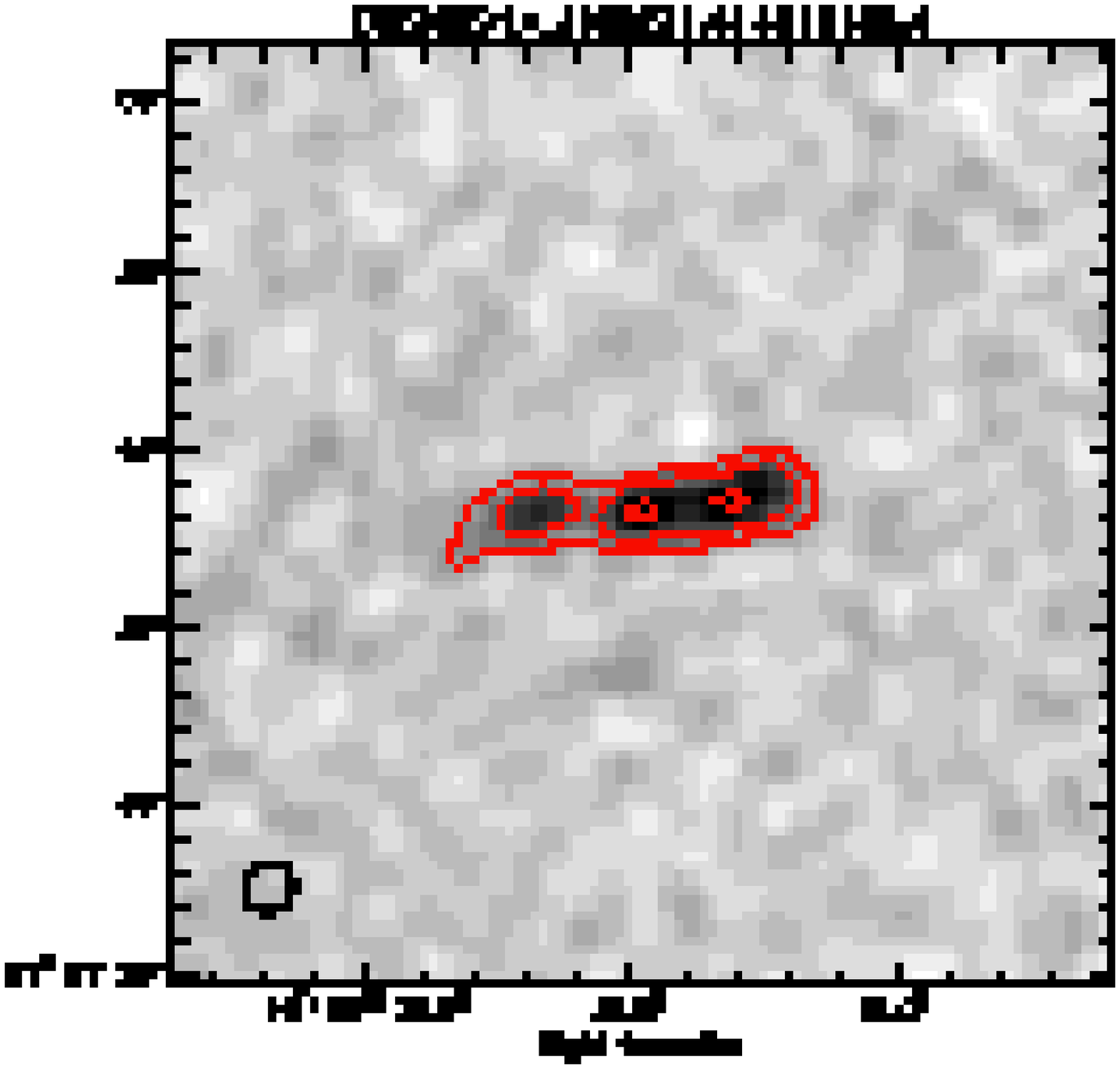}
                      \includegraphics[]{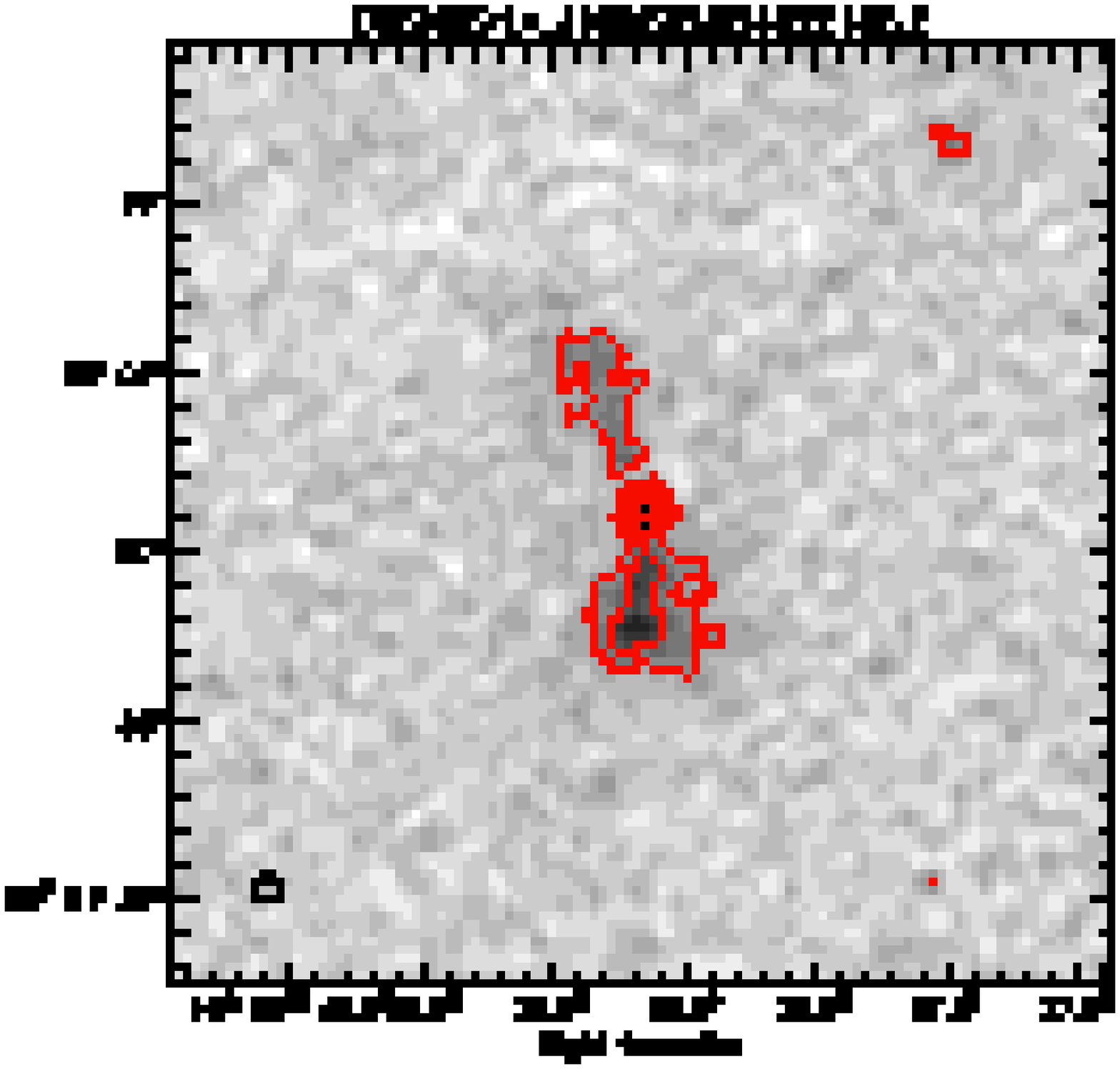}}
\clearpage
\end{center}

\clearpage

\begin{thebibliography}{}
     
\bibitem[Afonso et al.(2005)]{afo05} 
         Afonso, J., Georgakakis, A., Almeida, C., Hopkins, A.~M., Cram, 
         L.~E., Mobasher, B., \& Sullivan, M.\ 2005, \apj, 624, 135 
\bibitem[Afonso et al.(2006)]{afo06} 
         Afonso, J., Mobasher, B., Koekemoer, A., Norris, R.~P., \& Cram, 
         L.\ 2006, \aj, 131, 1216 
\bibitem[Aguirre et al.(2006)]{agu06}
        Aguirre, J.~E., et al. \ 2006, \apjs, this volume
\bibitem[Appleton et al.(2004)]{app04} 
         Appleton, P.~N., et  al.\ 2004, \apjs, 154, 147 
\bibitem[Aretxaga et al.(2005)]{are05} 
         Aretxaga, I., Hughes, D.~H., \& Dunlop, J.~S.\ 2005, 
         \mnras, 358, 1240 
\bibitem[Aussel et al.(2006)]{aus06}
        Aussel, H., et al. \ 2006, \apjs, this volume
\bibitem[Becker et al.(1995)]{bec95} 
         Becker, R.~H., White, R.~L., \& Helfand, D.~J.\ 1995, \apj, 450, 559
\bibitem[Benn et al.(1993)]{ben93} 
         Benn, C.~R., Rowan-Robinson, M., McMahon, R.~G., Broadhurst, T.~J., 
         \& Lawrence, A.\ 1993, \mnras, 263, 98
\bibitem[Bertin \& Arnouts(1996)]{ber96} 
         Bertin, E., \& Arnouts, S.\ 1996, \aaps, 117, 393 
\bibitem[Bertoldi et al.(2006)]{ber06}
        Bertoldi, F., et al. \ 2006, \apjs, this volume
\bibitem[Bondi et al.(2003)]{bon03} 
         Bondi, M., et al.\ 2003, \aap, 403, 857 
\bibitem[Capak et al.(2006)]{cap06}
        Capak, P., et al. \ 2006, \apjs, this volume
\bibitem[Carilli \& Barthel(1996)]{car96} 
        Carilli, C.~L., \& Barthel, P.~D.\ 1996, \aapr, 7, 1 
\bibitem[Carilli \& Yun(2000)]{car00} 
         Carilli, C.~L., \& Yun, M.~S.\ 2000, \apj, 530, 618 
\bibitem[Ciliegi et al.(1999)]{cil99} 
         Ciliegi, P., et al.\ 1999, \mnras, 302, 222 
\bibitem[Condon(1992)]{con92} 
         Condon, J.~J.\ 1992, \araa, 30, 575
\bibitem[Condon(1997)]{con97}
         Condon, J.~J.\ 1997, \pasp, 109, 166 
\bibitem[Condon et al.(1998)]{con98} 
         Condon, J.~J., Cotton, W.~D., Greisen, E.~W., Yin, Q.~F.,
         Perley, R.~A., Taylor, G.~B.,Broderick, J.~J.\ 1998, \aj, 115, 1693
\bibitem[Condon et al.(2003)]{con03} 
         Condon, J.~J., Cotton, W.~D., Yin, Q.~F., Shupe, D.~L., 
         Storrie-Lombardi, L.~J., Helou, G., Soifer, B.~T., \& Werner, M.~W.
         \ 2003, \aj, 125, 2411 
\bibitem[Ferguson et al.(2000)]{fer00} 
         Ferguson, H.~C., Dickinson, M., \& Williams, R.\ 2000, \araa, 38, 667
\bibitem[Fey et al.(2004)]{fey04} 
         Fey, A.~L., et al.\ 2004, \aj, 127, 3587 
\bibitem[Fomalont(1999)]{fom99} 
         Fomalont, E.~B.\ 1999, ASP Conf.~Ser.~180: Synthesis Imaging in 
         Radio Astronomy II, 180, 463 
\bibitem[Fomalont et al.(2006)]{fom06}
         Fomalont, E.~B., Kellermann, K.~I., Cowie, L.~L., Capak, P., 
         Barger, A.~J., Partridge, R.~B., Windhorst, R.~A., Richards, E.~A. 
         \ 2006, \apjs, in press (astro-ph/0607058)
\bibitem[Garrett(2002)]{gar02} 
         Garrett, M.~A.\ 2002, \aap, 384, L19 
\bibitem[Greisen(2003)]{gre03} 
         Greisen, E.~W.\ 2003, 
         Information Handling in Astronomy - Historical Vistas, 109 
\bibitem[Gruppioni et al.(1999)]{gru99} 
         Gruppioni, C., Mignoli, M., \& Zamorani, G.\ 1999, \mnras, 304, 199 
\bibitem[Haarsma et al.(2000)]{haa00} 
         Haarsma, D.~B., Partridge, R.~B., Windhorst, R.~A., \& Richards, 
         E.~A.\ 2000, \apj, 544, 641 
\bibitem[Hasinger et al.(2006)]{has06}
        Hasinger, G., et al. \ 2006, \apjs, this volume
\bibitem[Hopkins et al.(2003)]{hop03} 
         Hopkins, A.~M., Afonso, J., Chan, B., Cram, L.~E., Georgakakis, A., 
         \& Mobasher, B.\ 2003, \aj, 125, 465 
\bibitem[Hopkins et al.(1998)]{hop98} 
         Hopkins, A.~M., Mobasher, B., Cram, L., \& Rowan-Robinson, M.\ 
         1998, \mnras, 296, 839 
\bibitem[Huynh et al.(2005)]{huy05} 
         Huynh, M.~T., Jackson, C.~A., Norris, R.~P., \& Prandoni, I.\ 
         2005, \aj, 130, 1373 
\bibitem[Impey et al.(2006)]{imp06}
        Impey, C.~D. et al. \ 2006, \apjs, this volume
\bibitem[Ivezi{\'c} et al.(2002)]{ive02} 
         Ivezi{\'c}, {\v Z}., et al.\ 2002, \aj, 124, 2364 
\bibitem[Lilly et al.(2006)]{lil06}
        Lilly, S. et al. \ 2006, \apjs, this volume
\bibitem[Norris et al.(2005)]{nor05} 
         Norris, R.~P., et al.\ 2005, \aj, 130, 1358 
\bibitem[Prandoni et al.(2001)]{pra01} 
         Prandoni, I., Gregorini, L., Parma, P., de Ruiter, H.~R., Vettolani, 
         G., Wieringa, M.~H., \& Ekers, R.~D.\ 2001, \aap, 365, 392 
\bibitem[Richards(2000)]{ric00} 
         Richards, E.~A.\ 2000, \apj, 533, 611 
\bibitem[Roche et al.(2002)]{roc02} 
         Roche, N.~D., Lowenthal, J.~D., \& Koo, D.~C.\ 2002, \mnras, 330, 307 
\bibitem[de Ruiter et al.(1997)]{rui97} 
         de Ruiter, H.~R., et al.\ 1997, \aap, 319, 7 
\bibitem[Sadler et al.(2002)]{sad02} 
         Sadler, E.~M., et al.\ 2002, \mnras, 329, 227 
\bibitem[Sanders et al.(2006)]{san06}
        Sanders, D.~B., et al. \ 2006, \apjs, this volume
\bibitem[Schiminovich et al.(2006a)]{sch06}
        Schiminovich, D., et al. \ 2006a, \apjs, this volume
\bibitem[Schinnerer et al.(2004)]{sch04} 
         Schinnerer, E., et al.\ 2004, \aj, 128, 1974 
\bibitem[Scoville et al.(2006a)]{sco06}
        Scoville, N.~Z., et al. \ 2006a, \apjs, this volume
\bibitem[Scoville et al.(2006b)]{sco06b}
        Scoville, N.~Z., et al. \ 2006b, \apjs, this volume
\bibitem[Smol\v{c}i\'{c} et al.(2006)]{smo06}
        Smol\v{c}i\'{c}, V., et al. \ 2006, \apjs, this volume
\bibitem[Steidel et al.(1999)]{ste99} 
         Steidel, C.~C., Adelberger, K.~L., Giavalisco, M., Dickinson, M., 
         \& Pettini, M.\ 1999, \apj, 519, 1 
\bibitem[Taniguchi et al.(2006)]{tan06}
        Taniguchi, Y., et al. \ 2006, \apjs, this volume
\bibitem[Trump et al.(2006)]{tru06}
        Trump, J.~R., et al. \ 2006, \apjs, this volume
\bibitem[White et al.(1997)]{whi97} 
         White, R.~L., Becker, R.~H., Helfand, D.~J., \& Gregg, M.~D.
         \ 1997, \apj, 475, 479 
\bibitem[Windhorst et al.(1985)]{win85} 
         Windhorst, R.~A., Miley, G.~K., Owen, F.~N., Kron, R.~G., \& 
         Koo, D.~C.\ 1985, \apj, 289, 494 
\end{thebibliography}
 \end{document}